\documentclass[preprint,3p,sort&compress,final,times]{elsarticle}

\usepackage{amssymb}
\usepackage{placeins}
\usepackage{tikz}
\usepackage{ams math,cases}
\usepackage{siunitx}
\usepackage{mathtools}
\usepackage{array}
\usepackage[pagewise, mathlines]{lineno}
\usepackage{siunitx}
\definecolor{darkred}{rgb}{0.7,0,0}
\definecolor{darkgreen}{rgb}{0,0.5,0}
\definecolor{darkblue}{rgb}{0,0,0.7}
\definecolor{darkbrown}{rgb}{0.28,0.07,0.07}
\definecolor{black}{rgb}{0,0,0}
\definecolor{plotGreen}{rgb}{0,0.5,0}
\definecolor{plotRed}{rgb}{1.0,0,0}
\definecolor{plotBlue}{rgb}{0,0,1.0}
\definecolor{plotCyan}{rgb}{0,1.0,1.0}
\definecolor{plotGray}{gray}{0.25}

\definecolor{reviewer1color}{rgb}{0,0,0}
\definecolor{reviewer2color}{rgb}{0,0,0}
\definecolor{reviewer3color}{rgb}{0,0,0}

\newenvironment{reviewer1}{\color{reviewer1color}}{\ignorespacesafterend}
\newenvironment{reviewer2}{\color{reviewer2color}}{\ignorespacesafterend}
\newenvironment{reviewer3}{\color{reviewer3color}}{\ignorespacesafterend}

\newcommand{\reviewerone}[1]{{\color{reviewer1color} #1}}
\newcommand{\reviewertwo}[1]{{\color{reviewer2color} #1}}
\newcommand{\reviewerthree}[1]{{\color{reviewer3color} #1}}

\newdefinition{remark}{Remark}

\newcommand{\tabref}[1]{Table~\ref{#1}} % reference a table
\newcommand{\figref}[1]{Figure~\ref{#1}} % reference a figure
\newcommand{\secref}[1]{Section~\ref{#1}} % reference a section
\newcommand{\appendixref}[1]{\ref{#1}} % reference a remark

\newcommand{\normalized}[1]{\bar{#1}}

\begin{document}
%\linenumbers
\begin{frontmatter}



\journal{Journal of Computational Physics}
\title{A mimetic spectral element solver for the Grad-Shafranov equation}


\author[cst]{A.~Palha\corref{cor}}
\ead{a.palha@tue.nl}
\author[casa]{B.~Koren}
%\ead{b.koren@tue.nl}
\author[cst]{F.~Felici}
%\ead{f.felici@tue.nl}

\cortext[cor]{Corresponding author}

\address[cst]{Eindhoven University of Technology, Department of Mechanical Engineering, P.O. Box 513, 5600 MB Eindhoven, The Netherlands}

\address[casa]{Eindhoven University of Technology, Department of Mathematics and Computer Science, P.O. Box 513, 5600 MB Eindhoven, The Netherlands}

\begin{abstract}
	In this work we present a robust and accurate arbitrary order solver for the fixed-boundary plasma equilibria in toroidally axisymmetric geometries. To achieve this we apply the mimetic spectral element formulation presented in \cite{Palha2014} to the solution of the Grad-Shafranov equation. This approach combines a finite volume discretization with the mixed finite element method. In this way the discrete differential operators ($\nabla$, $\nabla\times$, $\nabla\cdot$) can be represented exactly and metric and all approximation errors are present in the constitutive relations. The result of this formulation is an arbitrary order method even on highly curved meshes. Additionally, the integral of the \reviewerone{toroidal current $J_{\phi}$} is exactly equal to the boundary integral of the poloidal field over the plasma boundary. This property can play an important role in the coupling between equilibrium and transport solvers. The proposed solver is tested on a varied set of plasma cross-sections (smooth and with an X-point) and also for a wide range of pressure and \reviewertwo{toroidal magnetic flux profiles}. Equilibria accurate up to machine precision are obtained. Optimal algebraic convergence rates of order $p+1$ and geometric convergence rates are shown for Soloviev solutions (including high Shafranov shifts), field-reversed configuration (FRC) solutions and spheromak analytical solutions. The robustness of the method is demonstrated for non-linear test cases, in particular on an equilibrium solution with a pressure pedestal.

\end{abstract}


\end{frontmatter}


\section{Introduction}
	The numerical computation of magnetohydrodynamic (MHD) equilibria plays a central role in the study of magnetically confined plasmas. In particular, MHD equilibria are used as input to complex algorithms capable of performing detailed simulations of MHD turbulence and stability, transport, heating, etc, see for example \cite{Sovinec2004,Lapillonne2009,Gorler2011,Brambilla1999,Czarny2008,Gruber1981,Fable2013a}. Other applications of MHD equilibrium computations that have been gaining an increasing attention are discharge scenario validation and control of tokamak reactors, see \cite{Humphreys2015} for an overview of current and future applications in control.  In the context of control and discharge scenario validation, MHD equilibria are typically used in coupled simulations with 1D transport codes, e.g. \cite{Hinton1976a,Hirshman1979a,Artaud2010b,Coster2010a,Parail2013a,Fable2013a}. With these applications in mind, the development of fast, robust and accurate MHD equilibrium solvers on arbitrary geometries has become an important and active topic of research.

\begin{reviewer1}
For plasmas in axisymmetric configuration, such as in tokamak devices, the MHD equilibrium  can be expressed in cylindrical coordinates $(r,z,\phi)$ by the Grad-Shafranov equation, see \cite{Grad1958,Shafranov1958}:
%%	\begin{equation}
%%		-\nabla \cdot \left(\frac{1}{\mu_{0}r}\nabla\psi\right) =  r \frac{\mathrm{d}P}{\mathrm{d}\psi} + \frac{1}{\mu_{0}r}f\frac{\mathrm{d}f}{\mathrm{d}\psi} \quad \mathrm{in}\quad\Omega_{p}\,. \label{eq::grad_shafranov}
%%	\end{equation}
	\begin{equation}
		-\frac{1}{\mu_{0} r}\frac{\partial^{2}\psi}{\partial r^{2}} + \frac{1}{\mu_{0} r^{2}}\frac{\partial \psi}{\partial r} -\frac{1}{\mu_{0} r}\frac{\partial^{2}\psi}{\partial z^{2}} =  r \frac{\mathrm{d}P}{\mathrm{d}\psi} + \frac{1}{\mu_{0}r}f\frac{\mathrm{d}f}{\mathrm{d}\psi} \quad \mathrm{in}\quad\Omega_{p}\,,  \label{eq::grad_shafranov}
	\end{equation}
	 where $\psi$ is the flux function, $f$ is related to the toroidal component of the magnetic flux, $P$ is the plasma pressure and where $\Omega_{p}=\Omega_{p}(\psi)$ denotes the plasma domain. For a complete derivation of this equation see for example the book by Goedbloed et al. \cite{Goedbloed2010Book}.
\end{reviewer1}
	 
	 The Grad-Shafranov equation, \eqref{eq::grad_shafranov}, is a non-linear elliptic partial differential equation. Its non-linear character stems from the non-linear dependence of $P$ and $f$ on the unknown flux function $\psi$ and from the fact that the plasma domain $\Omega_{p}$ is also an unknown, that in general can only be determined once the flux function is known. These two characteristics make the solution of this equation a challenging task.
	
	The literature on the numerical solution of MHD equilibria is extensive. A detailed review of different formulations and codes prior to 1991 is presented in \cite{Takeda1991} and a shorter review up to 1984 is given in \cite{Blum1984}. More recently, several other approaches have been proposed and are used by different research groups, e.g. CEDRES++ \cite{Heumann2015}, CHEASE \cite{Lutjens1996,Lutjens1992}, CREATE-NL+ \cite{Albanese2015}, ECOM \cite{Pataki2013,Lee2015}, EEC \cite{Li2014}, ESC \cite{Zakharov1999}, HELENA \cite{Helena1990}, NIMEQ \cite{Howell2014}, SPIDER \cite{IvanovSPIDER2005}, etc. 
	
	The solution of MHD equilibria can be grouped into two distinct classes: (i) fixed-boundary (e.g. CHEASE, ECOM, EEC, ESC, HELENA, NIMEQ, SPIDER) and (ii) free-boundary (e.g. CEDRES++, CREATE-NL+). In the fixed-boundary case the plasma domain is prescribed together with $\psi=\mathrm{constant}$ at its boundary, $\partial\Omega_{p}$. Equation \eqref{eq::grad_shafranov} is then used to find $\psi$ inside the plasma. The free-boundary approach requires the solution of \reviewerone{the MHD equilibrium} in an infinite domain with homogeneous boundary conditions, $\psi = 0$, at infinity and taking into consideration the current flowing in a set of external coils and the Grad-Shafranov equation in the plasma region, \eqref{eq::grad_shafranov}. In this situation, both the plasma domain and the flux function, $\psi$, are unknowns that need to be computed consistently. Both in the fixed- and free-boundary cases, the functions $P(\psi)$ and $f(\psi)$ are either prescribed or determined from transport codes (e.g. ASTRA \cite{ASTRA2002}, CORSICA \cite{CorsicaReport1997}, CRONOS \cite{Artaud2010b}, JETTO \cite{Cenacchi1988}, RAPTOR \cite{Felici2011}, TRANSP \cite{Budny1992}).
	
	The different schemes to numerically solve the MHD equilibrium problem can either compute the flux function, $\psi$, on a prescribed mesh in the $(r,z)$ coordinate system (Eulerian or direct solvers, e.g. CEDRES++, CHEASE, CREATE-NL+, ECOM, \cite{Imazawa2015,Jardin2004,Gourdain2006}) or employ a flux-based mesh and compute the physical coordinates $(r,z)$ from the plasma geometry and $\psi$ (Lagrangian or indirect solvers, e.g. EEC, ESC, \cite{Ludwig1997a,Jardin2010Book}). Regarding the numerical formulation, several different approaches have been proposed: finite element method (CEDRES++, CHEASE, CREATE-NL+, EEC, ESC, HELENA, NIMEQ, \cite{Jardin2004}), \reviewerone{spectral element or collocation method (ECOM, ESC, NIMEQ, \cite{Ludwig1997a})}, finite difference method (\cite{Gourdain2006}) and radial-basis function meshfree method (\cite{Imazawa2015}).
		
		In this article we present a new arbitrary order, fixed-boundary, Eulerian MHD equilibrium solver based on a mimetic spectral element method formulation, \cite{Palha2014}. High-order accuracy is obtained by (i) reformulating the Grad-Shafranov equation, \eqref{eq::grad_shafranov}, as a non-linear scalar Poisson equation with a non-uniform tensorial material property in the constitutive equation (see for example \cite{Neuman1977} for a derivation of Darcy's law), (ii) using a mixed finite element formulation for the equations containing metric and material properties and a finite volume formulation for the equations establishing topological relations, and (iii) using a particular set of arbitrary order finite element basis functions (edge basis functions, see \cite{Palha2014,gerritsma::edge_basis}). To our knowledge, the proposed approach results in the first spectrally accurate Grad-Shafranov solver capable of reconstructing the total plasma current on highly curved meshes. \reviewerone{An important characteristic of this solver is that it can account for arbitrary plasma shapes, including plasma shapes with an X-point. %Due to the geometric nature of the degrees of freedom used in this discretization, the regularity at the geometric axis is intrinsically accounted for. 
		Also, no restrictions on the flux functions $f(\psi)$ and $P(\psi)$ are required, allowing for the computation of a wide variety of equilibria.}
		
		Mimetic methods aim to preserve essential physical/mathematical structures in a discrete setting. Many of such structures are topological, i.e. independent of metric, and involve integral relations. For these reasons the mimetic method uses an integral formulation in order to preserve these properties at the discrete level. A general presentation of the mimetic formulation used in this work is given in \cite{Palha2014} and earlier work is presented in \cite{gerritsmaicosahom2012, palha:lssc2009, bouman::icosahom2009, palha::icosahom2009,Gerritsma}. For other mimetic formulations see the references in \cite{Palha2014}.
		
		The outline of this paper is as follows. In \secref{sec::numerical_method} we present the proposed numerical method. We start by introducing the Grad-Shafranov equation as a non-linear Poisson equation in \secref{sec::grad_shafranov_equation} and then in \secref{sec::poisson_iteration} the iterative procedure used is discussed. This is followed by the mimetic discretization of the linear Poisson equation in \secref{sec::poisson_solver}. We finalize the presentation of the method by applying the Poisson solver to the discretization of the Grad-Shafranov equation. In \secref{sec::numerical_test_cases} the proposed method is applied and tested on a varied set of test cases. We start by testing algebraic and geometric convergence rates on Soloviev (\secref{sec::test_cases_soloviev}), field-reversed (\secref{sec::test_cases_frc}) and spheromak (\secref{sec::test_cases_spheromak}) analytical solutions. We then show the robustness of the method for linear and non-linear eigenvalue test cases, \secref{sec::test_cases_linear_eigenvalue} and \secref{sec::test_cases_nonlinear_eigenvalue} respectively. In \secref{sec::conclusions} we conclude with a discussion of the merits and limitations of this solver and future extensions.

\section{Numerical method} \label{sec::numerical_method}
	\subsection{The Grad-Shafranov equation as a Poisson equation} \label{sec::grad_shafranov_equation}
		As stated before, the work presented here is focussed on the fixed-boundary solution of the Grad-Shafranov equation, which corresponds to a homogeneous Dirichlet problem given by:
%		\begin{equation}
%		\begin{dcases}
%			-\nabla \cdot \left(\frac{1}{\mu_{0}r}\nabla\psi\right) =  r \frac{\mathrm{d}P}{\mathrm{d}\psi} + \frac{1}{\mu_{0}r}f\frac{\mathrm{d}f}{\mathrm{d}\psi} & \mathrm{in} \quad \Omega_{p} \\
%			\psi = 0 & \mathrm{in} \quad \partial\Omega_{p}\,.
%		\end{dcases}
%		\end{equation}
		\begin{equation}
		\begin{dcases}
			-\frac{1}{\mu_{0} r}\frac{\partial^{2}\psi}{\partial r^{2}} + \frac{1}{\mu_{0} r^{2}}\frac{\partial \psi}{\partial r} -\frac{1}{\mu_{0} r}\frac{\partial^{2}\psi}{\partial z^{2}} =  r \frac{\mathrm{d}P}{\mathrm{d}\psi} + \frac{1}{\mu_{0}r}f\frac{\mathrm{d}f}{\mathrm{d}\psi} & \mathrm{in} \quad \Omega_{p}\,, \\
			\psi = 0 & \mathrm{on} \quad \partial\Omega_{p}\,.
		\end{dcases}  \label{eq::grad_shafranov_typical}
		\end{equation}
		Since we consider a fixed-boundary solution, $\Omega_{p}$ is known. In the same manner, both $P$ and $f$ are also given.
		
		Another form of this equation, that is typically presented in the literature, is 
		\begin{equation}
		\begin{dcases}
			-\nabla \cdot \left(\frac{1}{\mu_{0} r}\nabla\psi\right) =  r \frac{\mathrm{d}P}{\mathrm{d}\psi} + \frac{1}{\mu_{0}r}f\frac{\mathrm{d}f}{\mathrm{d}\psi} & \mathrm{in} \quad \Omega_{p}\,, \\
			\psi = 0 & \mathrm{on} \quad \partial\Omega_{p}\,.
		\end{dcases} \label{eq::poisson_typical}
		\end{equation}
		Although this expression is correct, we prefer to use an equivalent formulation that highlights the physical nature of the problem:
		\begin{equation}
		\begin{dcases}
			\nabla \times \left(\mathbb{K}\nabla\times\psi\right) =  J_{\phi}& \mathrm{in} \quad \Omega_{p}\,, \\
			\psi = 0 & \mathrm{on} \quad \partial\Omega_{p}\,,
		\end{dcases} \quad\mathrm{where}\quad
		\begin{dcases}
			\mathbb{K}:=\left[\begin{array}{cc}\frac{1}{\mu_{0}r} & 0 \\ 0 & \frac{1}{\mu_{0}r}\end{array}\right]\,, \\
			J_{\phi} := r \frac{\mathrm{d}P}{\mathrm{d}\psi} + \frac{1}{\mu_{0}r}f\frac{\mathrm{d}f}{\mathrm{d}\psi}\,,
		\end{dcases}
		\label{eq::non_linear_poisson}
		\end{equation}
		and with $\nabla\times\psi := \frac{\partial\psi}{\partial z}\vec{e}_{r} - \frac{\partial\psi}{\partial r}\vec{e}_{z}$. This form shows that this boundary-value problem can be seen as a non-linear vector Poisson problem in 2D with a non-uniform tensorial material property $\mathbb{K}$ in the constitutive relation. Additionally, the following two relations are explicitly expressed in \eqref{eq::non_linear_poisson}:
		\begin{equation}
			\mathbb{K}\nabla\times\psi = \vec{h}_{\mathrm{p}} = h_{r}\vec{e}_{r} + h_{z}\vec{e}_{z} \qquad\mathrm{and}\qquad \nabla\times\vec{h}_{\mathrm{p}} = J_{\phi}\,,
		\end{equation}
		with $\vec{h}_{\mathrm{p}}$ the poloidal component of the magnetic field. By posing the fixed-boundary Grad-Shafranov problem as a non-linear Poisson problem we can focus on the numerical solution of the more general Poisson problem and then substitute $\mathbb{K}$ and $J_{\phi}$ by the particular cases present in MHD equilibria.
		
	\subsection{Iterative solution of non-linear Poisson problem} \label{sec::poisson_iteration}
		The solution of the non-linear Poisson problem \eqref{eq::non_linear_poisson} requires an iterative procedure such as Newton's method or a more straightforward fixed-point iteration method. In this work we focus on the fixed-point iteration scheme. \reviewertwo{It is important to note that the fixed-point iteration procedure does not converge in all cases. In the future a Newton method will be required as a more robust solver.} Under some conditions \eqref{eq::non_linear_poisson} becomes an eigenvalue problem. In this situation a modification to the standard fixed-point iteration method is required. For this reason, we present first the non-eigenvalue case and then the eigenvalue one.
		
		\subsubsection{Non-eigenvalue case} \label{sec::non_eigenvalue_case}
		 With this simple method the updated value of the flux function, $\psi^{k+1}$, is computed by solving the Poisson problem with the right-hand side evaluated at the previous value $\psi^{k}$, that is $J_{\phi}(\psi^{k},r,z)$. This means that for each iteration $k$ a linear Poisson problem is solved:
		\begin{equation}
		\begin{dcases}
			\nabla \times \left(\mathbb{K}\nabla\times\psi^{k+1}\right) =  J_{\phi}(\psi^{k},r,z) & \mathrm{in} \quad \Omega_{p}\,, \\
			\psi = 0 & \mathrm{on} \quad \partial\Omega_{p}\,.
		\end{dcases} \label{eq::non_linear_poisson_iteration}
		\end{equation}
		The iterative procedure is stopped once the residual error satisfies
		\[
			\|\nabla \times \left(\mathbb{K}\nabla\times\psi^{k+1}\right)-J_{\phi}(\psi^{k+1},r,z)\|_{L^{s}} < \epsilon \ll 1\,,
		\]
		 with $\|\cdot\|_{L^{s}}$ the standard $s$-norm given by
		 \[
		 	\|f\|_{L^{s}} := \left(\int_{\Omega_{p}} |f|^{s}\,\mathrm{d}V\right)^{\frac{1}{s}}\,,
		 \]
		 and $s\in\mathbb{N}$.
		
		\subsubsection{Eigenvalue case} \label{sec::eigenvalue_case}
		In some situations the current profile $J_{\phi}$ has the form
		\begin{equation}
			J_{\phi}(\psi,r,z) = \tilde{J}_{\phi}(\psi,r,z)\,\psi\,.
		\end{equation}
		In this case $\psi=0$ is a trivial solution of this equation and therefore the iterative procedure outlined above needs to be adapted in order to recover the physically relevant solution. Under this condition, the non-linear Poisson problem \eqref{eq::non_linear_poisson} becomes an eigenvalue problem, see for example \cite{Pataki2013,Goedbloed1984,LoDestro1994,SAITOH2012, Kikuchi1984}. Below we outline the procedure presented in \cite{Pataki2013}, and followed in this work, for the solution of this eigenvalue problem.
		
		The main idea is that it is possible to rescale the flux function, $\normalized{\psi} := \psi\,\|\psi\|_{L^{s}}^{-1}$, and the total toroidal current, $\normalized{J}_{\phi} := J_{\phi}\,\sigma^{-1}$. Introducing these rescalings into $P$, $f$ and $J_{\phi}$ we get:
		\begin{equation}
			\frac{\mathrm{d}P}{\mathrm{d}\psi} = \frac{1}{\|\psi\|_{L^{s}}}\frac{\mathrm{d}P}{\mathrm{d}\normalized{\psi}}\,,\quad f\frac{\mathrm{d}f}{\mathrm{d}\psi} = \frac{1}{2}\frac{\mathrm{d}f^{2}}{\mathrm{d}\psi} = \frac{1}{\|\psi\|_{L^{s}}}\frac{\mathrm{d}f^{2}}{\mathrm{d}\normalized{\psi}} \quad \mathrm{and}\quad J_{\phi}(\psi,r,z) = \frac{\sigma}{\|\psi\|_{L^{s}}}\,\normalized{J}_{\phi}(\normalized{\psi},r,z)\,. \label{eq::normalized_psi}
		\end{equation}
		Using the rescaling \eqref{eq::normalized_psi} results in the following non-linear eigenvalue problem:
		\begin{equation}
			\begin{dcases}
				\nabla \times \left(\mathbb{K}\nabla\times\normalized{\psi}\right) = \normalized{\sigma}\,\normalized{J}_{\phi}& \mathrm{in} \quad \Omega_{p}\,, \\
				\normalized{\psi} = 0 & \mathrm{on} \quad \partial\Omega_{p}\,,
			\end{dcases} \label{eq::non_linear_eigenvalue}
			\end{equation}
		with $\normalized{\sigma}:= \frac{\sigma}{\|\psi\|_{L^{s}}^{2}}$ representing the eigenvalue. This form of the MHD equilibrium equation clearly shows that it is an eigenvalue problem on the eigenfunction $\normalized{\psi}$ and associated eigenvalue $\normalized{\sigma}$. This, as noted in \cite{Pataki2013,LoDestro1994}, demonstrates the well known scale-invariance property of the Grad-Shafranov equation under the transformation
		\begin{equation}
			(\psi,\sigma,r,z) \longrightarrow (\lambda\,\psi,\lambda^{2}\sigma,r,z)\,.
		\end{equation}
	
		The fixed-point iterative procedure to find the physically relevant eigenfunction of this non-linear eigenvalue problem can either be employed to find the normalized flux function solution or the solution that satisfies a specific total toroidal current. Here we use the approach for finding a normalized flux function solution as presented in \cite{Pataki2013}. This method computes in each iteration a new eigen-pair $(\normalized{\psi}^{k+1},\normalized{\sigma}^{k+1})$ by first using the previous eigen-pair $(\normalized{\psi}^{k},\normalized{\sigma}^{k})$ to solve the linear Poisson problem:
		\begin{equation}
			\begin{dcases}
				\nabla \times \left(\mathbb{K}\nabla\times\psi^{k+1}\right) =  \normalized{\sigma}^{k}\,\normalized{J}_{\phi}(\normalized{\psi}^{k},r,z)& \mathrm{in} \quad \Omega_{p}\,, \\
				\psi^{k+1} = 0 & \mathrm{on} \quad \partial\Omega_{p}\,.
			\end{dcases} \label{eq::linear_poisson_iteration}
			\end{equation}
			The new eigen-pair is computed in the following way:
			\begin{equation}
				\normalized{\psi}^{k+1} = \frac{\psi^{k+1}}{\|\psi^{k+1}\|_{L^{s}}} \quad\mathrm{and}\quad \normalized{\sigma}^{k+1} = \frac{\normalized{\sigma}^{k}}{\|\psi^{k+1}\|_{L^{s}}}\,. \label{eq::normalization_iteration_eigenvalue}
			\end{equation}
	
		Once showed that the iterative procedure to solve the non-linear Poisson problem \eqref{eq::non_linear_poisson} (and consequently the Grad-Shafranov problem \eqref{eq::grad_shafranov_typical}) relies heavily on the successive solution of a linear Poisson problem, we proceed with the discussion on the application of the mimetic spectral element discretization to the solution of the linear Poisson problem.
			
	\subsection{Poisson problem and its discrete solution} \label{sec::poisson_solver}
		In this section we present the application of the mimetic spectral element discretization to the solution of the linear Poisson problem, such as the one appearing in the iterative solution of the Grad-Shafranov problem, \eqref{eq::linear_poisson_iteration}:
		\begin{equation}
			\begin{dcases}
				\nabla\times\left(\mathbb{K}\nabla\times\psi\right) = J & \mathrm{in} \quad \Omega_{p}\,, \\
				\psi = \psi_{b} & \mathrm{on} \quad \partial\Omega_{p}\,.
			\end{dcases} \label{eq::poisson_linear_mimetic}
		\end{equation}
		Note that (i) $J=J(r,z)$ since it is a linear Poisson problem and (ii) $\psi_{b}\neq 0$. The general case of inhomogeneous boundary conditions is outlined since it will be necessary in the solution of the Soloviev test case. 
		
		We start by rewriting \eqref{eq::poisson_linear_mimetic} as a system of first order equations:
		\begin{equation}
			\begin{dcases}
				\mathbb{K}\nabla\times\psi = \vec{h}_{p}&\mathrm{in}\quad\Omega_{p}\,, \\
				\nabla\times \vec{h}_{p} = J & \mathrm{in}\quad\Omega_{p}\,, \\
				\psi = \psi_{b} & \mathrm{on} \quad \partial\Omega_{p}\,.
			\end{dcases} \label{eq::first_order_system}
		\end{equation}
		In this form, it is possible to separate the topological laws (exact) from the metric-dependent (approximate) ones. The second expression, $\nabla\times\vec{h}_{p} = J$, is a topological law (a circuital law in particular) that relates the flux integral of the current density $J$ through a surface $\mathcal{N}$ to the line integral of $\vec{h}_{p}$ over the boundary of $\mathcal{N}$:
		\begin{equation}
			\int_{\partial\mathcal{N}} \vec{h}_{p}\cdot\mathrm{d}\vec{l} = \int_{\mathcal{N}} J \,\mathrm{d}V\,, \label{eq::integral_form}
		\end{equation}
		where we have used Stokes' theorem to establish the relation: $\int_{\partial\mathcal{N}} \vec{h}_{p}\cdot\mathrm{d}\vec{l} = \int_{\mathcal{N}} \nabla\times\vec{h}_{p}\,\mathrm{d}V$. The integral form \eqref{eq::integral_form} highlights the topological, metric-free nature of this equation. On the other hand, the first equation, $\mathbb{K}\nabla\times\psi = \vec{h}_{p}$, is an approximate relation since it combines a topological relation, $\nabla\times\psi = \vec{b}_{p}$, with a constitutive one, $\mathbb{K} \vec{b}_{p} = \vec{h}_{p}$. Constitutive relations establish connections between different physical quantities by means of (inexact) physical constants and metric-dependent relations, as in this case where $\mathbb{K}$ is the metric-dependent tensorial material property. For these reasons, this equation has a local (metric-dependent) and approximate character opposed to the exact nature of the previously discussed topological law. For a detailed discussion of the nature of physical laws the authors advise the book by Tonti, \cite{BookTonti2013}.
		
		It turns out, see for example\cite{BookTonti2013,tonti1975formal,Tonti2014,mattiussi1997analysis,gerritsmaicosahom2012,bossavit_japan_computational_2,bossavit_japan_computational_1,bossavit_japan_computational_4,bossavit_japan_computational_5,Palha2014} for an extensive discussion, that it is highly desirable (and in some situations essential) to exactly satisfy the topological equations at the discrete level, while all approximation and interpolation errors can be included in the constitutive relations.
		
		With this objective in mind we will establish the mimetic discretization of the Poisson equation as expressed by the first order system \eqref{eq::first_order_system}. We start by recalling the standard inner product definitions for both scalar and vector valued functions,
		\begin{equation}
			 \langle g,f\rangle_{\Omega} := \int_{\Omega} g f \,\mathrm{d}V
			\qquad \mathrm{and} \qquad \langle \vec{u},\vec{v}\rangle_{\Omega} := \int_{\Omega} \vec{u} \cdot\vec{v} \,\mathrm{d}V\,, \label{eq::traditional_inner_product}
		\end{equation}
		the associated norms,
		\begin{equation}
			\|f\|_{L^{2}(\Omega)} := \langle f,f\rangle_{\Omega}^{\frac{1}{2}} \qquad \mathrm{and}\qquad \|\vec{u}\|_{L^{2}(\Omega)} := \langle \vec{u},\vec{u}\rangle_{\Omega}^{\frac{1}{2}}\,,
		\end{equation}
		\begin{reviewer3}
		and the function spaces $L^{2}(\Omega)$ and $H(\nabla\times,\Omega)$,
		\[
			L^{2}(\Omega) := \{f \mid \|f\|_{L^{2}(\Omega)}<\infty\} \qquad \mathrm{and} \qquad H(\nabla\times,\Omega):=\{ \vec{u} \in L^{2}(\Omega) \mid \nabla\times\vec{u} \in L^{2}(\Omega)\}\,.
		\]
		Note that the $L^{2}(\Omega)$ space is defined for both scalar and vector valued functions.
		
		The standard mixed finite element formulation starts by constructing the weak problem, see \cite{brezzi1991mixed},
		\begin{equation}
			\begin{dcases}
				\mathrm{Find} \,\, \psi\in L^{2}(\Omega_{p}) \,\, \mathrm{and}\,\, \vec{h}_{p}\in H(\nabla\times,\Omega_{p})\,\,\mathrm{such}\,\,\mathrm{that}  \\
				\langle \psi,\nabla\times(\mathbb{K}\,\vec{\varphi})\rangle_{\Omega_{p}} = \langle\vec{h}_{p},\vec{\varphi}\rangle_{\Omega_{p}}\,,\quad\forall \vec{\varphi}\in H(\nabla\times,\Omega_{p})\,, \\
				\langle\nabla\times\vec{h}_{p}, \phi\rangle_{\Omega_{p}} = \langle J,\phi\rangle_{\Omega_{p}}\,,\quad\forall \phi\in L^{2}(\Omega_{p})\,.
			\end{dcases} \label{eq::mixed_finite_element_weak_form}
		\end{equation}
		Note that we consider here homogeneous Dirichlet boundary conditions $\psi = \psi_{b} = 0$ on $\partial\Omega_{p}$, which are now imposed weakly. 
		\end{reviewer3}
		Well-posedness of this weak formulation can be found in any book on mixed finite elements, e.g. \cite{brezzi1991mixed}.
		
		This formulation has two shortcomings:
		\begin{enumerate}
			\item The discrete $\mathbb{K}\nabla\times$ operator will not be the adjoint of the discrete $\nabla\times$ operator, as opposed to the continuous case where both operators are adjoint. This can lead to the loss of self-adjointness and negative definiteness of the discrete Laplacian operator, $\nabla\times\mathbb{K}\nabla\times$, on general grids, leading to poor convergence properties, see for example \cite{HymanShashkovSteinberg97}.
			\item The second equation, representing the topological relation, is satisfied only \reviewerthree{approximately in curved geometries, not exactly. We will see that with the use of a proper set of basis functions it is possible to write this equation in a purely topological form, which is exact even on curved geometries.}
		\end{enumerate}
		
		In order to overcome these two aspects of the standard mixed finite element formulation we will present the mimetic spectral element discretization. We start by introducing an alternative inner product, as discussed in \cite{Palha2014,Rebelo2014,HymanShashkovSteinberg97}, and show that this results in a $\mathbb{K}\nabla\times$ operator  that is the adjoint of the $\nabla\times$ operator. Afterwards, we introduce the set of basis functions used to discretize the physical quantities and show that they result in an exact representation of the topological relation in \eqref{eq::first_order_system}.
		
		\subsubsection{The natural inner product}
			We can define an alternative inner product between two vector fields $\vec{u},\vec{v}\in H(\nabla\times,\Omega)$ using the material metric tensor $\mathbb{K}$, since it is symmetric and positive definite. This inner product, which we denote by \emph{natural} inner product, is defined by
			\begin{equation}
				\langle\vec{u},\vec{v}\rangle_{\mathbb{K},\Omega} := \int_{\Omega}\left(\mathbb{K}^{-1}\vec{u}\right)\cdot\vec{v}\,\mathrm{d}V = \int_{\Omega}\vec{u}\cdot\left(\mathbb{K}^{-1}\vec{v}\right)\,\mathrm{d}V\,. \label{eq::natural_inner_product_definition}
			\end{equation}
			If this inner product is used on the first equation in \eqref{eq::mixed_finite_element_weak_form} we obtain the following alternative equality
			\begin{equation}
				\langle \mathbb{K}\nabla\times\psi,\vec{\varphi}\rangle_{\mathbb{K},\Omega_{p}} = \langle\vec{h}_{p},\vec{\varphi}\rangle_{\mathbb{K},\Omega_{p}}\,, \quad\forall \vec{\varphi}\in H(\nabla\times,\Omega_{p})\,.
			\end{equation}
			Using the definition of the inner product, \eqref{eq::natural_inner_product_definition}, and integrating by parts we obtain:
			\begin{equation}
				\langle\mathbb{K}\nabla\times\psi,\vec{\varphi}\rangle_{\mathbb{K},\Omega_{p}} := \int_{\Omega_{p}}\left(\mathbb{K}^{-1}\mathbb{K}\nabla\times\psi\right)\cdot\vec{\varphi}\,\mathrm{d}V = \int_{\Omega_{p}}\psi\cdot\nabla\times\vec{\varphi}\,\mathrm{d}V  - \int_{\partial\Omega_{p}}\psi\vec{\varphi}\cdot\mathrm{d}\vec{l} = \langle\psi,\nabla\times\vec{\varphi}\rangle - \int_{\partial\Omega_{p}}\psi\vec{\varphi}\cdot\mathrm{d}\vec{l}\,.
			\end{equation}
			For homogeneous boundary conditions, $\psi_{b} = 0$, this expression shows that the natural inner product between vector valued functions, \eqref{eq::natural_inner_product_definition}, satisfies  the adjoint relation between $\mathbb{K}\nabla\times$ and $\nabla\times$. In order to enforce this adjoint relation at the discrete level we rewrite the mixed finite element formulation, \eqref{eq::mixed_finite_element_weak_form}, as:
			\begin{equation}
			\begin{dcases}
				\mathrm{Find} \,\, \psi\in L^{2}(\Omega_{p}) \,\, \mathrm{and}\,\, \vec{h}_{p}\in H(\nabla\times,\Omega_{p})\,\,\mathrm{such}\,\,\mathrm{that} \\
				\langle\psi,\nabla\times\vec{\varphi}\rangle_{\Omega_{p}} = \langle\vec{h}_{p},\vec{\varphi}\rangle_{\mathbb{K},\Omega_{p}}\,, \quad\forall \vec{\varphi}\in H(\nabla\times,\Omega_{p})\,, \\
				\langle\nabla\times\vec{h}_{p}, \phi\rangle_{\Omega_{p}} = \langle J,\phi\rangle_{\Omega_{p}}\,, \quad\forall \phi\in L^{2}(\Omega_{p})\,.
			\end{dcases} \label{eq::mixed_finite_element_weak_form_adjoint}
		\end{equation}
		\begin{reviewer2}
		Note that the Dirichlet boundary conditions $\psi = \psi_{b} = 0$ on $\partial\Omega_{p}$ are still imposed weakly. For inhomogeneous boundary conditions the boundary term $\int_{\partial\Omega_{p}}\psi\vec{\varphi}\cdot\mathrm{d}\vec{l}$ must be included.
		\end{reviewer2}
			
		\subsubsection{The finite dimensional basis functions} \label{sec::the_finite_dimensional_basis_functions}
			In this work we approximate $\psi$ and $\vec{h}_{p}$ by expanding them in two distinct families of tensor product polynomials of at most degree $p$ in $r$ and $z$ coordinates. First, two types of polynomials are introduced, one associated to nodal interpolation and the other associated to histopolation (see \cite{Robidoux2011,robidoux-polynomial,gerritsma::edge_basis} for an extensive discussion of histopolation and its relation to integral interpolation). Subsequently, these two types of polynomials will be combined to generate the two-dimensional polynomial basis functions used to discretize $\psi$ and $\vec{h}_{p}$.
			
			Consider the canonical interval $I=[-1,1]\subset\mathbb{R}$ and the Legendre polynomials, $L_{p}(\xi)$ of degree $p$ with $\xi\in I$. The $p+1$ roots, $\xi_{i}$, of the polynomial $(1-\xi^{2})\frac{\mathrm{d}L_{p}}{\mathrm{d}\xi}$ are called Gauss-Lobatto-Legendre (GLL) nodes and satisfy $-1 = \xi_{0} < \xi_{1} < \dots < \xi_{p-1} < \xi_{p} = 1$. Let $l^{p}_{i}(\xi)$ be the Lagrange polynomial of degree $p$ through the GLL nodes, such that
			\begin{equation}
				l^{p}_{i}(\xi_{j}) :=
				\begin{cases}
					1 & \mbox{if } i = j \\
					& \\
					0 & \mbox{if } i \neq j
				\end{cases}\,, \quad i,j = 0,\dots,p\,. \label{eq::nodal_basis_polynomials}
			\end{equation}
			The explicit form of these Lagrange polynomials is given by
			\begin{equation}
				l^{p}_{i}(\xi) = \prod_{\substack{k=0\\k\neq i}}^{p}\frac{\xi-\xi_{k}}{\xi_{i}-\xi_{k}}\,. \label{eq::lagrange_interpolants}
			\end{equation}
			Let $q(\xi)$ be a function defined on $I$ and $q_{i} = q(\xi_{i})$, then its expansion in terms of these polynomials, $q_{h}(\xi)$, is given by  
			\begin{equation}
				q_{h}(\xi) := \sum_{i=0}^{p}q_{i}l^{p}_{i}(\xi)\,. \label{eq::nodal_polynomial_expansion}
			\end{equation}
			By \eqref{eq::nodal_basis_polynomials} $q_{h}(\xi)$ is a polynomial interpolant of degree $p$ of $q(\xi)$. For this reason we denote the Lagrange polynomials in \eqref{eq::lagrange_interpolants} by \emph{nodal polynomials}.
			
			\begin{reviewer2}
			Before introducing the second set of basis polynomials that will be used in this work it is important to introduce the reader to the concept of \emph{histopolant}. Given a histogram, a histopolant is a function whose integrals over the cells (or bins) of the histogram are equal to the area of the corresponding bars of the histogram, see \figref{fig::example_histopolation}. If the histopolant is a polynomial we say that it is a \emph{polynomial histopolant}. In the same way as a polynomial interpolant that passes exactly through $p+1$ points has degree $p$, a polynomial that exactly histopolates a histogram with $p+1$ bins has polynomial degree $p$. Consider now a function $g(x)$ and its associated integrals over a set of cells $[a_{j-1},a_{j}]$, $g_{j} = \int_{a_{j-1}}^{a_{j}} g(x)\mathrm{d}x$, with $ a_{0}, < \dots < a_{j} < \dots < a_{p}$. The set of integral values $g_{j}$ and cells $[a_{j-1},a_{j}]$ can be seen as a histogram. As mentioned before, it is possible to construct a histopolant of this histogram. This histopolant will be an approximating function of $g$ that has the particular property of having the same integral over the cells $[a_{j-1},a_{j}]$ as $g$. In the same way as an interpolant exactly reconstructs the original function at the interpolating points, a histopolant exactly reconstructs the integral of the original function over the cells.
			
			\begin{figure}[!ht]
				\begin{center}
					\includegraphics[width=0.3\textwidth]{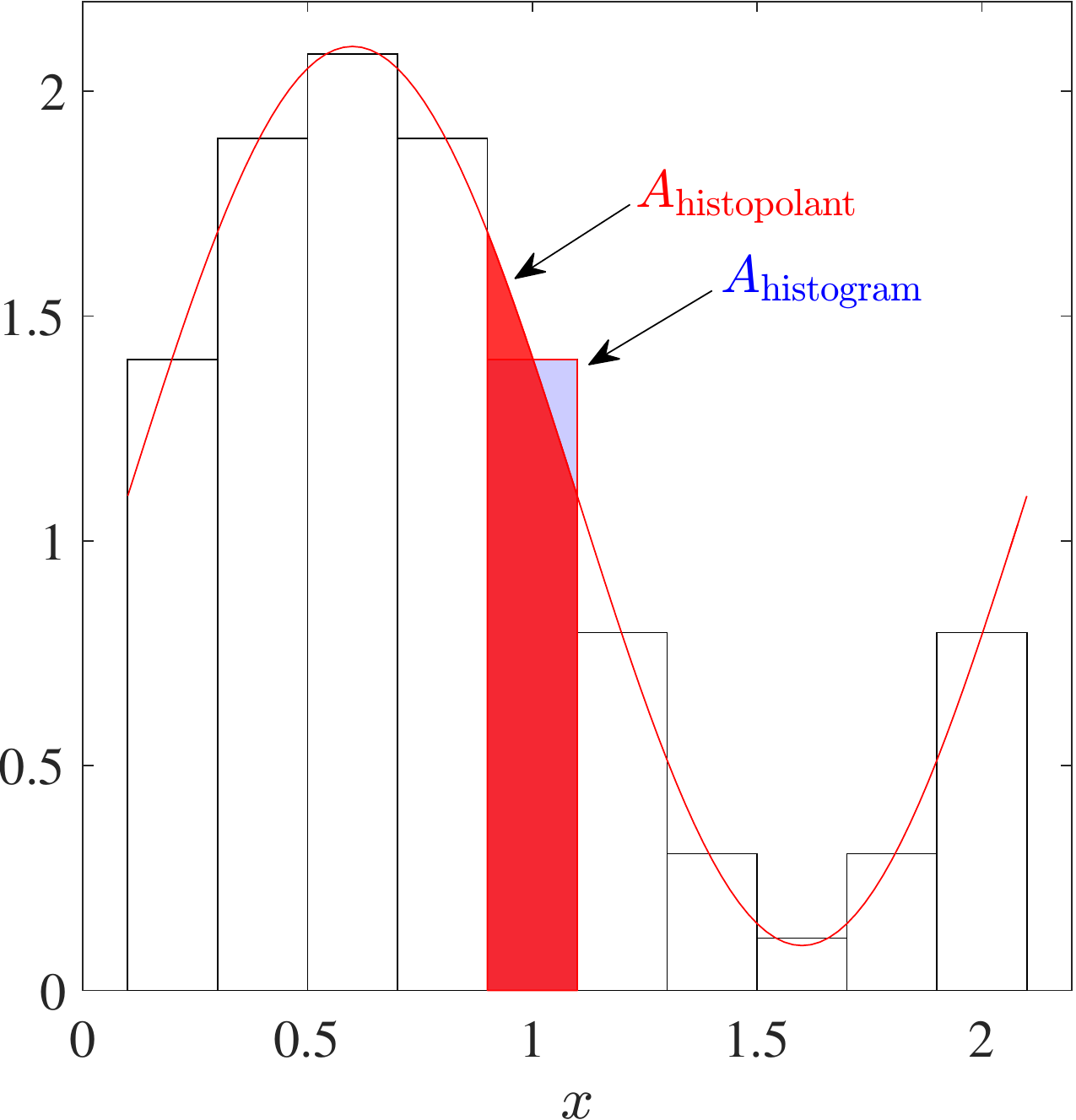} 
					\caption{\reviewertwo{Histogram and an example of a histopolant (red curve). By definition, the integral of the histopolant over each cell (or bin) $A_{\mathrm{histopolant}}$ is equal to the area of the corresponding bar of the histogram $A_{\mathrm{histogram}}$.}}
					\label{fig::example_histopolation}
				\end{center}
			\end{figure}
			
			\end{reviewer2}
			
			Using the nodal polynomials we can define another set of basis polynomials, $e^{p}_{i}(\xi)$, as
			\begin{equation}
				e^{p}_{i}(\xi) := - \sum_{k=0}^{i-1}\frac{\mathrm{d}l^{p}_{k}(\xi)}{\mathrm{d}\xi}\,, \qquad i=1,\dots,p\,. \label{eq::histopolant_polynomials_definition}
			\end{equation}
			\reviewerthree{These polynomials $e^{p}_{i}(\xi)$ have polynomial degree $p-1$ and satisfy,}
			\begin{equation}
				\int_{\xi_{j-1}}^{\xi_{j}}e^{p}_{i}(\xi)\,\mathrm{d}\xi = 
				\begin{cases}
					1 & \mbox{if } i=j \\
					& \\
					0 & \mbox{if } i\neq j
				\end{cases}\,, \quad i,j = 1,\dots,p\,. \label{eq::histopolant_polynomials_properties}
			\end{equation}
			\begin{reviewer2}
			The proof that the polynomials $e^{p}_{i}(\xi)$ have degree $p-1$ follows directly from the fact that their definition \eqref{eq::histopolant_polynomials_definition} involves a linear combination of the derivative of polynomials of degree $p$. The proof of \eqref{eq::histopolant_polynomials_properties} results from the properties of $l_{k}^{p}(\xi)$. Using  \eqref{eq::histopolant_polynomials_definition} the integral of $e^{p}_{i}(\xi)$ becomes
			\[
				\int_{\xi_{j-1}}^{\xi_{j}}e^{p}_{i}(\xi)\,\mathrm{d}\xi =  - \int_{\xi_{j-1}}^{\xi_{j}}\sum_{k=0}^{i-1}\frac{\mathrm{d}l^{p}_{k}(\xi)}{\mathrm{d}\xi} = - \sum_{k=0}^{i-1}\int_{\xi_{j-1}}^{\xi_{j}}\frac{\mathrm{d}l^{p}_{k}(\xi)}{\mathrm{d}\xi} = - \sum_{k=0}^{i-1} \left(l^{p}_{k}(\xi_{j}) - l^{p}_{k}(\xi_{j-1})\right) = - \sum_{k=0}^{i-1} \left(\delta_{k,j} -\delta_{k,j-1}\right)\,.
			\]
			It is straightforward to see that 
			\[
				- \sum_{k=0}^{i-1} \left(\delta_{k,j} -\delta_{k,j-1}\right) = \begin{cases}
					1 & \mbox{if } i=j \\
					& \\
					0 & \mbox{if } i\neq j
				\end{cases}\,, \quad i,j = 1,\dots,p\,.
			\]
			For more details see \cite{Robidoux2011,robidoux-polynomial,gerritsma::edge_basis}.
			\end{reviewer2}
			
			Let $g(\xi)$ be a function defined on $I$ and $g_{i} = \int_{\xi_{i-1}}^{\xi_{i}}g(\xi)\,\mathrm{d}\xi$, then its expansion in terms of these polynomials, $g_{h}(\xi)$, is given by
			\begin{equation}
				g_{h}(\xi) = \sum_{i=1}^{p}g_{i}e^{p}_{i}(\xi)\,.
			\end{equation}
			By \eqref{eq::histopolant_polynomials_properties} we have $\int_{\xi_{i-1}}^{\xi_{i}}g_{h}(\xi)\,\mathrm{d}\xi = g_{i}$ and \reviewertwo{therefore} $g_{h}(\xi)$ is a polynomial histopolant of degree $p-1$ of $g(\xi)$. For this reason we denote the polynomials in \eqref{eq::histopolant_polynomials_definition} by \emph{histopolant polynomials}.
			
\begin{figure}
				\begin{center}
					\includegraphics[width=0.4\textwidth]{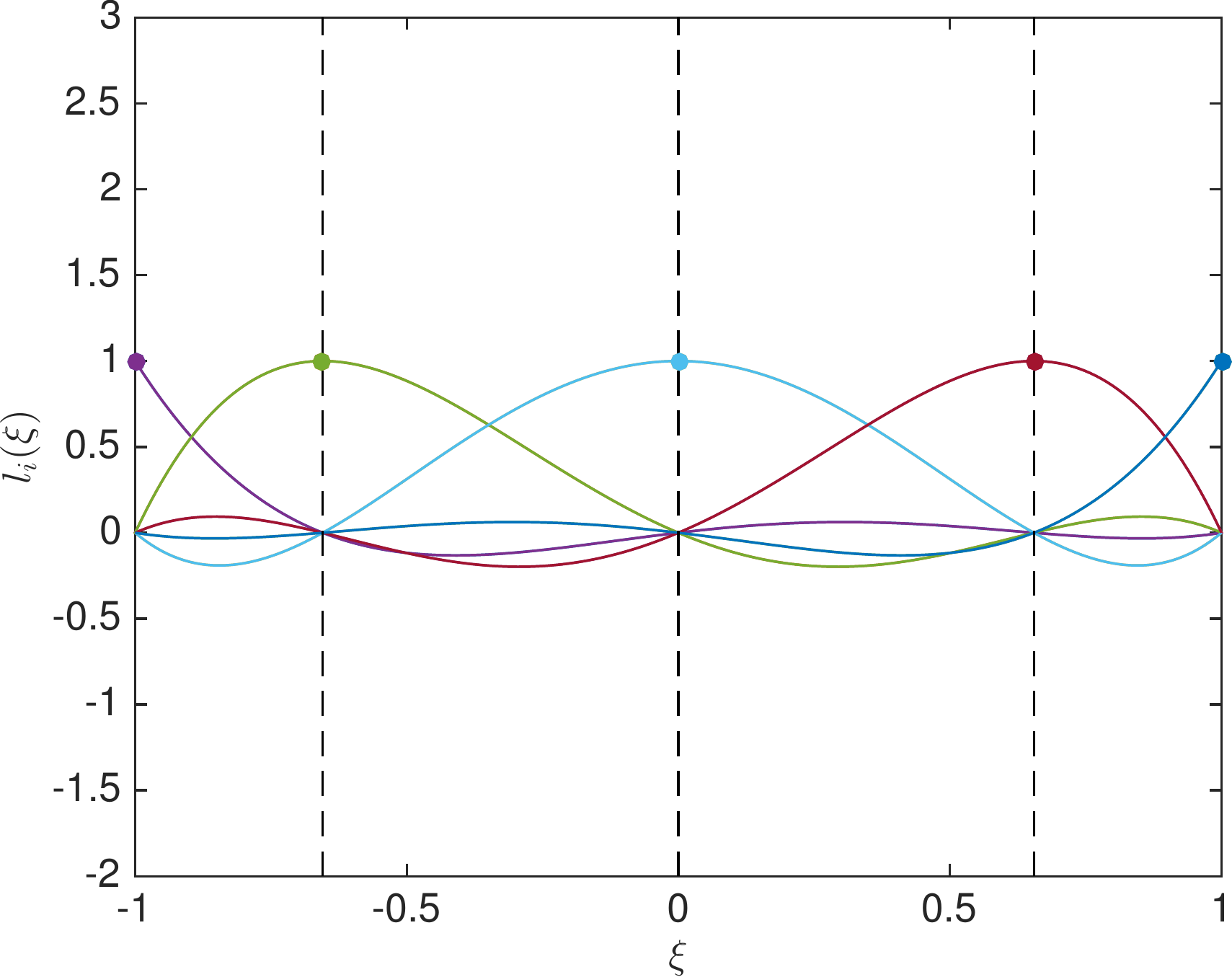} \hspace{1cm}
					\includegraphics[width=0.4\textwidth]{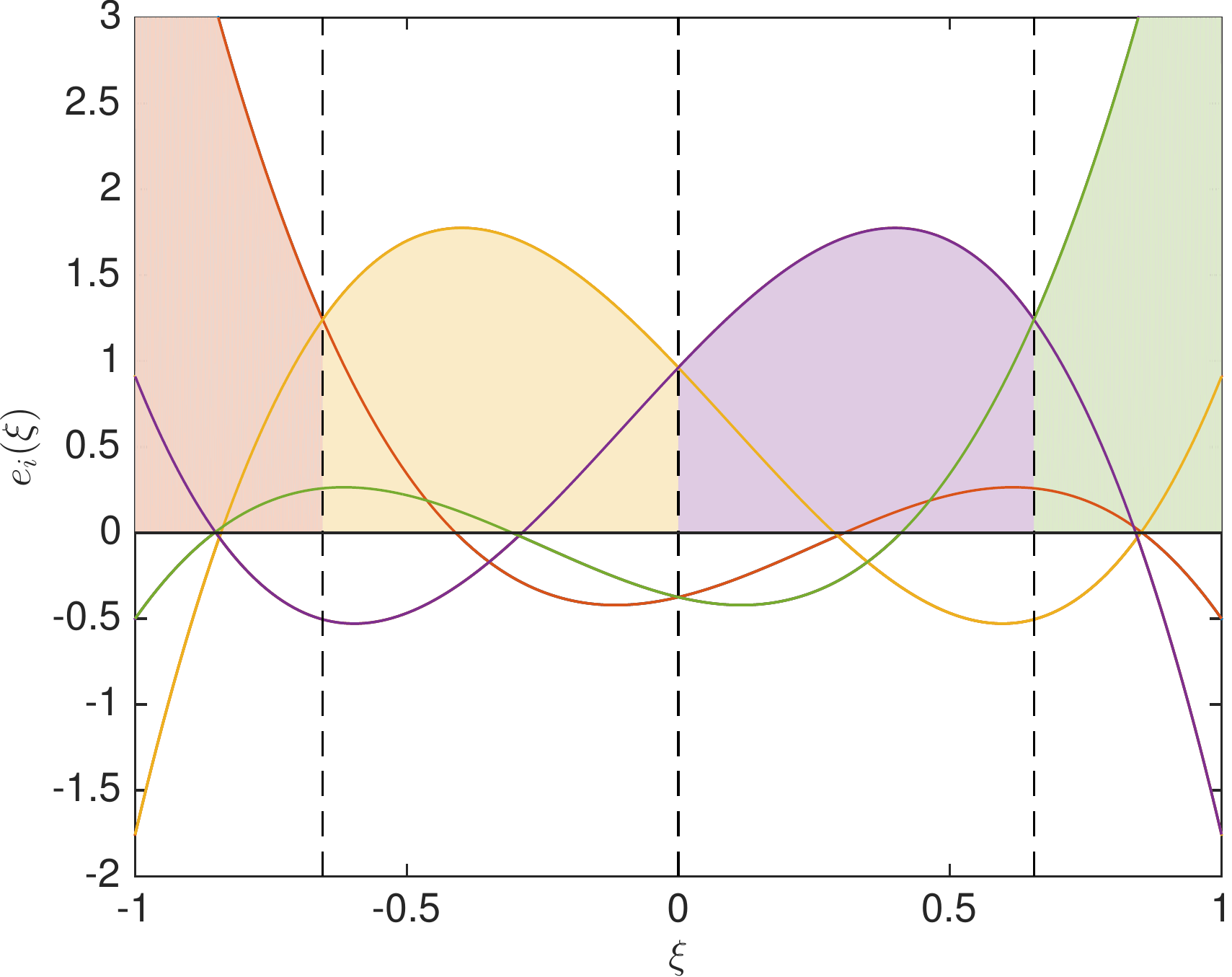}
					\caption{Basis polynomials associated to $p=4$. Left: nodal polynomials, the value of the basis polynomial at the corresponding node is one and on the other nodes is zero. Right: histopolant polynomials, the integral of the basis polynomials over the corresponding shaded area evaluates to one and to zero on the others.}
					\label{fig::basis_polynomials}
				\end{center}
			\end{figure}

			It can be shown, \cite{robidoux-polynomial,gerritsma::edge_basis}, that if $q(\xi)$ is expanded in terms of nodal polynomials, as in \eqref{eq::nodal_polynomial_expansion}, then the expansion of its derivative $\frac{\mathrm{d}q(\xi)}{\mathrm{d}\xi}$ in terms of histopolant polynomials is
			\begin{equation}
				\left(\frac{\mathrm{d}q(\xi)}{\mathrm{d}\xi}\right)_{h} = \sum_{i=1}^{p} \left(\int_{\xi_{i-1}}^{\xi_{i}}\frac{\mathrm{d}q(\xi)}{\mathrm{d}\xi}\mathrm{d}\xi\right)e^{p}_{i}(\xi) = \sum_{i=1}^{p} \left(q(\xi_{i}) - q(\xi_{i-1})\right)e^{p}_{i}(\xi) = \sum_{i=1}^{p} \left(q_{i} - q_{i-1}\right)e^{p}_{i}(\xi) = \sum_{i=1,j=0}^{p}\mathsf{E}^{1,0}_{i,j}q_{j}e^{p}_{i}(\xi)\,,
			\end{equation} 
			where $\mathsf{E}^{1,0}_{i,j}$ are the coefficients of the $p\times(p+1)$ matrix $\boldsymbol{\mathsf{E}}^{1,0}$
			\begin{equation}
				\boldsymbol{\mathsf{E}}^{1,0} := 
					\left(
						\begin{array}{cccccc} 
							-1 & 1 & 0 & 0 &  \dots & 0  \\ 
							0 & -1 & 1 & 0 & \ddots & 0 \\
							\vdots &    &  \ddots & \ddots  &  & \vdots \\
							0 & \ddots & 0 & -1 & 1 & 0 \\
							0 & \dots & 0 & 0 & -1 & 1
						\end{array}
					\right)\,,
			\end{equation}
			and the following identity holds
			\begin{equation}
				\left(\frac{\mathrm{d}q(\xi)}{\mathrm{d}\xi}\right)_{h} = \frac{\mathrm{d}q_{h}(\xi)}{\mathrm{d}\xi}\,.
			\end{equation}
			%Note that the $p+1$ polynomials $\frac{\mathrm{d}l^{p}_{i}}{\mathrm{d}\xi}$ do \emph{not} form a basis, since if $g(\xi) = 0$ is expanded in terms of $\frac{\mathrm{d}l^{p}_{i}}{\mathrm{d}\xi}$ it does \emph{not} imply that the coefficients of the expansion are zero. On the other hand, the $p$ polynomials $e^{p}_{i}(\xi)$ \emph{do} form a basis, since if $g(\xi) = 0$ is expanded in terms of $e^{p}_{i}(\xi)$ it \emph{does} imply that the coefficients of the expansion are zero.
			For an example of the basis polynomials corresponding to $p=4$ see \figref{fig::basis_polynomials}.
			
			Combining histopolant polynomials we can construct the polynomial basis functions used to discretize $\psi$ on quadrilaterals. Consider the canonical interval $I=[-1,1]$, the canonical square $Q=I\times I\subset \mathbb{R}^{2}$, the histopolant polynomials \eqref{eq::histopolant_polynomials_definition}, $e^{p}_{i}(\xi)$ of degree $p-1$, and take $\xi,\eta\in I$. Then a set of two-dimensional basis polynomials, $\omega^{p}_{k}(\xi,\eta)$, can be constructed as the tensor product of the one-dimensional ones
			\begin{equation}
				\omega^{p}_{k}(\xi,\eta) := e^{p}_{i}(\xi)\,e^{p}_{j}(\eta), \qquad i,j=1,\dots,p, \quad k = j + (i-1)p\,. \label{eq::volume_polynomials_definition}
			\end{equation} 
			These polynomials, $\omega^{p}_{k}(\xi,\eta)$, have degree $p-1$ in each variable and satisfy, see \cite{gerritsma::edge_basis,Palha2014},
			\begin{equation}
				\int_{\xi_{i-1}}^{\xi_{i}}\int_{\eta_{j-1}}^{\eta_{j}}\omega^{p}_{k}(\xi,\eta)\,\mathrm{d}\xi\mathrm{d}\eta = 
				\begin{cases}
					1 & \mbox{if }k=(i-1)p + j \\
					 & \\
					0 & \mbox{if } k\neq (i-1)p + j
				\end{cases}\,, \quad
				i,j = 1,\dots,p, \quad k = 1,\dots,p^{2}\,. \label{eq::volume_polynomials_properties}
			\end{equation} 
			\reviewerthree{Where, as before, $\xi_{i}$ and $\eta_{i}$ with $i=0,\dots,p$ are the Gauss-Lobatto-Legendre (GLL) nodes.} Let $\psi(\xi,\eta)$ be a function defined on $Q$ and $\psi_{k} = \int_{\xi_{i-1}}^{\xi_{i}}\int_{\eta_{j-1}}^{\eta_{j}}\psi(\xi,\eta)\,\mathrm{d}\xi\mathrm{d}\eta$ with $k = j + (i-1)p$, then its expansion in terms of these polynomials, $\psi_{h}(\xi,\eta)$, is given by
			\begin{equation}
				\psi_{h}(\xi,\eta) = \sum_{k=1}^{p^{2}}\psi_{k}\omega_{k}^{p}(\xi,\eta)\,. \label{eq::volume_polynomials_expansion}
			\end{equation}
			By \eqref{eq::volume_polynomials_properties} we have $\int_{\xi_{i-1}}^{\xi_{i}}\int_{\eta_{j-1}}^{\eta_{j}}\psi_{h}(\xi,\eta)\,\mathrm{d}\xi\mathrm{d}\eta = \psi_{k}$ for $k = j + (i-1)p$, with $i,j=1,\dots,p$. Additionally, $\psi_{h}(\xi,\eta)$ is a two-dimensional histopolant of $\psi(\xi,\eta)$ of degree $p-1$ in each variable. For this relation between the expansion coefficients and volume integration (in two dimensions volumes are surfaces) we denote the polynomials in \eqref{eq::volume_polynomials_definition} by \emph{volume polynomials}. Moreover, these basis polynomials satisfy $\omega^{p}_{k}(\xi,\eta)\in L^{2}(Q)$.
			
			In a similar way, but combining nodal polynomials with histopolant polynomials, we can construct the polynomial basis functions used to discretize $\vec{h}_{p}$ on quadrilaterals. Consider the nodal polynomials \eqref{eq::lagrange_interpolants}, $l^{p}_{i}(\xi)$ of degree $p$, the histopolant polynomials \eqref{eq::histopolant_polynomials_definition}, $e^{p}_{i}(\xi)$ of degree $p-1$, and take $\xi,\eta\in I$. A set of two-dimensional basis polynomials, $\vec{\epsilon}^{p}_{k}(\xi,\eta)$, can be constructed as the tensor product of the one-dimensional ones
			\begin{equation}
				\vec{\epsilon}^{p}_{k}(\xi,\eta):=
				\begin{cases}
					e_{i}^{p}(\xi)l_{j}^{p}(\eta)\,\vec{e}_{\xi} & \mbox{if } k < p(p+1), \quad \mbox{with } i=1,\dots,p, \quad j=0,\dots,p, \quad k = (i-1)(p+1) + j\,, \\
					& \\
					l_{i}^{p}(\xi)e_{j}^{p}(\eta)\,\vec{e}_{\eta} & \mbox{if } k \geq p(p+1), \quad \mbox{with } i=0,\dots,p, \quad j=1,\dots,p, \quad k = ip + j + p(p+1)-1\,.
				\end{cases} \label{eq::edge_polynomials_definition}
			\end{equation}
			These polynomials, $\vec{\epsilon}^{p}_{k}(\xi,\eta)$, have degree $p-1$ in $\xi$ and $p$ in $\eta$ if $k<p(p+1)$. If $k\geq p(p+1)$ then the degree in $\xi$ is $p$ and the degree in $\eta$ is $p-1$. It is possible to show, see \cite{gerritsma::edge_basis,Palha2014}, that these polynomials satisfy
			\begin{equation}
				\int_{\xi_{i-1}}^{\xi_{i}} \vec{\epsilon}_{k}^{p}(\xi,\eta_{j})\cdot\vec{e}_{\xi}\,\mathrm{d}\xi =
				\begin{cases}
					1 & \mbox{if } k = (i-1)(p+1) + j \\
					 & \\
					0 & \mbox{if } k \neq (i-1)(p+1) + j
				\end{cases}\,,\quad\mbox{with}\quad 
				\begin{cases}
					i = 1,\dots,p, \\ 
					j = 0, \dots, p, \\
					k = 0,\dots,2p(p+1)-1\,,
				\end{cases}
				\label{eq::edge_polynomials_properties_xi}
			\end{equation}
			and
			\begin{equation}
				\int_{\eta_{j-1}}^{\eta_{j}} \vec{\epsilon}_{k}^{p}(\xi_{i},\eta)\cdot\vec{e}_{\eta}\,\mathrm{d}\eta =
				\begin{cases}
					1 & \mbox{if } k = ip + j + p(p+1) - 1\\
					 & \\
					0 & \mbox{if } k \neq ip + j + p(p+1) - 1
				\end{cases}\,,\quad\mbox{with}\quad 
				\begin{cases}
					i = 0,\dots,p, \\ 
					j = 1, \dots, p, \\
					k = 0,\dots,2p(p+1)-1\,. 
				\end{cases}
				\label{eq::edge_polynomials_properties_eta}
			\end{equation}
			Let $\vec{h}_{p}(\xi,\eta)$ be a vector valued function defined on $Q$ and
			\begin{equation}
				h_{p,k} = 
				\begin{dcases}
					\reviewerthree{\int_{\xi_{i-1}}^{\xi_{i}} \vec{h}_{p}(\xi,\eta_{j})\cdot\vec{e}_{\xi}\,\mathrm{d}\xi\,} & \mbox{if } k < p(p+1), \quad \mbox{with} \quad \begin{cases} i=1,\dots,p, \\ j=0,\dots,p, \\ k = (i-1)(p+1) + j\,, \end{cases} \\
					& \\
					\int_{\eta_{j-1}}^{\eta_{j}} \vec{h}_{p}(\xi_{i},\eta)\cdot\vec{e}_{\eta}\,\mathrm{d}\eta & \mbox{if } k \geq p(p+1), \quad \mbox{with} \quad \begin{cases} i=0,\dots,p, \\ j=1,\dots,p, \\ k = ip + j + p(p+1)-1\,. \end{cases}
				\end{dcases} \label{eq::edge_basis_expansion_coefficientes}
			\end{equation}
			then its expansion in terms of these polynomials, $\vec{h}_{p,h}(\xi,\eta)$, is given by
			\begin{equation}
				\vec{h}_{p,h}(\xi,\eta) = \sum_{k=0}^{2p(p+1)-1} h_{p,k}\vec{\epsilon}^{p}_{k}(\xi,\eta)\,, \label{eq::expansion_edge_polynomials}
			\end{equation}
			Using \eqref{eq::edge_polynomials_properties_xi} we have
			\begin{equation}
				\int_{\xi_{i-1}}^{\xi_{i}} \vec{h}_{p,h}(\xi,\eta_{j})\cdot\vec{e}_{\xi}\,\mathrm{d}\xi = h_{p,k}\,,\,\, \mbox{ for }\,\, k = (i-1)(p+1) + j\,,\,\,\mbox{ with } \,\, i=1,\dots,p \,\,\mbox{ and }\,\, j=0,\dots,p\,.
			\end{equation} 
			In a similar way, using \eqref{eq::edge_polynomials_properties_eta} we have
			\begin{equation}
				\int_{\eta_{j-1}}^{\eta_{j}} \vec{h}_{p,h}(\xi_{i},\eta)\cdot\vec{e}_{\eta}\,\mathrm{d}\eta = h_{p,k}\,,\,\, \mbox{ for }\,\, k = ip + j + p(p+1)-1\,,\,\, \mbox{ with }\,\, i=0,\dots,p\,\,\mbox{ and }\,\, j=1,\dots,p\,.
			\end{equation} 
			The expansion $\vec{h}_{p,h}(\xi,\eta)$ is a two-dimensional edge histopolant (interpolates line integrals) of $\vec{h}_{p}(\xi,\eta)$ with degrees $p-1$ in $\xi$ and $p$ in $\eta$ along the $\xi$ component and degrees $p$ in $\xi$ and $p-1$ in $\eta$ along the $\eta$ component. Since the coefficients of this expansion are edge (or line) integrals, we denote the polynomials in \eqref{eq::edge_polynomials_definition} by \emph{edge polynomials}. Additionally, these basis polynomials satisfy $\vec{\epsilon}_{k}^{p}(\xi,\eta)\in H(\nabla\times,\Omega)$, see \cite{gerritsma::edge_basis,Palha2014}.
			
			It can be shown, \cite{gerritsma::edge_basis,Palha2014}, that if $\vec{h}_{p}(\xi,\eta)$ is expanded in terms of edge polynomials, as in \eqref{eq::expansion_edge_polynomials}, then the expansion of $\nabla\times\vec{h}_{p}$ in terms of the volume polynomials, \eqref{eq::volume_polynomials_definition}, is
			\begin{align}
				\left(\nabla\times\vec{h}_{p}(\xi,\eta)\right)_{h} &\stackrel{\phantom{\eqref{eq::edge_basis_expansion_coefficientes}}}{=} \sum_{i,j=1}^{p}\left(\int_{\xi_{i-1}}^{\xi_{i}}\int_{\eta_{j-1}}^{\eta_{j}}\nabla\times\vec{h}_{p}(\xi,\eta)\,\mathrm{d}\xi\mathrm{d}\eta\right)\omega_{j+(i-1)p}^{p}(\xi,\eta) \nonumber\\
				 &\stackrel{\phantom{\eqref{eq::edge_basis_expansion_coefficientes}}}{=}\begin{aligned}[t]
				 		&\sum_{i,j=1}^{p}\left(\int_{\xi_{i-1}}^{\xi_{i}}\vec{h}_{p}(\xi,\eta_{j-1})\cdot\vec{e}_{\xi}\,\mathrm{d}\xi + \int_{\eta_{j-1}}^{\eta_{j}}\vec{h}_{p}(\xi_{i},\eta)\cdot\vec{e}_{\eta}\,\mathrm{d}\eta\right. \nonumber \\ 
						&\qquad\qquad\left.- \int_{\xi_{i-1}}^{\xi_{i}}\vec{h}_{p}(\xi,\eta_{j})\cdot\vec{e}_{\xi}\,\mathrm{d}\xi - \int_{\eta_{j-1}}^{\eta_{j}}\vec{h}_{p}(\xi_{i-1},\eta)\cdot\vec{e}_{\eta}\,\mathrm{d}\eta \right)\omega_{j+(i-1)p}^{p}(\xi,\eta)
					\end{aligned} \nonumber\\
				&\stackrel{\eqref{eq::edge_basis_expansion_coefficientes}}{=} \sum_{k=1}^{p^{2}} \left(h_{p,k+(k\,\mathrm{div} p)} + h_{p,k+p(p+2)} - h_{p,k+(k\,\mathrm{div} p) +1 } - h_{p,k+p(p+1)} \right)\omega_{k}^{p}(\xi,\eta)\nonumber \\
				&\stackrel{\phantom{\eqref{eq::edge_basis_expansion_coefficientes}}}{=} \sum_{k=1,j=0}^{p^{2},2p(p+1)-1} \mathsf{E}^{2,1}_{k,j} h_{p,j} \omega_{k}^{p}(\xi,\eta)\,, \label{eq::edge_basis_expansion_curl}
			\end{align}
			where $\mathsf{E}^{2,1}_{k,j}$ are the coefficients of the $p^{2}\times 2p(p+1)$ matrix $\boldsymbol{\mathsf{E}}^{2,1}$ and are defined as
			\begin{equation}
				\mathsf{E}_{k,j}^{2,1} :=
				\begin{cases}
					1  & \mbox{if } j = k + (k\,\mathrm{div} p)\,, \\
					1  & \mbox{if } j = k + p(p+2)\,, \\
					-1 & \mbox{if } j = k + (k\,\mathrm{div} p) + 1\,, \\
					-1 & \mbox{if } j = k + p(p+1)\,, \\
					0  & \mbox{otherwise}\,.
				\end{cases}
			\end{equation}
			The following identity holds
			\begin{equation}
				\left(\nabla\times\vec{h}_{p}(\xi,\eta)\right)_{h} = \nabla\times\vec{h}_{p,h}\,. \label{eq::edge_basis_commuting_projection}
			\end{equation}
			Here, $(k\,\mathrm{div} p)$ denotes integer division, that is division in which the fractional part (remainder) is discarded, e.g. $(5\,\mathrm{div}\,3) = 1$, $(5\,\mathrm{div}\,2) = 2$ and $(5\,\mathrm{div}\,6) = 0$.
			
\begin{figure}[!hb]
				\begin{center}
					\includegraphics[width=0.3\textwidth]{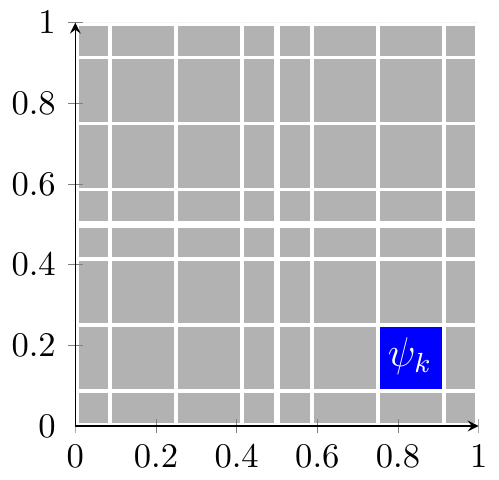} \hspace{1cm}
					\includegraphics[width=0.3\textwidth]{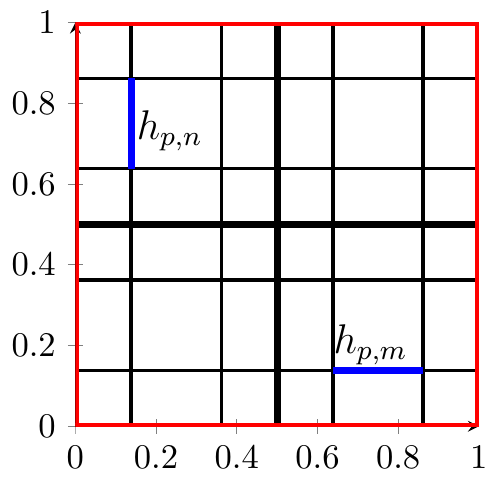}
					\caption{Visual representation of the coefficients of the polynomial expansions using volume polynomials (left) and edge polynomials (right). These coefficients correspond to \reviewerthree{geometric degrees of freedom}. For volume polynomials these \reviewerthree{geometric degrees of freedom} are associated to surfaces of the spectral element mesh and for edge polynomials they are associated to edges of the spectral element mesh.}
					\label{fig::basis_polynomials_dofs}
				\end{center}
			\end{figure}

			A graphical representation of the coefficients in the polynomial expansions, \eqref{eq::volume_polynomials_expansion} and \eqref{eq::expansion_edge_polynomials}, is shown in \figref{fig::basis_polynomials_dofs}.
			\FloatBarrier
			
		\subsubsection{Discrete representation of topological laws}
			The finite dimensional polynomial basis functions presented in \secref{sec::the_finite_dimensional_basis_functions} enable us to exactly satisfy the topological laws at a discrete level. Consider the topological law in \eqref{eq::mixed_finite_element_weak_form_adjoint}
			\begin{equation}
				\langle\nabla\times\vec{h}_{p},\phi\rangle_{\Omega_{p}} = \langle J, \phi\rangle_{\Omega_{p}},\qquad \forall \phi \in L^{2}(\Omega_{p})\,.
			\end{equation}
			If $\vec{h}_{p}$ and $J$ are substituted by their polynomial expansions as in \eqref{eq::expansion_edge_polynomials} and \eqref{eq::volume_polynomials_expansion} respectively, using \eqref{eq::edge_basis_expansion_curl} the following expression is obtained
			\begin{equation}
				\langle\sum_{k=1,j=0}^{p^{2},2p(p+1)-1} \mathsf{E}^{2,1}_{k,j} h_{p,j} \omega_{k}^{p}(\xi,\eta),\phi\rangle_{\Omega_{p}} = \langle \sum_{k=1}^{p^{2}} J_{k}\omega_{k}^{p}(\xi,\eta), \phi\rangle_{\Omega_{p}},\qquad \forall \phi \in L^{2}(\Omega_{p})\,.
			\end{equation}
			By linearity of the inner product this can be rewritten as
			\begin{equation}
				\sum_{k=1,j=0}^{p^{2},2p(p+1)-1} \mathsf{E}^{2,1}_{k,j} h_{p,j} \langle\omega_{k}^{p}(\xi,\eta),\phi\rangle_{\Omega_{p}} =  \sum_{k=1}^{p^{2}} J_{k}\langle\omega_{k}^{p}(\xi,\eta), \phi\rangle_{\Omega_{p}},\qquad \forall \phi \in L^{2}(\Omega_{p})\,.
			\end{equation}
			\reviewerthree{Since the basis elements $\omega_{k}^{p}$ form a basis, we have that $\sum a_{k}\omega_{k}^{p} = 0$ is equivalent to $a_{k}=0$, therefore we can write it in a purely topological fashion (independent of metric and material properties)}
			%term $\langle\omega_{k}^{p}(\xi,\eta), \phi\rangle_{\Omega_{p}}$ appears on both sides of this equation, 
			\begin{equation}
				\sum_{k=1,j=0}^{p^{2},2p(p+1)-1} \mathsf{E}^{2,1}_{k,j} h_{p,j} =  \sum_{k=1}^{p^{2}} J_{k}\,.
			\end{equation}
			In matrix notation, this equation takes the more compact form
			\begin{equation}
				\boldsymbol{\mathsf{E}}^{2,1} \boldsymbol{h}_{p} = \boldsymbol{J}\,, \label{eq::discrete_topological_law}
			\end{equation}
			where $\boldsymbol{h}_{p}$ denotes the row vector with $j$-th entry equal to $h_{p,j}$ and $\boldsymbol{J}$ represents the row vector with $j$-th entry equal to $J_{j}$.
			
			The key factor that allows the construction of a purely topological law at the discrete level, \eqref{eq::discrete_topological_law}, is the appropriate definition of a set of basis functions such that \eqref{eq::edge_basis_expansion_curl} and \eqref {eq::edge_basis_commuting_projection} are satisfied.
			
		\subsubsection{The discrete system of equations}
			Having defined a natural inner product, \eqref{eq::natural_inner_product_definition}, having constructed a set of basis functions with which to discretize the unknown physical quantities $\psi$ and $\vec{h}_{p}$, \eqref{eq::volume_polynomials_definition} and \eqref{eq::edge_polynomials_definition}, and having derived the equivalent discrete topological law, \eqref{eq::discrete_topological_law}, we can now revisit the weak formulation of the Poisson problem, \eqref{eq::mixed_finite_element_weak_form_adjoint}.
			
			Consider the computational domain $\Omega_{p}$ in the $(r,z)$-plane and its tessellation $\mathcal{T}(\Omega_{p})$ consisting of $N$ arbitrary quadrilaterals (possibly curved), $\Omega_{m}$ with $m=1,\dots,N$.  We assume that all quadrilateral elements $\Omega_{m}$ can be obtained from a map $\Phi_{m}:(\xi,\eta)\in Q\rightarrow (r,z)\in\Omega_{m}$. Then the pushforward $\Phi_{m,*}$ maps functions in the reference element $Q$ to functions in the physical element $\Omega_{m}$, see for example \cite{abraham_diff_geom,frankel}. For this reason it suffices to explore the analysis on the reference domain $Q$.
			
			\begin{remark}
				If a differential geometry formulation was used, the physical quantities would be represented by differential $k$-forms and the map $\Phi_{m}:(\xi,\eta)\in Q\rightarrow (r,z)\in\Omega_{m}$ would generate a pullback, $\Phi^{*}_{m}$, mapping $k$-forms in physical space, $\Omega_{m}$, to $k$-forms in the reference element, $Q$.
			\end{remark}
			
			Let $U$ and $V$ be finite-dimensional polynomial spaces such that $U=\mathrm{span}\{\omega_{1}^{p},\dots,\omega_{p^{2}}^{p}\}\subset L^{2}(Q)$ and $V=\mathrm{span}\{\vec{\epsilon}_{0}^{p},\dots,\vec{\epsilon}_{2p(p+1)-1}^{p}\}\subset H(\nabla\times,Q)$. The mimetic formulation for the discrete solution of \eqref{eq::first_order_system} on $Q$ is given by
			\begin{equation}
				\begin{dcases}
					\mathrm{Find} \,\, \psi_{h}\in U \,\, \mathrm{and}\,\, \vec{h}_{p,h}\in V\,\,\mathrm{such}\,\,\mathrm{that:} \\
					\langle\psi_{h},\nabla\times\vec{\epsilon}_{j}^{p}\rangle_{Q}  = \langle\vec{h}_{p,h},\vec{\epsilon}_{j}^{p}\rangle_{\mathbb{K},Q}\,, \quad j=0,\dots,2p(p+1)-1\,, \\
					\langle\nabla\times\vec{h}_{p}, \omega_{j}^{p}\rangle_{Q} = \langle J,\omega_{j}^{p}\rangle_{Q}\,, \quad j=1,\dots,p^{2}\,.
				\end{dcases} \label{eq::mixed_finite_element_weak_form_mimetic_discrete}
			\end{equation}
			Well-posedness of this discrete formulation follows directly from the fact that $V\subset H(\nabla\times,Q)$, $U\subset L^{2}(Q)$ and $\mathcal{Z}_{h}:=\{\vec{v}_{h}\in V \mid \nabla\times\vec{v}_{h}\} \subset \mathcal{Z}:=\{\vec{v}\in H(\nabla\times,Q) \mid \nabla\times\vec{v}\}$. For a more detailed discussion see \cite{Palha2014}.
			
			Substituting $\psi_{h}$, $\vec{h}_{p,h}$ by their expressions and $J$ by its expansion, as in \eqref{eq::expansion_edge_polynomials}, \eqref{eq::volume_polynomials_expansion} and \eqref{eq::expansion_edge_polynomials} respectively, and using both the linearity of the inner product and \eqref{eq::edge_basis_expansion_curl} we obtain
			\begin{equation}
				\begin{dcases}
					\mathrm{Find} \,\, \psi_{h}\in U \,\, \mathrm{and}\,\, \vec{h}_{p,h}\in V\,\,\mathrm{such}\,\,\mathrm{that:} \\
					\sum_{i=1,k=1}^{p^{2}}\mathsf{E}_{k,j}^{2,1}\psi_{i}\langle\omega_{i}^{p},\omega_{k}^{p}\rangle_{Q}  = \sum_{k=0}^{2p(p+1)-1}h_{p,k}\langle\vec{\epsilon}_{k}^{p},\vec{\epsilon}_{j}^{p}\rangle_{\mathbb{K},Q}\,, \quad j=0,\dots,2p(p+1)-1\,, \\
					\sum_{i=0,k=1}^{2p(p+1)-1,p^{2}}\mathsf{E}_{k,i}^{2,1}h_{p,i}\langle\omega_{k}^{p}, \omega_{j}^{p}\rangle_{Q} = \sum_{k=1}^{p^{2}}J_{k}\langle \omega_{k}^{p},\omega_{j}^{p}\rangle_{Q}\,, \quad j=1,\dots,p^{2} \,.
				\end{dcases}
			\end{equation}
			Note that we have used $\nabla\times\vec{\epsilon}_{j}^{p} = \sum_{k=1}^{p^{2}}\mathsf{E}_{k,j}\omega_{k}^{p}$, a special case of  \eqref{eq::edge_basis_expansion_curl}. This formulation takes a more compact form in matrix notation
			\begin{equation}
				\begin{dcases}
					\mathrm{Find} \,\, \boldsymbol{\psi}\in \mathbb{R}^{p^{2}} \,\, \mathrm{and}\,\, \boldsymbol{h}_{p}\in \mathbb{R}^{2p(p+1)}\,\,\mathrm{such}\,\,\mathrm{that:} \\
					\left(\boldsymbol{\mathsf{E}}^{2,1}\right)^{\intercal}\boldsymbol{\mathsf{N}}\,\boldsymbol{\psi} = \boldsymbol{\mathsf{M}}\,\boldsymbol{h}_{p}\,,\\
					\boldsymbol{\mathsf{N}}\boldsymbol{\mathsf{E}}^{2,1}\boldsymbol{h}_{p} = \boldsymbol{\mathsf{N}}\,\boldsymbol{J}\,,
				\end{dcases}
			\end{equation}
			with 
			\begin{equation}
				\mathsf{N}_{i,j} := \langle\omega_{i}^{p},\omega_{j}^{p}\rangle_{Q}\quad \mbox{and} \quad \mathsf{M}_{i,j} := \langle\vec{\epsilon}_{i}^{p},\vec{\epsilon}_{j}^{p}\rangle_{\mathbb{K},Q} \,. \label{eq::mimetic_poisson_algebraic_system}
			\end{equation}
			As expected, for homogeneous Dirichlet boundary conditions, $\psi_{b} = 0$, the system of algebraic equations is symmetric.
		
	\subsection{Grad-Shafranov problem and its discrete solution} \label{sec::grad_shafranov_solver}
		The discrete solution of the Grad-Shafranov problem \eqref{eq::non_linear_poisson} can be directly obtained combining \eqref{eq::mimetic_poisson_algebraic_system} with the iterative procedures described in \secref{sec::non_eigenvalue_case} (non-eigenvalue case) and \secref{sec::eigenvalue_case} (eigenvalue case).
		
		\subsubsection{Non-eigenvalue case}
			For the non-eigenvalue case we replace in \eqref{eq::non_linear_poisson_iteration} the continuous operators by their discrete versions \eqref{eq::mimetic_poisson_algebraic_system}. Hence, for each iteration $k$ the following algebraic system of equations is solved
			\begin{equation}
			\begin{dcases}
				\mathrm{Find} \,\, \boldsymbol{\psi}^{k+1}\in \mathbb{R}^{p^{2}} \,\, \mathrm{and}\,\, \boldsymbol{h}_{p}^{k+1}\in \mathbb{R}^{2p(p+1)}\,\,\mathrm{such}\,\,\mathrm{that:} \\
				\left(\boldsymbol{\mathsf{E}}^{2,1}\right)^{\intercal}\boldsymbol{\mathsf{N}}\,\boldsymbol{\psi}^{k+1} = \boldsymbol{\mathsf{M}}\,\boldsymbol{h}_{p}^{k+1}\,,\\
				\boldsymbol{\mathsf{N}}\boldsymbol{\mathsf{E}}^{2,1}\boldsymbol{h}_{p}^{k+1} = \boldsymbol{\mathsf{N}}\,\boldsymbol{J}^{k}\,,
			\end{dcases} 
			\end{equation}
			with 
			\begin{equation}
				J^{k}_{n} = \int_{\xi_{i-1}}^{\xi_{i}}\int_{\eta_{j-1}}^{\eta_{j}} J_{\phi}(\psi_{h}^{k}(r,z),r,z)\,\mathrm{d}\xi\mathrm{d}\eta, \qquad n=j+(i-1)p,\quad i,j=1,\dots,p\,.
			\end{equation}
		As before, the iterative procedure is stopped once the residual between two consecutive iterations satisfies
		\begin{equation}
			\|\nabla\times\vec{h}_{p,h}^{k+1} - J_{h}^{k+1}\|_{L^{s}} < \epsilon\,, \label{eq::stopping_criterium}
		\end{equation}
		with $\epsilon \ll 1$ and $s \in \mathbb{N}$.
		
		\subsubsection{Eigenvalue case}
			In a similar way, for the eigenvalue case we just replace in \eqref{eq::linear_poisson_iteration} and \eqref{eq::normalization_iteration_eigenvalue} the continuous operators by their discrete versions \eqref{eq::mimetic_poisson_algebraic_system}. Therefore, for each iteration $k$ a new eigen-pair $(\normalized{\boldsymbol{\psi}}^{k+1},\sigma^{k+1})$ is computed by solving the algebraic system of equations
			\begin{equation}
				\begin{dcases}
					\mathrm{Find} \,\, \boldsymbol{\psi}^{k+1}\in \mathbb{R}^{p^{2}} \,\, \mathrm{and}\,\, \boldsymbol{h}_{p}^{k+1}\in \mathbb{R}^{2p(p+1)}\,\,\mathrm{such}\,\,\mathrm{that:} \\
					\left(\boldsymbol{\mathsf{E}}^{2,1}\right)^{\intercal}\boldsymbol{\mathsf{N}}\,\boldsymbol{\psi}^{k+1} = \boldsymbol{\mathsf{M}}\,\boldsymbol{h}_{p}^{k+1}\,,\\
					\boldsymbol{\mathsf{N}}\boldsymbol{\mathsf{E}}^{2,1}\boldsymbol{h}_{p}^{k+1} = \sigma^{k}\boldsymbol{\mathsf{N}}\,\boldsymbol{J}^{k}\,,
				\end{dcases} 
			\end{equation}
			and then normalizing
			\begin{equation}
				\normalized{\boldsymbol{\psi}}^{k+1} = \frac{\boldsymbol{\psi}^{k+1}}{\|\psi^{k+1}_{h}(\xi,\eta)\|_{L^{s}}} \quad\mathrm{and}\quad \sigma^{k+1} = \frac{\sigma^{k}}{\|\psi^{k+1}_{h}(\xi,\eta)\|_{L^{s}}}\,,
			\end{equation}
			with 
			\begin{equation}
				J^{k}_{n} = \int_{\xi_{i-1}}^{\xi_{i}}\int_{\eta_{j-1}}^{\eta_{j}} J_{\phi}(\normalized{\psi}_{h}^{k}(r,z),r,z)\,\mathrm{d}\xi\mathrm{d}\eta, \qquad n=j+(i-1)p,\quad i,j=1,\dots,p\,.
			\end{equation}
		As before, the iterative procedure is stopped once the residual  satisfies \eqref{eq::stopping_criterium}.
			
\section{Numerical test cases} \label{sec::numerical_test_cases}
	In order to assess the accuracy and robustness of the proposed Grad-Shafranov solver we apply it to the solution of different test cases. We start by analyzing the convergence properties of the solver on different grids by solving equilibria where exact solutions are known, \secref{sec::convergence_tests}. We then proceed to address the case where the computational domain coincides with the plasma domain, \secref{sec::plasma_boundary}. We first present results for the case of smooth plasma boundaries for both linear, \secref{sec::test_cases_linear_eigenvalue}, and non-linear eigenvalue problems, \secref{sec::test_cases_nonlinear_eigenvalue}. We finalize with an application to an X-point configuration, \secref{sec::test_cases_x_point}.
	
	The choice of the specific test cases addressed here was made with the objective of comparing the proposed solver with results from recently published high order solvers, namely \cite{Pataki2013, Lee2015,Howell2014}. Therefore most of the test cases discussed here are also presented either in \cite{Howell2014} or \cite{Pataki2013}. A summary of the different profile models used in the test cases is presented in \tabref{tab::summary_profile_models}.
	\begin{table}[htp]
		\caption{A summary of the profile models used in the test cases.}
		\begin{center}
			\begin{tabular}{p{5cm} p{3cm} p{4cm}}
				\hline
				Equilibria & $f$ model & $P$ model \\
				\hline
				Soloviev & $f(\psi) = f_{0}$ & $P(\psi) = C \psi + P_{0}$ \\
				FRC &  $f(\psi) = 0$ & $P(\psi) = P_{0} + P_{1}\frac{\psi^{2}}{\psi_{0}^{2}}$\\
				Spheromak & $f(\psi) = f_{0}\psi$ & $P(\psi) = P_{0}\frac{\psi}{\psi_{0}}$ \\
				Linear eigenvalue & $f(\psi) = 0$ & $P(\psi) = \frac{C_{1}+C_{2}r^{2}}{2\mu_{0}}\psi^{2} + P_{0}$ \\
				non-linear eigenvalue &  $f(\psi) = 0$ & $P(\psi) = \frac{C_{1} + C_{2}\psi^{2}}{\mu_{0}}\left(1-e^{-\frac{\psi^{2}}{\eta}}\right)$\\
				\reviewerone{X-point (Soloviev)} &  $f(\psi) = \sqrt{2\mu_{0}A\psi}$ & $P(\psi) = (1-A)\psi$ \\
				\reviewerone{X-point (linear eigenvalue)}  & $f(\psi) = 0$ & $P(\psi) = \frac{C_{1}+C_{2}r^{2}}{2\mu_{0}}\psi^{2} + P_{0}$ \\
				\reviewerone{X-point (non-linear eigenvalue)} & $f(\psi) = 0$ & $P(\psi) = \frac{C_{1} + C_{2}\psi^{2}}{\mu_{0}}\left(1-e^{-\frac{\psi^{2}}{\eta}}\right) $\\
				\hline
			\end{tabular}
		\end{center}
		\label{tab::summary_profile_models}
	\end{table}%

	\subsection{Convergence tests}\label{sec::convergence_tests}
		The convergence tests focus mainly on both $h$- (mesh refinement) and $p$- (increase of polynomial degree of basis functions) convergence studies. We intend to show optimal convergence rates of order $p+1$ for $h$-refinement and exponential convergence rates for $p$-refinement. Another important aspect we intend to show is the robustness of the present method with respect to highly curved meshes. For this reason we will compare the convergence rates on orthogonal meshes with the ones on highly curved meshes. We will use three plasma configurations where exact solutions are known: (i) Soloviev solution in \secref{sec::test_cases_soloviev}, (ii) field-reversed configuration (FRC) in \secref{sec::test_cases_frc} and (iii) spheromak configuration in \secref{sec::test_cases_spheromak}.
		
		\subsubsection{Soloviev test case} \label{sec::test_cases_soloviev}
			The first test case corresponds to a Soloviev solution of the Grad-Shafranov equation. This solution is one of the most widely used analytical solutions of the Grad-Shafranov equation. This special case corresponds to a constant model $f(\psi) = f_{0}$ combined with a linear model $P(\psi) = C\psi + P_{0}$. This leads to the following Grad-Shafranov equation
			\begin{equation}
				\begin{dcases}
					\nabla\times\left(\mathbb{K}\nabla\times\psi\right) = \frac{r}{\mu_{0}} & \mbox{in} \quad \Omega_{p}\,, \\
					\psi = \psi_{a} & \mbox{on}\quad\partial\Omega_{p}\,,
				\end{dcases} \label{eq::soloviev_test_case}
			\end{equation}
			where we have set $C=\frac{1}{\mu_{0}}$. \reviewerthree{Note that since we are solving the Grad-Shafranov equation on a rectangular domain, here $\Omega_{p}$ stands for a rectangular domain, which includes the plasma.}
			
			We outline here the methodology presented in \cite{Cerfon2010,Pataki2013} to obtain physically relevant solutions to this equation. The analytical solution, $\psi_{a}(r,z)$, is constructed by adding to the particular solution $\frac{r^{4}}{8}$ a linear combination of three homogeneous solutions,
			\begin{equation}
				\psi_{a}(r,z) = \frac{r^{4}}{8} + d_{1} + d_{2}r^{2} + d_{3}\left(r^{4} - 4r^{2}z^{2}\right)\,. \label{eq::soloviev_test_case_analytical_solution}
			\end{equation}
			The coefficients $d_{i}$ are determined by imposing a physically relevant shape to the contour line $\psi_{a}(r,z) = 0$, the plasma cross section. The shape of the plasma is parameterized by three variables, see \cite{Cerfon2010,Pataki2013} for a detailed discussion: inverse aspect-ratio $\epsilon$, elongation $\kappa$, and triangularity $\delta$. The imposed boundary conditions correspond to the following equations
			\begin{equation}
				\begin{dcases}
					\psi_{a}(1\pm\epsilon,0) = 0 & \mbox{Lateral shape conditions,} \\
					\psi_{a}(1-\delta\epsilon,\kappa\epsilon) = 0 & \mbox{Vertical shape condition,}
				\end{dcases}
			\end{equation}
			which can be expressed as
			\begin{equation}
				\left(
					\begin{array}{ccc}
						1 & (1+\epsilon)^{2} & (1+\epsilon)^{4} \\
						1 & (1-\epsilon)^{2} & (1-\epsilon)^{4} \\
						1 & (1-\delta\epsilon)^{2} & (1-\delta\epsilon)^{4} -4(1-\delta\epsilon)^{2}\kappa^{2}\epsilon^{2}
					\end{array}
				\right) 
				\left(
					\begin{array}{c}
						d_{1} \\
						d_{2} \\
						d_{3}
					\end{array}
				\right) = -\frac{1}{8}
				\left(
					\begin{array}{c}
						(1+\epsilon)^{4} \\
						(1-\epsilon)^{4} \\
						(1-\delta\epsilon)^{4}
					\end{array}
				\right)\,. \label{eq:soloviev_parameters}
			\end{equation}
			
			The numerical method proposed in this work is applied to the solution of two Soloviev solution cases, as in \cite{Pataki2013}. One corresponds to an ITER-like cross section where $\epsilon=0.32$, $\kappa=1.7$ and $\delta=0.33$. The other solution corresponds to an NTSX-like cross section where $\epsilon=0.78$, $\kappa=2.0$ and $\delta=0.35$.
			
\begin{figure}[!ht]
				\begin{center}
				\includegraphics[height=0.271\textwidth]{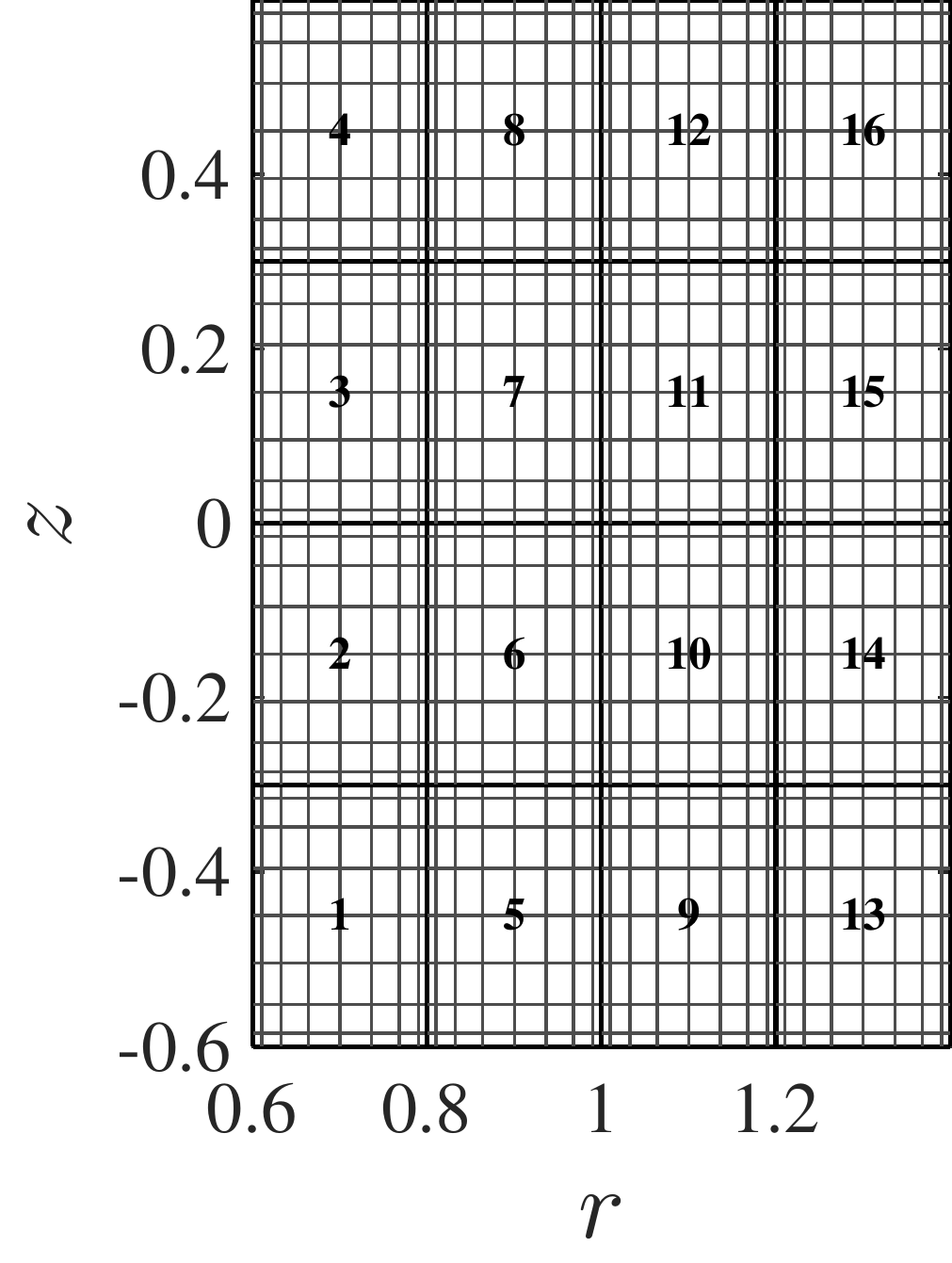}
				\includegraphics[height=0.271\textwidth]{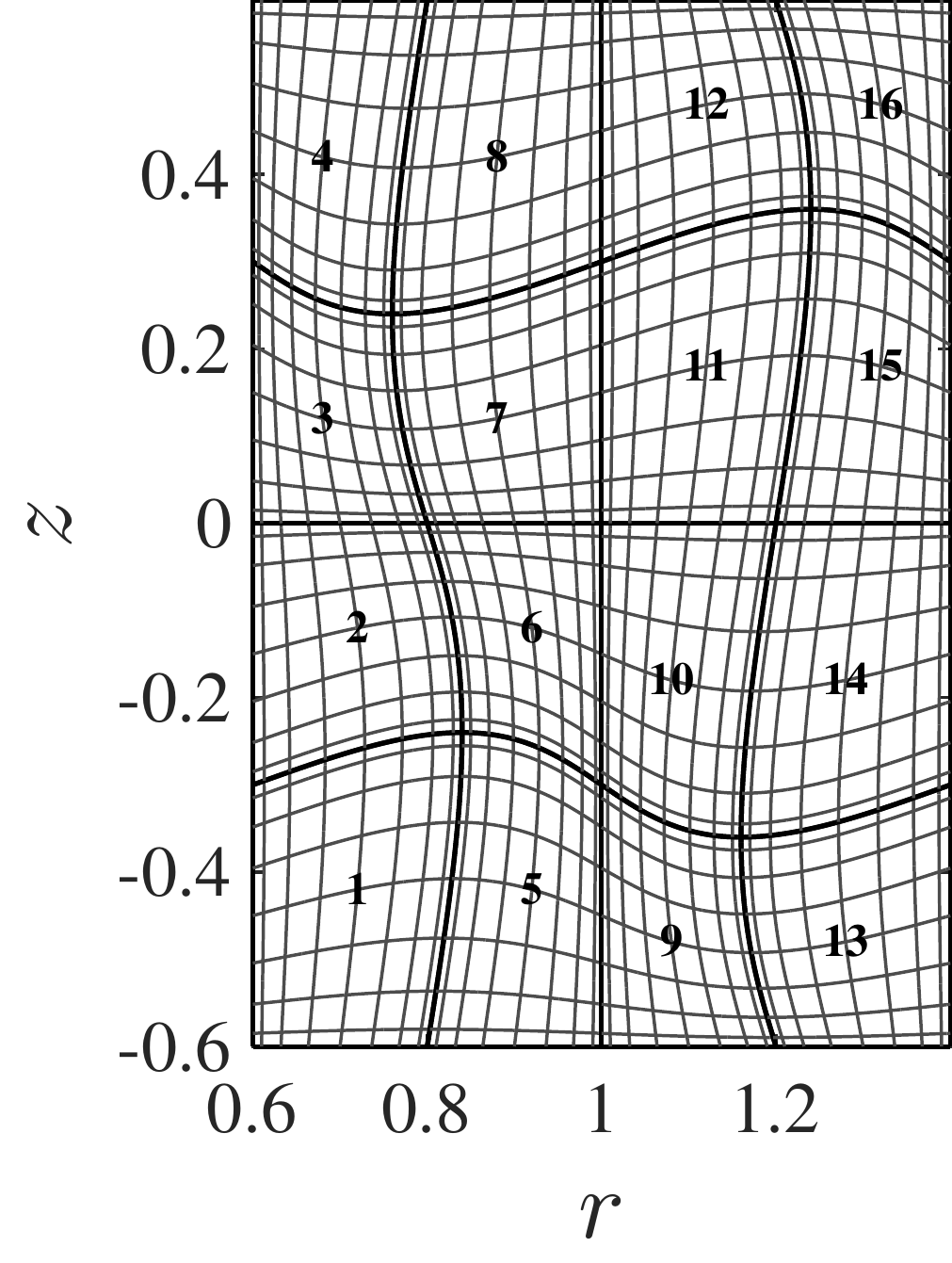}
                                 \includegraphics[height=0.271\textwidth]{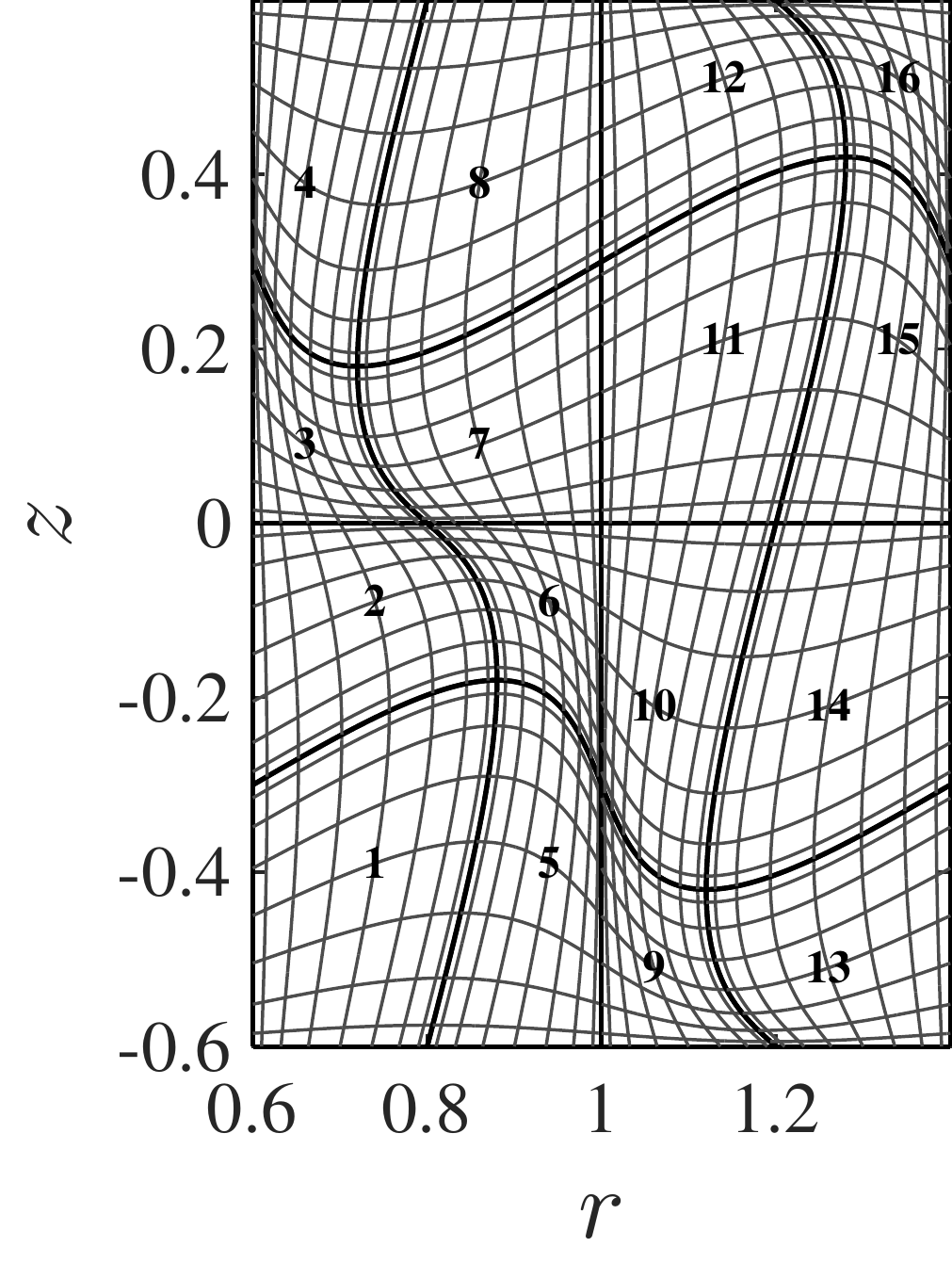}
				\includegraphics[height=0.271\textwidth]{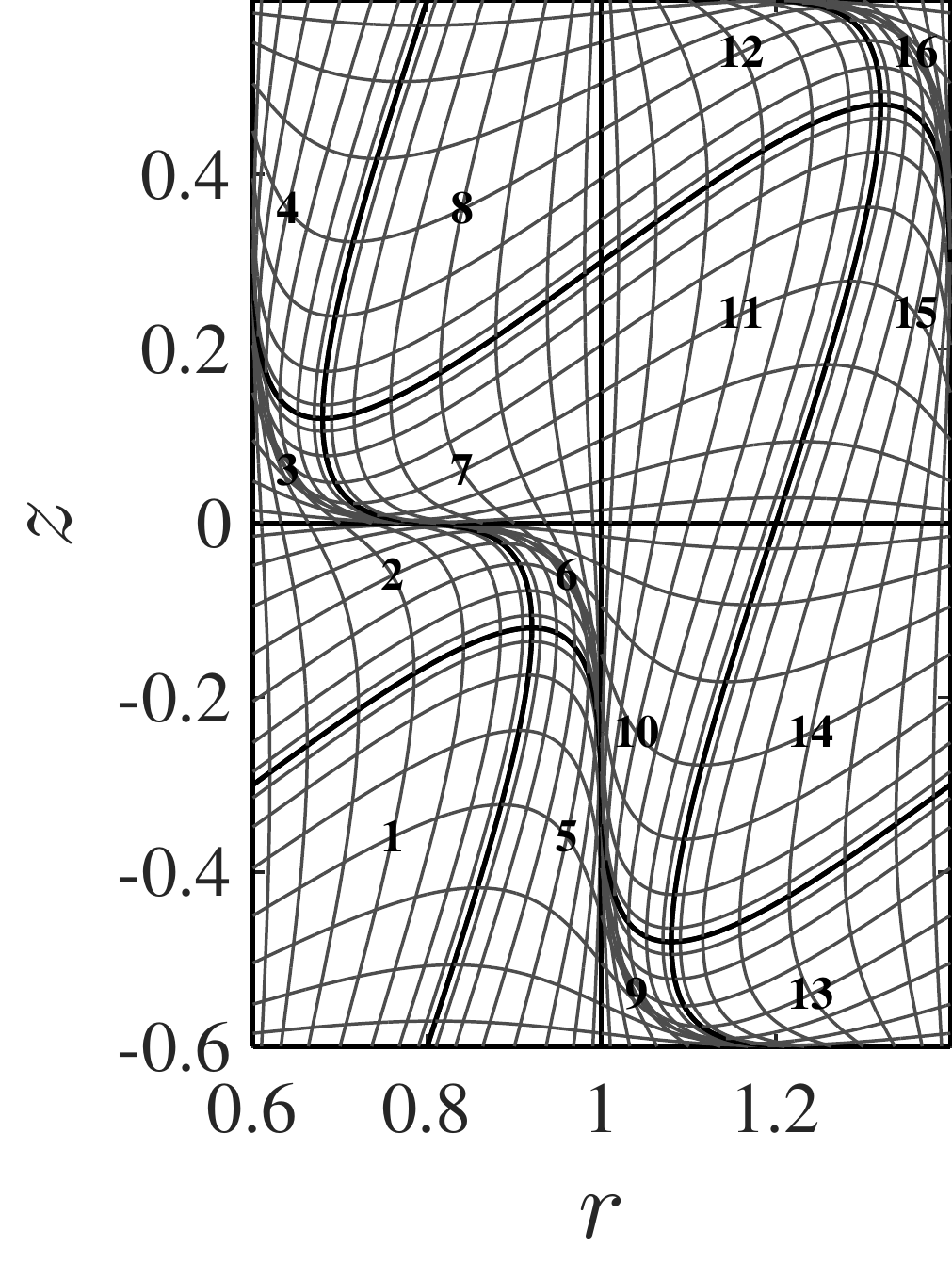}
				\end{center}
				\caption{Example of mesh deformation generated by \eqref{eq::mesh_deformation_mapping} for $c=0.0,0.1,0.2,0.3$ used in the ITER-like case. Mesh corresponding to $4\times4$ elements of polynomial degree $p=8$.}
				\label{fig::soloviev_test_case_iter_mesh_examples}
			\end{figure}

			For this test case we consider meshes obtained by the transformation of the unit square $[-1,1]^{2}$ to curvilinear coordinates on the domain $[r_{-},r_{+}]\times[z_{-},z_{+}]$ by the mapping given as $(r,z) = \Phi(\xi,\eta)$, with
			\begin{equation}
			\begin{dcases}
				r(\xi,\eta) & = \left(\xi + c\sin(\pi\xi)\sin(\pi\eta) + 1\right)\frac{r_{+}-r_{-}}{2}\,, \\
				z(\xi,\eta) & = \left(\eta + c\sin(\pi\xi)\sin(\pi\eta) + 1\right)\frac{z_{+}-z_{-}}{2}\,.
			\end{dcases} \label{eq::mesh_deformation_mapping}
			\end{equation}
			\figref {fig::soloviev_test_case_iter_mesh_examples} shows the meshes generated by \eqref{eq::mesh_deformation_mapping} with increasing values of $c$, for the solution domain used in the ITER-like case.
			
			The mimetic spectral element method is capable of very accurately reproducing the analytical solution for both the ITER and the NSTX cases even on highly deformed meshes, see \figref{fig::soloviev_test_case_iter} and \figref{fig::soloviev_test_case_ntsx} respectively.
			
			\begin{figure}[!ht]
				\begin{center}
				\includegraphics[height=0.271\textwidth]{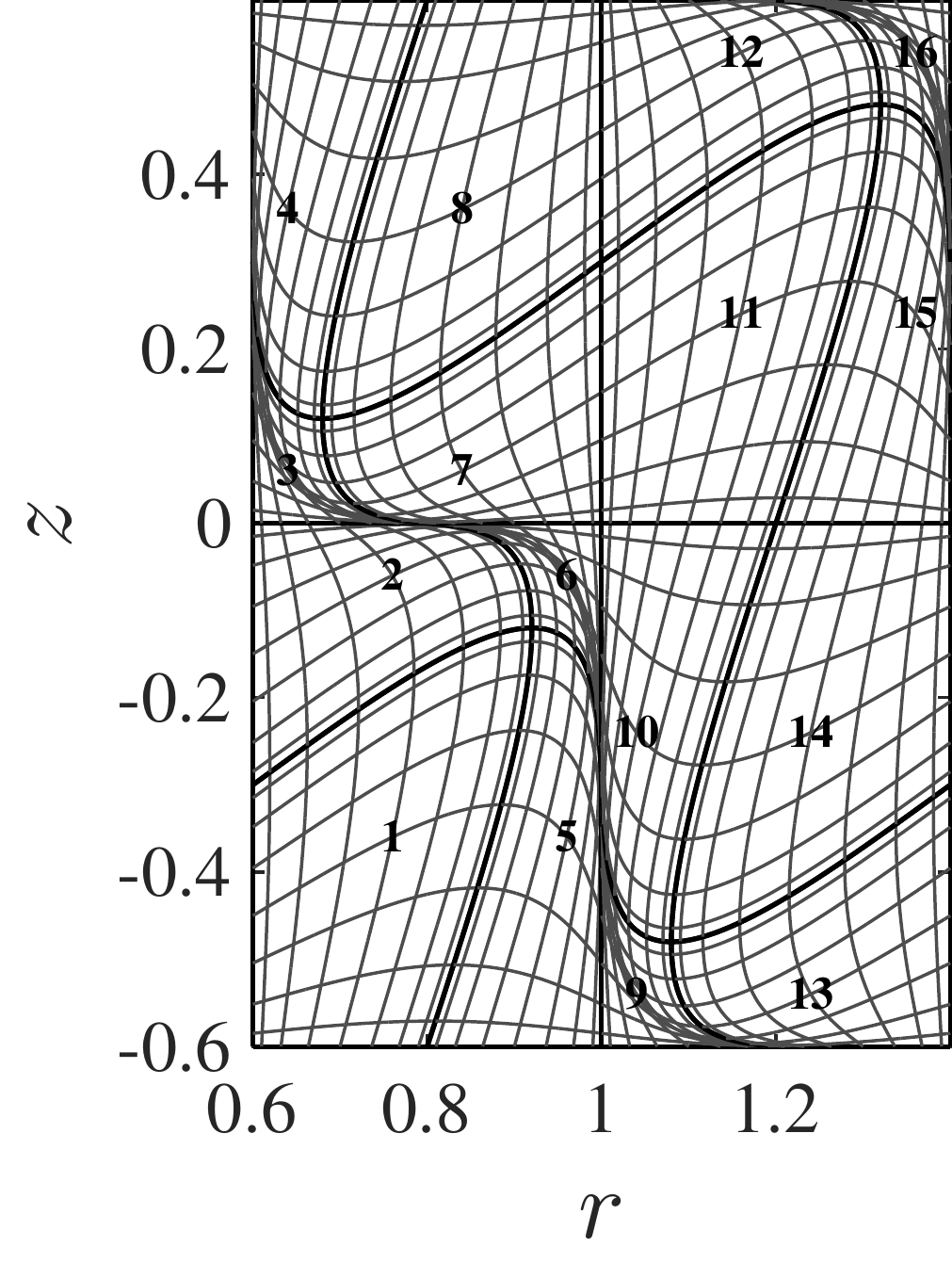}
				\includegraphics[height=0.28\textwidth]{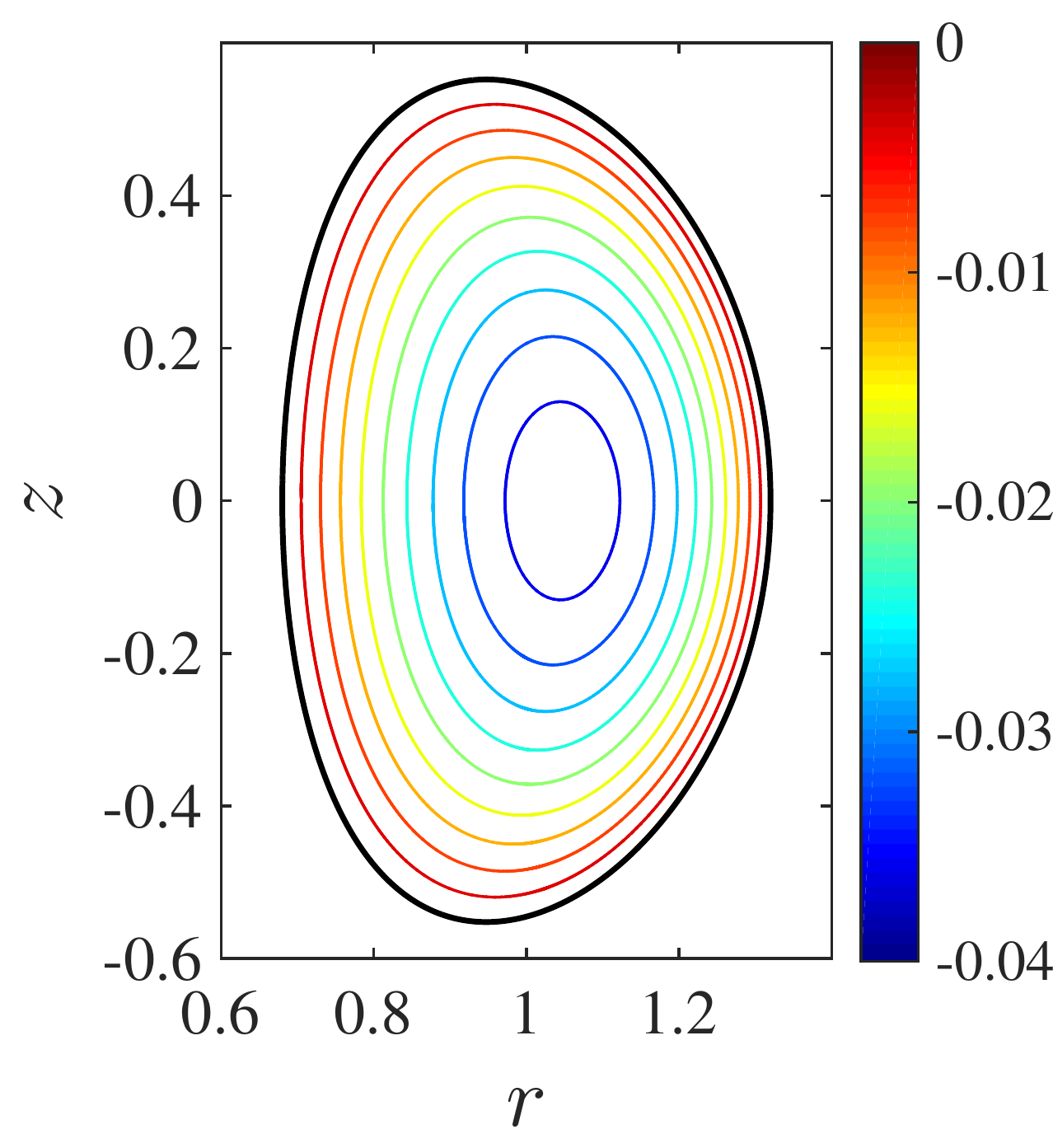}
				\includegraphics[height=0.271\textwidth]{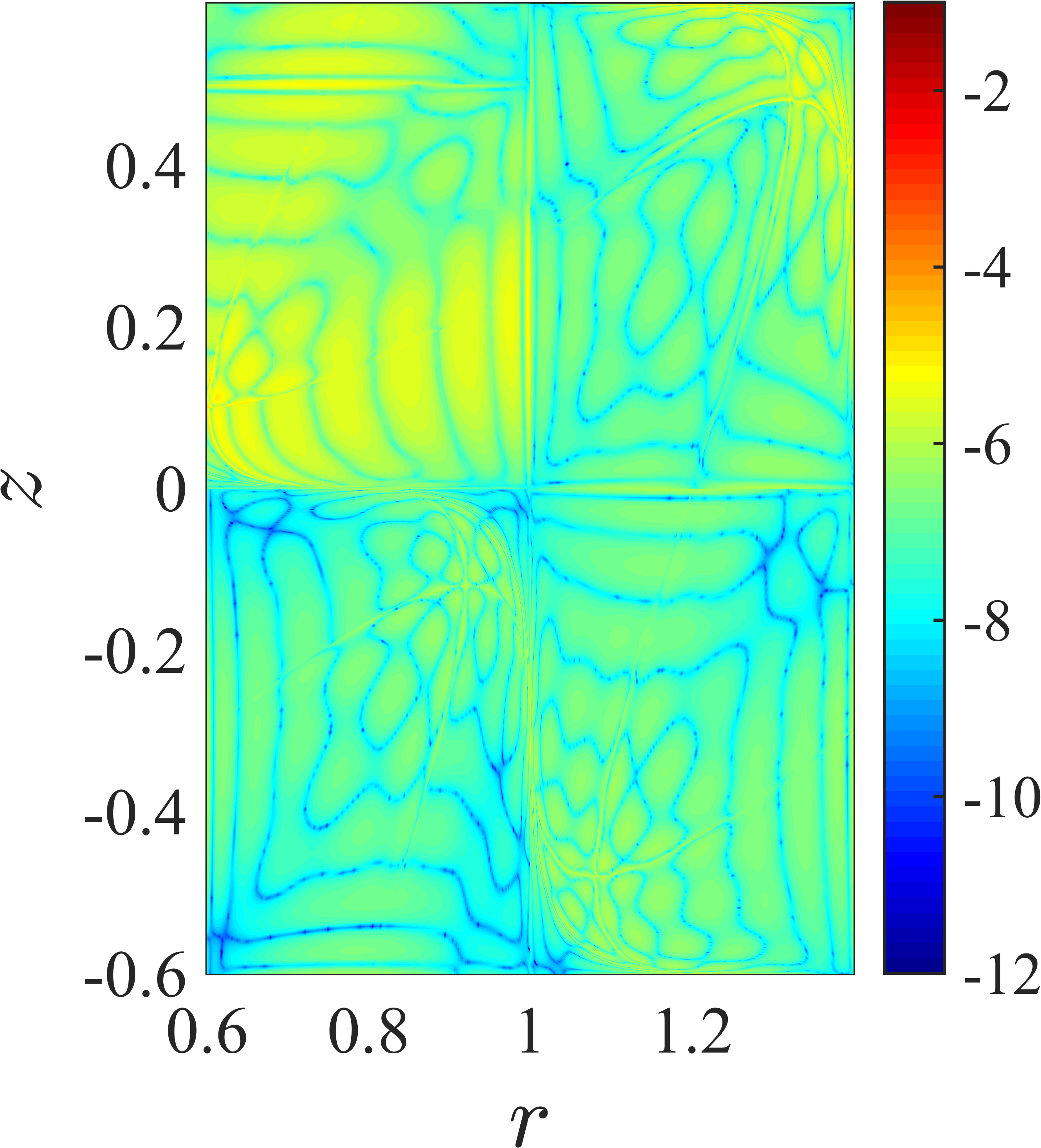}
				\end{center}
				\caption{\reviewerone{Numerical solution of the Soloviev test case, \eqref{eq::soloviev_test_case}, for ITER parameters, $\epsilon = 0.32$, $\kappa = 1.7$, $\delta=0.33$. From left to right: (i) computational mesh, with curvature parameter $c=0.3$, $4\times 4$ elements of polynomial degree $p=8$, %(ii) analytical solution \eqref{eq::soloviev_test_case_analytical_solution}, $\psi_{a}$, 
				(ii) numerical solution using the mesh in (i), $\psi_{h}$, and (iii) logarithmic error between the analytical solution and the numerical one, $\log_{10}|\psi_{a}-\psi_{h}|$.}}
				\label{fig::soloviev_test_case_iter}
			\end{figure}
			
			\begin{figure}[!ht]
				\begin{center}
				\includegraphics[height=0.295\textwidth]{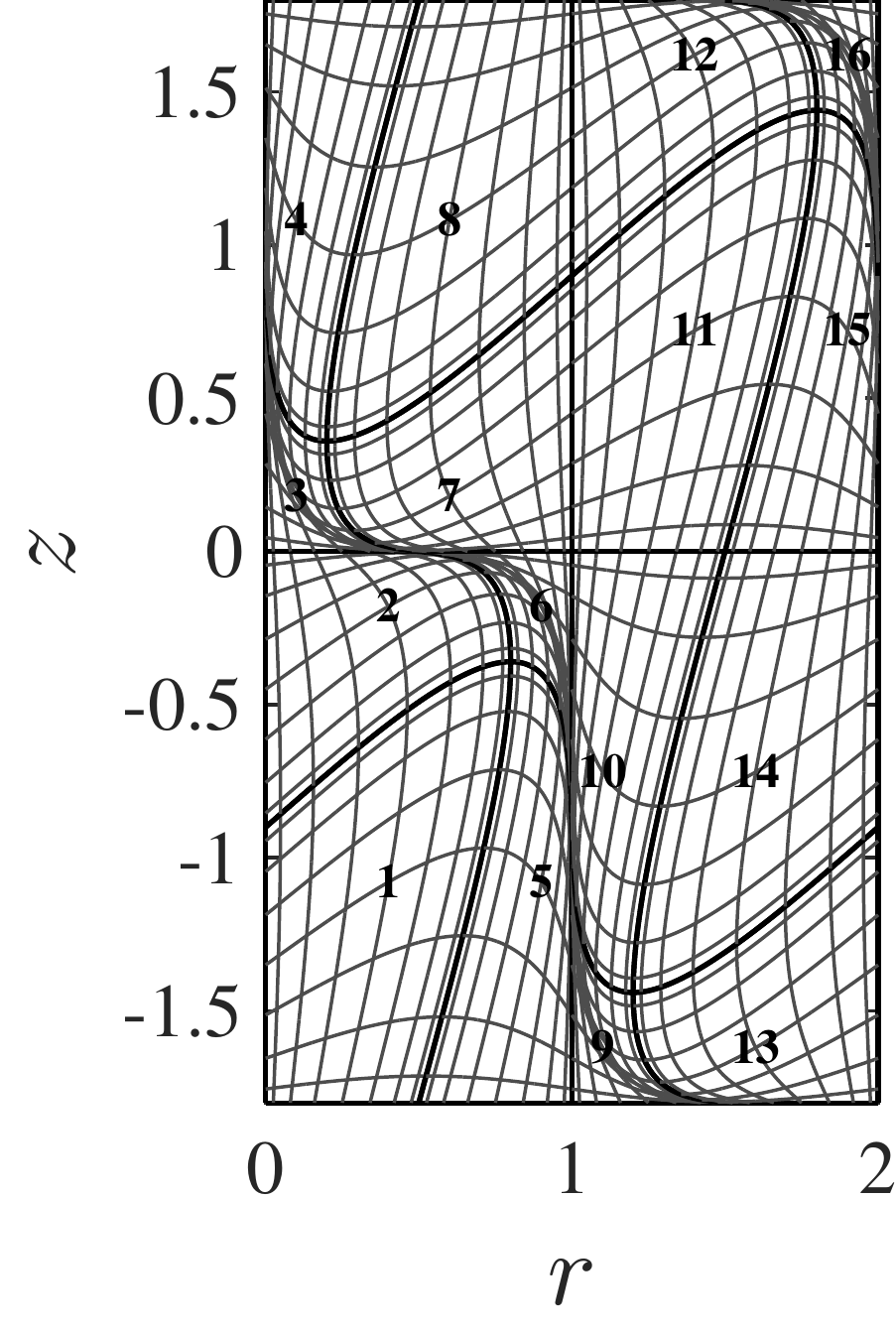}
				\includegraphics[height=0.305\textwidth]{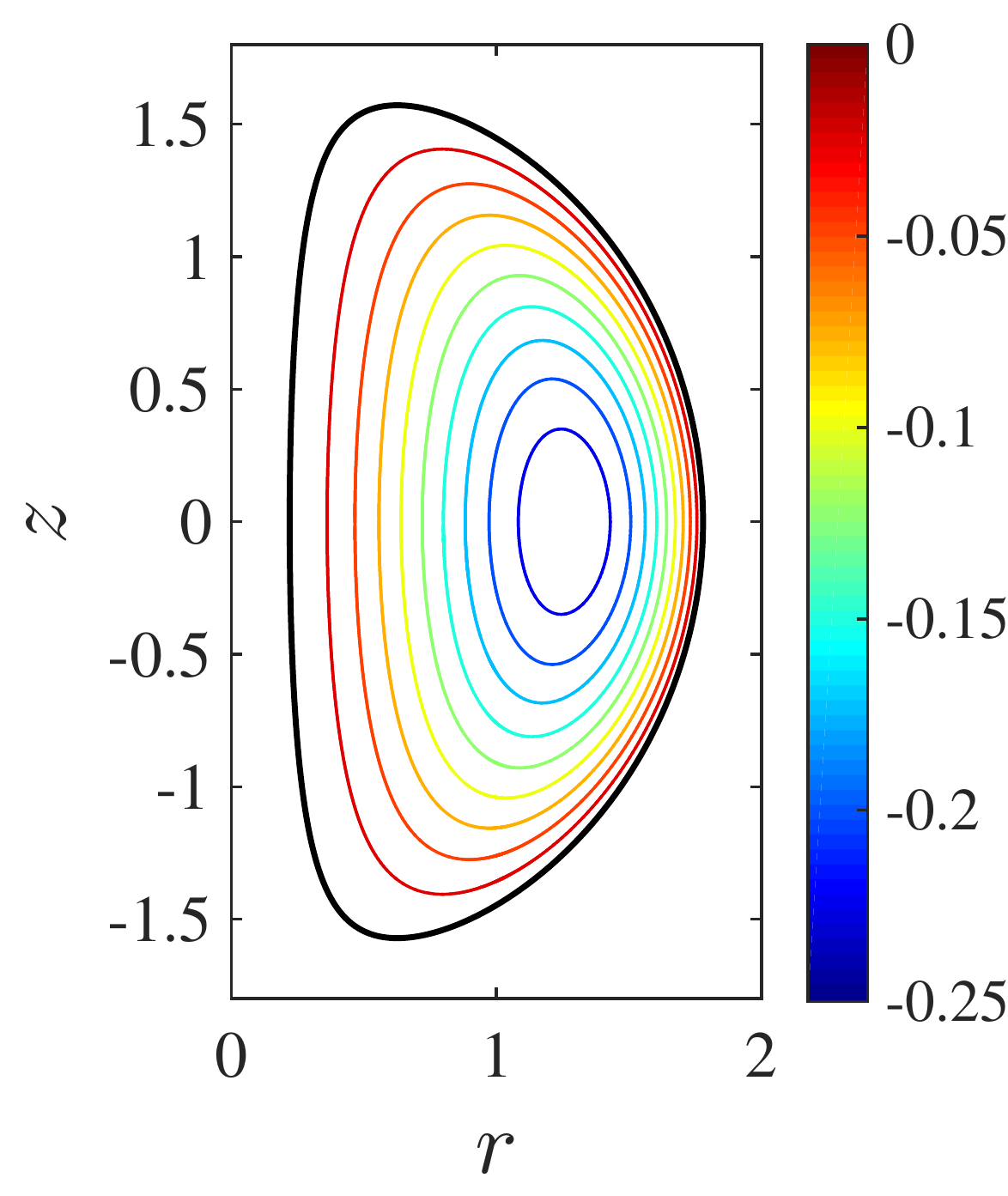}
				\includegraphics[height=0.295\textwidth]{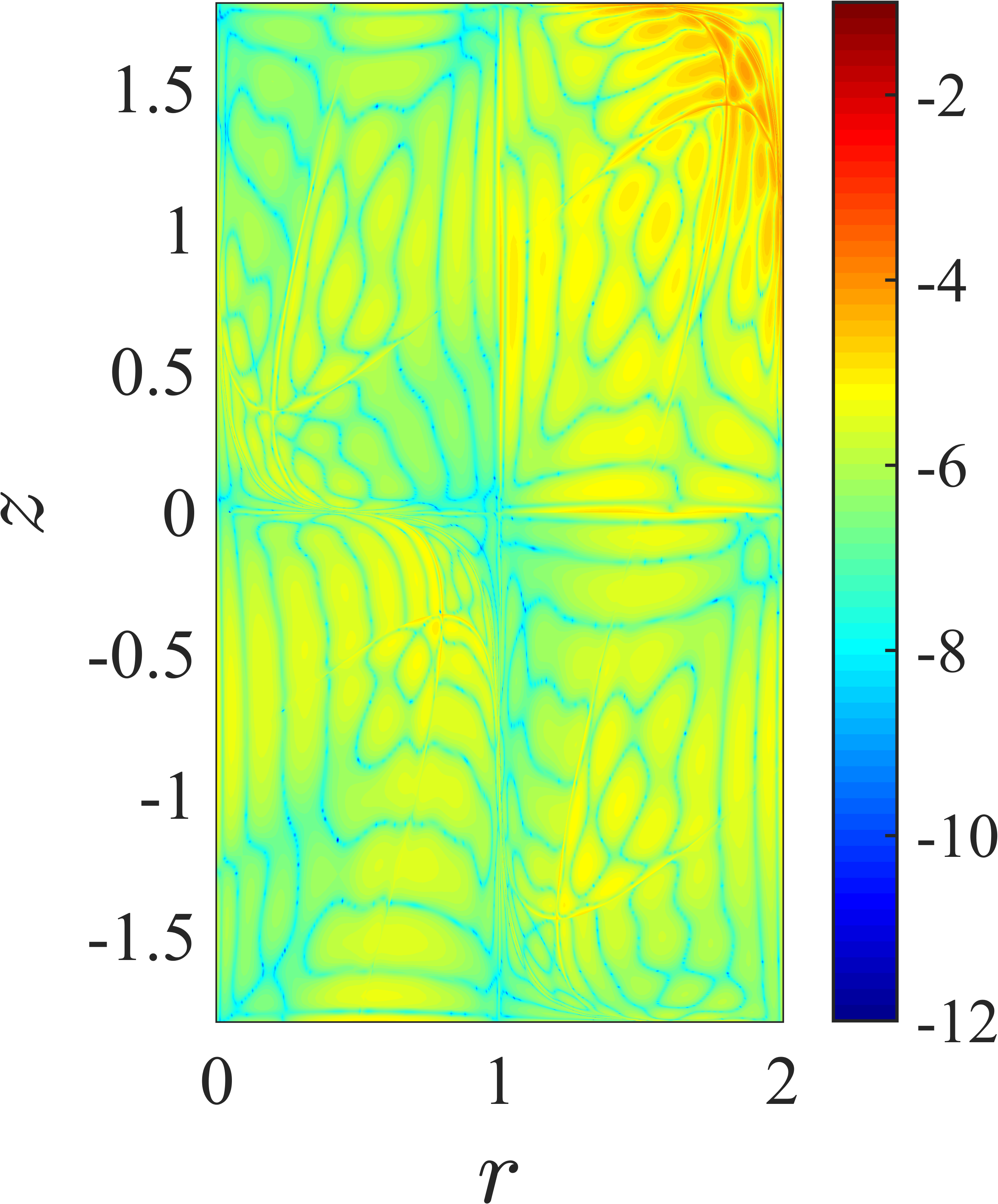}
				\end{center}
				\caption{\reviewerone{Numerical solution of the Soloviev test case, \eqref{eq::soloviev_test_case}, for NSTX parameters, $\epsilon = 0.78$, $\kappa = 2.0$, $\delta=0.35$. From left to right: (i) computational mesh, with curvature parameter $c=0.3$, $4\times 4$ elements of polynomial degree $p=8$, %(ii) analytical solution \eqref{eq::soloviev_test_case_analytical_solution}, $\psi_{a}$,
				(ii) numerical solution using the mesh in (i), $\psi_{h}$, and (iii) logarithmic error between the analytical solution and the numerical one, $\log_{10}|\psi_{a}-\psi_{h}|$.}}
				\label{fig::soloviev_test_case_ntsx}
			\end{figure}
			
\begin{figure}[htb]
				\begin{center}
				\includegraphics[height=0.295\textwidth]{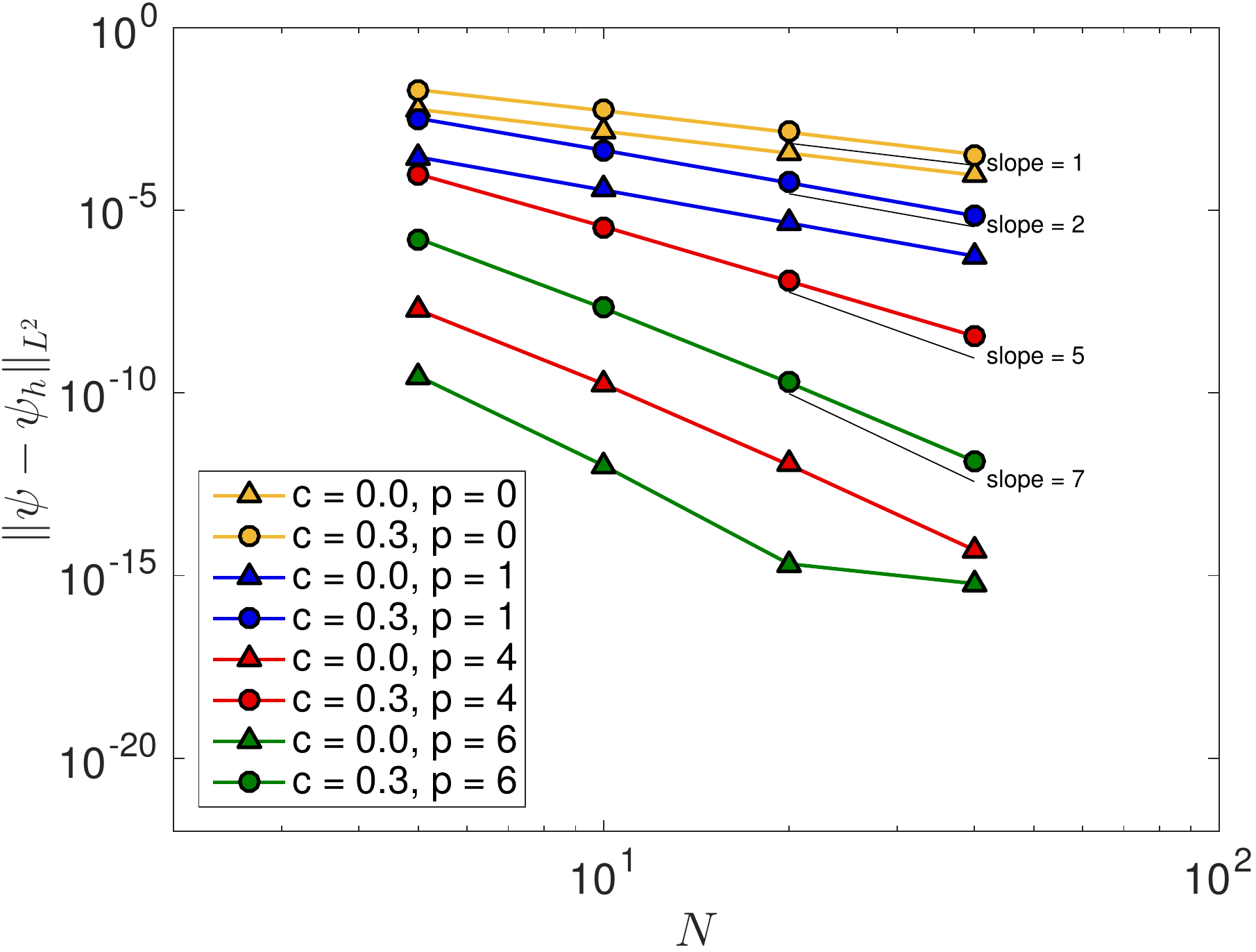} \hspace{1cm}
				\includegraphics[height=0.295\textwidth]{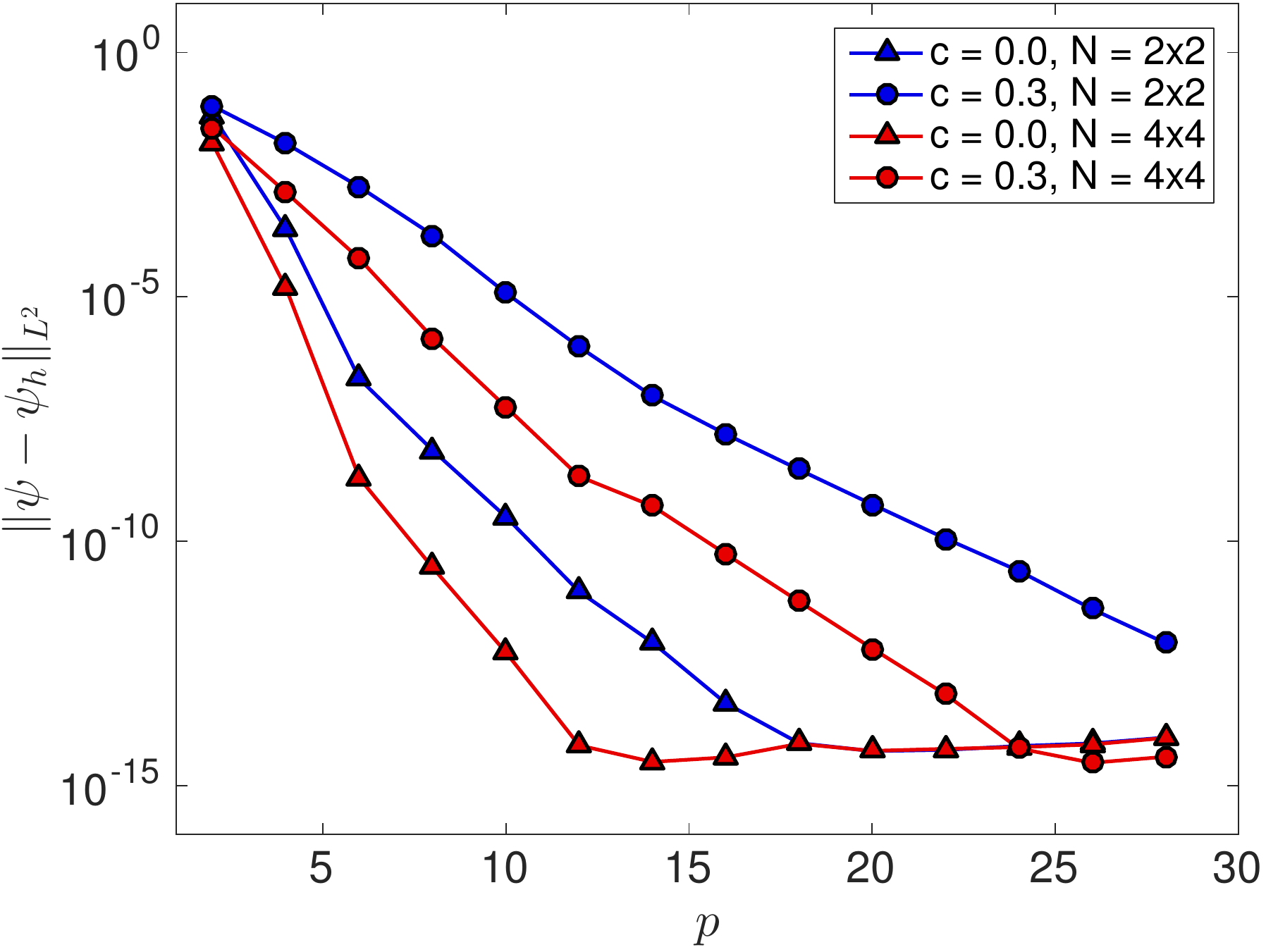}
				\end{center}
				\caption{Convergence plots for the numerical solution of the Soloviev test case \eqref{eq::soloviev_test_case} for ITER parameters, $\epsilon = 0.32$, $\kappa = 1.7$, $\delta=0.33$. Left: $h$-convergence plots. Right: $p$-convergence plots. Convergence plots computed for meshes with deformation, $c=0.0$ and $c=0.3$.}
				\label{fig::soloviev_test_case_convergence}
			\end{figure}

			The convergence tests for $h$-refinement and $p$-refinement show very good convergence rates. For $h$-convergence, \figref{fig::soloviev_test_case_convergence} left, we can see that the mimetic spectral element solver preserves high convergence rates very close to $p+1$ even on highly deformed meshes. Regarding $p$-convergence, \figref{fig::soloviev_test_case_convergence} right, we observe the same robustness of the method, with convergence rates maintaining their exponential character on highly deformed meshes. Furthermore, it is possible to observe convergence to machine accuracy.
			
			\FloatBarrier
			
		\subsubsection{FRC test case} \label{sec::test_cases_frc}
			The second test case corresponds to a field-reversed configuration (FRC). This solution is characterised by zero toroidal field $f(\psi) = 0$ combined with a quadratic model for pressure $P(\psi) = P_{0} + P_{1}\frac{\psi^{2}}{\psi_{0}^{2}}$, where $P_{0}$ and $P_{1}$ are free constants and $\psi_{0}$ is the flux at the magnetic axis, also a free parameter. This corresponds to the following Grad-Shafranov equation
			\begin{equation}
				\begin{dcases}
					\nabla\times\left(\mathbb{K}\nabla\times\psi\right) = 2rP_{1}\frac{\psi}{\psi_{0}^{2}} & \mbox{in} \quad \Omega_{p}\,, \\
					\psi = 0 & \mbox{on}\quad\partial\Omega_{p}\,,
				\end{dcases} \label{eq::frc_test_case}
			\end{equation}
			where $\Omega_{p} = [0,L_{r}]\times[0,L_{z}]$.
			
			An extensive derivation of the analytical solution under these conditions is presented in \cite{Parks2003}. Here we outline this procedure. The Grad-Shafranov equation can be written in terms of dimensionless variables $\bar{r}=\frac{r}{L_{r}}$ and $\bar{z}=\frac{z}{L_{z}}$ and the analytical solution obtained by the method of separation of variables such that
			\begin{equation}
				\psi_{a}(\bar{r},\bar{z}) = W(\bar{r})T(\bar{z})\,.
			\end{equation}
			
			\begin{reviewer3}
			A solution for $T(\bar{z})$ that satisfies the boundary conditions is given by $T(\bar{z}) = \sin(\pi\bar{z})$. Introducing the change of variables $\zeta = \frac{\boldsymbol{\mathrm{i}}L_{r}^{2}}{\psi_{0}}\sqrt{2\mu_{0}P_{1}}\bar{r}^{2}$ the equation for $W(\bar{z})$ becomes
			\begin{equation}
				\frac{\mathrm{d}^{2}W}{\mathrm{d}\zeta^{2}} + \left(\frac{\boldsymbol{\mathrm{i}}k}{\zeta} - \frac{1}{4}\right) W = 0\,, \label{eq::whittaker}
			\end{equation}
			with $k=\frac{\pi^{2}}{4L_{z}^{2}\sqrt{d}}$, $d=\frac{2\mu_{0}P_{1}}{\psi_{0}^{2}}$ and $\boldsymbol{\mathrm{i}} = \sqrt{-1}$. Equation \eqref{eq::whittaker} is Whittaker's equation with $m=\frac{1}{2}$. The solution that is real-valued and zero on the z-axis ($\zeta=0$) is given by $W(\zeta)=-\boldsymbol{\mathrm{i}}CM_{\boldsymbol{\mathrm{i}}k,\frac{1}{2}}(\zeta)$, with $C$ a real number and $M_{ik,\frac{1}{2}}(\zeta)$ Whittaker's $M$ function. By imposing the radial boundary condition, $W(\boldsymbol{\mathrm{i}}\sqrt{d}\bar{r}^{2})=0$, the lowest physically realistic value for $L_{r}$ is determined by finding the first non-zero root of $-\boldsymbol{\mathrm{i}}M_{\boldsymbol{\mathrm{i}}k,\frac{1}{2}}(\zeta)$, $\zeta_{1}$. The location of the magnetic O-point on the midplane $\zeta=\zeta_{0}$ corresponds to the maximum of $-\boldsymbol{\mathrm{i}}M_{\boldsymbol{\mathrm{i}}k,\frac{1}{2}}(\zeta)$ in the interval $[0,\zeta_{1}]$. The analytical solution is then given by
			\begin{equation}
				\psi_{a}(r,z) = \psi_{0}\frac{M_{\boldsymbol{\mathrm{i}}k,\frac{1}{2}}(\boldsymbol{\mathrm{i}}\sqrt{d}r^{2})}{M_{\boldsymbol{\mathrm{i}}k,\frac{1}{2}}(\boldsymbol{\mathrm{i}}\sqrt{d}r_{0}^{2})}\,\sin\left(\frac{\pi}{L_{z}z}\right)\,, \label{eq::frc_test_case_analytical_solution}
			\end{equation}
			with $r_{0} = \zeta_{0}^{\frac{1}{2}}d^{-\frac{1}{4}}$ and $L_{r} = \zeta_{1}^{\frac{1}{2}}d^{-\frac{1}{4}}$.
			\end{reviewer3}
			
			The mimetic spectral element method is applied to the FRC solution with parameters $\psi_{0}=0.1$, $L_{z}=1.0$, $\mu_{0} P_{1} = 0.277$, $r_{0} = 0.7670524200738164$ and $L_{r} = 1.0367471722606991$, as in \cite{Howell2014}. As in the Soloviev test case, \secref{sec::test_cases_soloviev}, we use the same mapping to generate the mesh deformation, \eqref{eq::mesh_deformation_mapping}, on the domain $[0,L_{r}]\times[0,L_{z}]$. 
			
\begin{figure}[!ht]
				\begin{center}
				\includegraphics[height=0.21\textwidth]{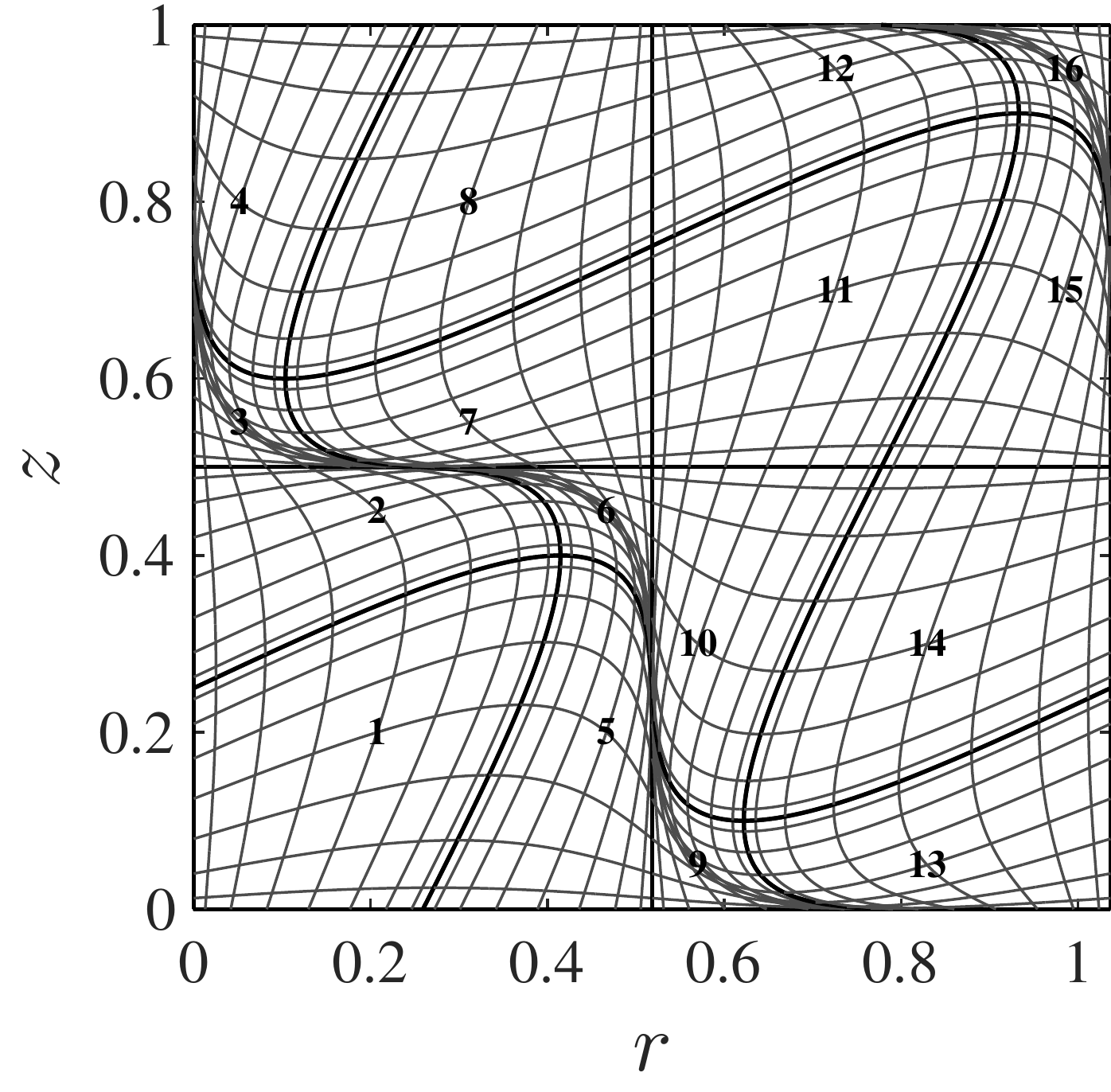}
				\includegraphics[height=0.21\textwidth]{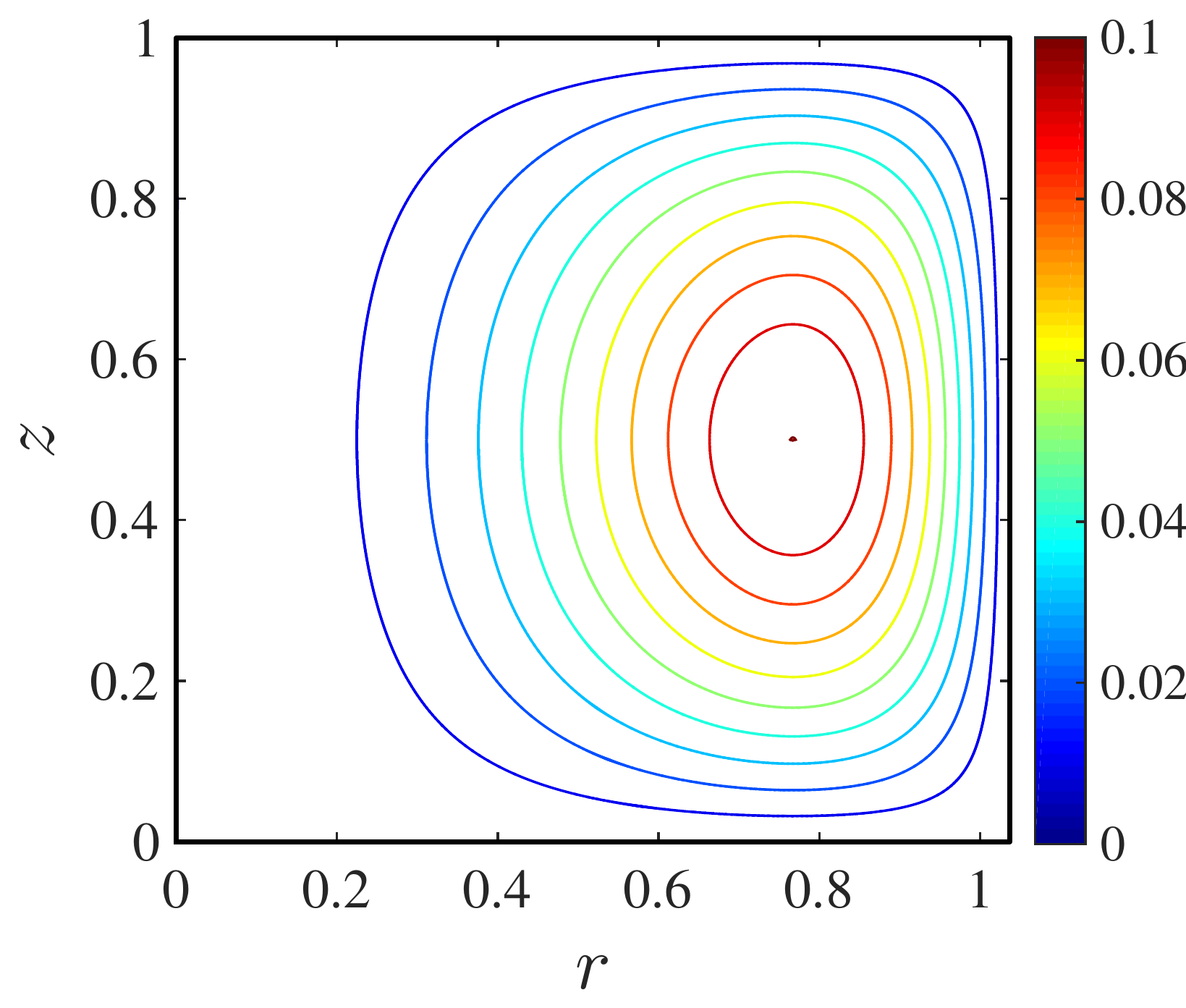}
				\includegraphics[height=0.21\textwidth]{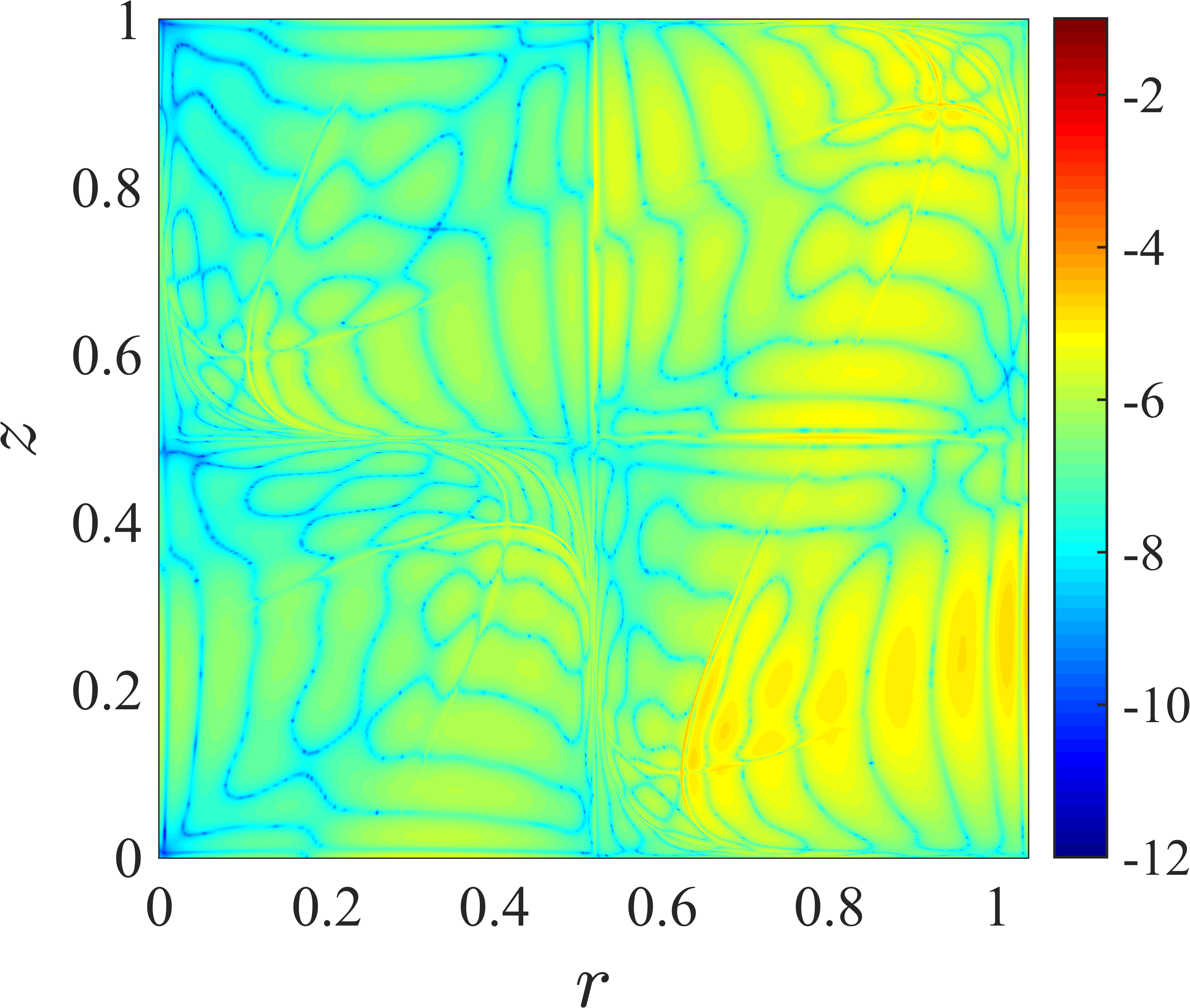}
				\end{center}
				\caption{\reviewerone{Numerical solution of the FRC test case, \eqref{eq::frc_test_case}, with $\psi_{0}=0.1$, $L_{z}=1.0$, $\mu_{0} P_{1} = 0.277$, $r_{0} = 0.7670524200738164$ and $L_{r} = 1.0367471722606991$. From left to right: (i) computational mesh, with curvature parameter $c=0.3$, $4\times 4$ elements of polynomial degree $p=8$, %(ii) analytical solution \eqref{eq::frc_test_case_analytical_solution}, $\psi_{a}$, 
				(ii) numerical solution using the mesh in (i), $\psi_{h}$, and (iii) logarithmic error between the analytical solution and the numerical one, $\log_{10}|\psi_{a}-\psi_{h}|$.}}
				\label{fig::frc_test_case}
			\end{figure}

			We see that the proposed method  accurately computes the FRC solution, showing small errors even on a highly deformed mesh, see \figref{fig::frc_test_case}.
			
\begin{figure}[!ht]
				\begin{center}
				\includegraphics[height=0.295\textwidth]{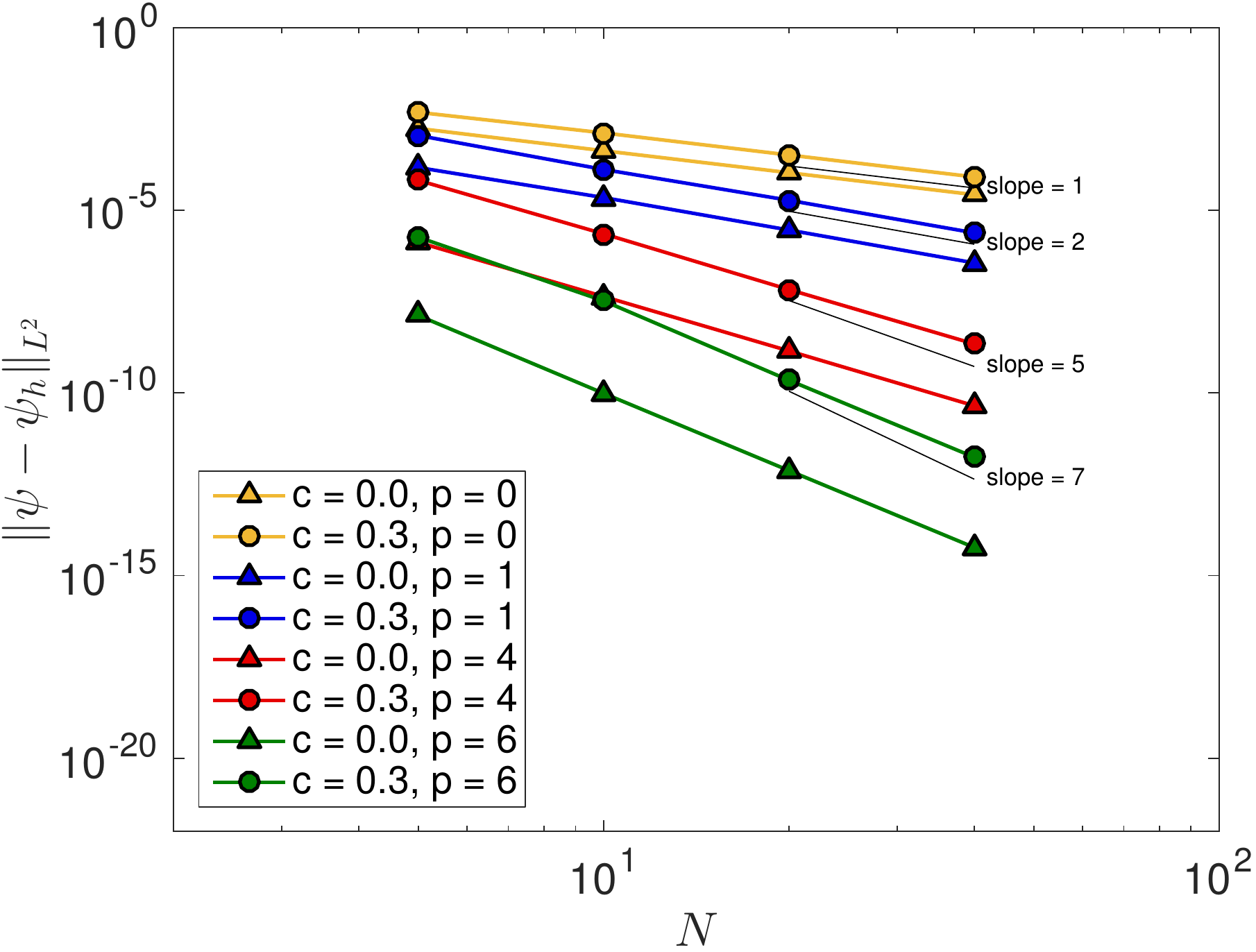} \hspace{1cm}
				\includegraphics[height=0.295\textwidth]{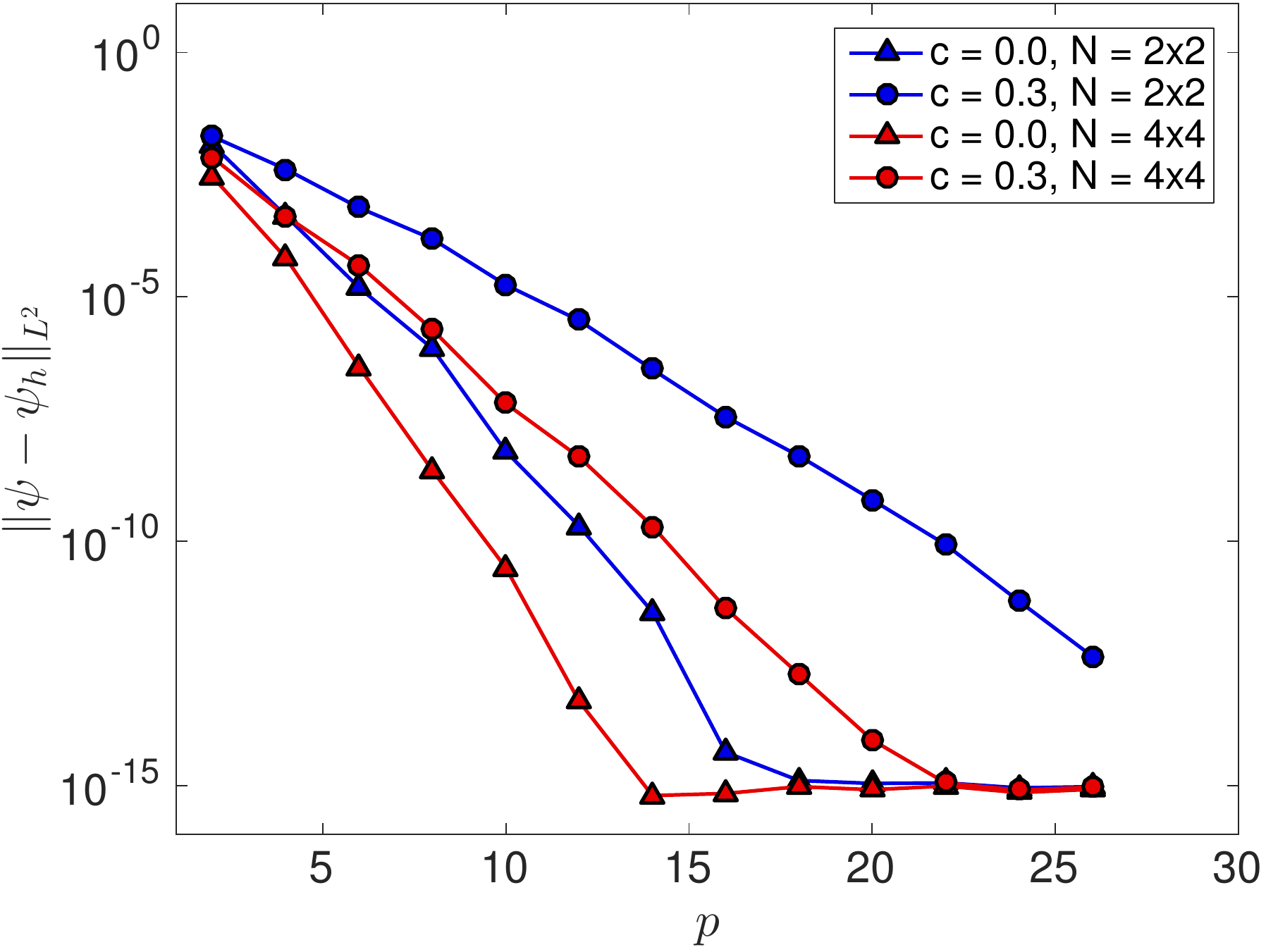}
				\end{center}
				\caption{Convergence plots for the numerical solution of the FRC test case \eqref{eq::frc_test_case}, with $\psi_{0}=0.1$, $L_{z}=1.0$, $\mu_{0} P_{1} = 0.277$, $r_{0} = 0.7670524200738164$ and $L_{r} = 1.0367471722606991$. Left: $h$-convergence plots. Right: $p$-convergence plots. Convergence plots computed for meshes with deformation, $c=0.0$ and $c=0.3$.}
				\label{fig::frc_test_case_convergence}
			\end{figure}

			The convergence tests for the FRC test case confirm the results obtained for the Soloviev solution. We can observe high $h$-convergence rates very close to order $p+1$, \figref{fig::frc_test_case_convergence} left. Similar behaviour can be seen for $p$-convergence with the method converging exponentially fast for both straight and curved meshes, \figref{fig::frc_test_case_convergence} right. Also for this test case machine accuracy is achieved.
			
			\begin{reviewer1}
			As pointed out in \cite{Parks2003}, FRC equilibria are often very elongated. In order to study the dependence of the error in the numerical solution with respect to the plasma elongation $\frac{L_{z}}{L_{r}}$, we have computed the error as a function of the elongation for a mesh with $20\times 20$ elements of polynomial degree $p=1$  (\figref{fig::frc_elongation_dependence} left) and for a mesh with $4\times 4$ elements of polynomial degree $p=5$ (\figref{fig::frc_elongation_dependence} right). As can be seen, the error increases slowly with the elongation (blue line). Since the area of the plasma increases with the elongation, a better comparison is the error normalized by the area of the plasma (red line). The normalized error shows a very small variation with the elongation of the plasma. As an example, we present in \figref{fig::frc_test_case_elongation_10} the numerical solution for a plasma with elongation 10.87.
			
			\begin{figure}[!ht]
				\begin{center}
				\includegraphics[height=0.295\textwidth]{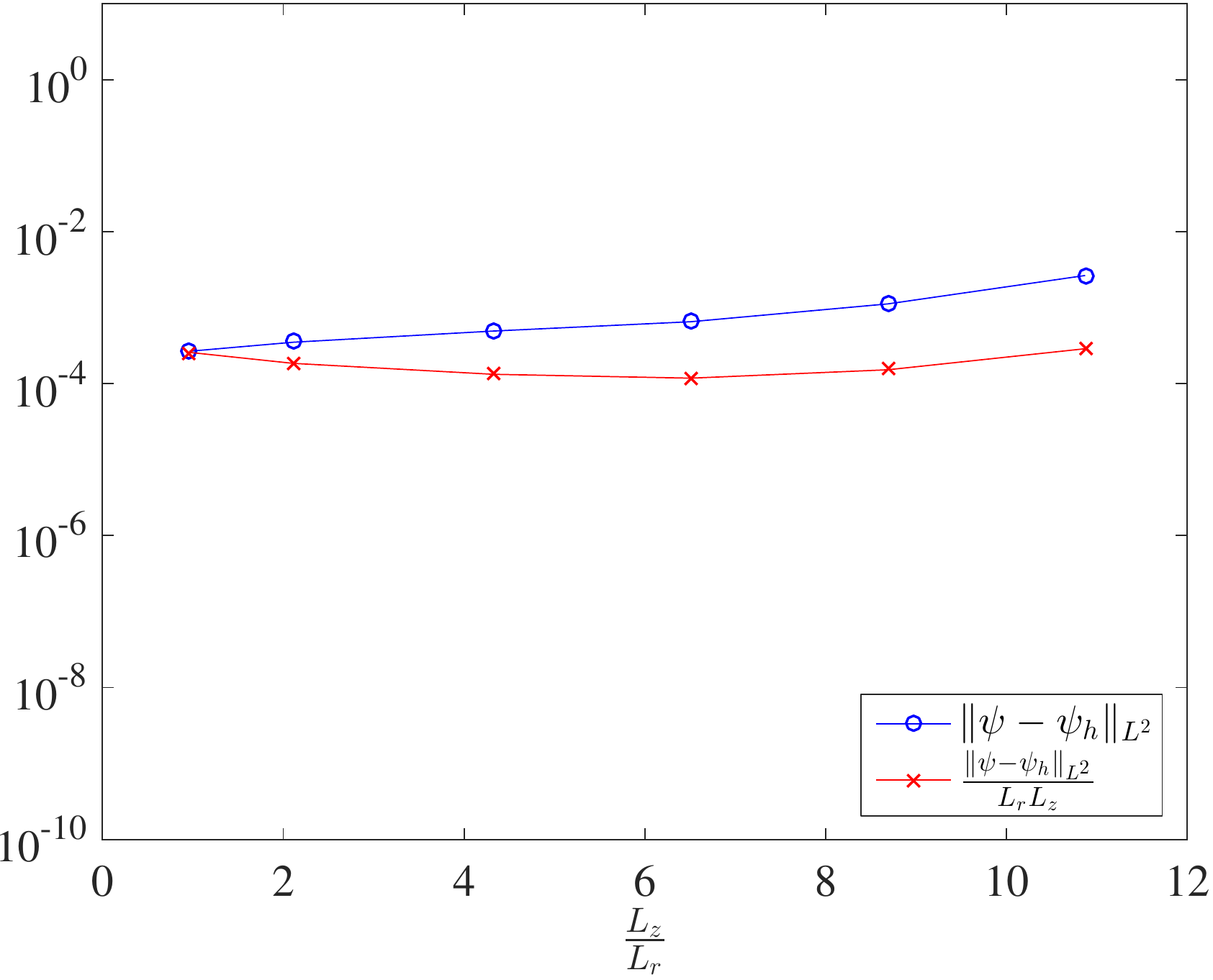} \hspace{1cm}
				\includegraphics[height=0.295\textwidth]{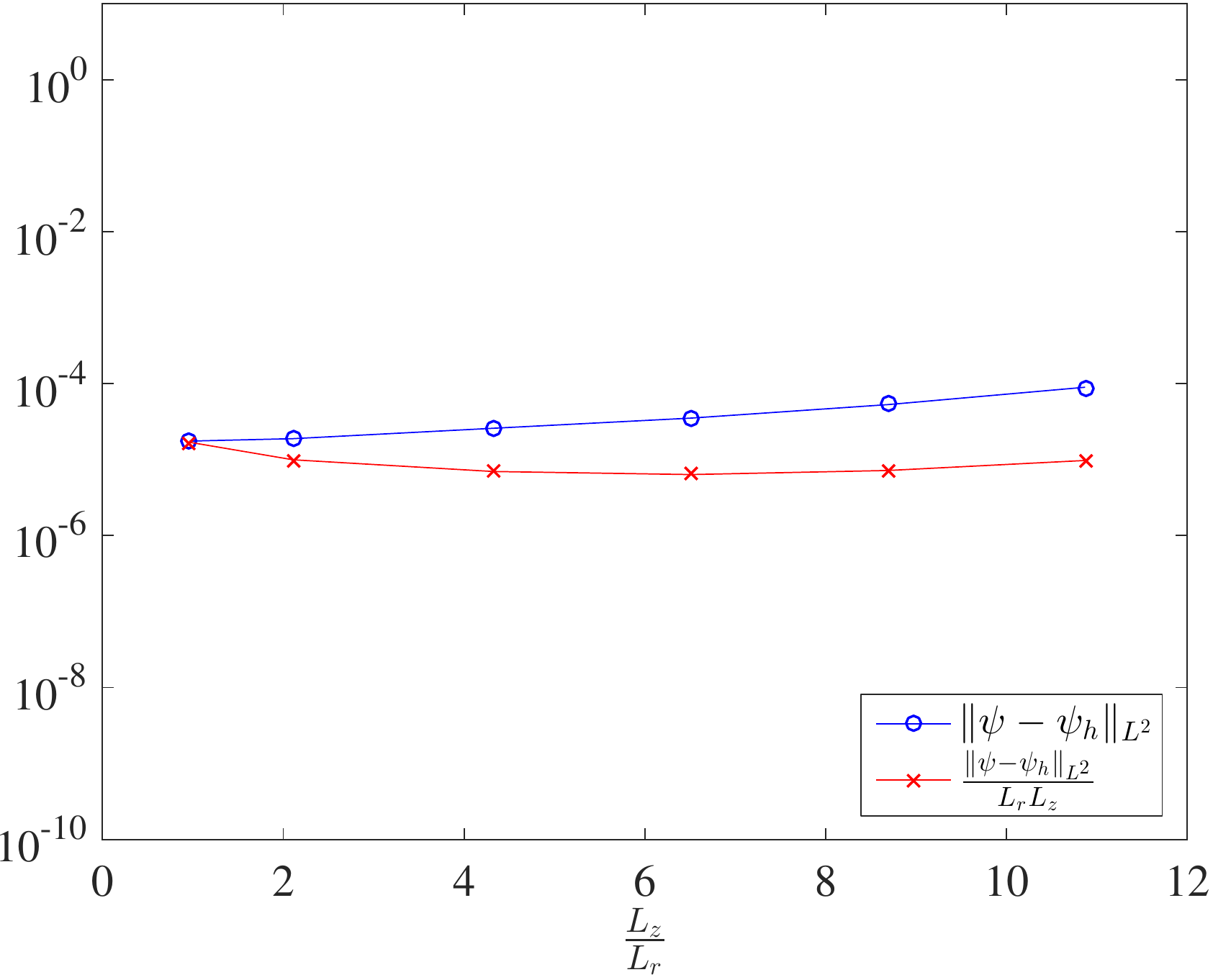}
				\end{center}
				\caption{\reviewerone{Error for the numerical solution of the FRC test case \eqref{eq::frc_test_case}, with $\psi_{0}=0.1$, $L_{z}=[1.0,2.0,4.0,6.0,8.0,10.0]$ and $\mu_{0} P_{1} = 0.277$. Left: mesh with $20\times 20$ elements of polynomial degree $p=1$. Right: mesh with $4\times 4$ elements of polynomial degree $p=5$.}}
				\label{fig::frc_elongation_dependence}
			\end{figure}
			
			\begin{figure}[!ht]
				\begin{center}
				\includegraphics[width=0.5\textwidth]{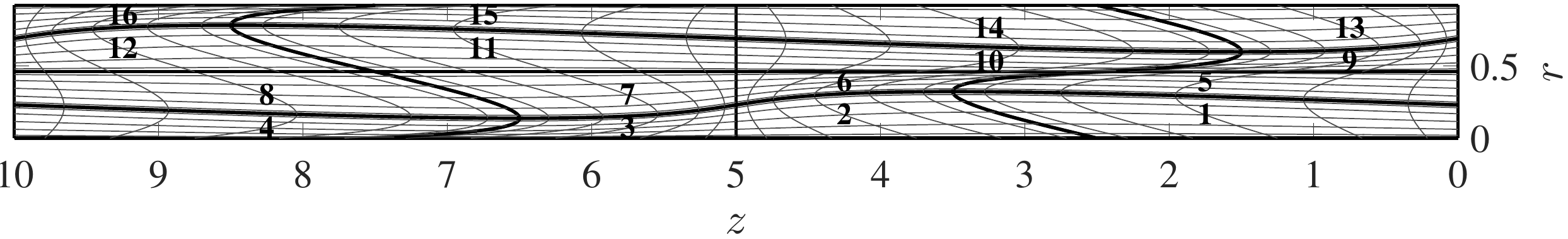}\vspace{0.5cm}
				\includegraphics[width=0.50\textwidth]{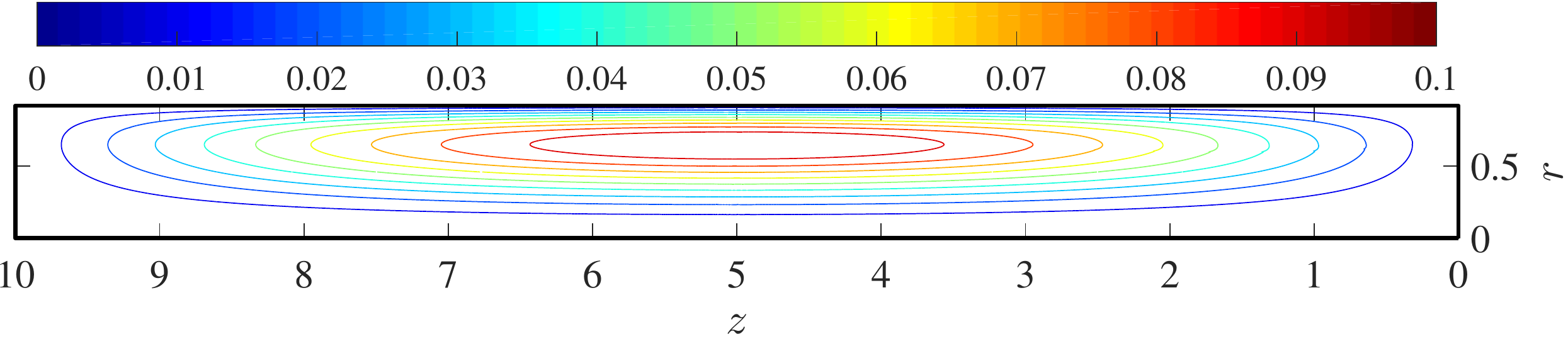}\vspace{0.5cm}
				\includegraphics[width=0.50\textwidth]{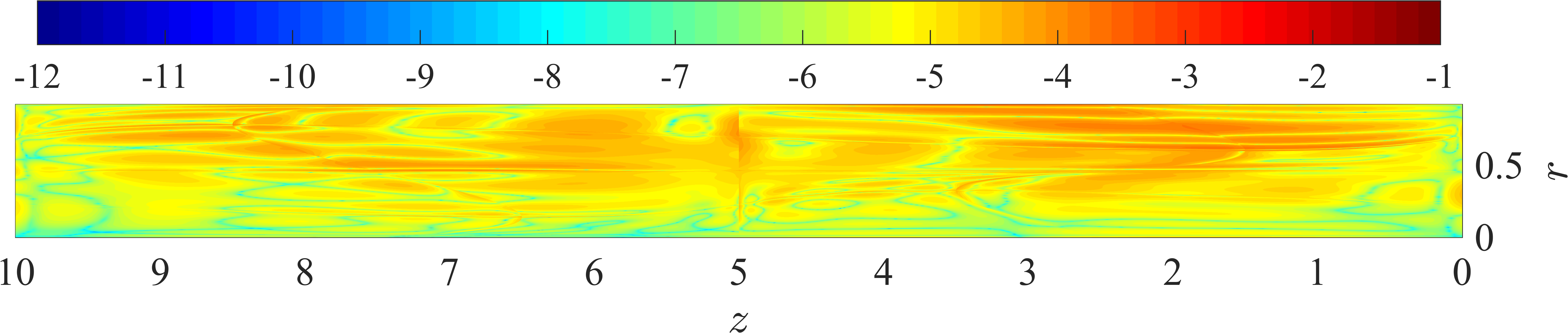}
				\end{center}
				\caption{\reviewerone{Numerical solution of the FRC test case, \eqref{eq::frc_test_case}, with $\psi_{0}=0.1$, $L_{z}=10.0$, $\mu_{0} P_{1} = 0.277$, $r_{0} = 0.6508076929394377$ and $L_{r} = 0.9199638879139966$. From left to right: (i) computational mesh, with curvature parameter $c=0.2$, $4\times 4$ elements of polynomial degree $p=5$, %(ii) analytical solution \eqref{eq::frc_test_case_analytical_solution}, $\psi_{a}$, 
				(ii) numerical solution using the mesh in (i), $\psi_{h}$, and (iii) logarithmic error between the analytical solution and the numerical one, $\log_{10}|\psi_{a}-\psi_{h}|$.}}
				\label{fig::frc_test_case_elongation_10}
			\end{figure}
			
			\end{reviewer1}

			\FloatBarrier
			
		\subsubsection{Spheromak test case} \label{sec::test_cases_spheromak}
			The third test case used to determine the $h$- and $p$-convergence properties of the proposed method is the cylindrical spheromak configuration. This case is obtained by a linear toroidal field $f(\psi) = f_{0}\psi$ together with a linear pressure profile $P(\psi) = P_{0}\frac{\psi}{\psi_{0}}$, where $f_{0}$ and $P_{0}$ are constants and $\psi_{0}$ is the flux at the magnetic axis, also a free parameter. With these models the Grad-Shafranov problem becomes
			\begin{equation}
				\begin{dcases}
					\nabla\times\left(\mathbb{K}\nabla\times\psi\right) = \frac{P_{0}r}{\psi_{0}} -\frac{f_{0}^{2}\psi}{\mu_{0}r} & \mbox{in} \quad \Omega_{p}\,, \\
					\psi = 0 & \mbox{on}\quad\partial\Omega_{p}\,,
				\end{dcases} \label{eq::spheromak_test_case}
			\end{equation}
			where $\Omega_{p} = [0,L_{r}]\times[0,L_{z}]$.
			
			Here we simply outline the derivation of the analytical solution as presented in \cite{Bellan2002}. The methodology to obtain the analytical solution is very similar to the one used in the FRC case, \secref{sec::test_cases_frc}. The first step is to multiply \eqref{eq::spheromak_test_case} by $r$ and rewrite \eqref{eq::spheromak_test_case} in terms of the normalized quantities $\normalized{\psi} = \frac{\psi}{\psi_{0}}$, $\normalized{r}=\frac{r}{r_{0}}$, $\normalized{z}=\frac{z}{r_{0}}$, $\normalized{f_{0}} = f_{0} r_{0}$, $\normalized{k} = k r_{0}=r_{0}\frac{\pi}{L_{z}}$ and with respect to $\chi = \normalized{\psi} + \beta\,\normalized{r}^{2}\normalized{f_{0}}^{-2}$, where $\beta = \pi^{2}\mu_{0}P_{0}r_{0}^{4}\psi_{0}^{-2}$ and $r_{0}$ is the radial coordinate of the plasma's axis:
			\begin{equation}
				\normalized{r}\frac{\partial}{\partial \normalized{r}}\left(\frac{1}{\normalized{r}}\frac{\partial\chi}{\partial\normalized{r}}\right) + \frac{\partial^{2}\chi}{\partial\normalized{z}} +\normalized{f_{0}}\,\chi = 0\,.
			\end{equation}
			The solution can now be obtained by the method of separation of variables such that
			\begin{equation}
				\chi(\normalized{r},\normalized{z}) = \normalized{r}g(\normalized{r})\sin(\normalized{k}\normalized{z})\,.
			\end{equation}
			Introducing the change of variables $x^{2} = (\normalized{f_{0}}^{2} - \normalized{k}^{2})\normalized{r}^{2}$ the equation for $g(x)$ becomes:
			\begin{equation}
				x^{2}\frac{\mathrm{d}^{2}g}{\mathrm{d}x^{2}} + x\frac{\mathrm{d}g}{\mathrm{d}x} + (x^{2}-1) g=0\,,
			\end{equation}
			which is Bessel's equation with $n=1$. The physically relevant solution is then
			\begin{equation}
				\psi_{a}(r,z) = \psi_{0} \frac{\chi_{11}}{\chi_{01}L_{r}} \frac{B_{1}(\frac{\chi_{11}r}{L_{r}})}{B_{1}(\chi_{01})} r\sin\left(\frac{\pi r}{L_{z}}\right) - \frac{\pi^{2}\mu_{0}P_{0}r}{f_{0}\psi_{0}^{2}}\,,\label{eq::spheromak_test_case_analytical_solution}
			\end{equation}
			with $B_{n}$ the $n$-th Bessel function of the first kind, $\chi_{ij}$ the $j$-th positive zero of Bessel function $B_{i}$. The constant $f_{0}$ is then given by $f_{0}=\sqrt{\chi_{11}^{2}L_{r}^{-2} + k^{2}}$.
			
			For this test case we apply the numerical method developed in this article to the spheromak solution with parameters $\psi_{0} = 0.1$, $L_{r} = 1.0$, $L_{z} = 1.0$, $P_{0} = 0.0$ and $f_{0} = 4.954954595474438$, as in \cite{Howell2014}.
			
\begin{figure}[!ht]
				\begin{center}
				\includegraphics[height=0.21\textwidth]{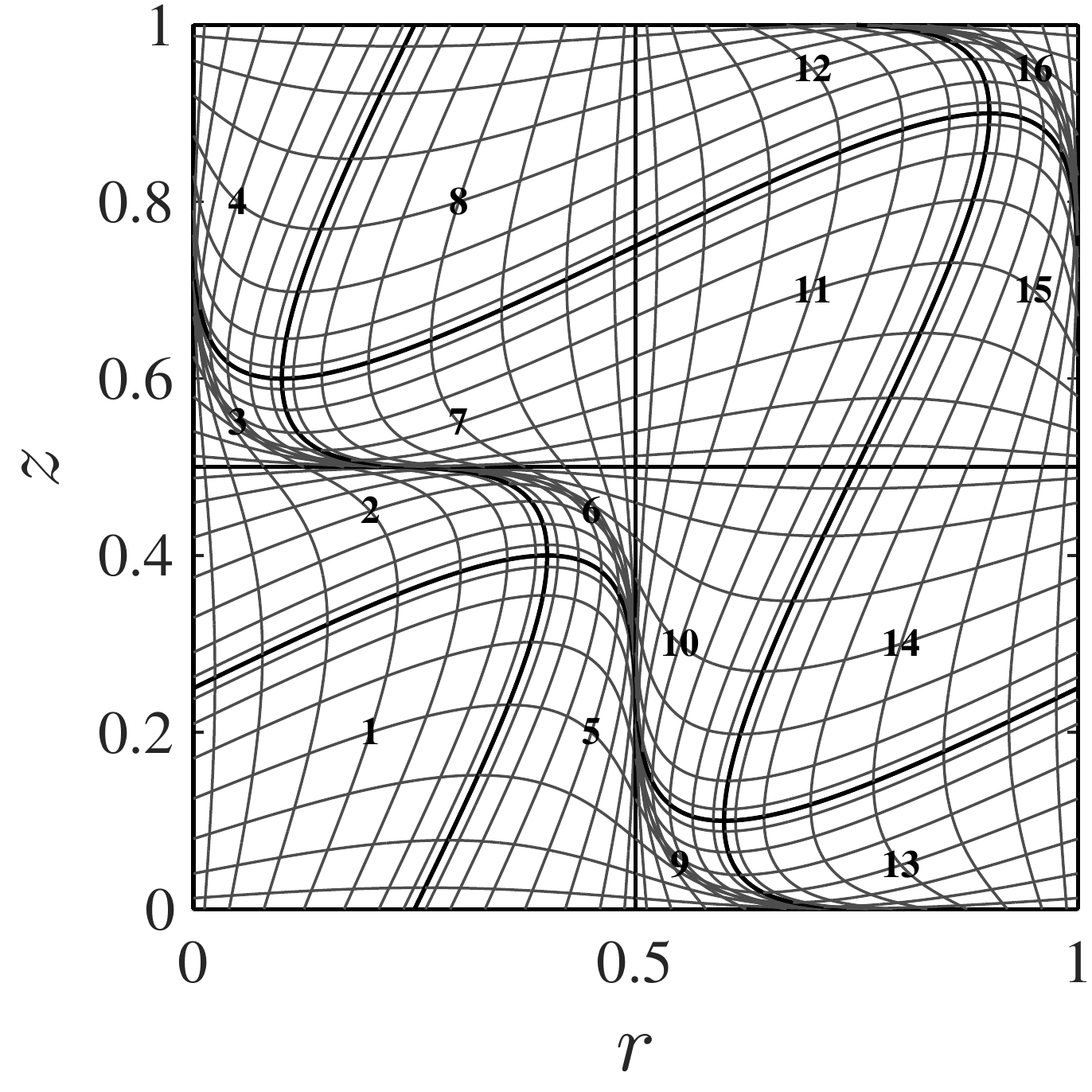}
				\includegraphics[height=0.21\textwidth]{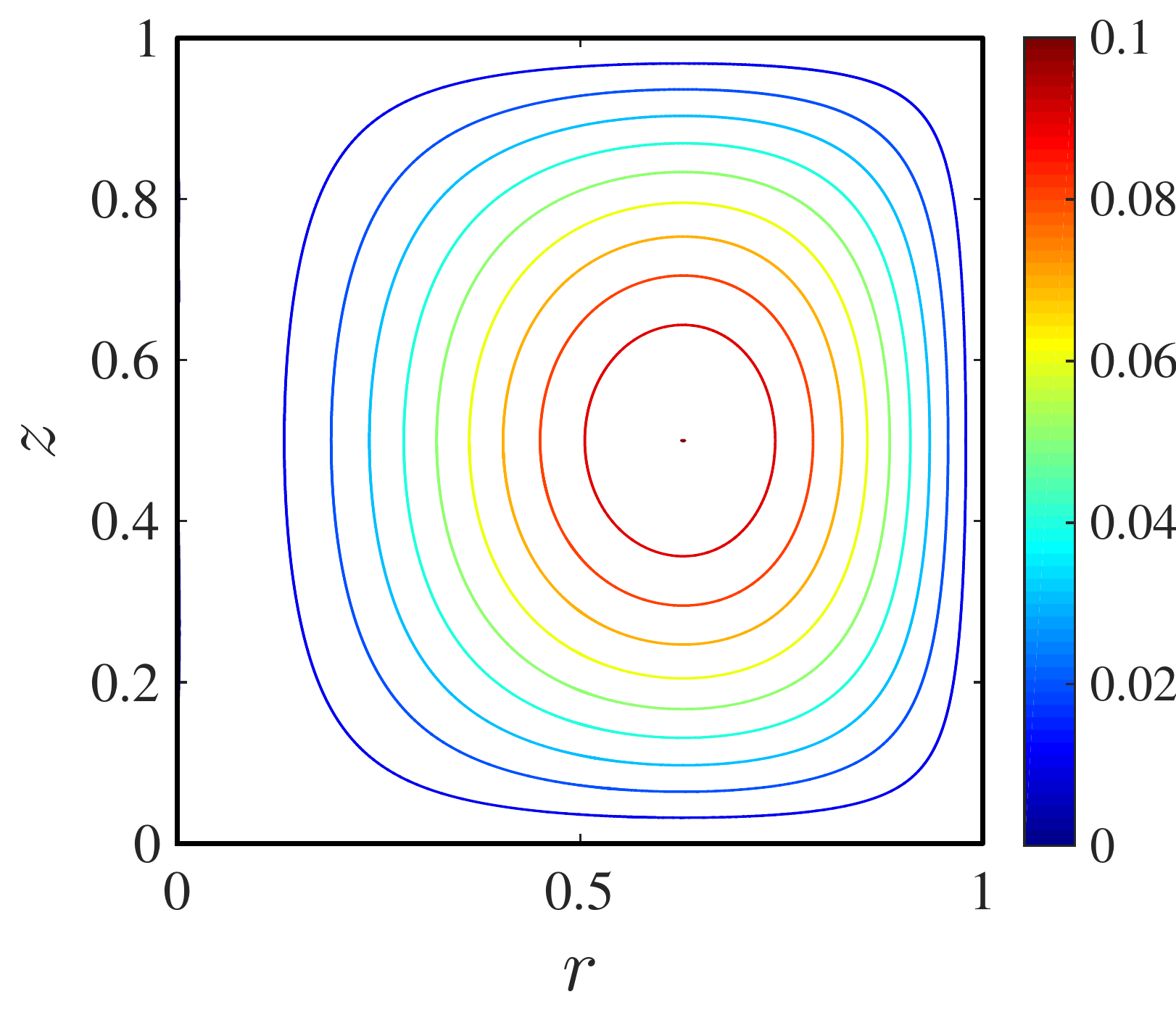}
				\includegraphics[height=0.21\textwidth]{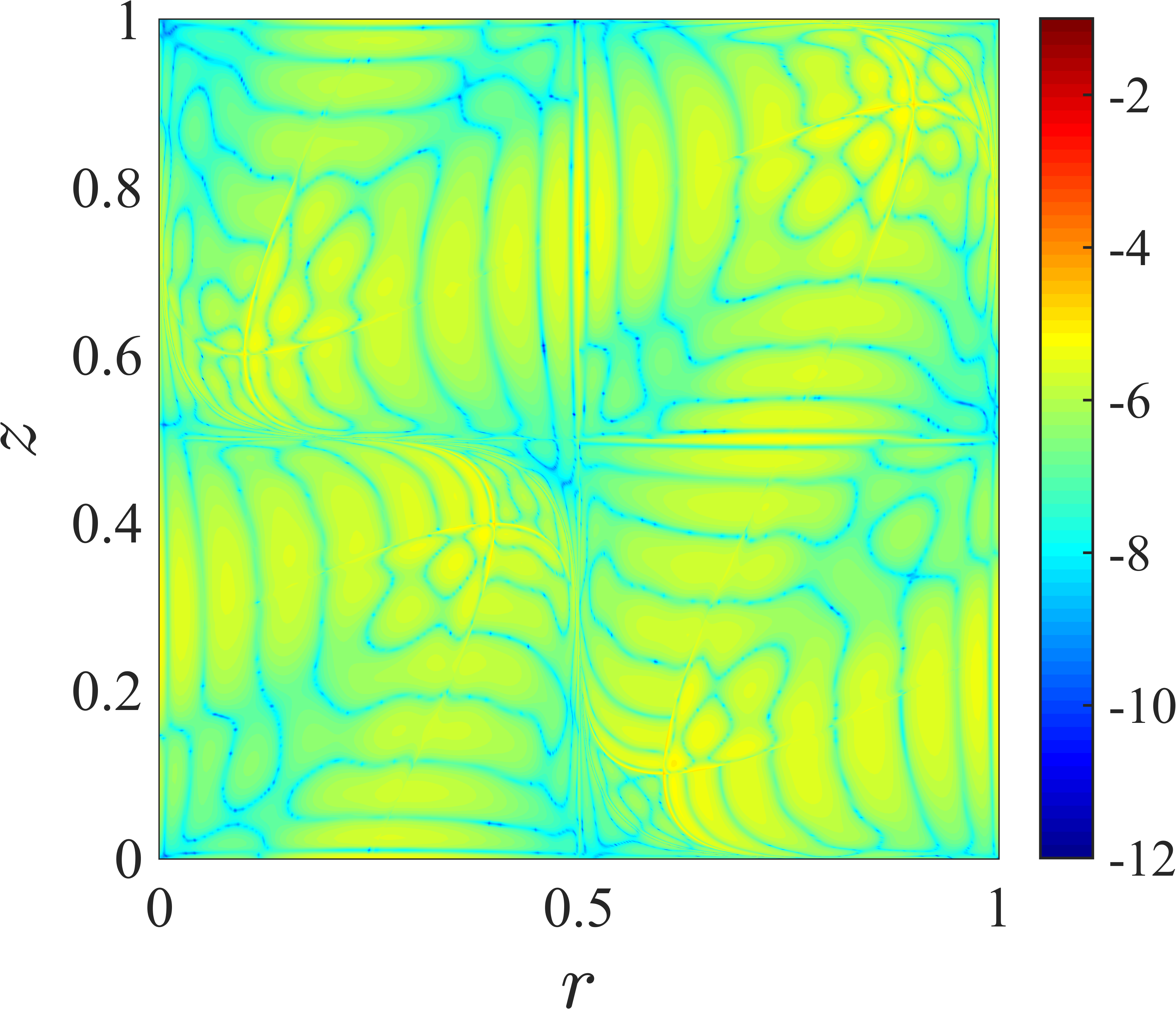}
				\end{center}
				\caption{\reviewerone{Numerical solution of the spheromak test case, \eqref{eq::spheromak_test_case}, with $\psi_{0}=0.1$, $L_{r}=1.0$, $L_{z} = 1.0$ and $f_{0}=4.954954595474438$. From left to right: (i) computational mesh, with curvature parameter $c=0.3$, $4\times 4$ elements of polynomial degree $p=8$, %(ii) analytical solution \eqref{eq::spheromak_test_case_analytical_solution}, $\psi_{a}$, 
				(ii) numerical solution using the mesh in (i), $\psi_{h}$, and (iii) logarithmic error between the analytical solution and the numerical one, $\log_{10}|\psi_{a}-\psi_{h}|$.}}
				\label{fig::spheromak_test_case}
			\end{figure}

\begin{figure}[!ht]
				\begin{center}
				\includegraphics[height=0.295\textwidth]{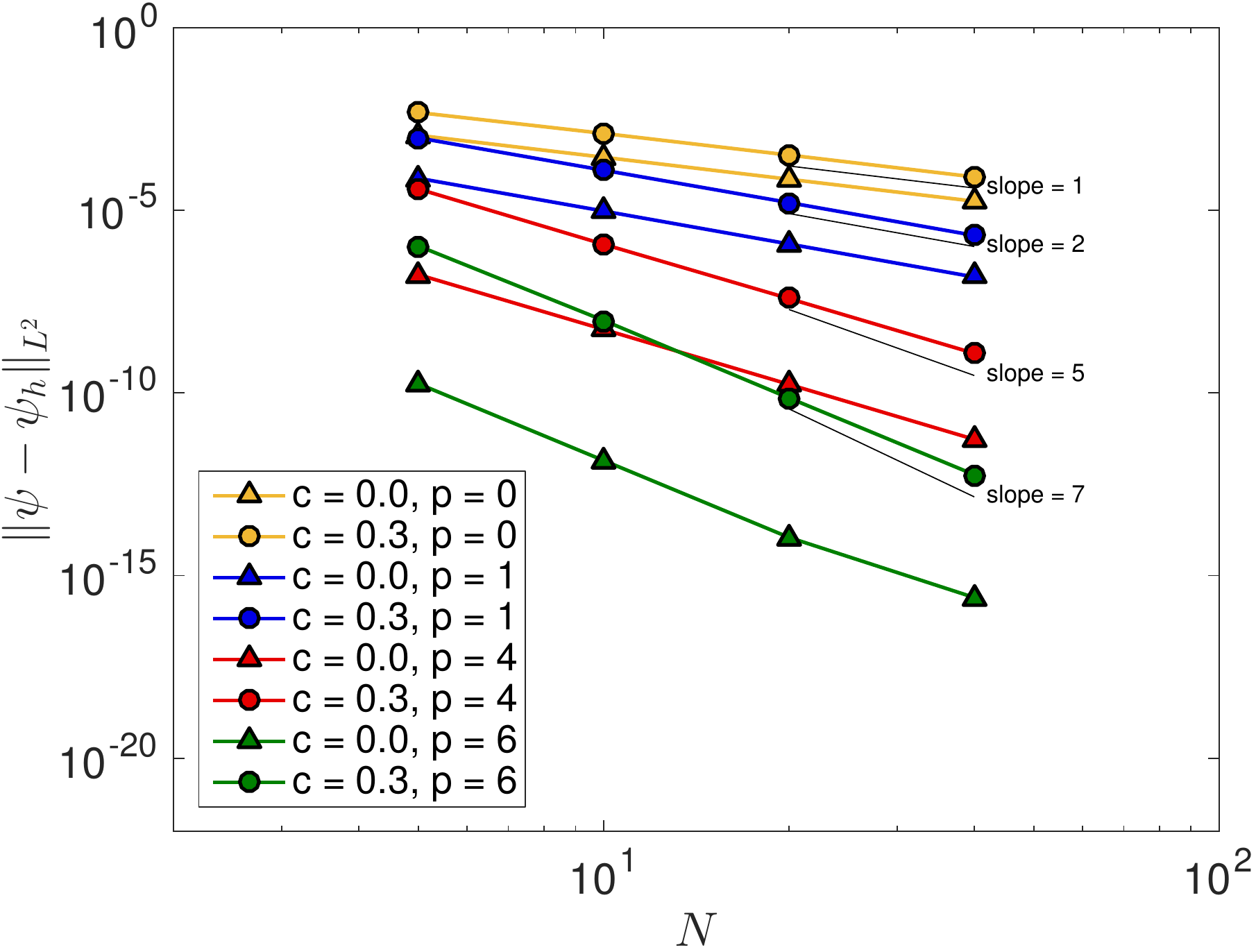} \hspace{1cm}
				\includegraphics[height=0.295\textwidth]{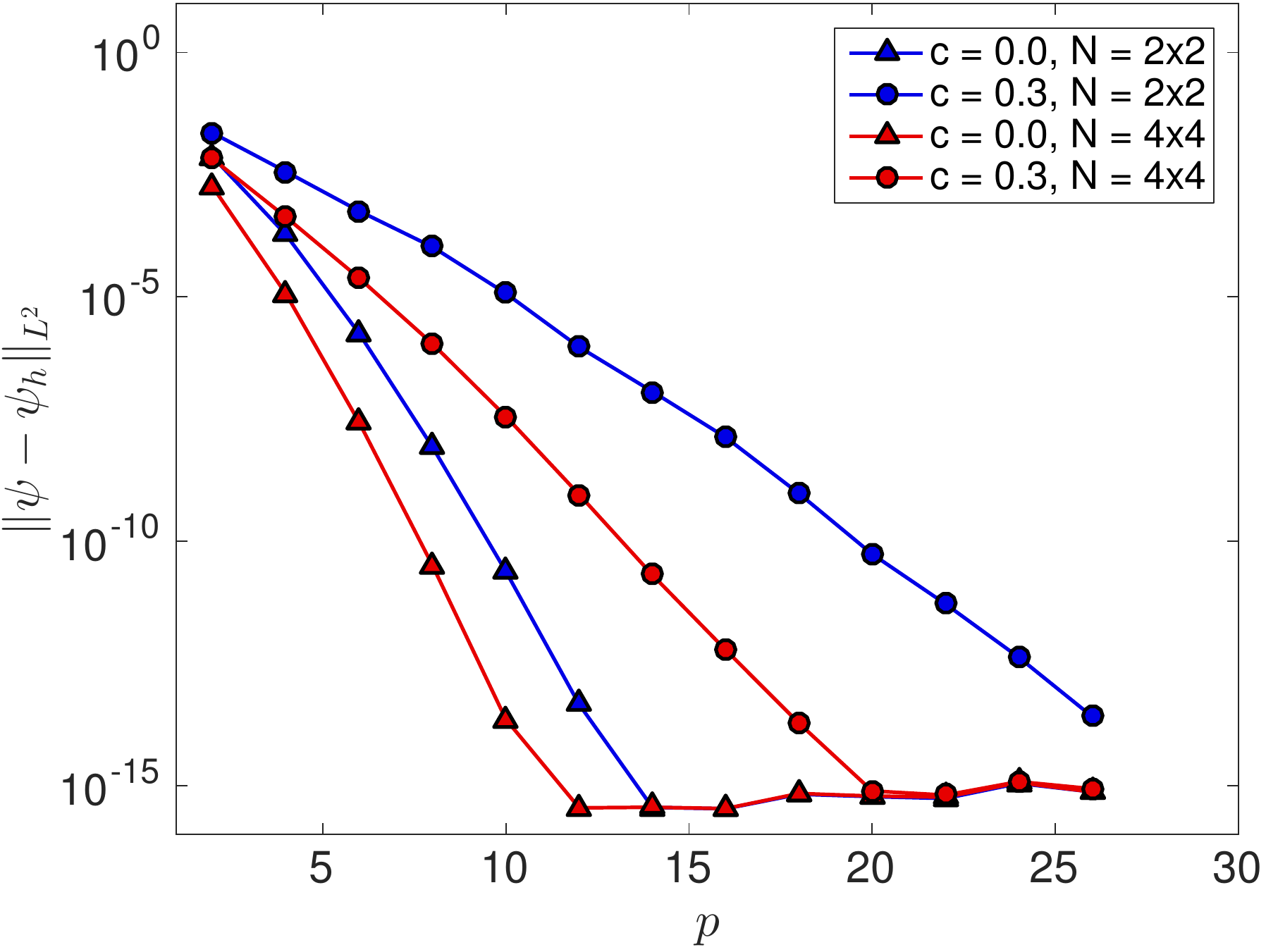}
				\end{center}
				\caption{Convergence plots for the numerical solution of the spheromak test case \eqref{eq::frc_test_case}, with $\psi_{0}=0.1$, $L_{r}=1.0$, $L_{z} = 1.0$ and $f_{0}=4.954954595474438$. Left: $h$-convergence plots. Right: $p$-convergence plots. Convergence plots computed for meshes with deformation, $c=0.0$, $c=0.1$, $c=0.2$ and $c=0.3$.}
				\label{fig::spheromak_test_case_convergence}
			\end{figure}

			As can be seen in \figref{fig::spheromak_test_case}, the mimetic spectral element method is capable of accurately reproducing the results on a highly curved mesh obtained with the mapping \eqref{eq::mesh_deformation_mapping}, on the domain $[0,L_{r}]\times[0,L_{z}]$. The convergence tests for the spheromak test case confirm the results obtained previously. We can observe high $h$-convergence rates close to order $p+1$, \figref{fig::spheromak_test_case_convergence} left. Similar behaviour can be seen for $p$-convergence with the method converging exponentially fast for both straight and curved meshes, \figref{fig::spheromak_test_case_convergence} right. Machine accuracy is also achieved in this test case.
			
			\FloatBarrier
			
	\subsection{Curved plasma boundary} \label{sec::plasma_boundary}
		In this section we assess the ability of the proposed numerical model to solve fixed boundary problems where the prescribed plasma boundary is curved. \reviewerone{For this purpose the method is tested first on the Soloviev solutions presented in \secref{sec::test_cases_soloviev} (now with curved plasma boundary), then on both linear and non-linear eigenvalue problems, and finally we test it on a plasma with an X-point.}
		
		In all cases, the plasma boundary is prescribed parametrically such that $r_{b} = \gamma_{r}(s)$ and $z_{b} = \gamma_{z}(s)$ with $s\in[s_{-},s_{+}]$. The computational mesh is prescribed %in two ways: (i) using a peudo-cylindrical coordinate system, as in \figref{fig::linear_eigenvalue_shape_1_test_case} top left, or (ii) 
		\reviewerone{using a transfinite mapping, see \cite{gordon::transfinite_mapping}}.%, as in \figref{fig::linear_eigenvalue_shape_1_test_case} bottom left.
		 
		\begin{reviewer1}
		\subsubsection{Soloviev test case} \label{sec::test_cases_curved_soloviev}
			In order to assess the accuracy of the proposed method for the solution of a curved boundary plasma shape, we first apply it to the Soloviev problem introduced in \secref{sec::test_cases_soloviev}. For this case the plasma boundary is given by the equation
			\begin{equation}
				\frac{r^{4}}{8}+d_{1}+d_{2}r^{2} + d_{3}\left(r^{4} - 4r^{2}z^{2}\right) = 0,
			\end{equation}
			with $d_{1},d_{2},d_{3}$ as in \eqref{eq:soloviev_parameters}.
			
			In \figref{fig::soloviev_test_case_iter} an example solution for ITER parameters is presented and in \figref{fig::soloviev_test_case_ntsx} another example solution for NSTX parameters is shown. As can be seen, both solutions are well reconstructed by the proposed method. The NSTX case has larger errors due to the more deformed underlying mesh. As was seen in \secref{sec::test_cases_soloviev}, a higher mesh deformation leads to higher errrors.
			
			\begin{figure}[!ht]
				\begin{center}
				\includegraphics[height=0.271\textwidth]{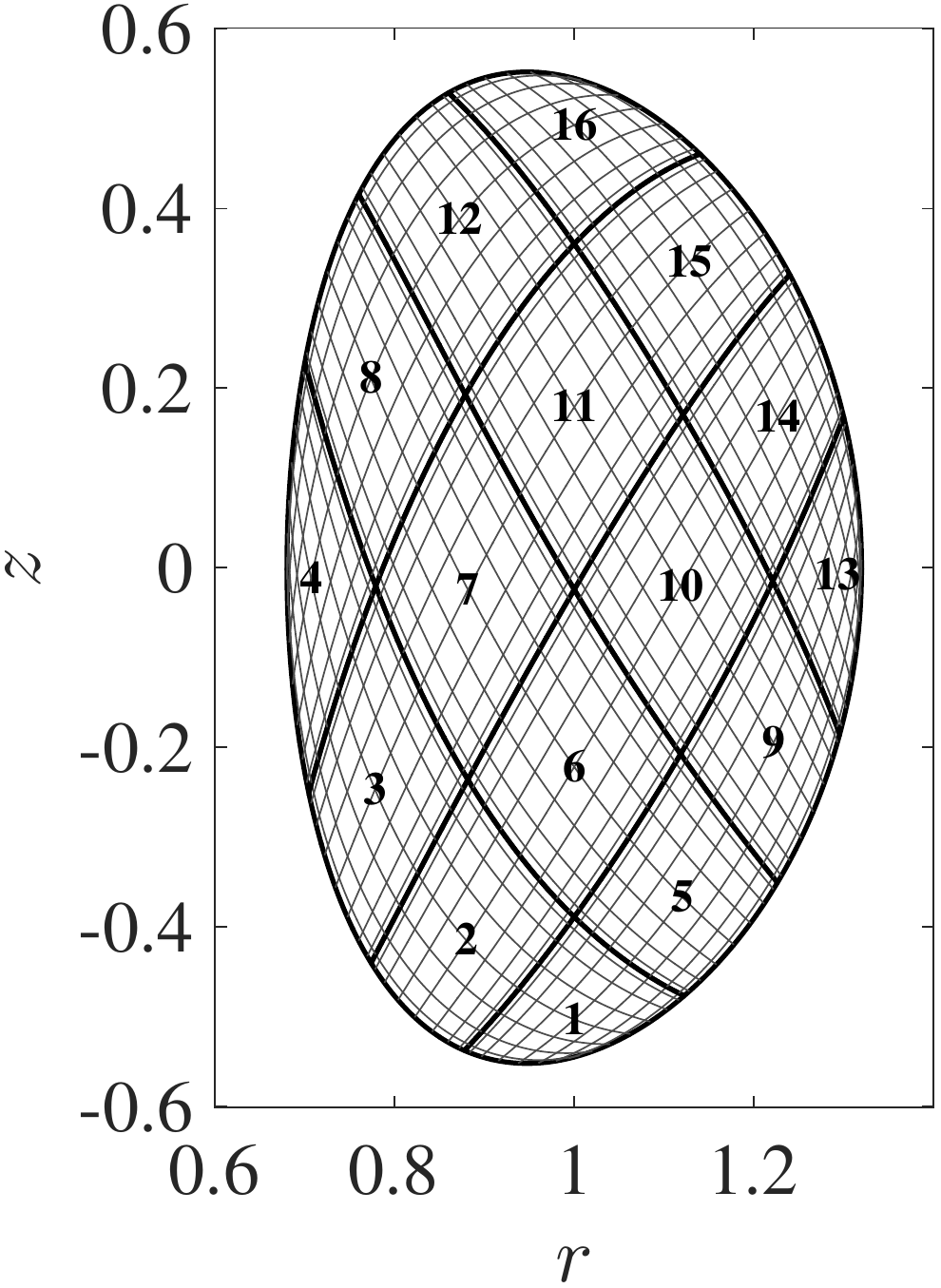}\hspace{0.75cm}
				\includegraphics[height=0.271\textwidth]{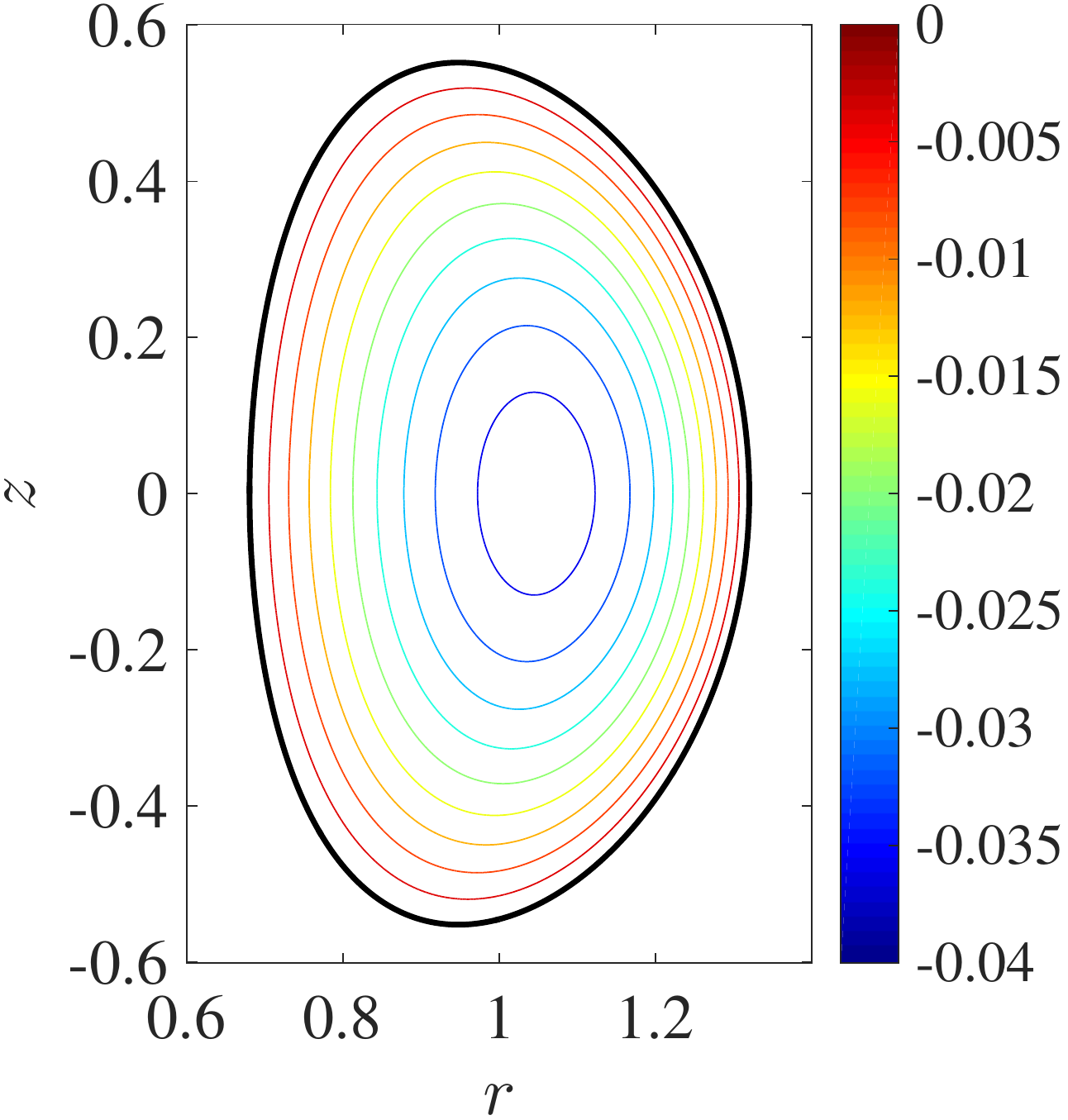}\hspace{0.75cm}
				\includegraphics[height=0.271\textwidth]{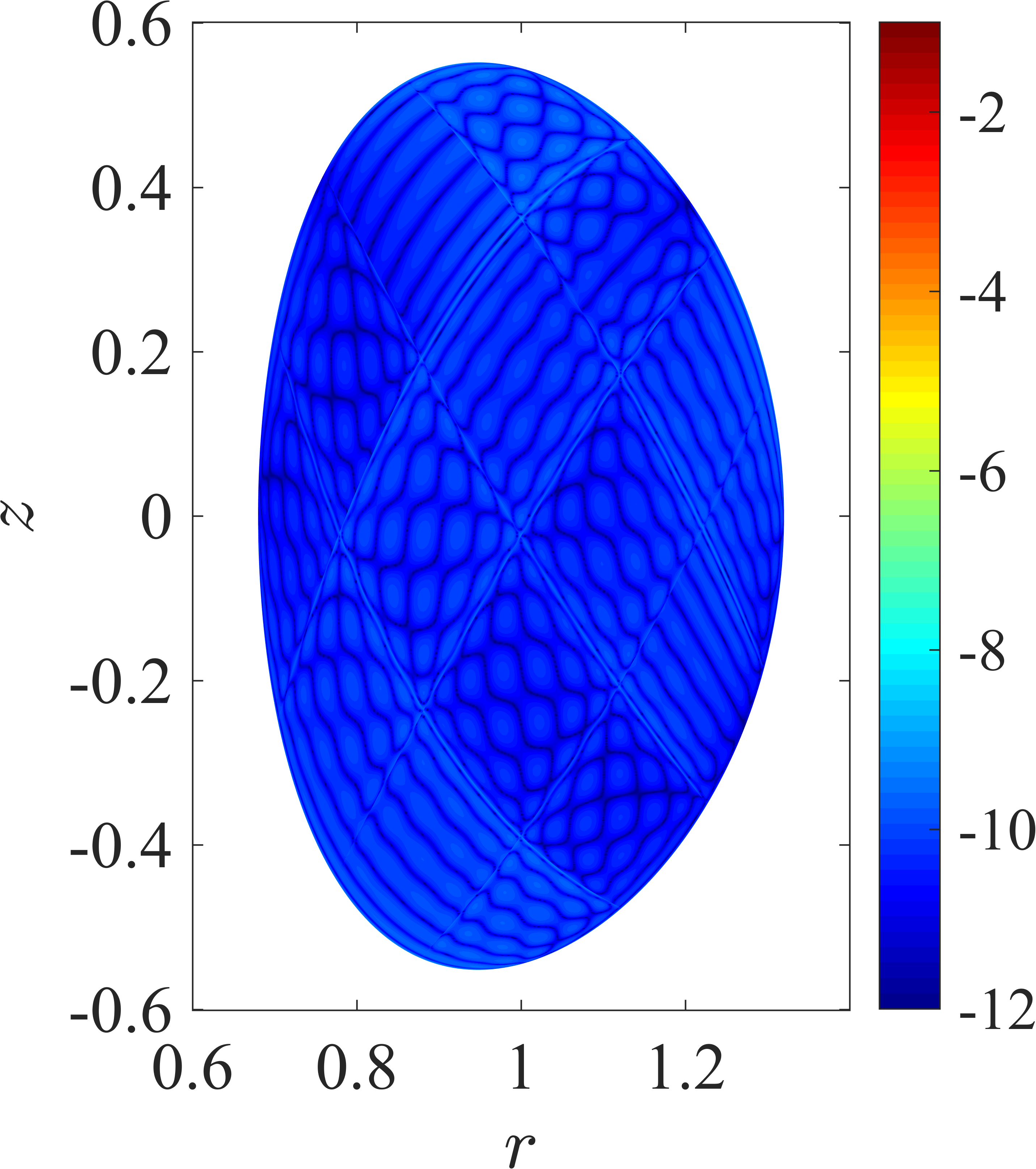}
				\end{center}
				\caption{\reviewerone{Numerical solution of the Soloviev test case, \eqref{eq::soloviev_test_case}, for ITER parameters, $\epsilon = 0.32$, $\kappa = 1.7$, $\delta=0.33$. From left to right: (i) computational mesh with $4\times 4$ elements of polynomial degree $p=8$, %(ii) analytical solution \eqref{eq::soloviev_test_case_analytical_solution}, $\psi_{a}$, 
				(ii) numerical solution using the mesh in (i), $\psi_{h}$, and (iii) logarithmic error between the analytical solution and the numerical one, $\log_{10}|\psi_{a}-\psi_{h}|$.}}
				\label{fig::soloviev_test_case_iter}
			\end{figure}
			
			\begin{figure}[!ht]
				\begin{center}
				\includegraphics[height=0.271\textwidth]{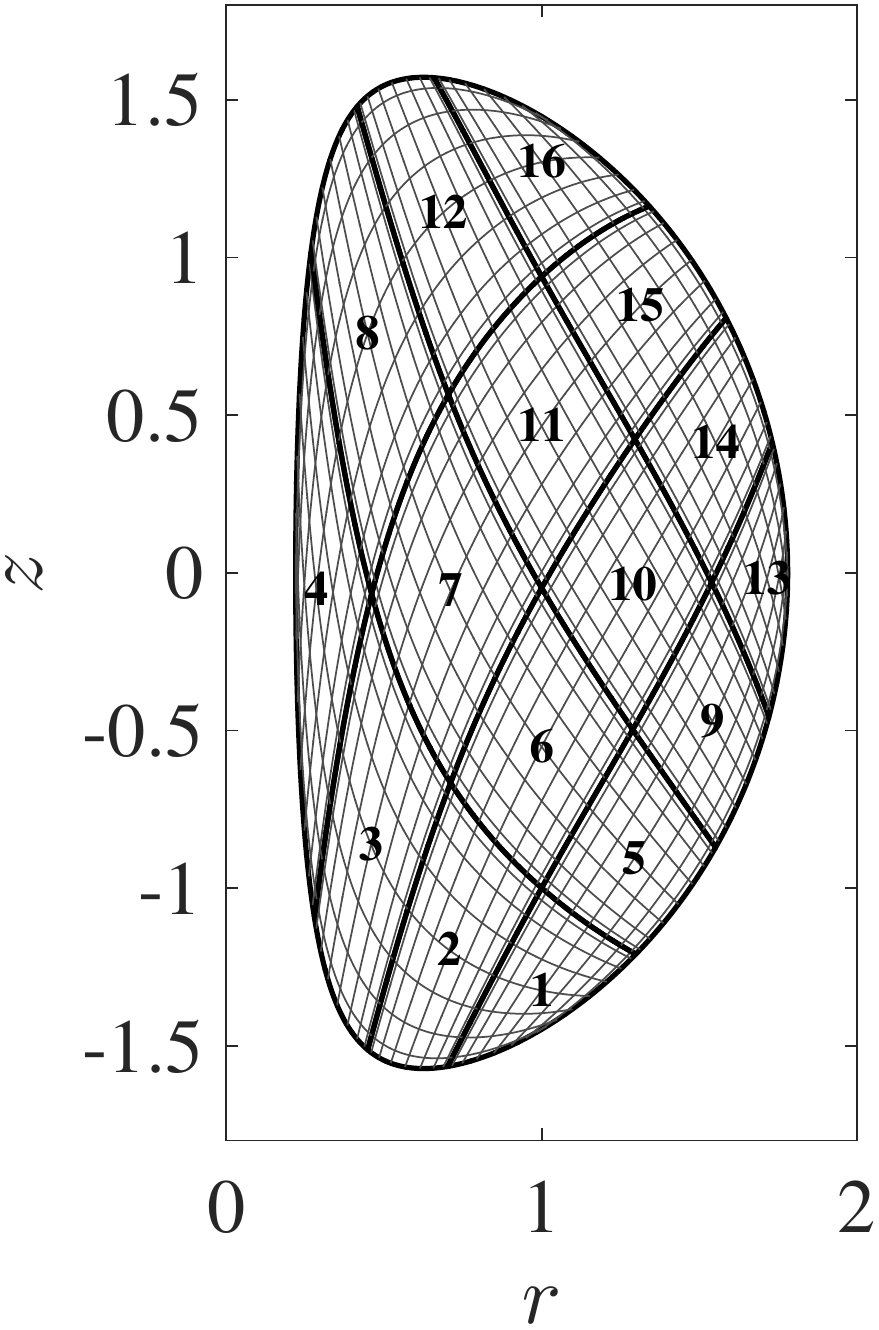}\hspace{0.75cm}
				\includegraphics[height=0.271\textwidth]{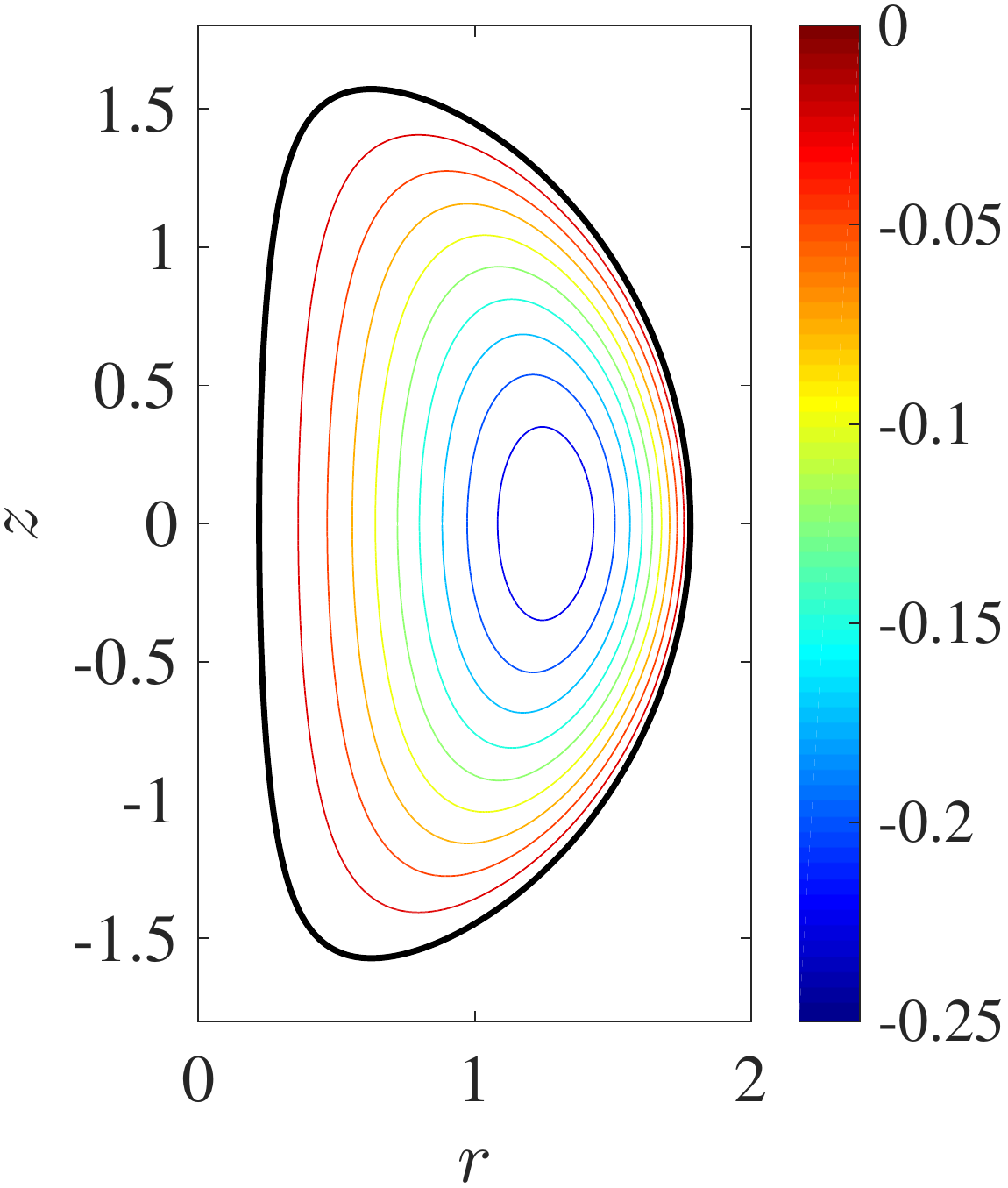}\hspace{0.75cm}
				\includegraphics[height=0.271\textwidth]{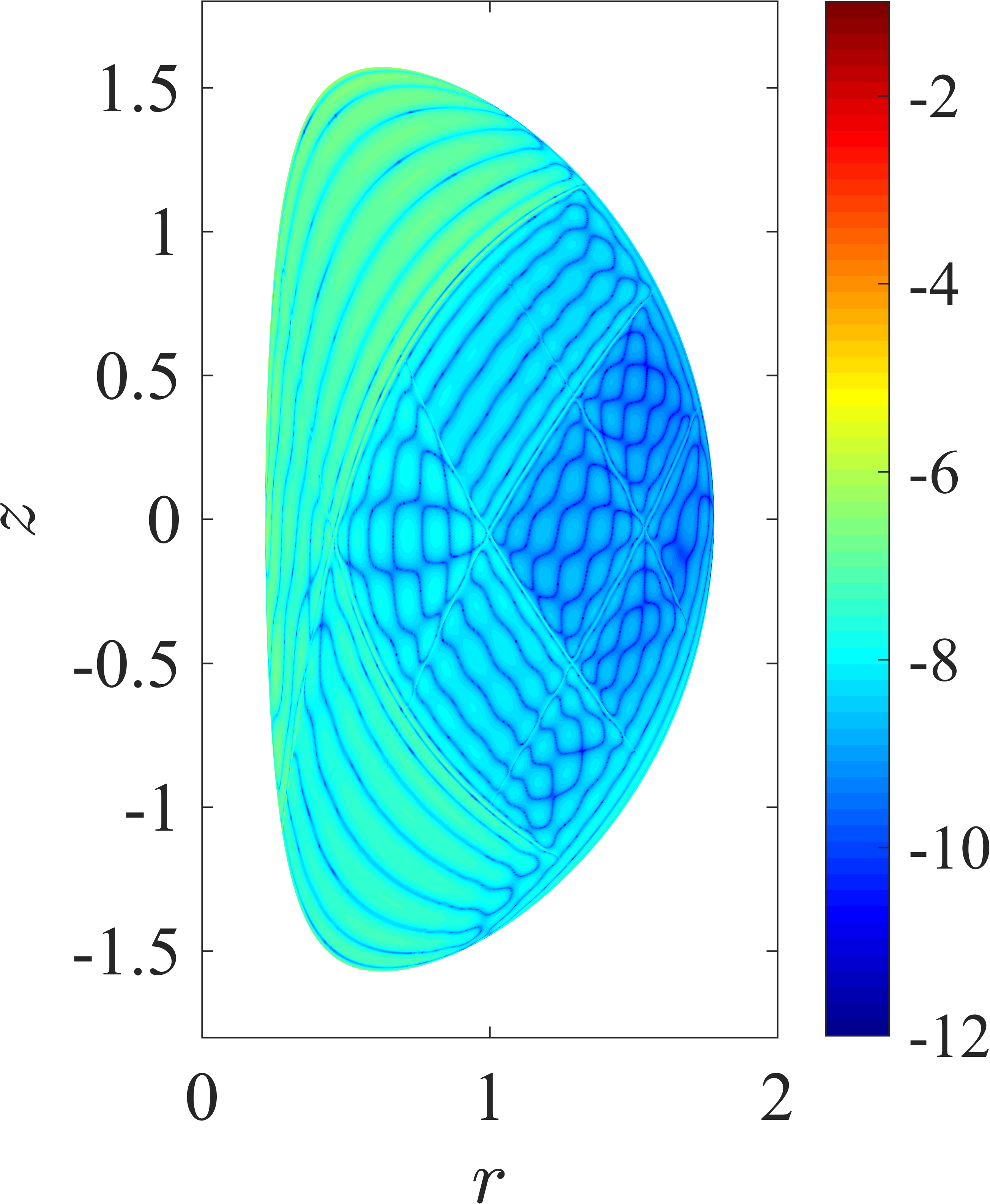}
				\end{center}
				\caption{\reviewerone{Numerical solution of the Soloviev test case, \eqref{eq::soloviev_test_case}, for NSTX parameters, $\epsilon = 0.78$, $\kappa = 2.0$, $\delta=0.35$. From left to right: (i) computational mesh with $4\times 4$ elements of polynomial degree $p=8$, %(ii) analytical solution \eqref{eq::soloviev_test_case_analytical_solution}, $\psi_{a}$,
				(ii) numerical solution using the mesh in (i), $\psi_{h}$, and (iii) logarithmic error between the analytical solution and the numerical one, $\log_{10}|\psi_{a}-\psi_{h}|$.}}
				\label{fig::soloviev_test_case_ntsx}
			\end{figure}
			
			The NSTX case convergence tests for $h$-refinement and $p$-refinement show very good convergence rates both for $\psi_{h}$ and for $\vec{h}_{h}$. For $h$-convergence, \figref{fig::curved_soloviev_test_case_convergence} left and \figref{fig::curved_soloviev_test_case_convergence_h} left, we can see that the mimetic spectral element solver preserves high convergence rates of $p+1$. Regarding $p$-convergence, \figref{fig::curved_soloviev_test_case_convergence} right and \figref{fig::curved_soloviev_test_case_convergence_h} right, we observe the same robustness of the method, with convergence rates maintaining their exponential character. As noted before, it is possible to obtain convergence to machine accuracy.
			
			\begin{figure}[htb]
				\begin{center}
				\includegraphics[height=0.295\textwidth]{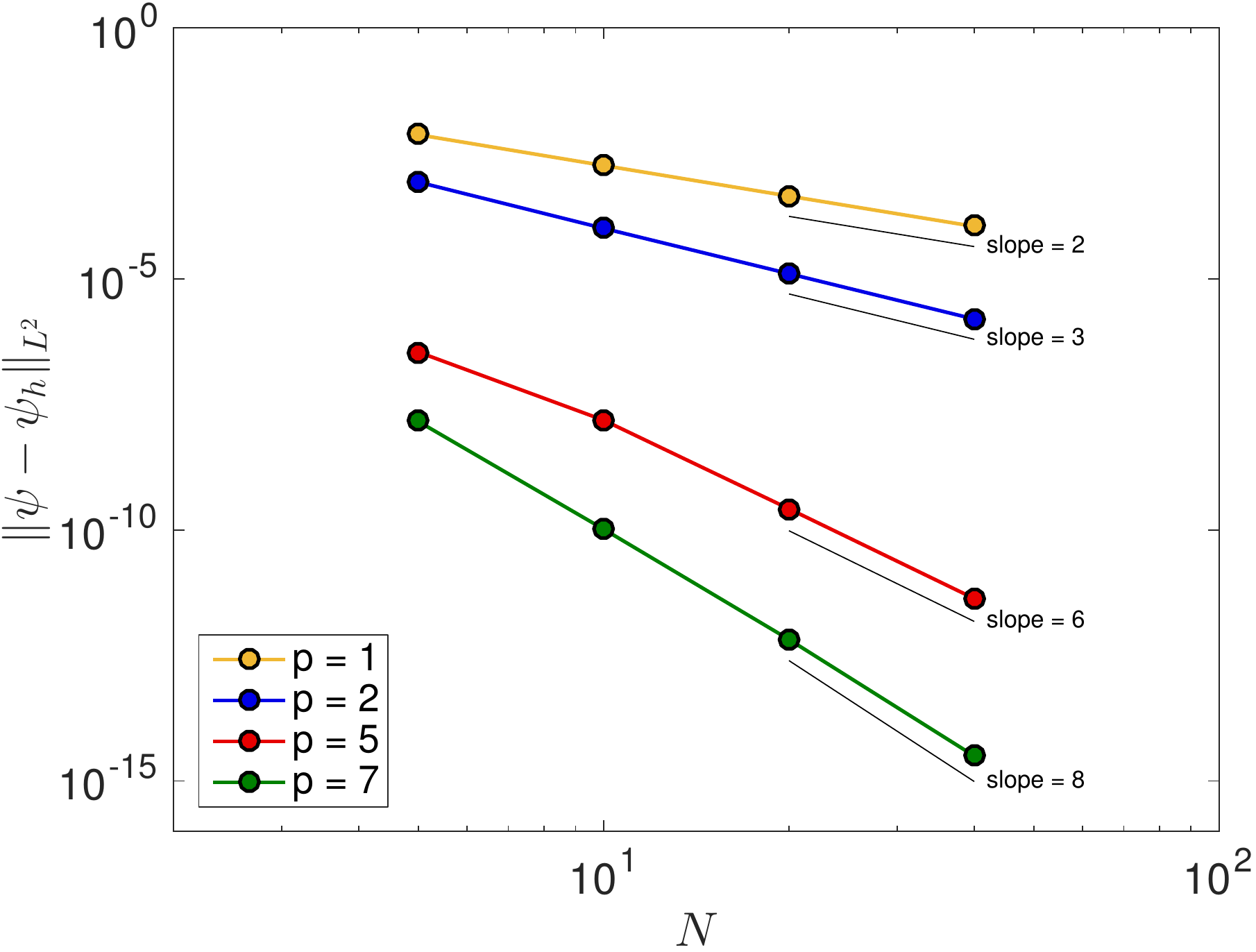} \hspace{1cm}
				\includegraphics[height=0.295\textwidth]{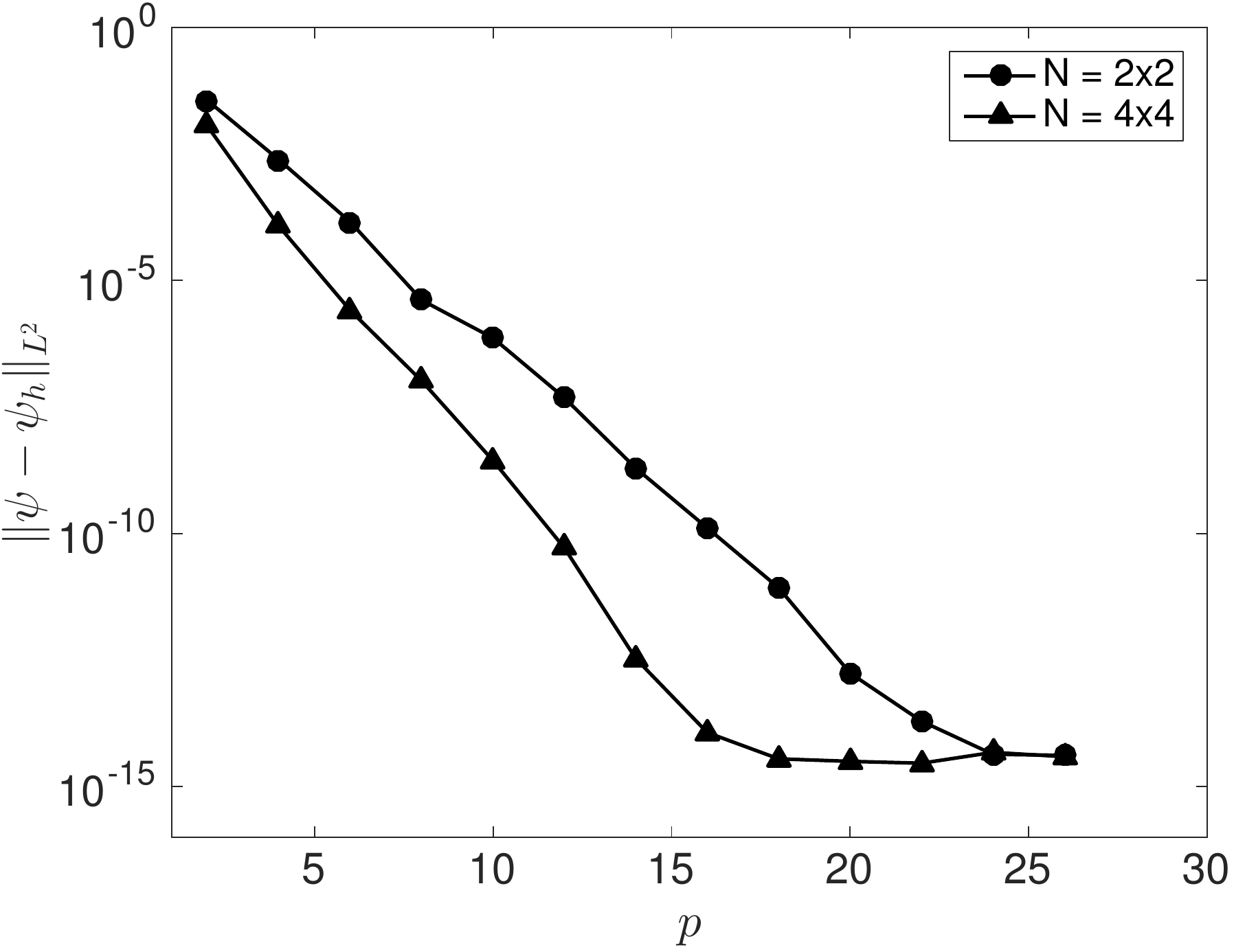}
				\end{center}
				\caption{\reviewerone{Convergence plots for the numerical solution of $\psi(r,z)$ of the Soloviev test case \eqref{eq::soloviev_test_case} for NSTX parameters, $\epsilon = 0.78$, $\kappa = 2.0$, $\delta=0.35$. Left: $h$-convergence plots. Right: $p$-convergence plots.}}
				\label{fig::curved_soloviev_test_case_convergence}
			\end{figure}
			
			\begin{figure}[htb]
				\begin{center}
				\includegraphics[height=0.295\textwidth]{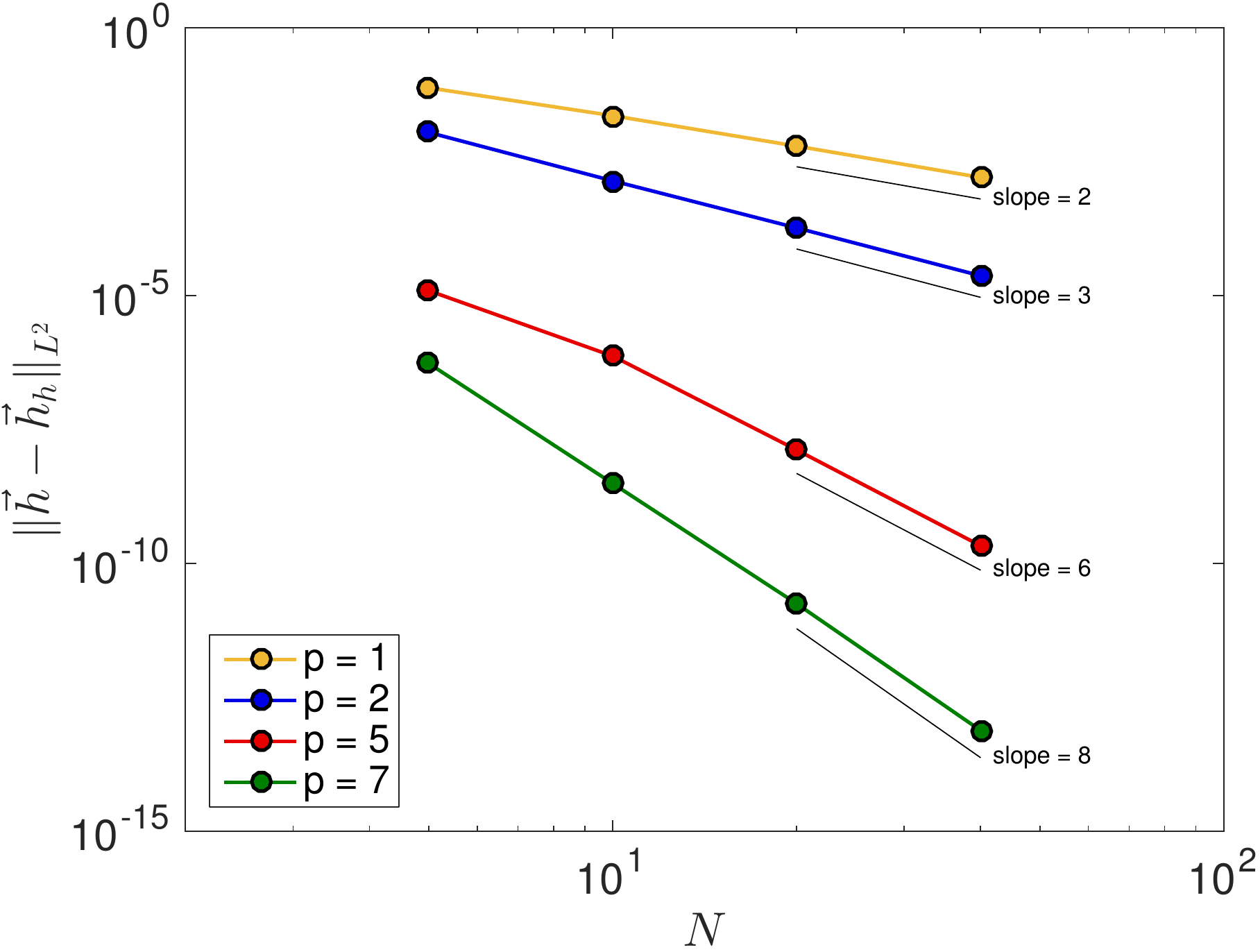} \hspace{1cm}
				\includegraphics[height=0.295\textwidth]{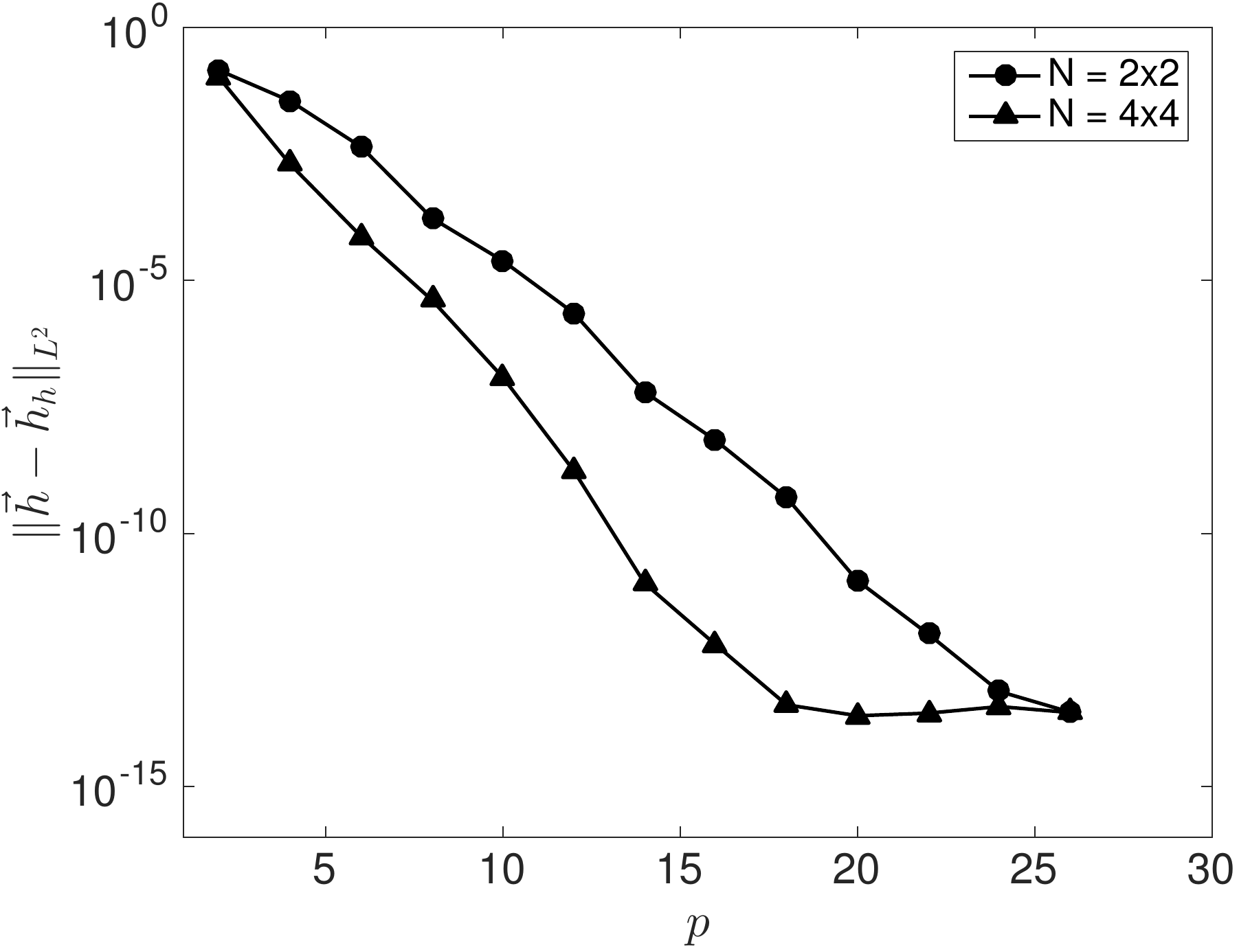}
				\end{center}
				\caption{\reviewerone{Convergence plots for the numerical solution of $\vec{h}_{h}(r,z)$ of the Soloviev test case \eqref{eq::soloviev_test_case} for NSTX parameters, $\epsilon = 0.78$, $\kappa = 2.0$, $\delta=0.35$. Left: $h$-convergence plots. Right: $p$-convergence plots.}}
				\label{fig::curved_soloviev_test_case_convergence_h}
			\end{figure}
			
			A final test consisted in computing the error between the flux integral $\int_{\Omega_{p}} J_{h}\mathrm{d}V$ and the contour integral  $\int_{\partial\Omega_{p}}\vec{h}_{h}\mathrm{d}\vec{l}$. We show the results for the NSTX test case in \figref{fig::curved_soloviev_test_case_convergence_integral}, where it is possible to see that the two quantities are identical up to machine precision. This solver can reconstruct the integral values up to machine precision, independently of the number of elements and polynomial degree of the basis functions.
			
			\begin{figure}[htb]
				\begin{center}
				\includegraphics[height=0.295\textwidth]{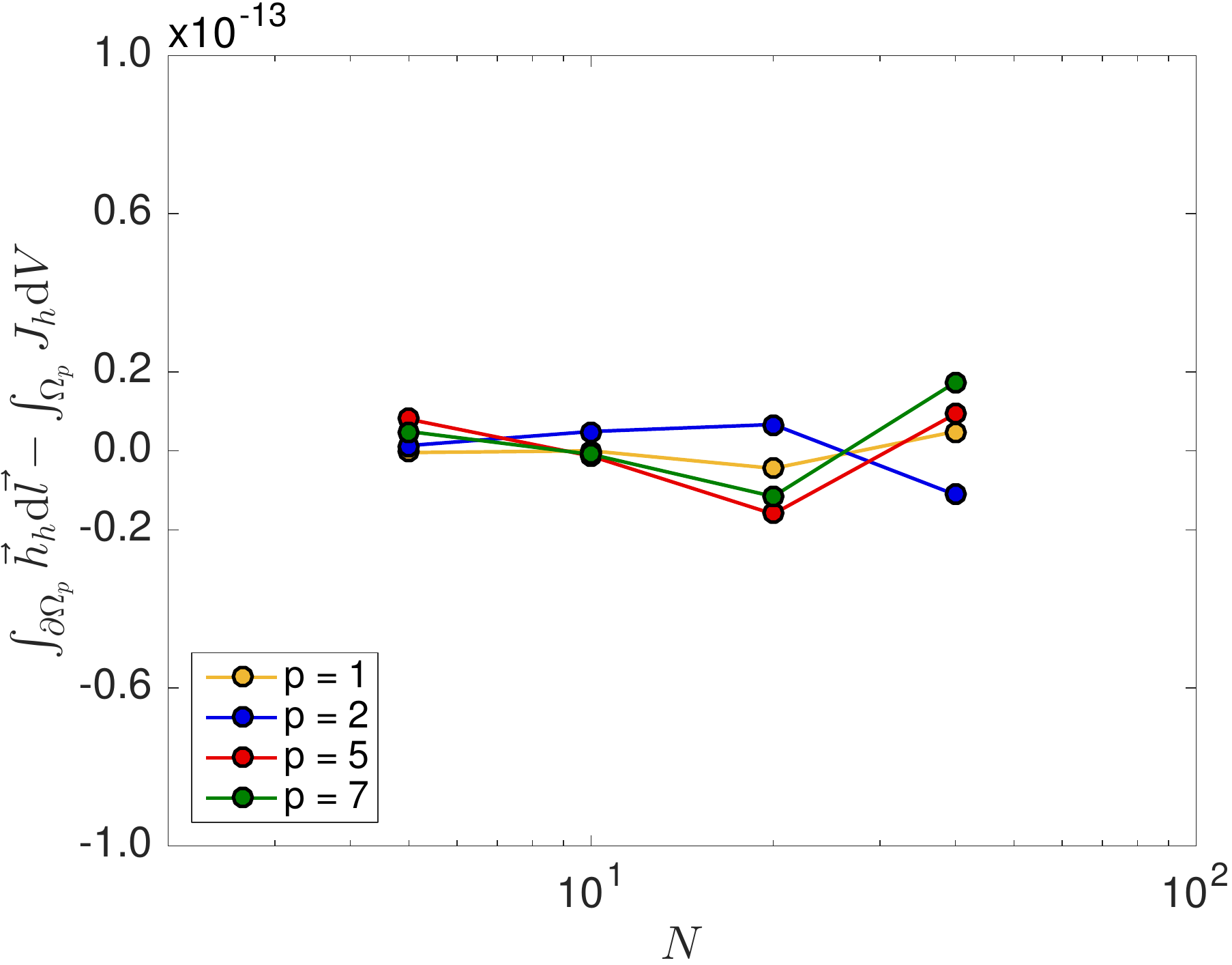} \hspace{1cm}
				\includegraphics[height=0.295\textwidth]{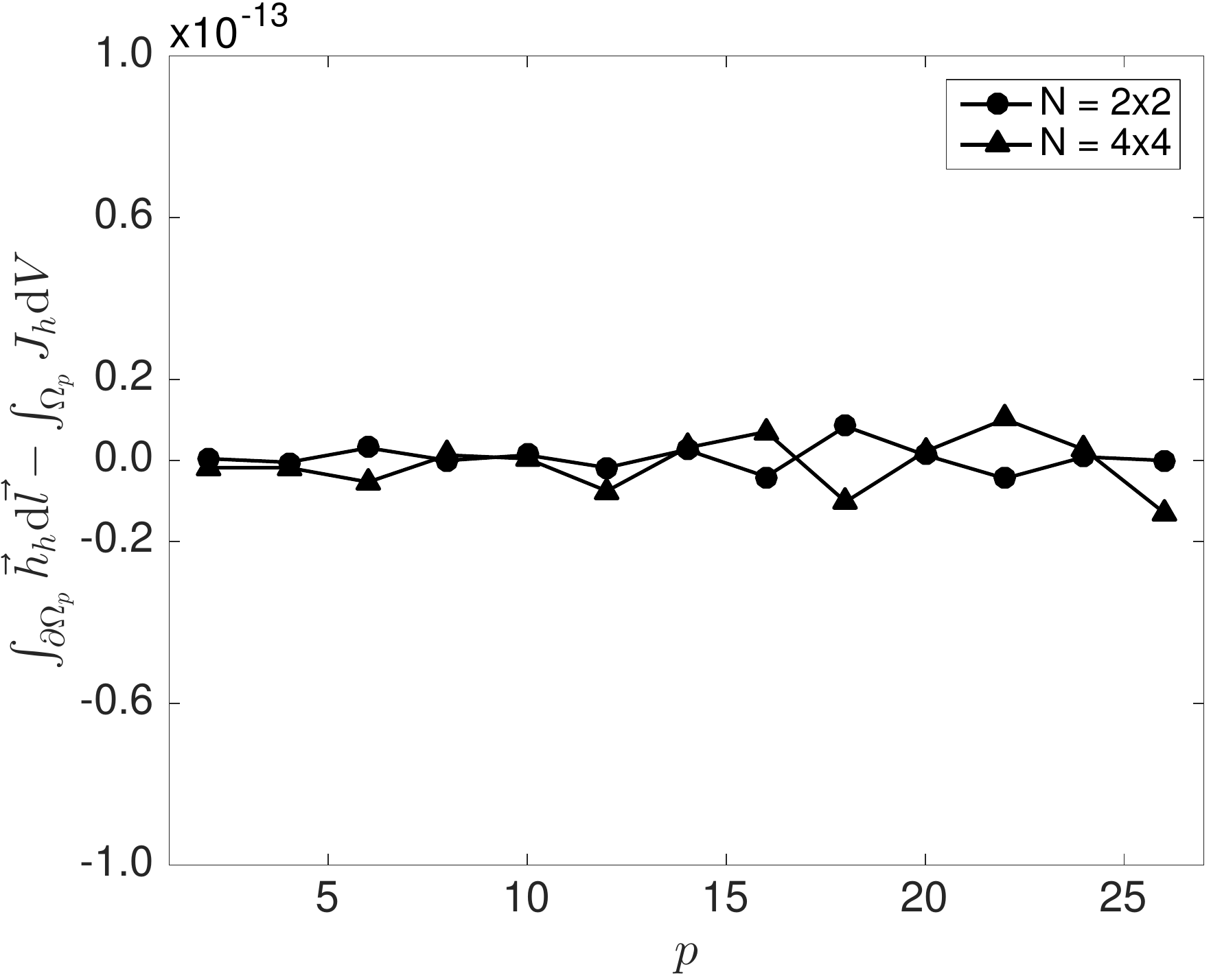}
				\end{center}
				\caption{\reviewerone{Error between the flux integral $\int_{\Omega_{p}} J_{h}\mathrm{d}V$ and the contour integral  $\int_{\partial\Omega_{p}}\vec{h}_{h}\mathrm{d}\vec{l}$, for the Soloviev test case \eqref{eq::soloviev_test_case} with NSTX parameters, $\epsilon = 0.78$, $\kappa = 2.0$, $\delta=0.35$. Left: $h$-convergence plots. Right: $p$-convergence plots.}}
				\label{fig::curved_soloviev_test_case_convergence_integral}
			\end{figure}
			
		\FloatBarrier	
		\end{reviewer1}

		\subsubsection{Linear eigenvalue problem} \label{sec::test_cases_linear_eigenvalue}
			For this fixed boundary test case we consider the plasma configuration used in \cite{Pataki2013}, corresponding to the models $f(\psi) = 0$ and $P(\psi)= \frac{C_{1}+C_{2}r^{2}}{2\mu_{0}}\,\psi^{2} + P_{0}$. This leads to the following Grad-Shafranov problem
			\begin{equation}
				\begin{dcases}
					\nabla\times\left(\mathbb{K}\nabla\times\psi\right) = \left(\frac{C_{1}}{r} +C_{2}r\right)\frac{\psi}{\mu_{0}} & \mbox{in} \quad \Omega_{p}\,, \\
					\psi = 0 & \mbox{on}\quad\partial\Omega_{p}\,,
				\end{dcases} \label{eq::linear_eigenvalue_test_case}
			\end{equation}
			with the plasma boundary $\Omega_{p}$ given by
			\begin{equation}
				\begin{dcases}
					r(s) = 2 + \epsilon \cos(s + \alpha \sin s)  & \mbox{with}\quad s\in[0,2\pi[\,,\\
					z(s) = \epsilon\kappa \sin s & \mbox{with}\quad s\in[0,2\pi[ \,. 
				\end{dcases} \label{eq::plasma_shape}
			\end{equation}
			
			\begin{reviewer1}
			The coefficients $C_{i}$ determine the Shafranov shift of the solutions. Several different values were tested but here we only present the results corresponding to $C_{1} = -1.0$ and $C_{2} = 2.0$, the highest Shafranov shift used in \cite{Pataki2013}. Two  plasma shapes have been tested: (i) $\epsilon = 0.32$, $\kappa = 1.7$ and $\delta=0.33$ (ITER shape, \figref{fig::linear_eigenvalue_iter_test_case}), and (ii) $\epsilon = 0.78$, $\kappa = 2.0$ and $\delta=0.45$ (NSTX shape, \figref{fig::linear_eigenvalue_nstx_test_case}).
			
			Since analytical solutions are not available for these test cases, it is not possible to compute the exact error associated to the numerical approximation. In order to assess the accuracy of the method in these cases, we compute and analyse the following error term
			\begin{equation}
				\mathrm{E} = \left|\frac{\nabla\times\vec{h}_{h} - \sigma\,J(r,z,\psi_{h})}{\sigma\,J(r,z,\psi_{h})}\right|.
				\label{eq:nonlinear_error}
			\end{equation}
			
			We can see, \figref{fig::linear_eigenvalue_iter_test_case} and \figref{fig::linear_eigenvalue_nstx_test_case}, that the method presented in this article can accurately reproduce the plasma contour lines on both meshes, even for highly elongated plasmas. The error is substantially reduced when the polynomial degree of the basis functions is increased, as has already been seen in the previous test cases.
			
			\begin{figure}[!ht]
				\begin{center}
				\includegraphics[height=0.271\textwidth]{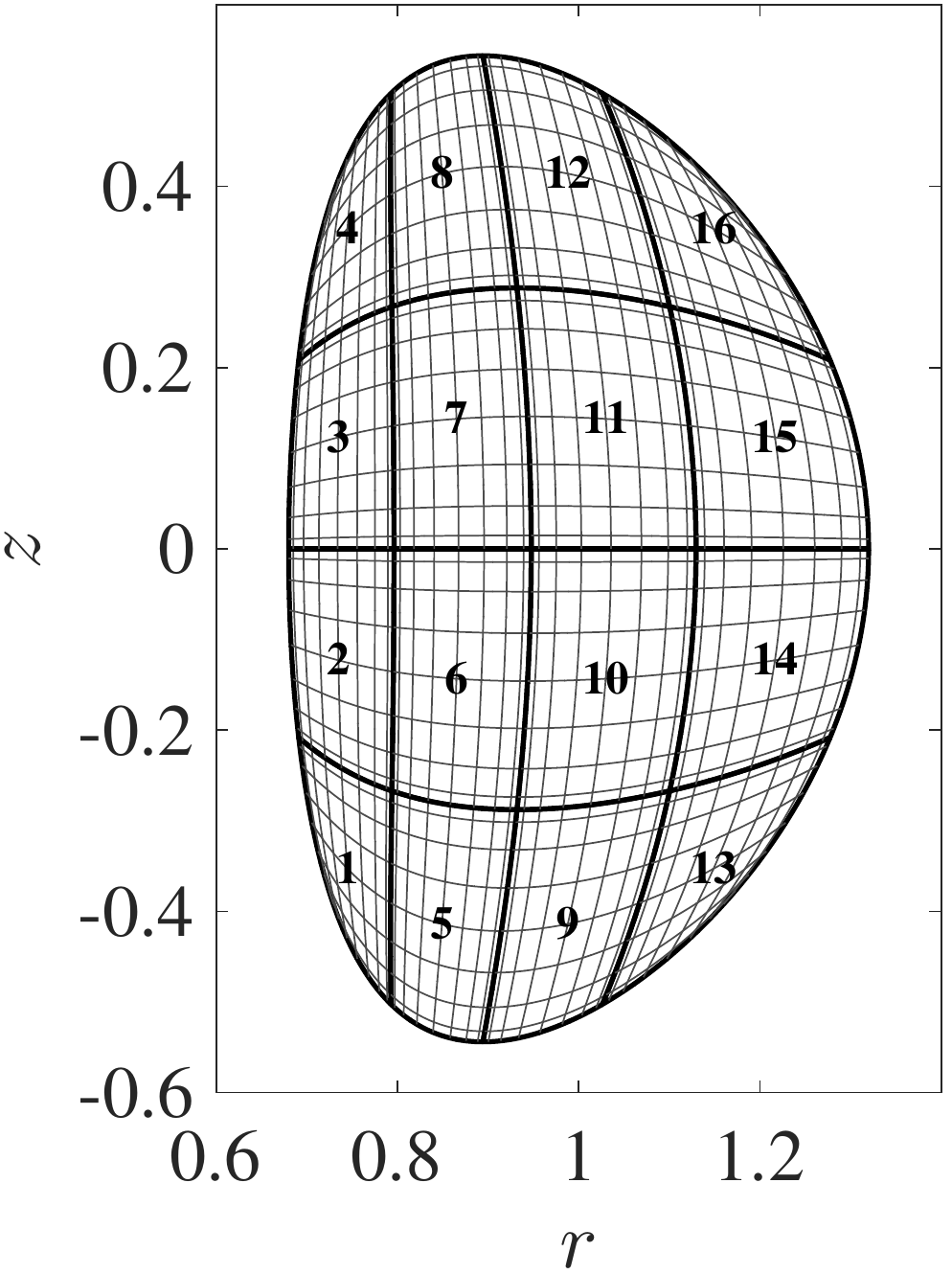}\hspace{0.75cm}
				\includegraphics[height=0.271\textwidth]{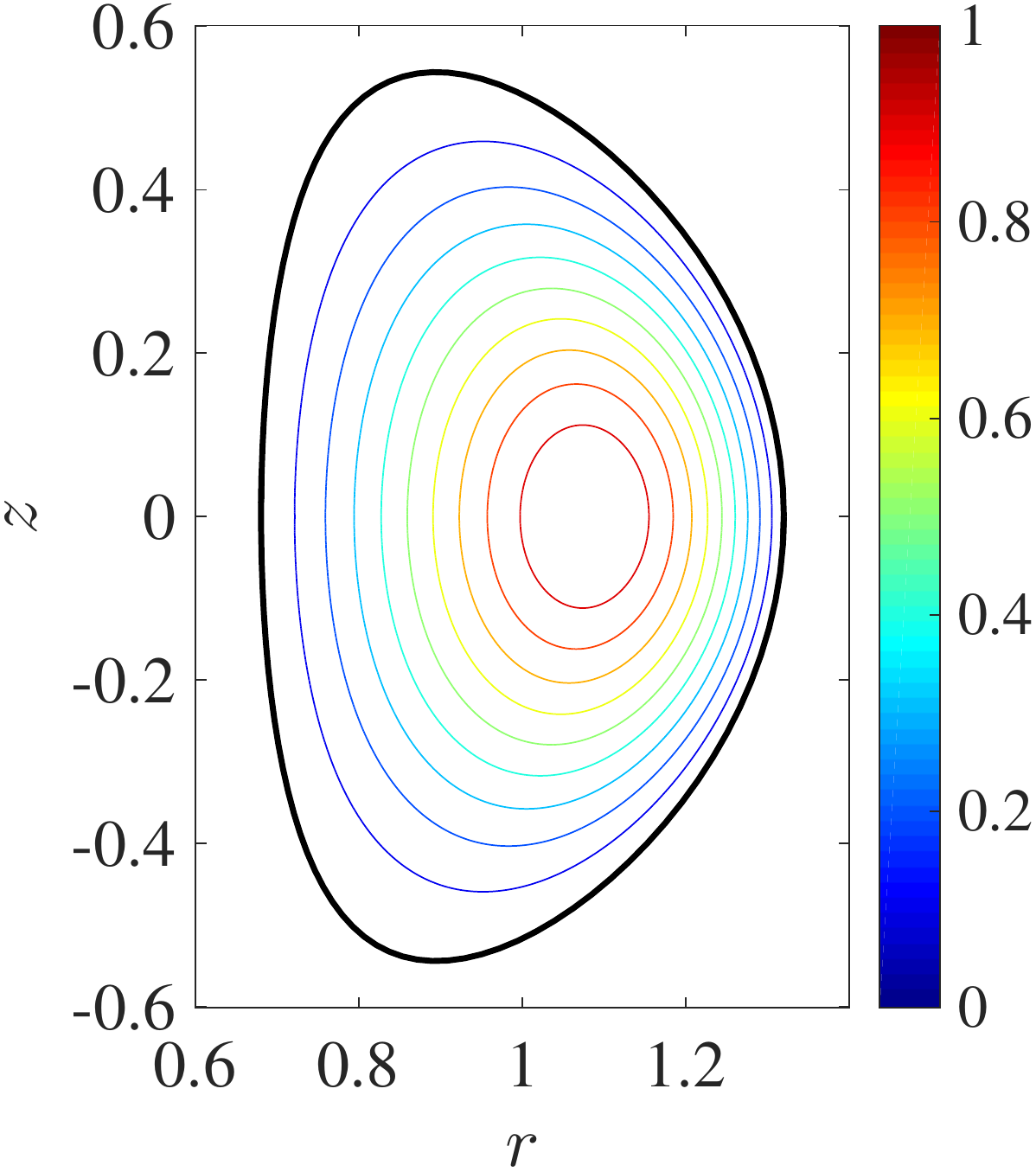}\hspace{0.75cm}
				\includegraphics[height=0.271\textwidth]{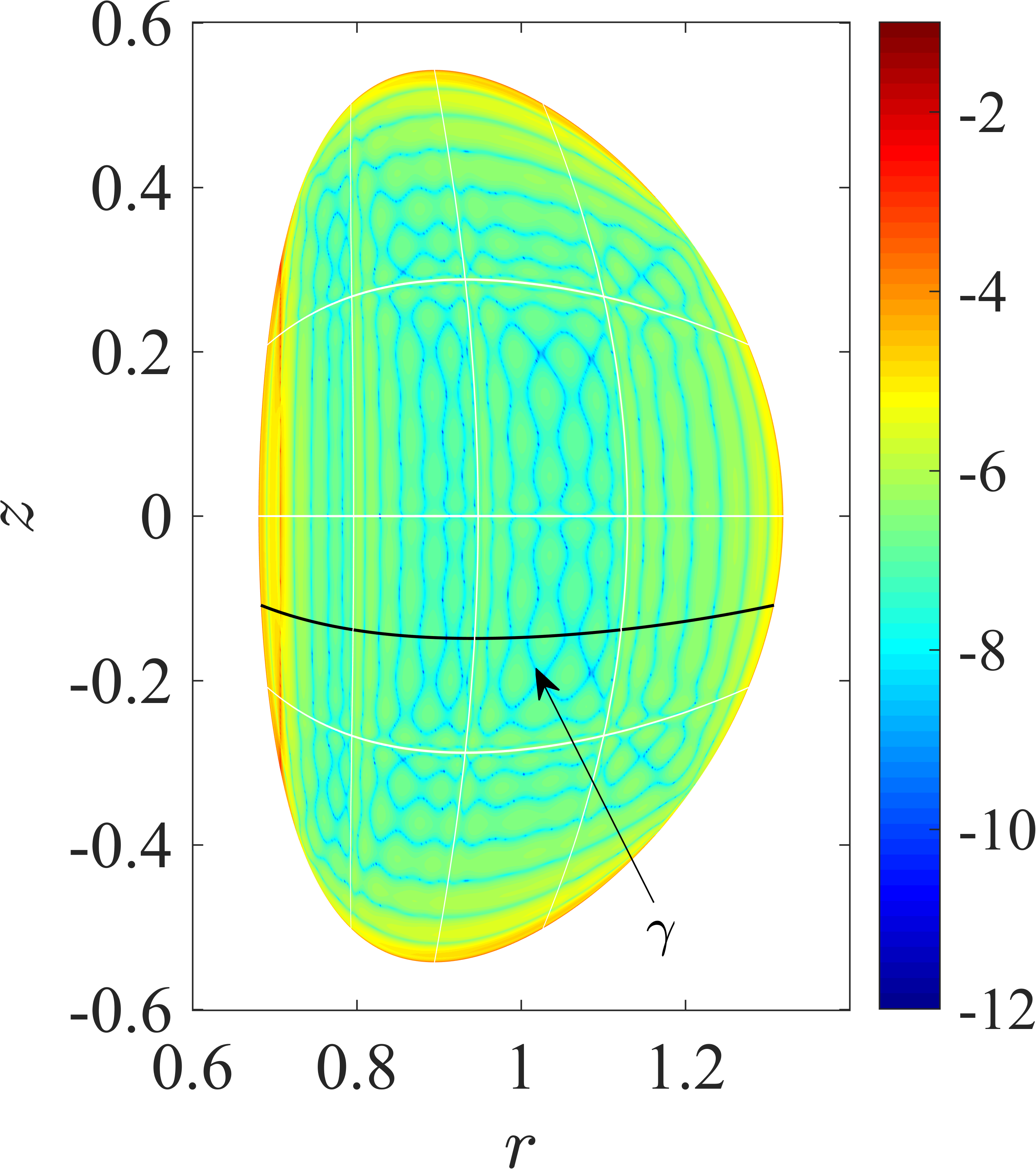}\\
				\includegraphics[height=0.271\textwidth]{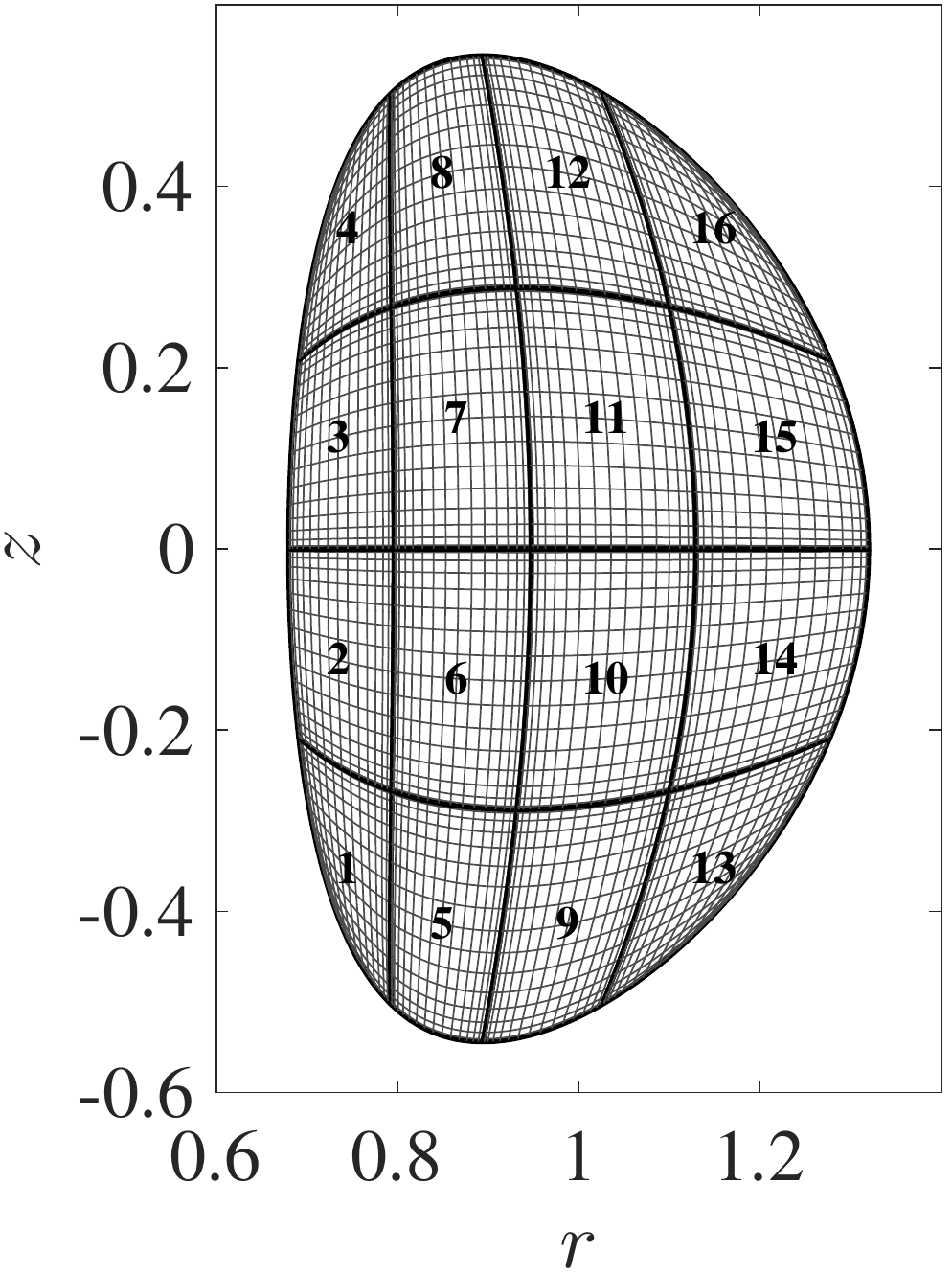}\hspace{0.75cm}
				\includegraphics[height=0.271\textwidth]{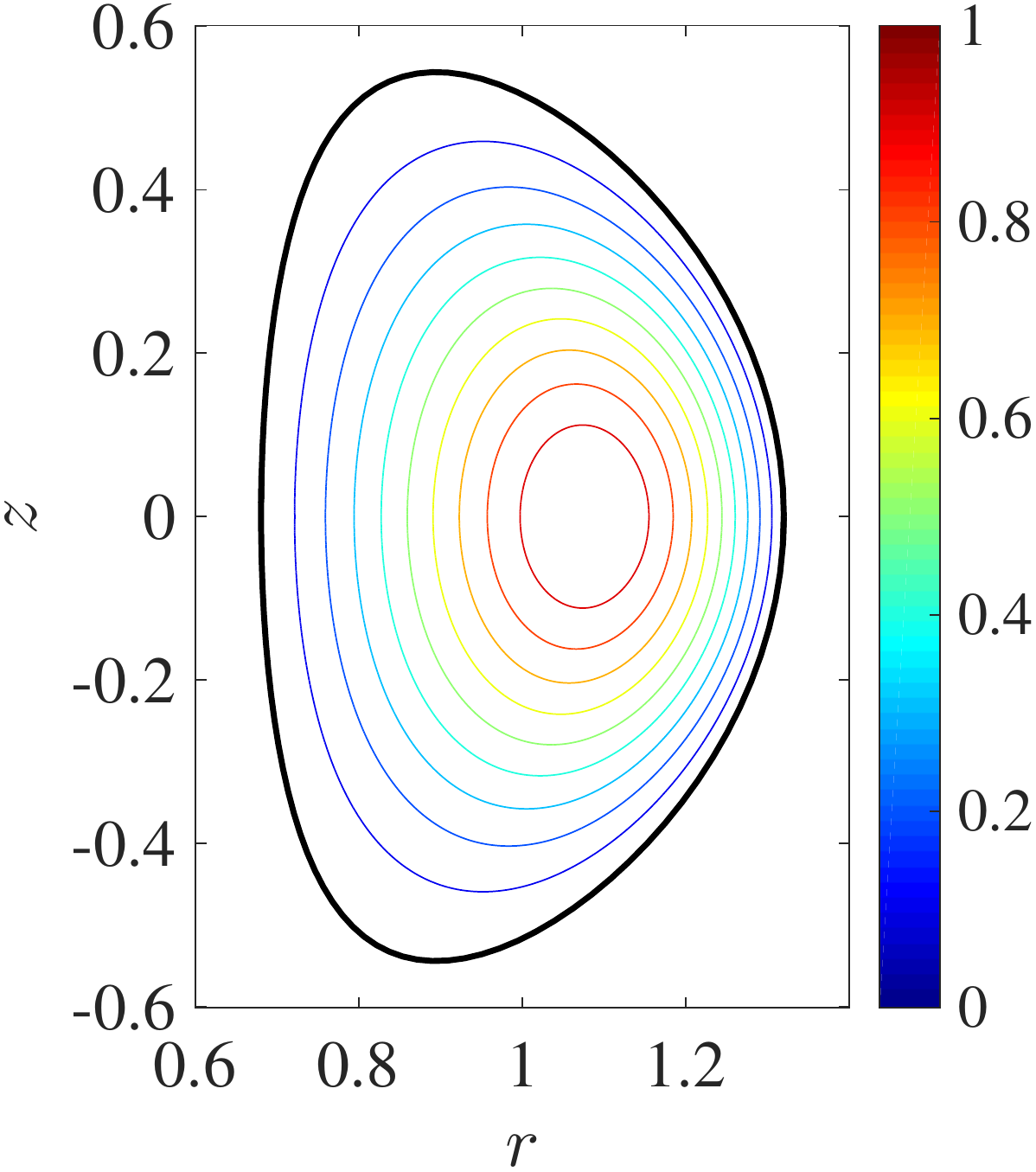}\hspace{0.75cm}
				\includegraphics[height=0.271\textwidth]{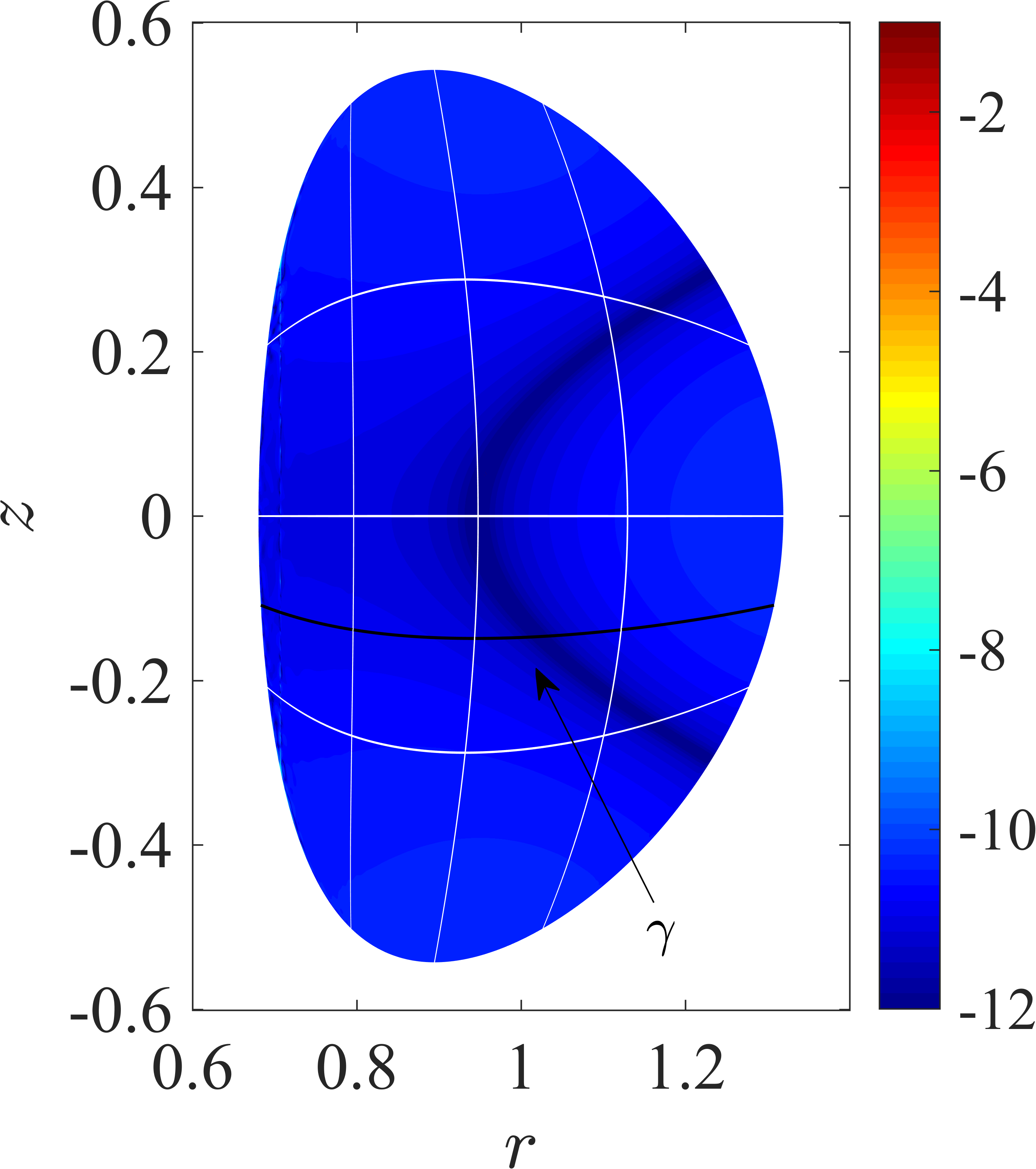}\\
				\end{center}
				\caption{\reviewerone{Numerical solution of the linear eigenvalue ITER test case, \eqref{eq::linear_eigenvalue_test_case}, with $|\psi_{0}|=1.0$, $\epsilon=0.32$, $\kappa = 1.7$, $\delta = 0.33$ and $C_{1}=-1.0$ and $C_{2}=2.0$ (high Shafranov shift). Top: mesh of $4\times 4$ elements and elements of polynomial degree $p=8$. Bottom: mesh of $4\times 4$ elements and elements of polynomial degree $p=16$. From left to right: computational mesh, numerical solution, and error as given by \eqref{eq:nonlinear_error}.}}
				\label{fig::linear_eigenvalue_iter_test_case}
			\end{figure}
			
			\begin{figure}[!ht]
				\begin{center}
				\includegraphics[height=0.3\textwidth]{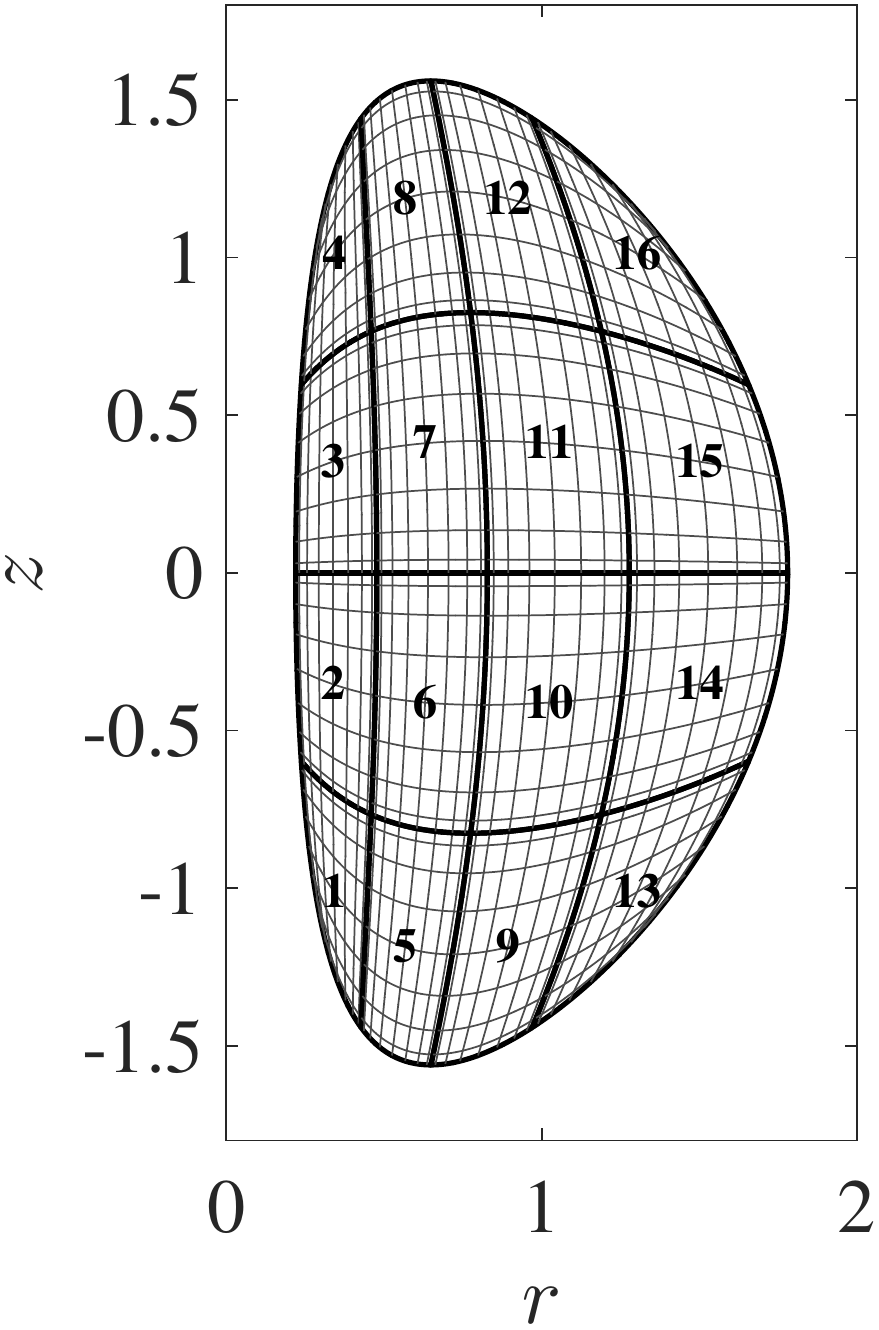}\hspace{0.75cm}
				\includegraphics[height=0.3\textwidth]{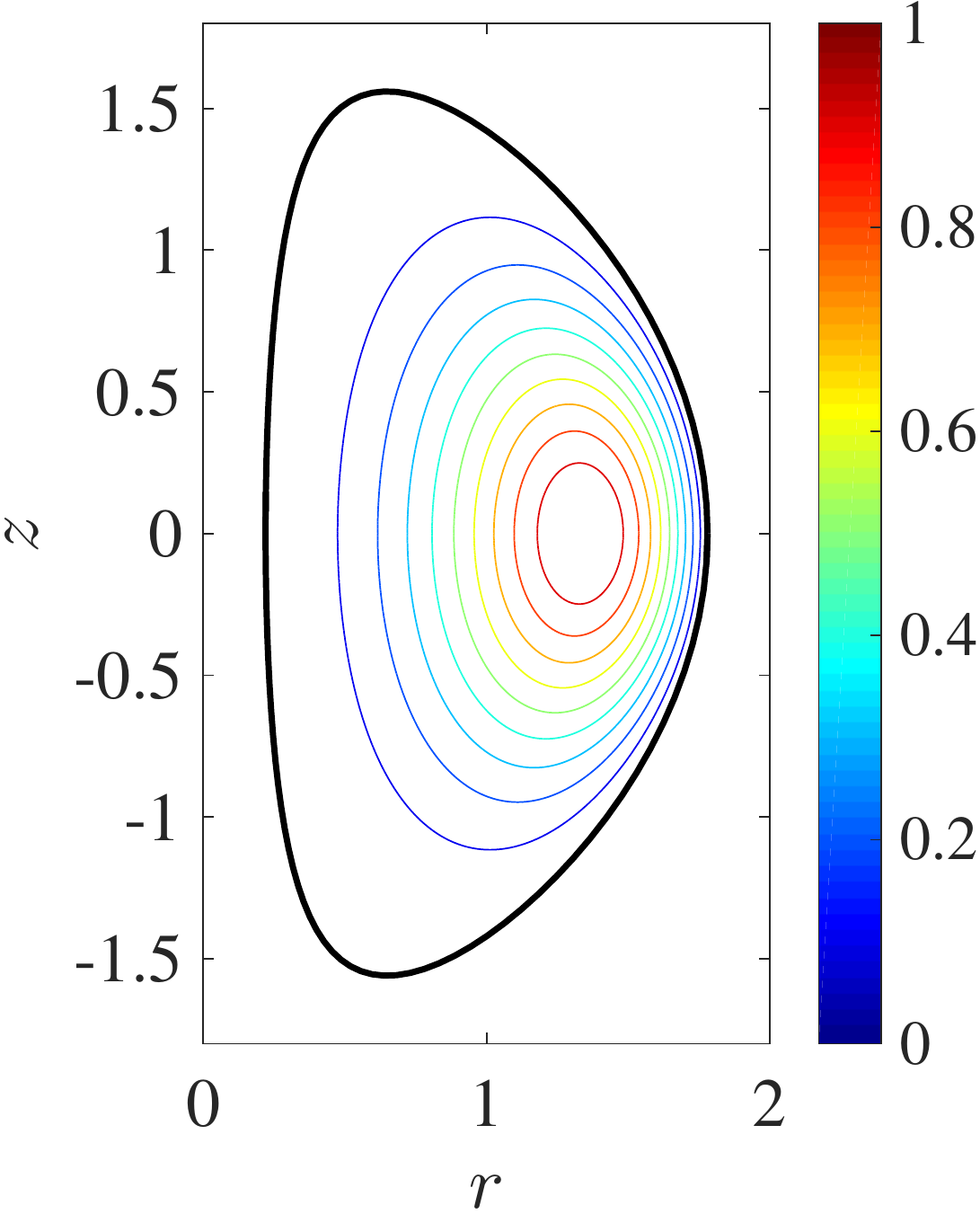}\hspace{0.75cm}
				\includegraphics[height=0.3\textwidth]{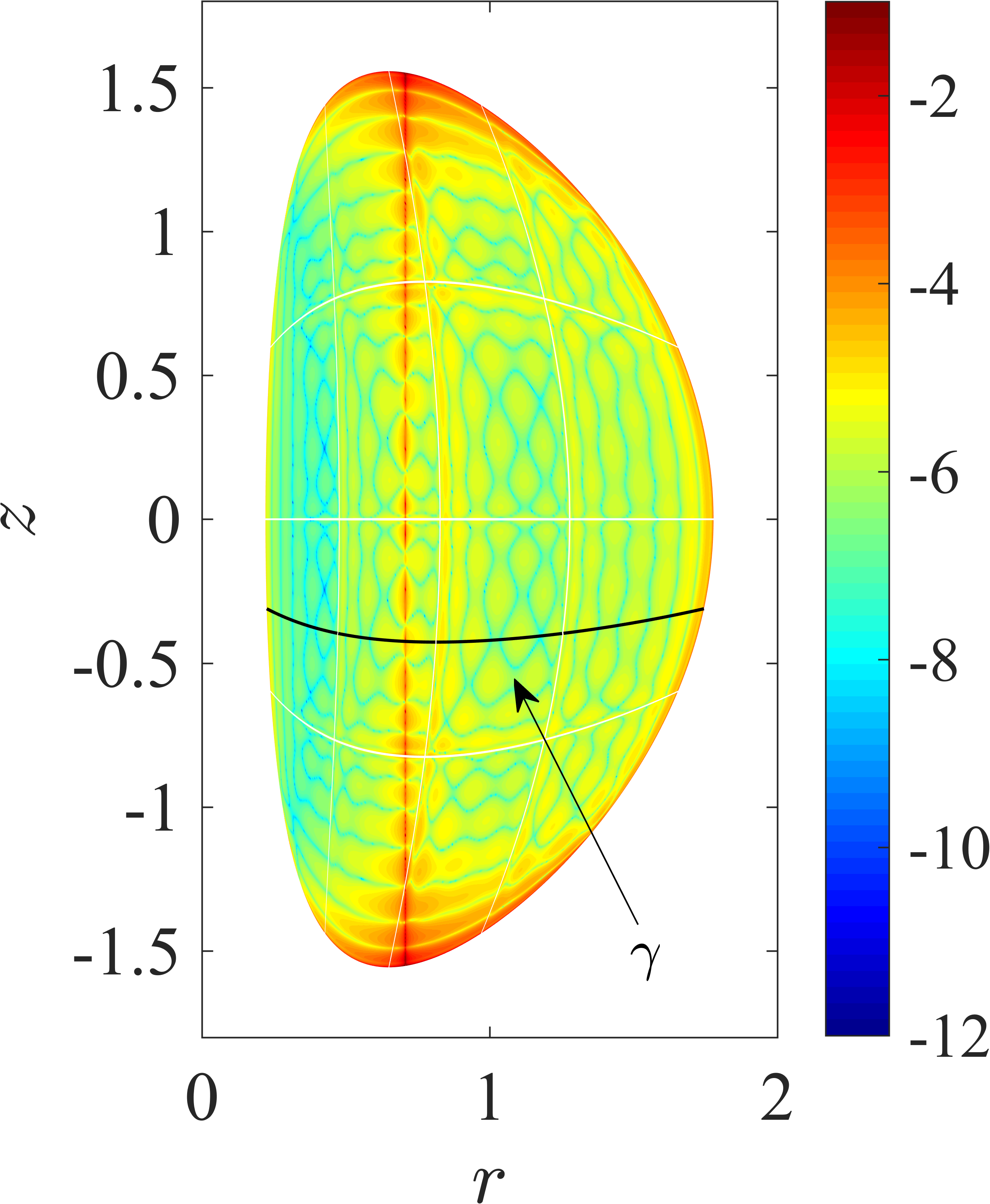}\\
				\includegraphics[height=0.3\textwidth]{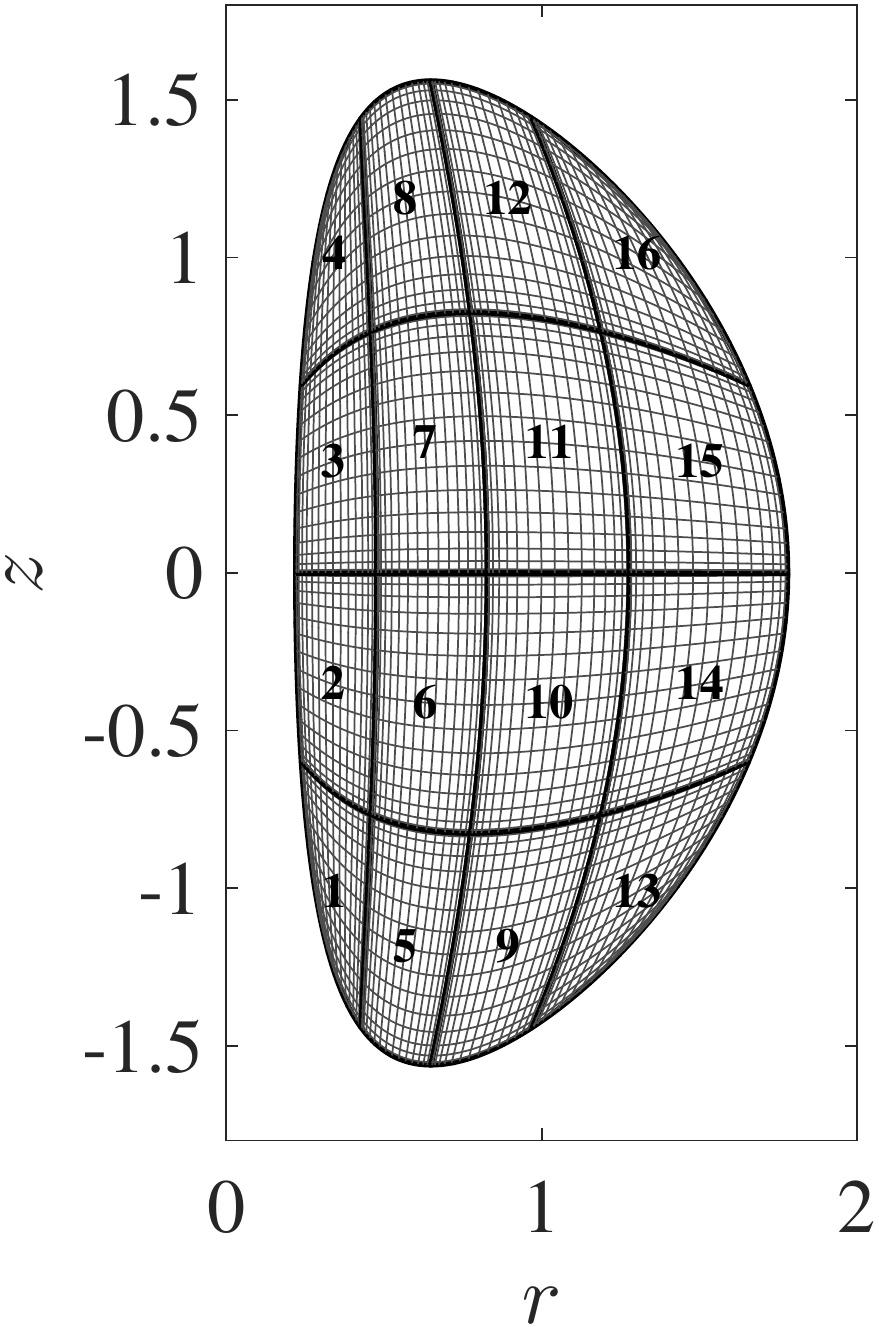}\hspace{0.75cm}
				\includegraphics[height=0.3\textwidth]{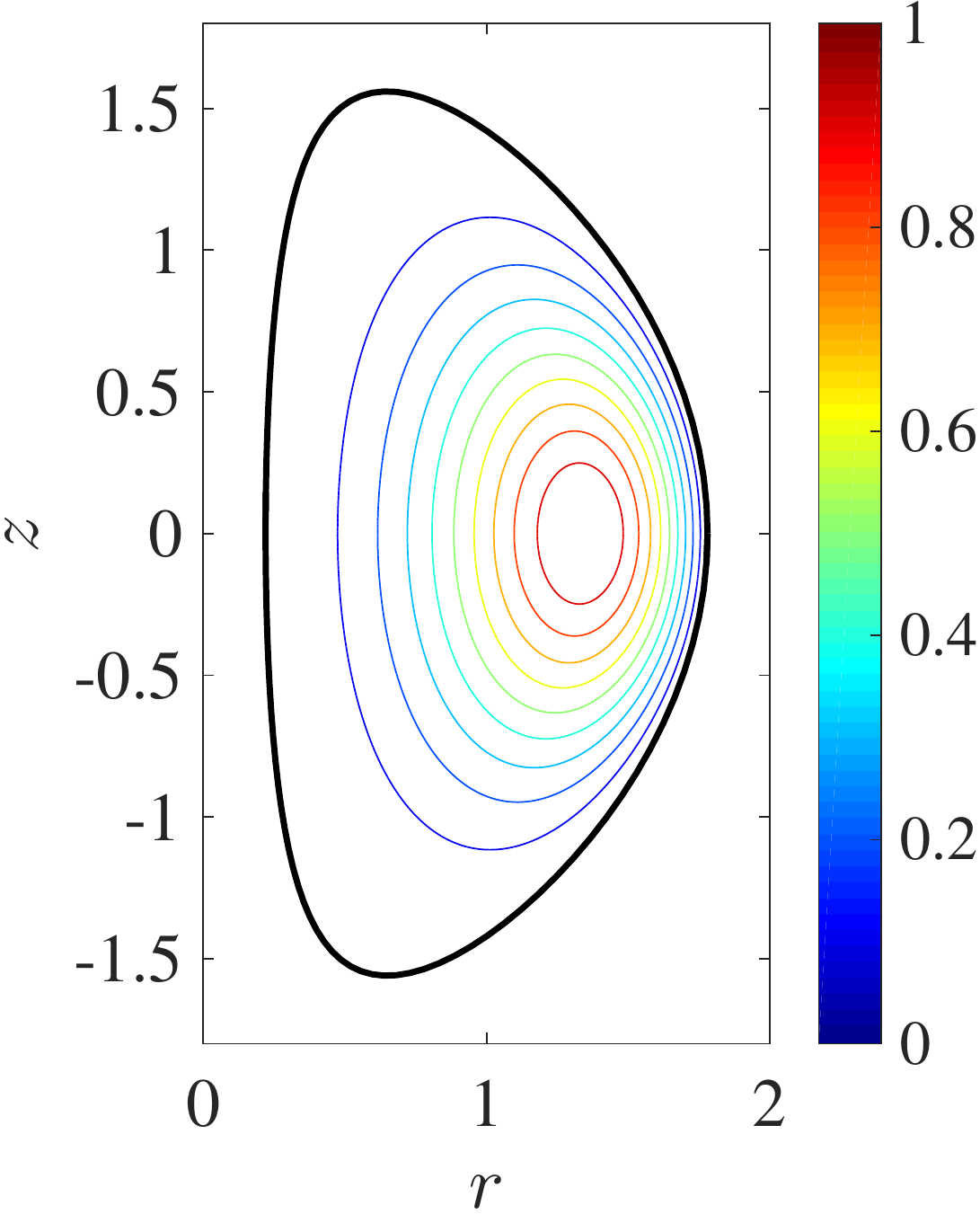}\hspace{0.75cm}
				\includegraphics[height=0.3\textwidth]{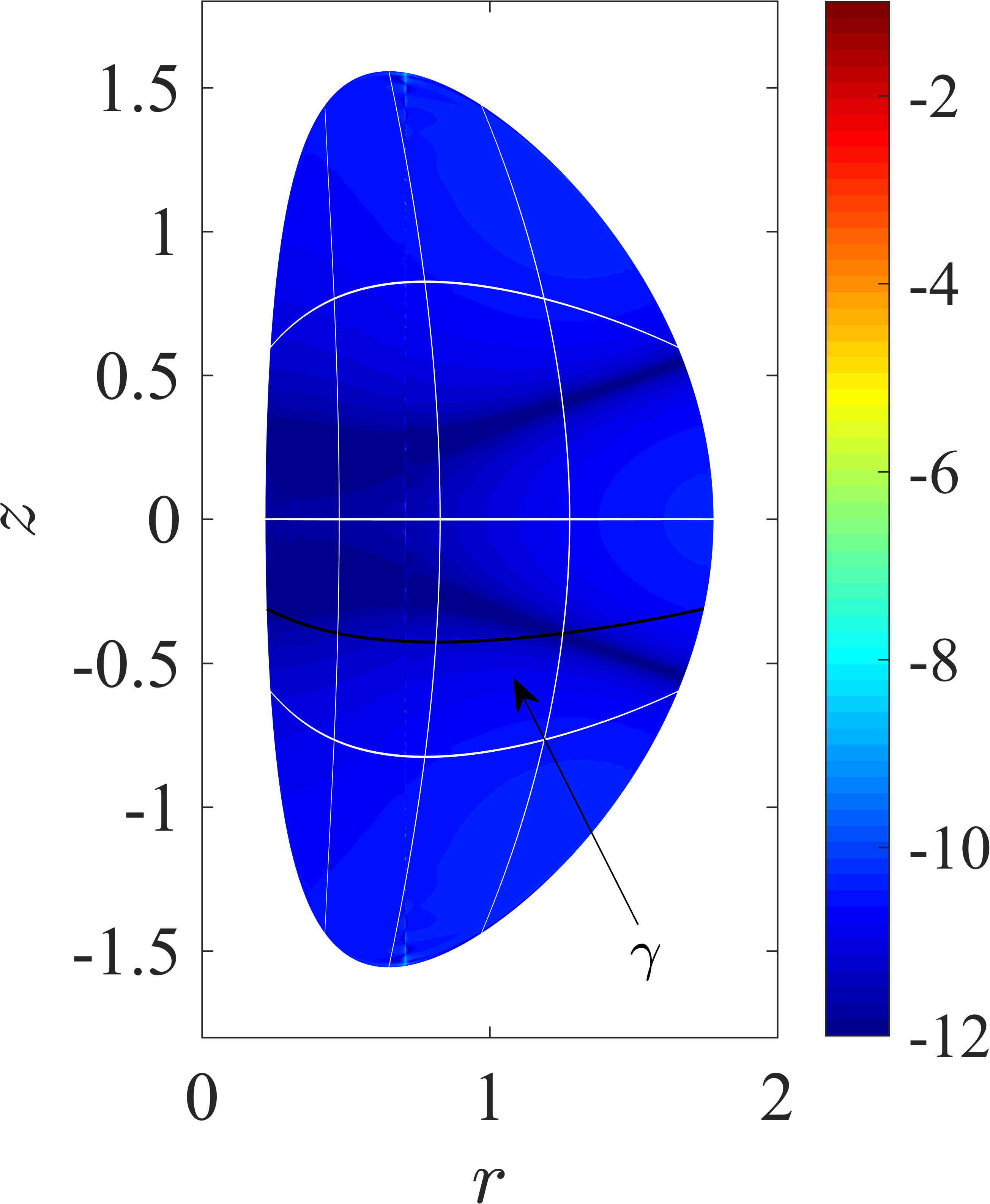}\\
				\end{center}
				\caption{\reviewerone{Numerical solution of the linear eigenvalue NSTX test case, \eqref{eq::linear_eigenvalue_test_case}, with $|\psi_{0}|=1.0$, $\epsilon=0.78$, $\kappa = 2.0$, $\delta = 0.45$ and $C_{1}=-1.0$ and $C_{2}=2.0$ (high Shafranov shift). Top: mesh of $4\times 4$ elements and elements of polynomial degree $p=8$. Bottom: mesh of $4\times 4$ elements and elements of polynomial degree $p=16$. From left to right: computational mesh, numerical solution, and error as given by \eqref{eq:nonlinear_error}.}}
				\label{fig::linear_eigenvalue_nstx_test_case}
			\end{figure}
			
			For the ease of comparison, the error \eqref{eq:nonlinear_error} along the line $\gamma$ (see \figref{fig::linear_eigenvalue_iter_test_case} for ITER test case and \figref{fig::linear_eigenvalue_nstx_test_case} for NSTX test case) is shown in \figref{fig::linear_eigenvalue_nstx_test_case_error_along_line}.
			
			\begin{figure}[!ht]
				\begin{center}
				\includegraphics[width=0.395\textwidth]{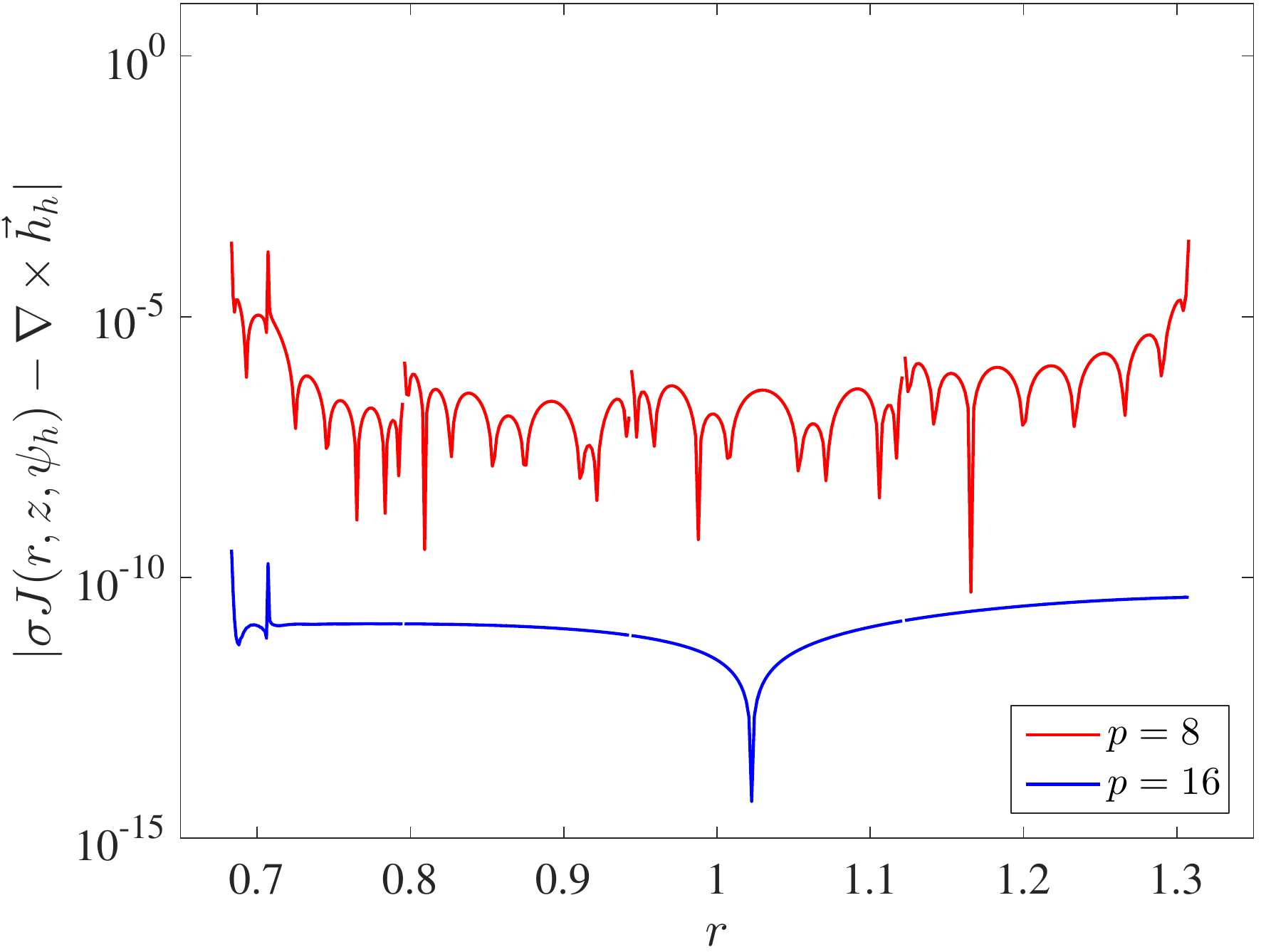}\hspace{0.5cm}
				\includegraphics[width=0.4\textwidth]{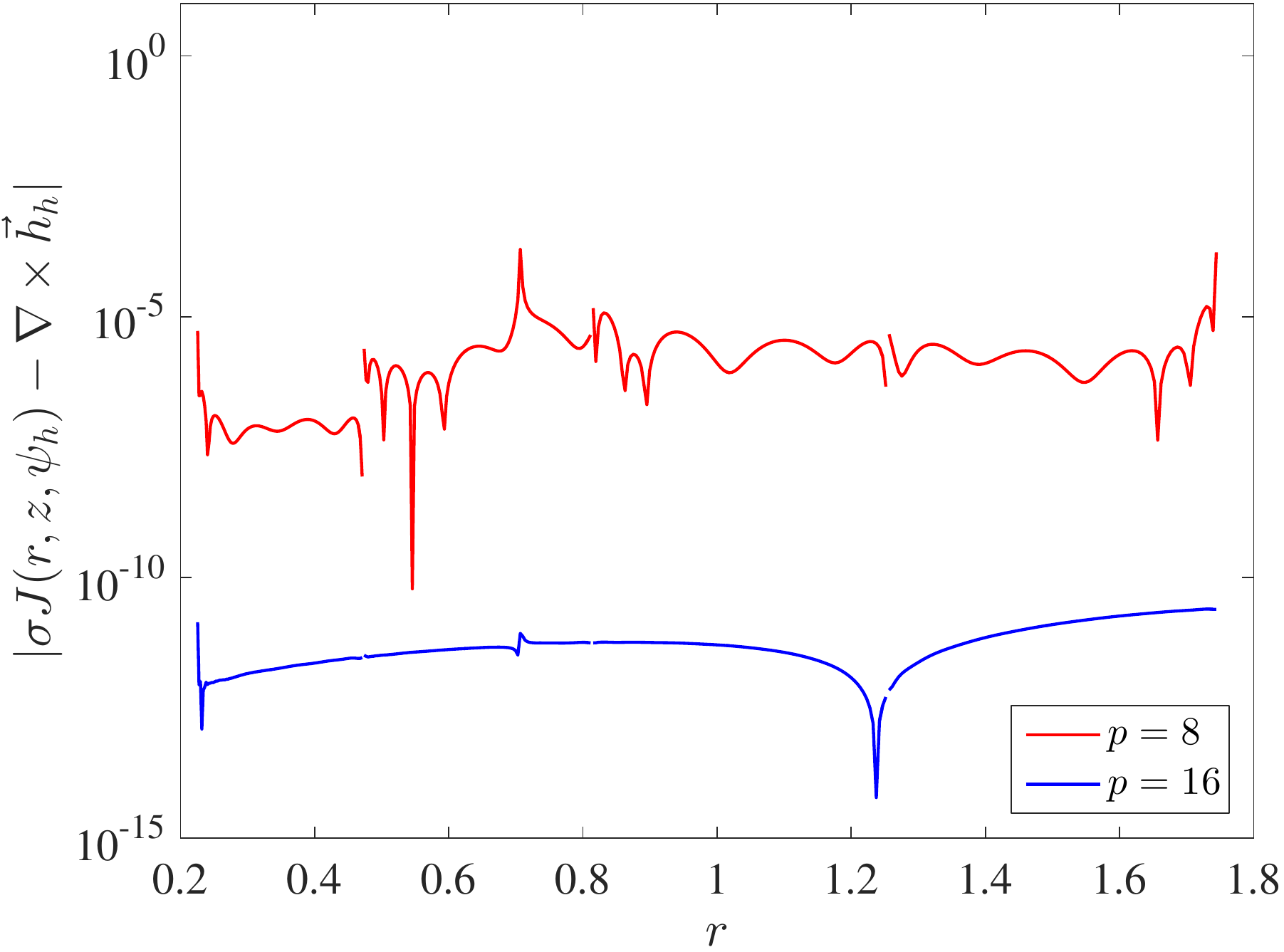}
				\end{center}
				\caption{\reviewerone{Error, as given by \eqref{eq:nonlinear_error}, along the curve $\gamma$ (see \figref{fig::linear_eigenvalue_iter_test_case} for ITER test case and \figref{fig::linear_eigenvalue_nstx_test_case} for NSTX test case). Left: ITER test case. Right:  NSTX test case.}}
				\label{fig::linear_eigenvalue_nstx_test_case_error_along_line}
			\end{figure}		
			
			\end{reviewer1}
			\FloatBarrier
			
		\subsubsection{Non-linear eigenvalue problem} \label{sec::test_cases_nonlinear_eigenvalue}
			\begin{reviewer1}
			To further assess the proposed numerical model on a fully non-linear problem, we apply it to the non-linear test case presented in \cite{Pataki2013}. This test case corresponds to a configuration with a pressure pedestal and consists of the following models:
			\begin{equation}
				f(\psi) = 0 \qquad\mbox{and}\qquad P(\psi) = \frac{C_{1} + C_{2}\psi^{2}}{\mu_{0}}\left(1-e^{-\frac{\psi^{2}}{\eta}}\right)\,, \label{eq::non-linear_pedestal_models}
			\end{equation}
			which lead to the following non-linear Grad-Shafranov problem
			\begin{equation}
				\begin{dcases}
					\nabla\times\left(\mathbb{K}\nabla\times\psi\right) = 2r\left[C_{2}\psi\left(1-e^{-\frac{\psi^{2}}{\eta}}\right) + \frac{C_{1}+C_{2}\psi^{2}}{\eta}\,\psi\eta^{-\frac{\psi^{2}}{\eta}}\right] & \mbox{in} \quad \Omega_{p}\,, \\
					\psi = 0 & \mbox{on}\quad\partial\Omega_{p}\,,
				\end{dcases} \label{eq::nonlinear_eigenvalue_test_case}
			\end{equation}
			with the plasma boundary $\Omega_{p}$ given by \eqref{eq::plasma_shape}.
			
			For the coefficients $C_{i}$ we use the same values as in \cite{Pataki2013}, $C_{1}=0.8$ and $C_{2} = 0.2$, and we set $\eta=0.1$. ITER (\figref{fig::nonlinear_eigenvalue_iter_test_case}) and NSTX (\figref{fig::nonlinear_eigenvalue_nstx_test_case}) plasma shapes, as in \secref{sec::test_cases_linear_eigenvalue}, are tested.
 			
			As can be seen in \figref{fig::nonlinear_eigenvalue_iter_test_case} and \figref{fig::nonlinear_eigenvalue_nstx_test_case}, an increase in the polynomial degree of the basis functions leads to a reduction of the error in the numerical solution. 
			\begin{figure}[!ht]
				\begin{center}
				\includegraphics[height=0.271\textwidth]{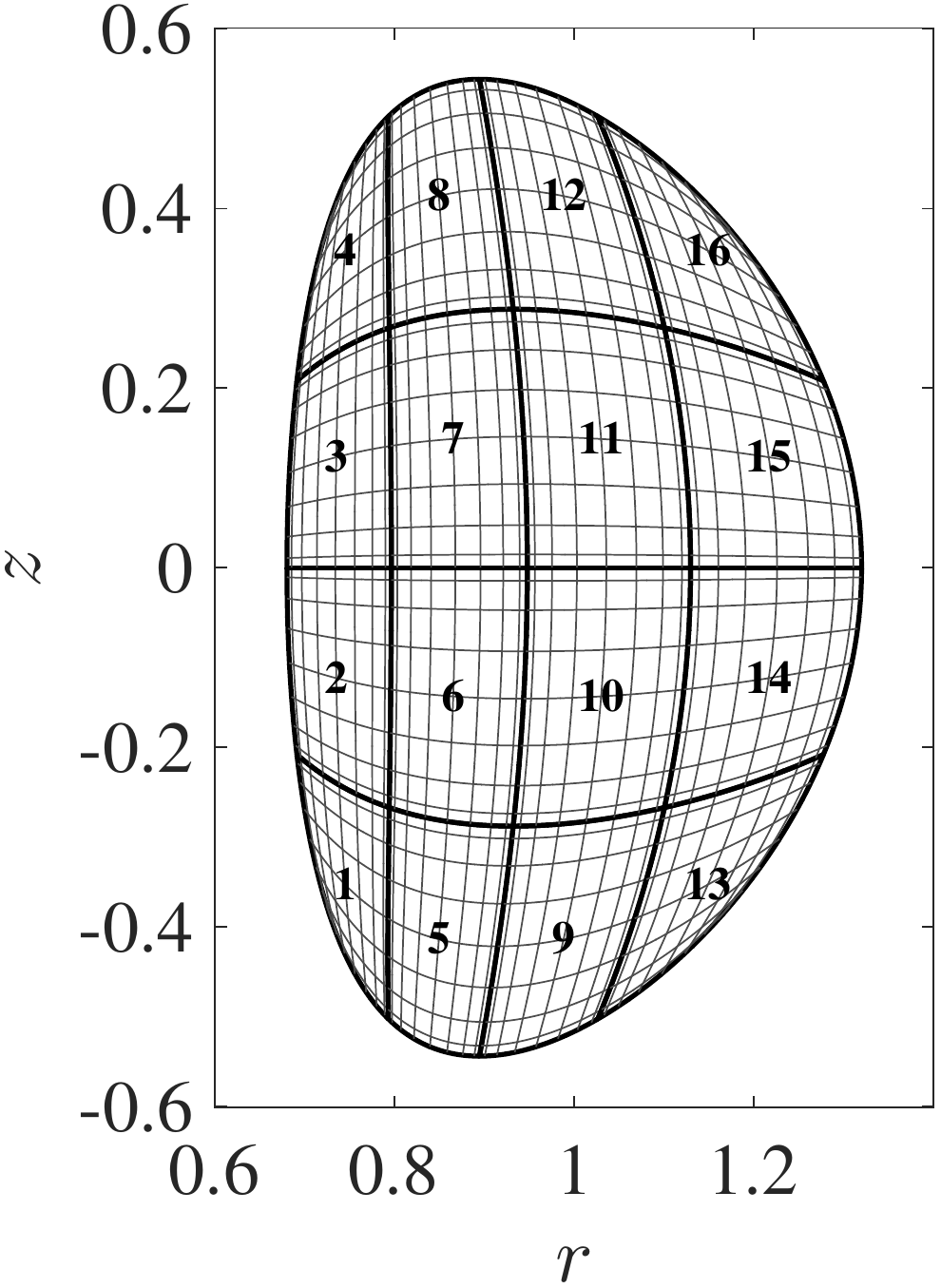}\hspace{0.75cm}
				\includegraphics[height=0.271\textwidth]{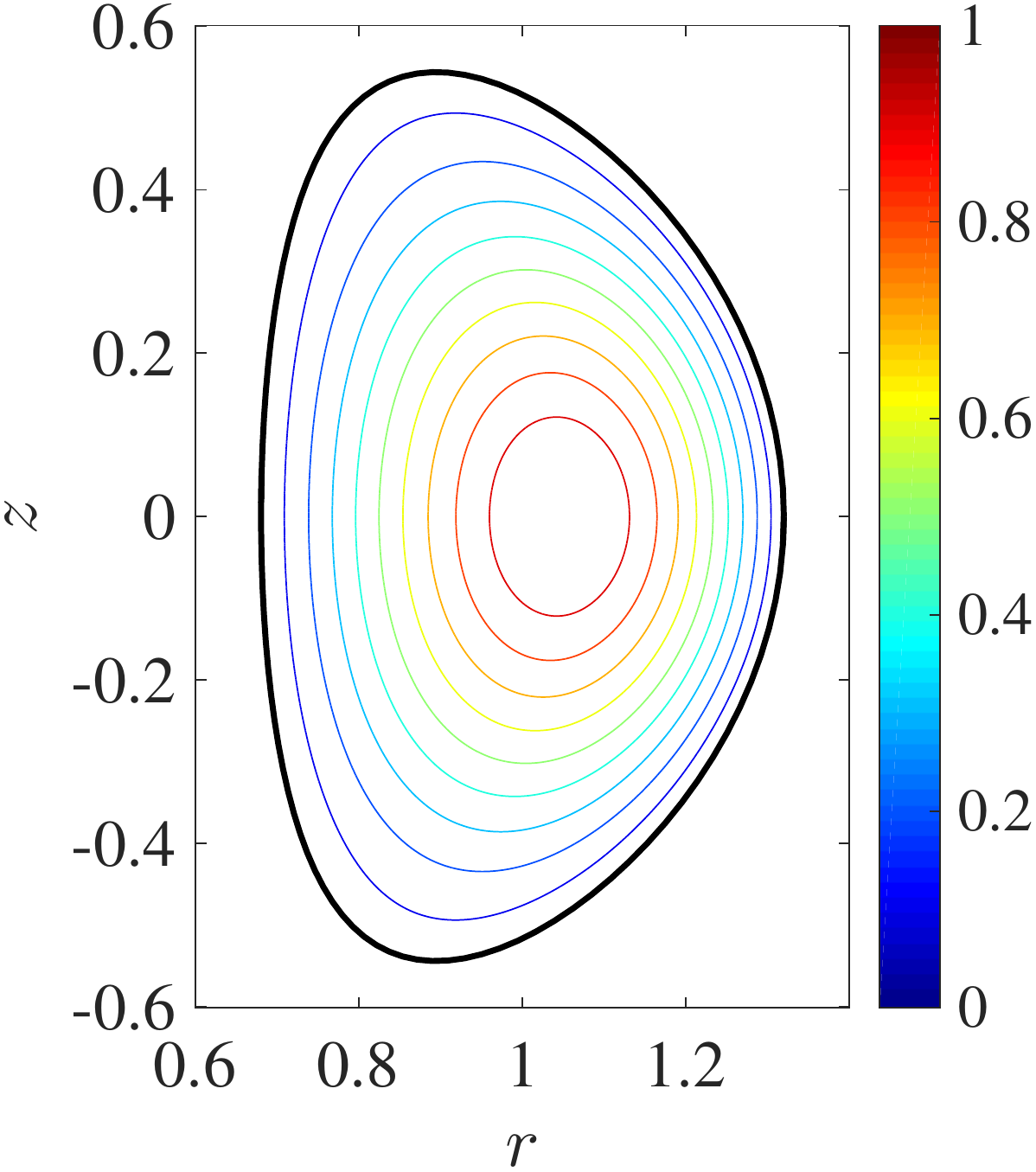}\hspace{0.75cm}
				\includegraphics[height=0.271\textwidth]{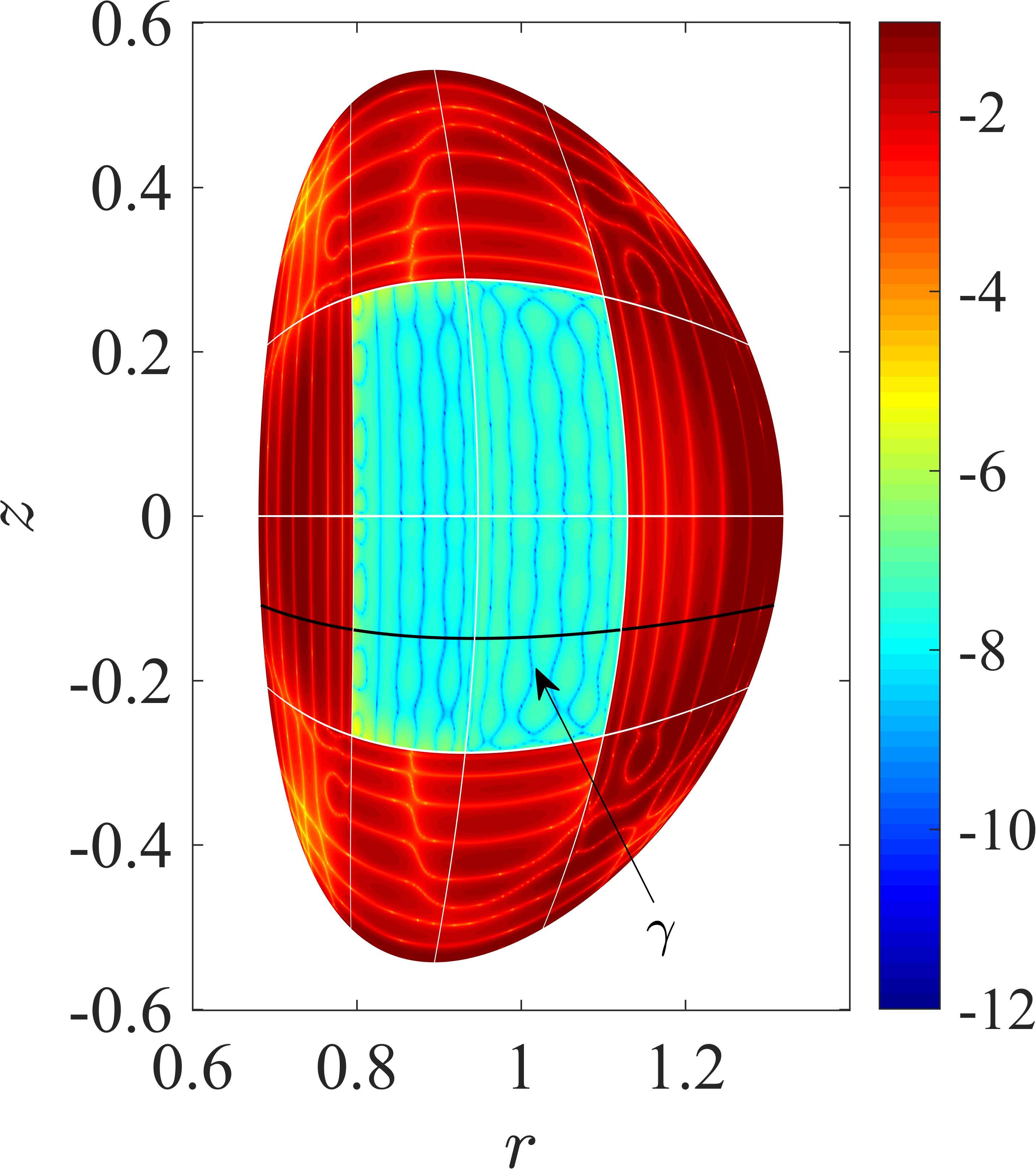}\\
				\includegraphics[height=0.271\textwidth]{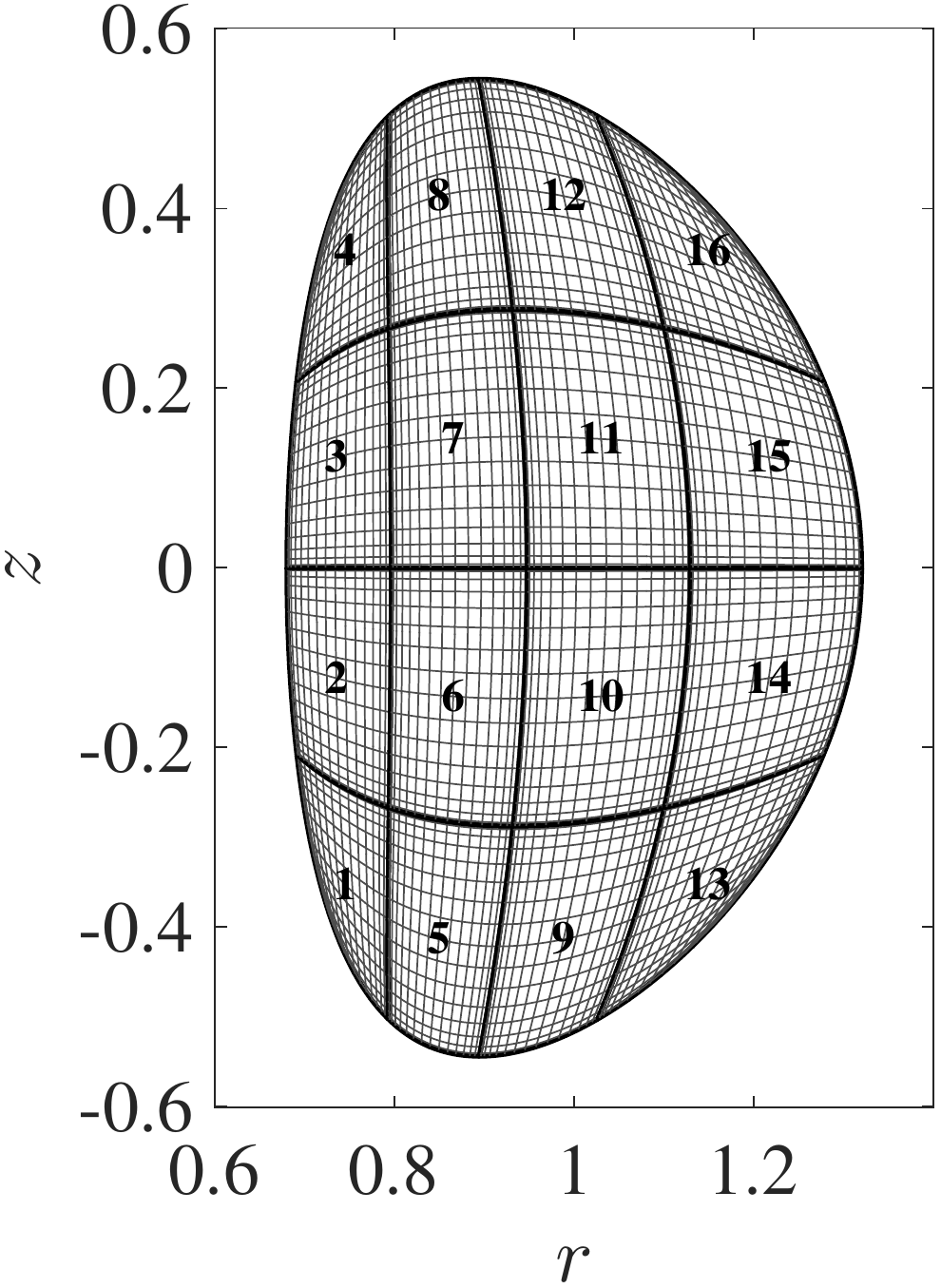}\hspace{0.75cm}
				\includegraphics[height=0.271\textwidth]{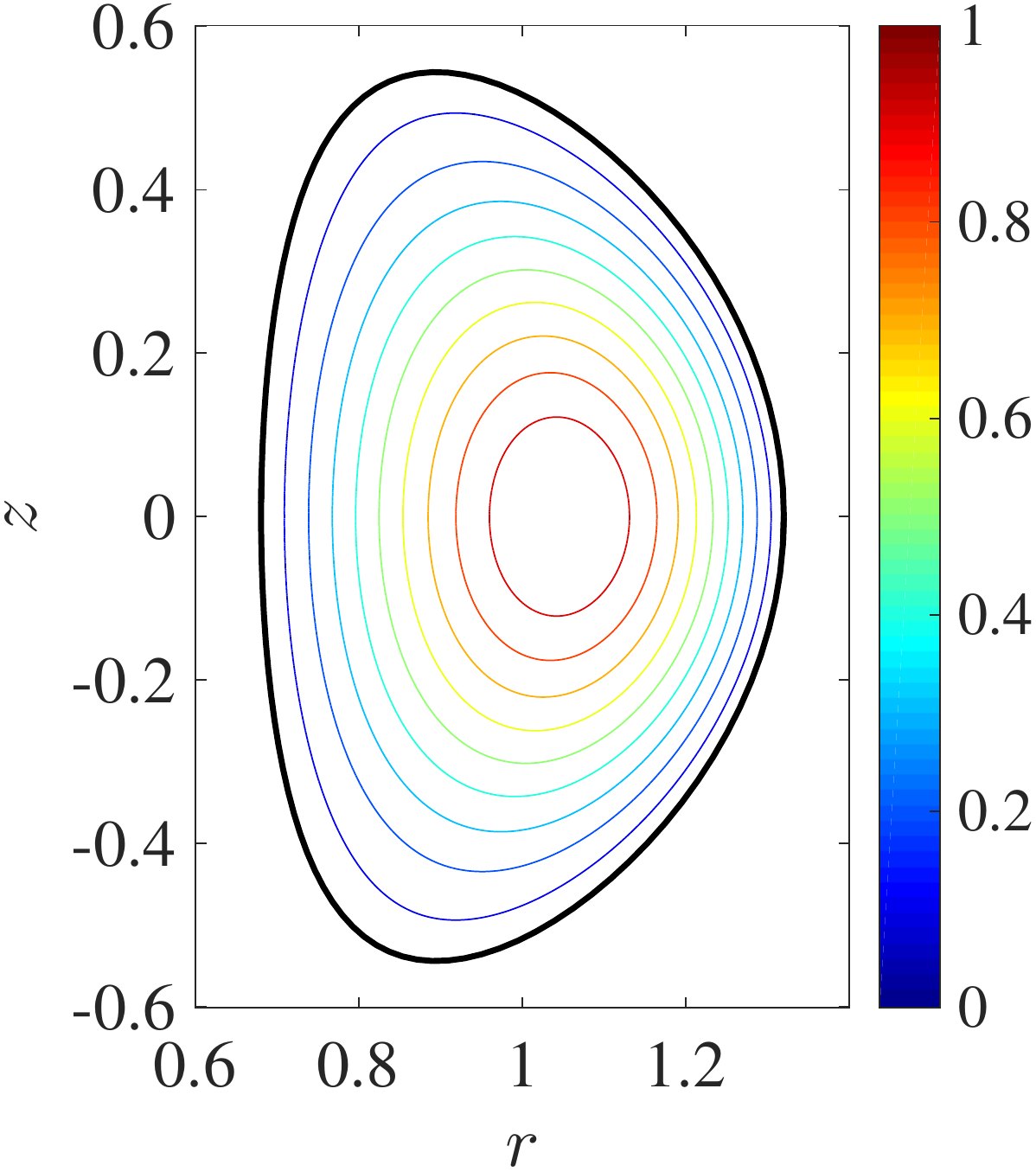}\hspace{0.75cm}
				\includegraphics[height=0.271\textwidth]{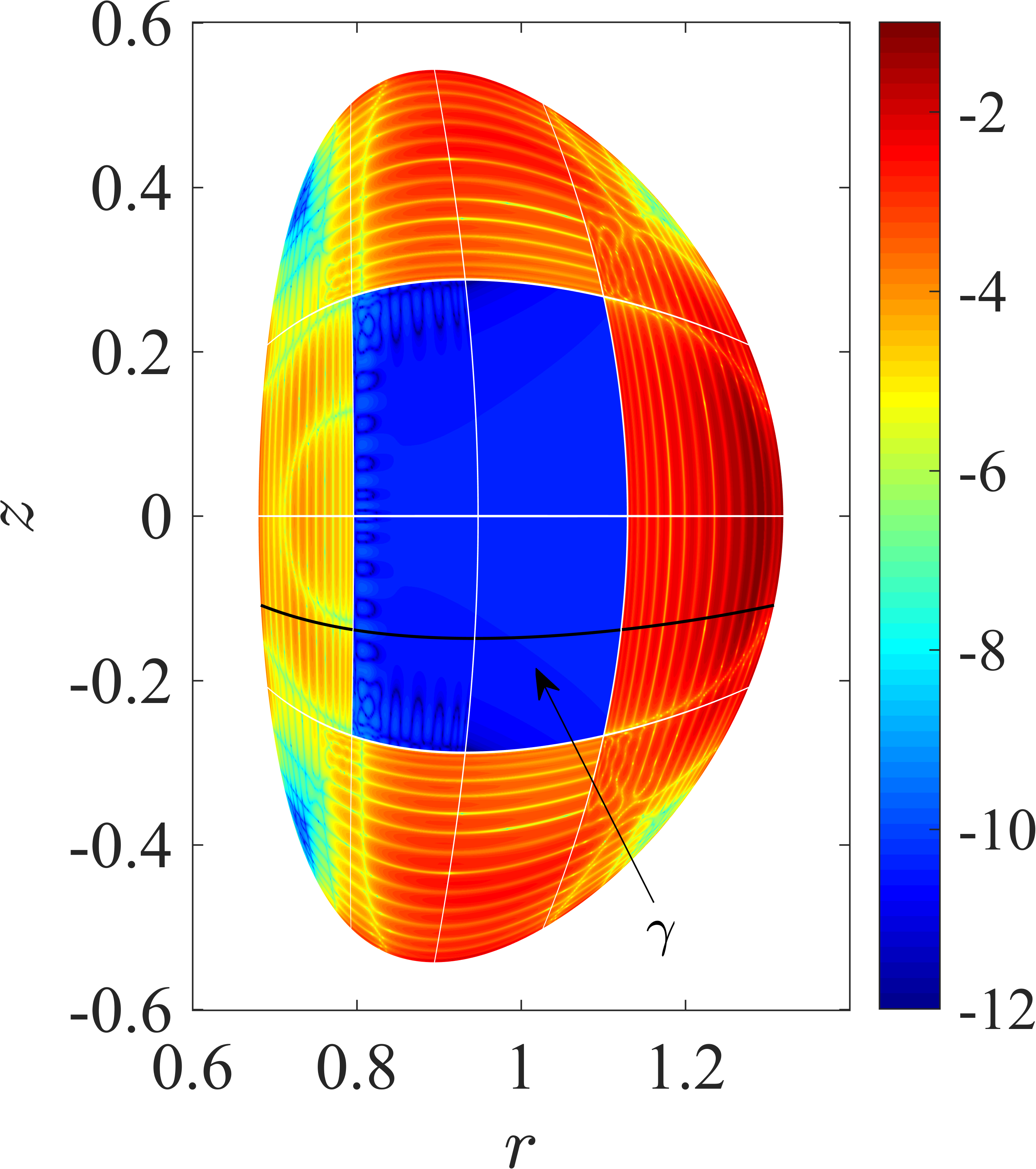}\\
				\end{center}
				\caption{\reviewerone{Numerical solution of the non-linear eigenvalue ITER test case, \eqref{eq::nonlinear_eigenvalue_test_case}, with $|\psi_{0}|=1.0$, $\epsilon=0.32$, $\kappa = 1.7$, $\delta = 0.33$ and $C_{1}=0.8$ and $C_{2}=0.2$ and $\eta = 0.1$. Top: mesh of $4\times 4$ elements and elements of polynomial degree $p=8$. Bottom: mesh of $4\times 4$ elements and elements of polynomial degree $p=16$. From left to right: computational mesh, numerical solution, and error as given by \eqref{eq:nonlinear_error}.}}
				\label{fig::nonlinear_eigenvalue_iter_test_case}
			\end{figure}
			
			\begin{figure}[!ht]
				\begin{center}
				\includegraphics[height=0.3\textwidth]{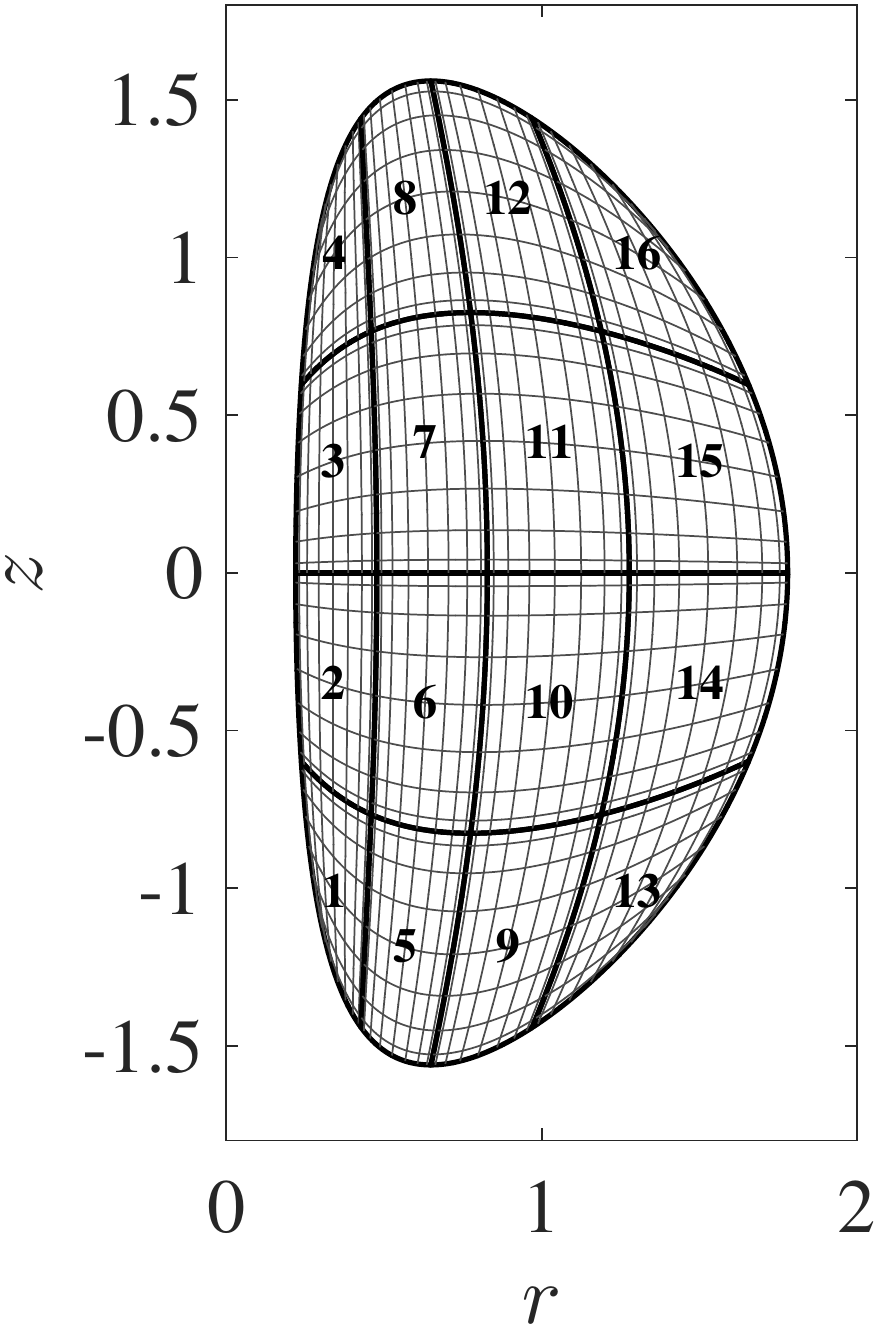}\hspace{0.75cm}
				\includegraphics[height=0.3\textwidth]{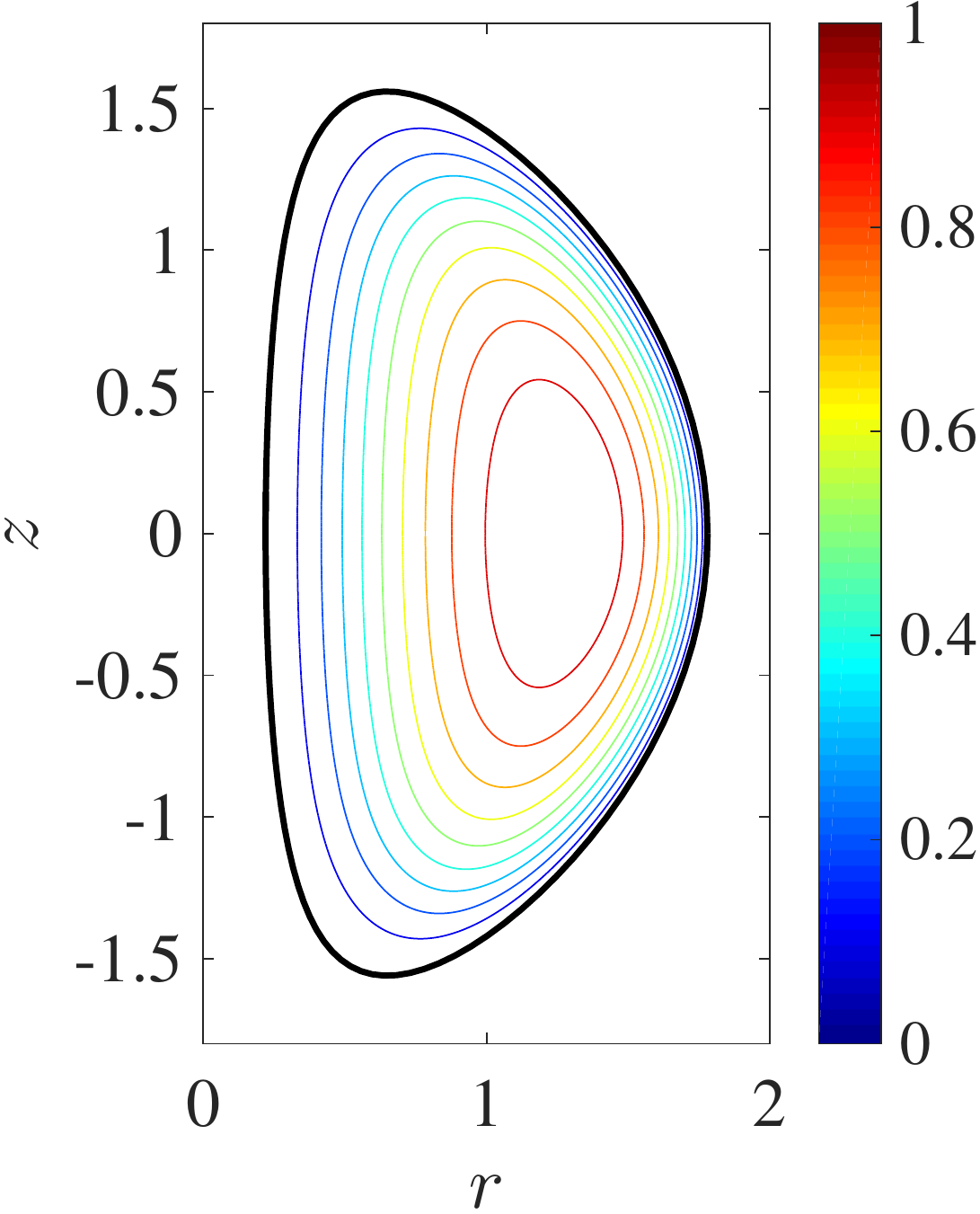}\hspace{0.75cm}
				\includegraphics[height=0.3\textwidth]{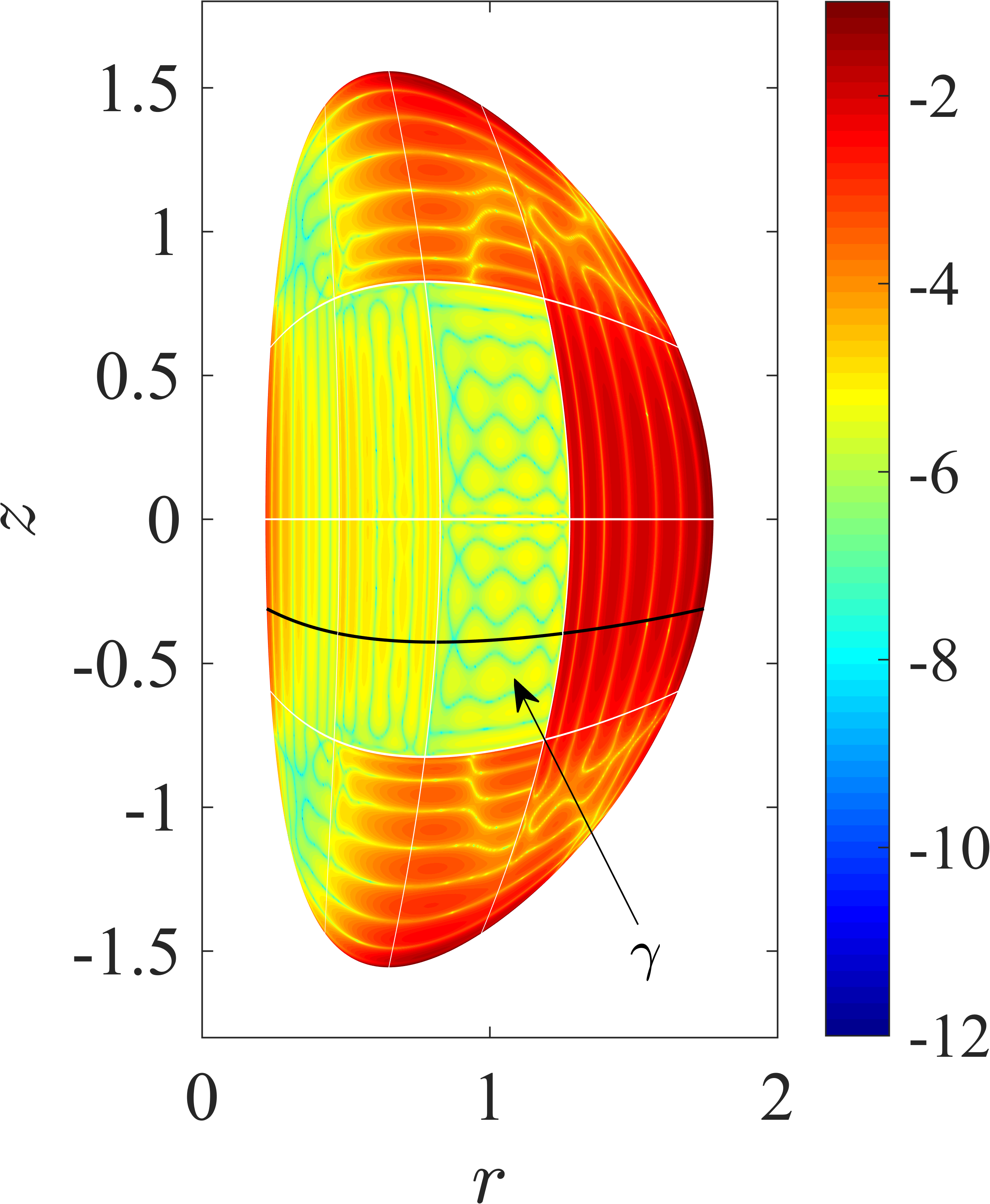}\\
				\includegraphics[height=0.3\textwidth]{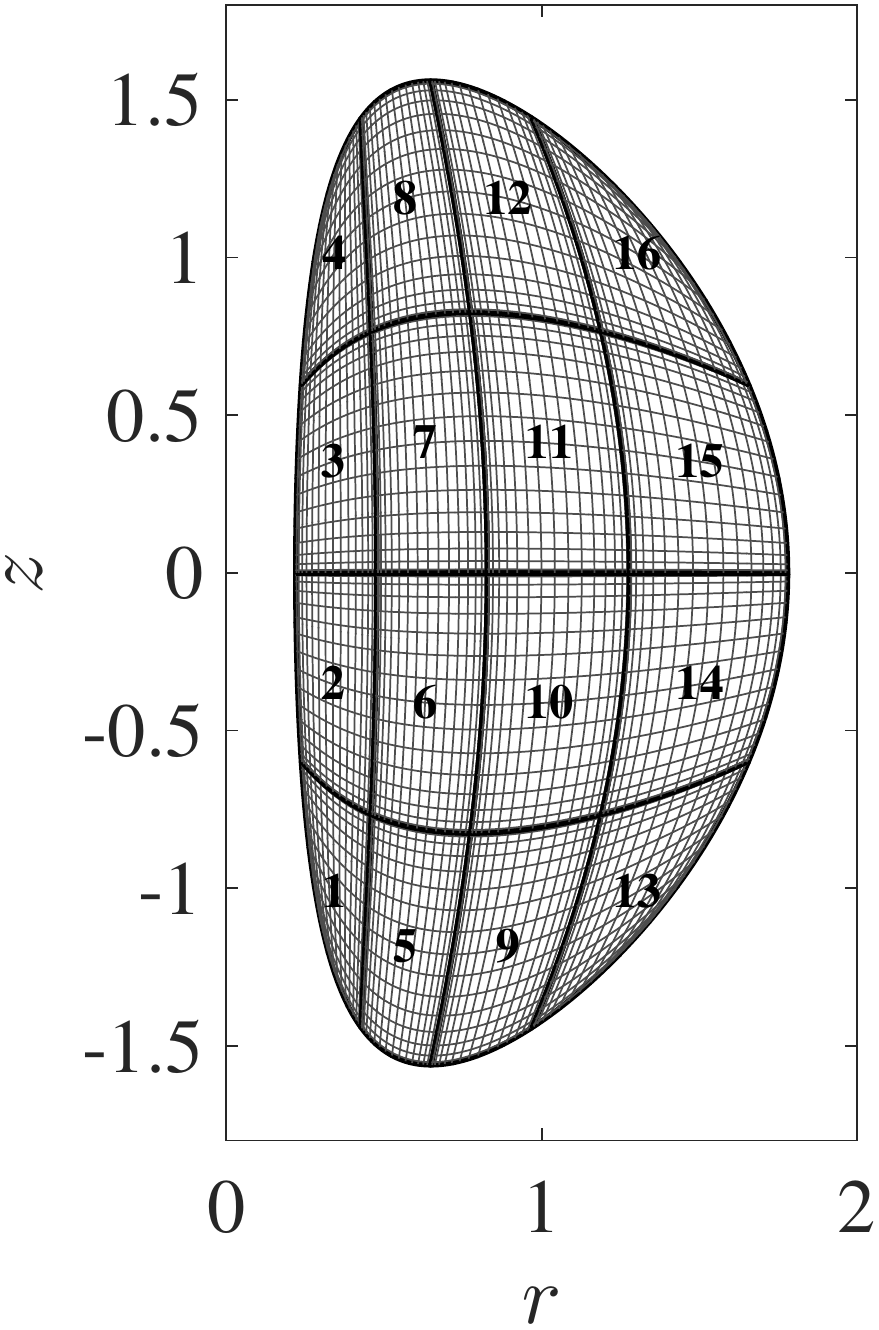}\hspace{0.75cm}
				\includegraphics[height=0.3\textwidth]{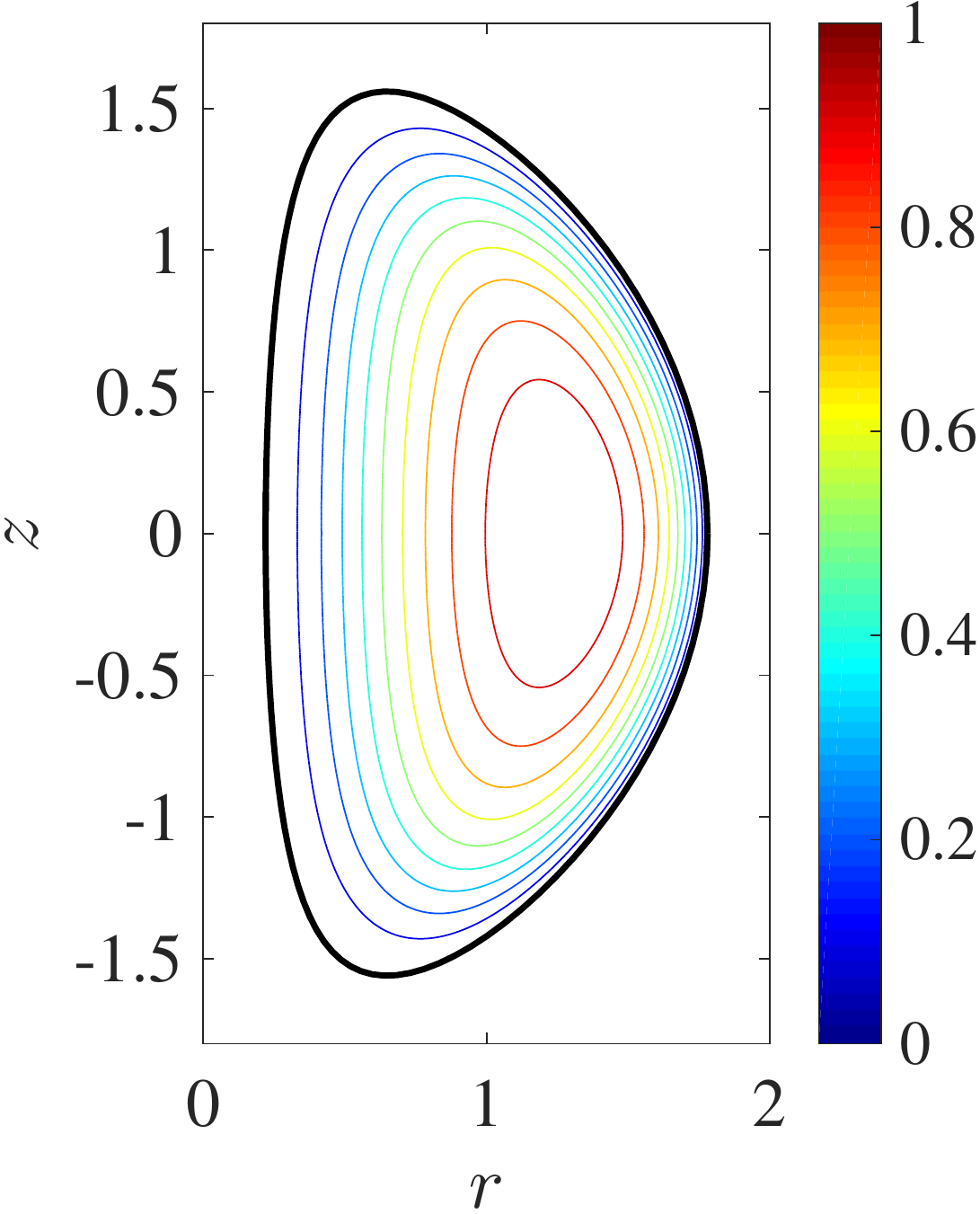}\hspace{0.75cm}
				\includegraphics[height=0.3\textwidth]{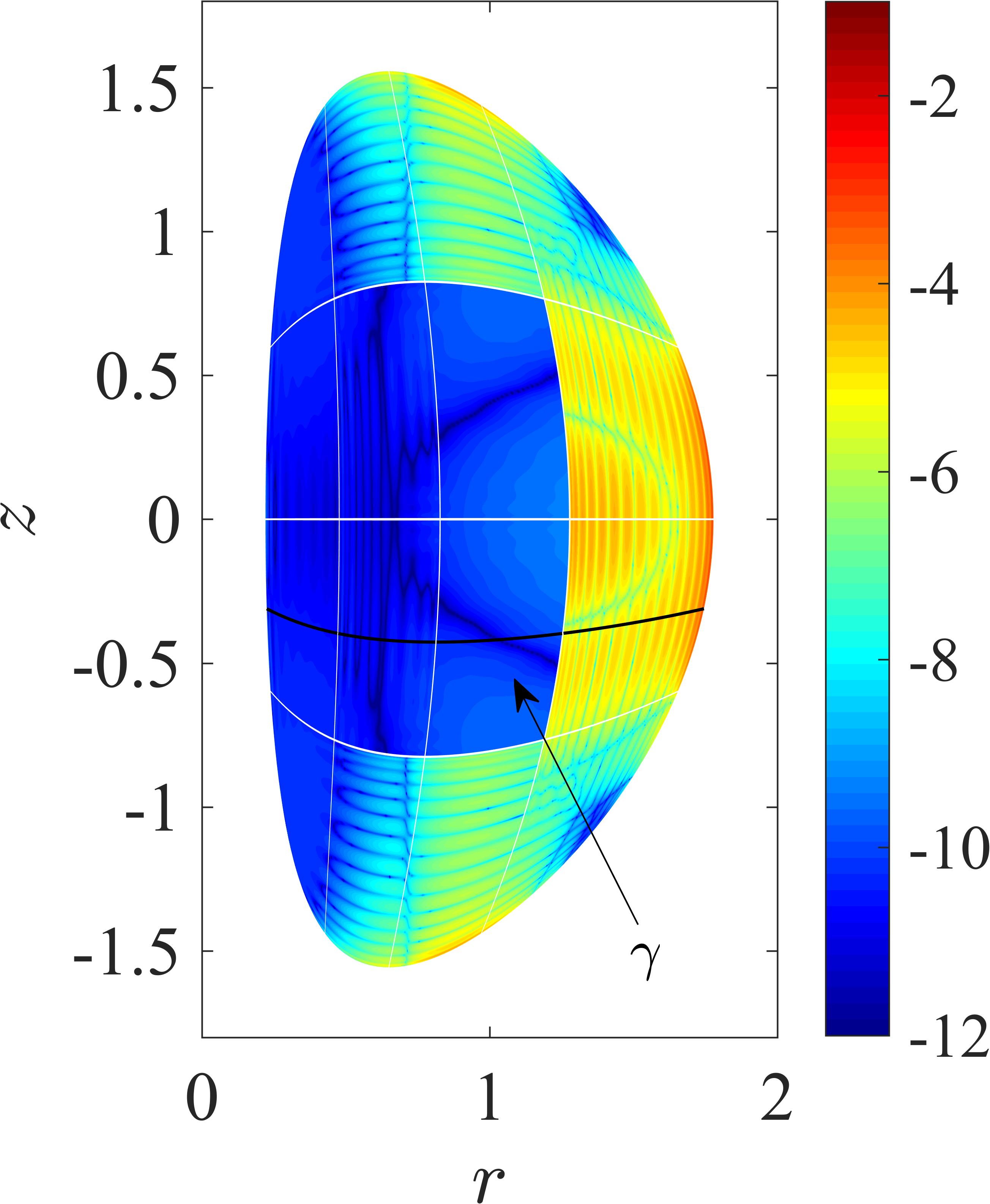}\\
				\end{center}
				\caption{\reviewerone{Numerical solution of the non-linear eigenvalue NSTX test case, \eqref{eq::nonlinear_eigenvalue_test_case}, with $|\psi_{0}|=1.0$, $\epsilon=0.78$, $\kappa = 2.0$, $\delta = 0.45$, $C_{1}=0.8$, $C_{2}=0.2$ and $\eta = 0.1$. Top: mesh of $4\times 4$ elements and elements of polynomial degree $p=8$. Bottom: mesh of $4\times 4$ elements and elements of polynomial degree $p=16$. From left to right: computational mesh, numerical solution, and error as given by \eqref{eq:nonlinear_error}.}}
				\label{fig::nonlinear_eigenvalue_nstx_test_case}
			\end{figure}
			
			An interesting aspect of this test case is that for these meshes the elements at the boundary show a substantially larger error than the interior elements. This behaviour can be explained by analysing $\nabla\times\vec{h}_{h}(r,z)$ and $\sigma J(r,z,\psi_{h})$. In \figref{fig::nonlinear_eigenvalue_nstx_test_case_J_along_line} we compare $\sigma J(r,z,\psi_{h})$ to $\nabla\times\vec{h}_{h}(r,z)$ for $p=8,16$ along the line $\gamma$ (\figref{fig::nonlinear_eigenvalue_iter_test_case} for ITER case and \figref{fig::nonlinear_eigenvalue_nstx_test_case} for NSTX case). As can be seen, for the ITER shape there is a sharp variation of $\sigma J(r,z,\psi_{h})$ close to the edge of the plasma. This sharp variation cannot be accurately recovered with $p=8$ for large elements, as used here. Nevertheless, with an increase of the polynomial degree to $p=16$, we can see a much better agreement.
			
			\begin{figure}[!ht]
				\begin{center}
				\includegraphics[width=0.4\textwidth]{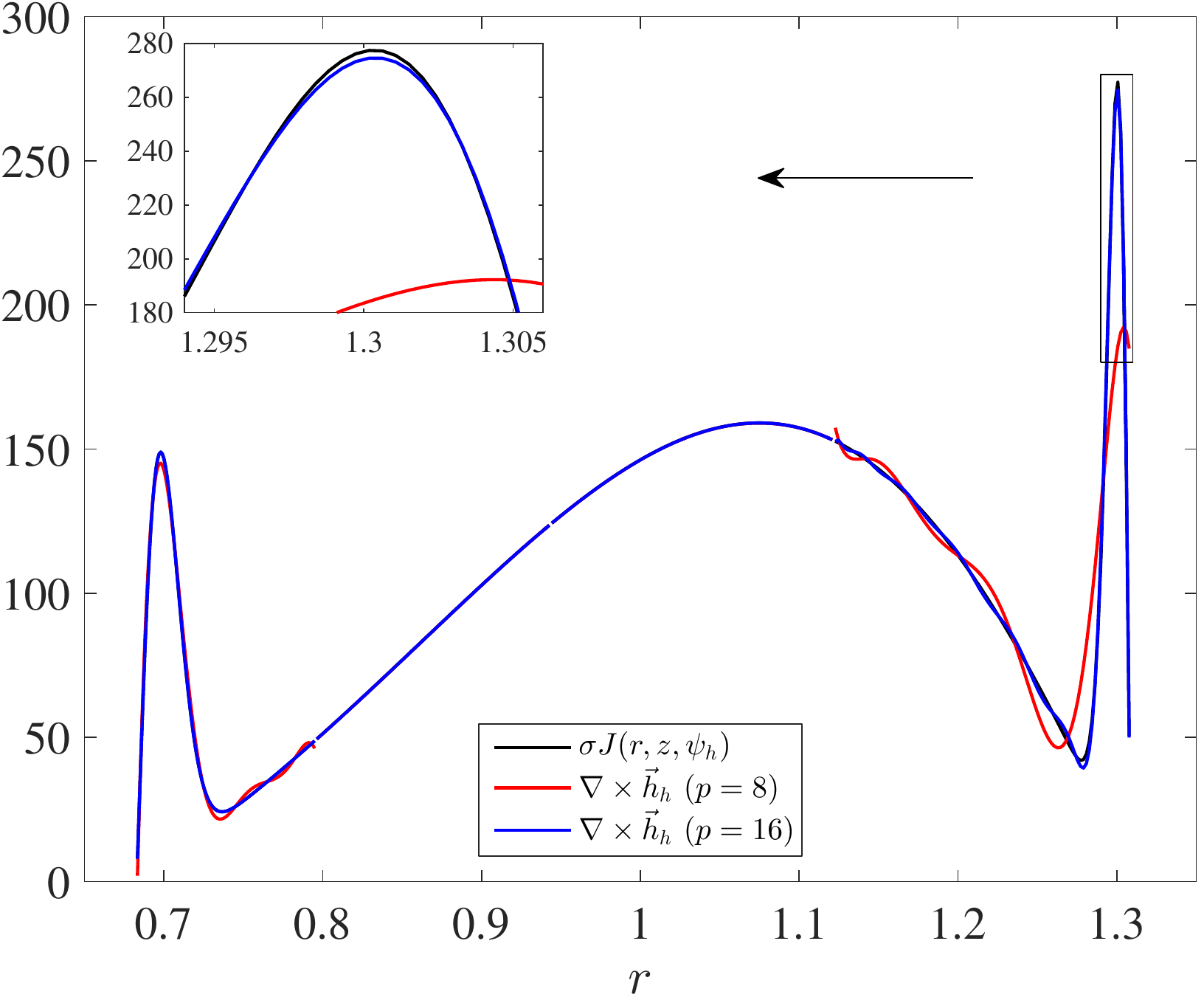}\hspace{0.5cm}
				\includegraphics[width=0.395\textwidth]{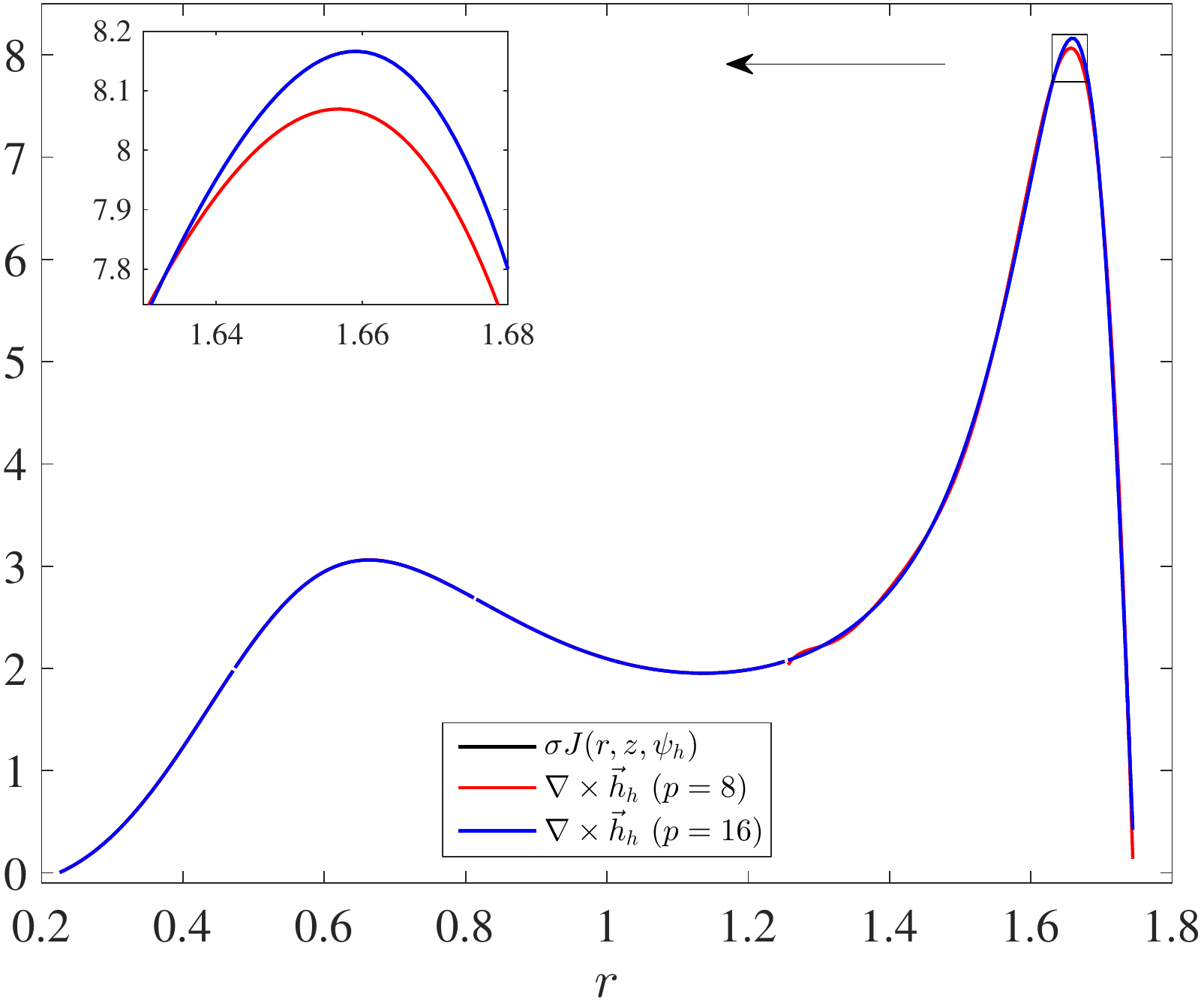}
				\end{center}
				\caption{\reviewerone{Comparison between $\sigma J(r,z,\psi_{h})$ and $\nabla\times\vec{h}_{h}(r,z)$ for $p=8,16$ along the line $\gamma$ (\figref{fig::nonlinear_eigenvalue_iter_test_case} for ITER case and \figref{fig::nonlinear_eigenvalue_nstx_test_case} for NSTX case). Left: ITER test case. Right:  NSTX test case.}}
				\label{fig::nonlinear_eigenvalue_nstx_test_case_J_along_line}
			\end{figure}
			
			\begin{figure}[!ht]
				\begin{center}
				\includegraphics[width=0.4\textwidth]{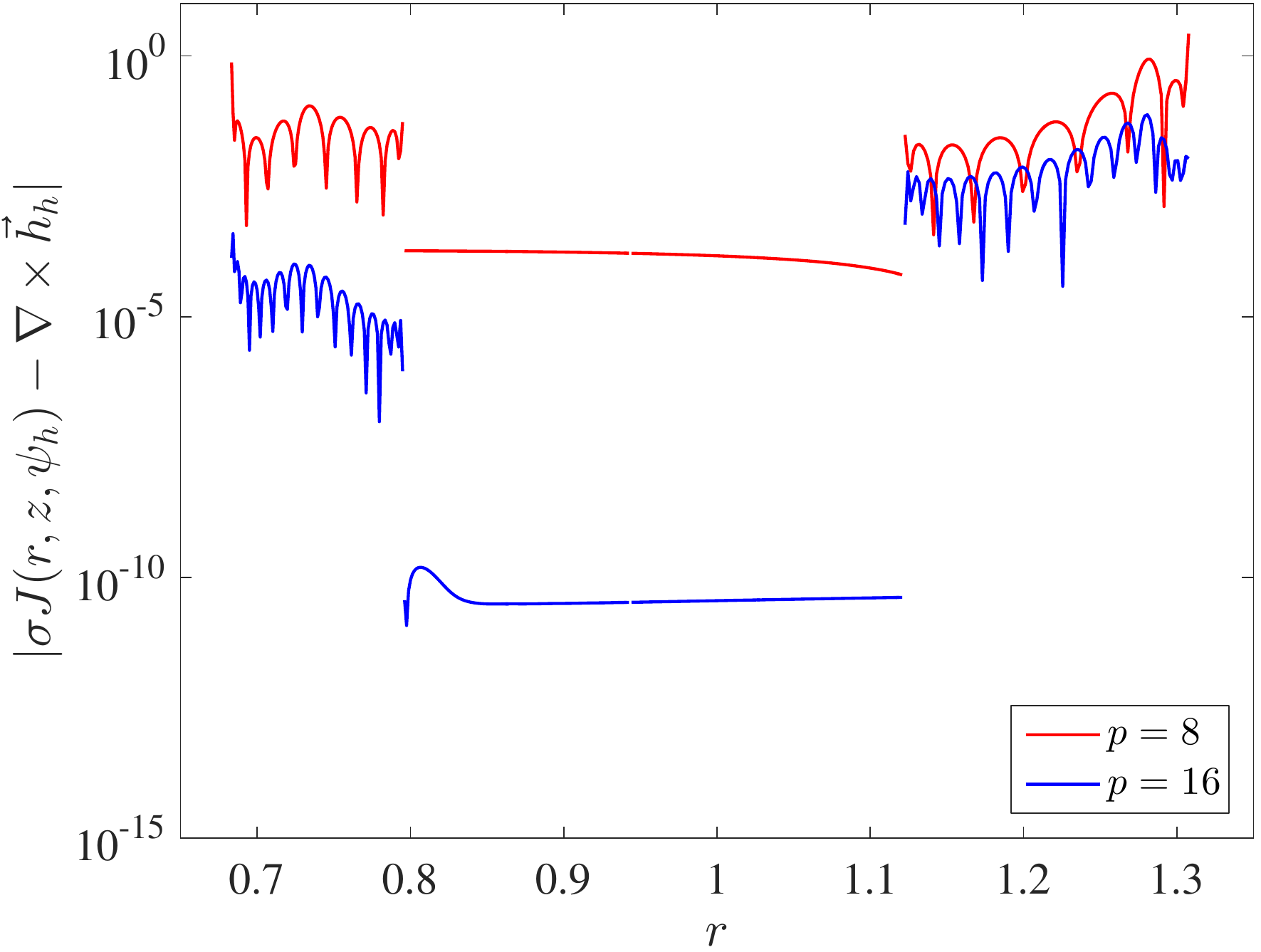}\hspace{0.5cm}
				\includegraphics[width=0.4\textwidth]{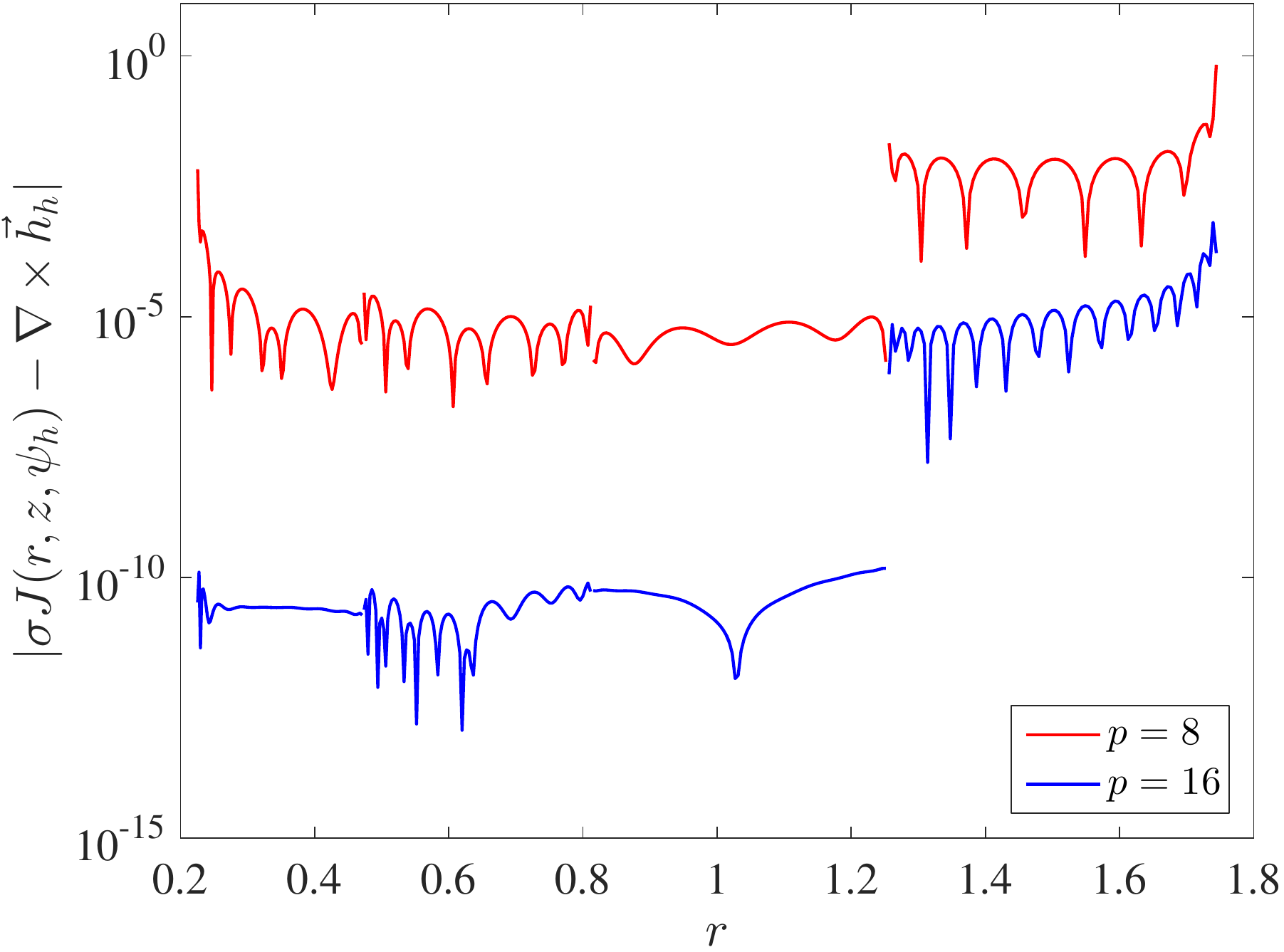}
				\end{center}
				\caption{\reviewerone{Error, as given by \eqref{eq:nonlinear_error}, along the curve $\gamma$ (see \figref{fig::nonlinear_eigenvalue_iter_test_case} for ITER test case and \figref{fig::nonlinear_eigenvalue_nstx_test_case} for NSTX test case). Left: ITER test case. Right:  NSTX test case.}}
				\label{fig::nonlinear_eigenvalue_nstx_test_case_J_along_line}
			\end{figure}
			
			\end{reviewer1}
			\FloatBarrier
		\begin{reviewer1}	
		\subsection{Plasma shape with an X-point} \label{sec::test_cases_x_point}
			The final test cases correspond to a plasma with an X-point, \reviewerthree{leading to a plasma boundary with a sharp corner}. This plasma configuration will be assessed for a Soloviev problem, the linear eigenvalue problem introduced in \secref{sec::test_cases_linear_eigenvalue} and the non-linear eigenvalue problem of \secref{sec::test_cases_nonlinear_eigenvalue}.
			
			\subsubsection{Plasma shape definition}
				The shape of the plasma considered here is the up-down asymmetric ITER-like configuration presented in \cite{Cerfon2010}, see \figref{fig::xpoint_plasma_shape}. This shape is given by the equation
				\begin{equation}
					\frac{r^{4}}{8} + A \left(\frac{r^{2}}{2}\ln r - \frac{r^{4}}{8}\right) + \sum_{k=1}^{12}c_{k}\, \psi_{k} (r,z) = 0, \label{eq:x_point_plasma_shape}
				\end{equation}
				with the functions $\psi_{i}(r,z)$ and the coefficients $c_{i}$ as in \appendixref{ap:x_point}.
				
				\begin{figure}[!ht]
					\begin{center}
					\includegraphics[height=0.295\textwidth]{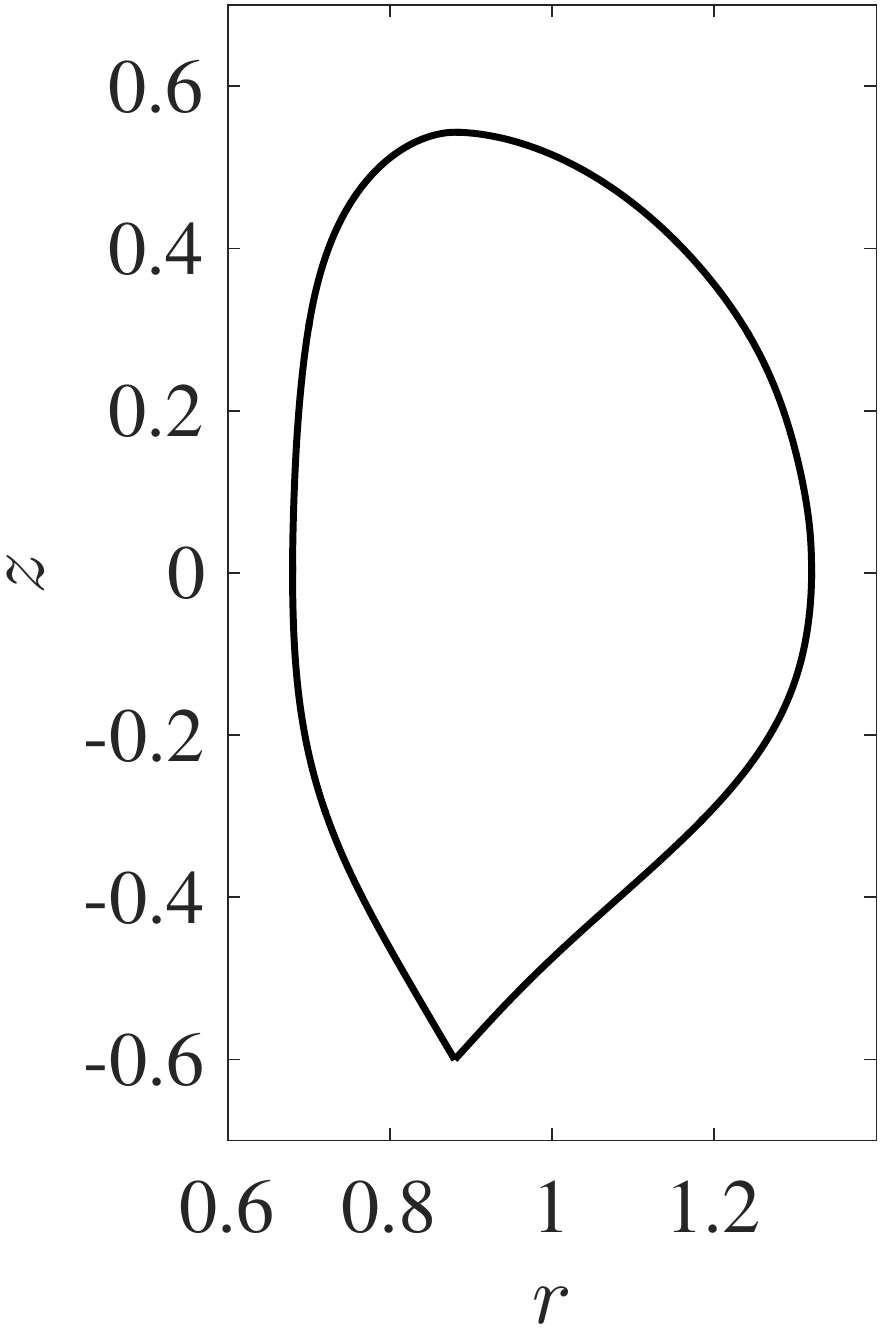}
					\end{center}
					\caption{\reviewerone{Plasma shape with an X-point, corresponding to the up-down asymmetric ITER-like configuration presented in \cite{Cerfon2010}.}}
					\label{fig::xpoint_plasma_shape}
				\end{figure}
				
			\subsubsection{Soloviev problem}
				The first test case for the X-point plasma corresponds to a Soloviev solution of the Grad-Shafranov equation. This special case corresponds to $f(\psi) = \sqrt{2\mu_{0}A\psi}$ combined with a linear model $P(\psi) = (1-A)\psi$, for a more detailed discussion see \cite{Cerfon2010}. This leads to the following Grad-Shafranov equation
			\begin{equation}
				\begin{dcases}
					\nabla\times\left(\mathbb{K}\nabla\times\psi\right) = (1-A)r + \frac{A}{r} & \mbox{in} \quad \Omega_{p}\,, \\
					\psi = 0 & \mbox{on}\quad\partial\Omega_{p}\,,
				\end{dcases} \label{eq::soloviev_test_case_x_point}
			\end{equation}
			where we will consider $A=-0.155$.
			
			The analytical solution, $\psi_{a}(r,z)$, is constructed by adding to the particular solution $\frac{r^{4}}{8} + A \left(\frac{r^{2}}{2}\ln r - \frac{r^{4}}{8}\right)$ a linear combination of twelve homogeneous solutions, $\psi_{i}(r,z)$ with $i=1,\dots,12$:
			\begin{equation}
				\psi_{a}(r,z) =\frac{r^{4}}{8} + A \left(\frac{r^{2}}{2}\ln r - \frac{r^{4}}{8}\right) + \sum_{k=1}^{12}c_{k}\, \psi_{k} (r,z)\,, \label{eq::soloviev_test_case_analytical_solution_x_point}
			\end{equation}
			with the homogeneous solutions and the coefficients $c_{i}$ as given in \appendixref{ap:x_point}. The methodology to obtain this solution is similar to the one presented in \secref{sec::test_cases_soloviev} and is fully detailed in \cite{Cerfon2010}.
			
			In \figref{fig::soloviev_test_case_xpoint} an example solution with $4\times 4$ elements of polynomial degree $p=8$ is presented. As can be seen, the solution is well reconstructed by the proposed method.
			
			\begin{figure}[!ht]
				\begin{center}
				\includegraphics[height=0.271\textwidth]{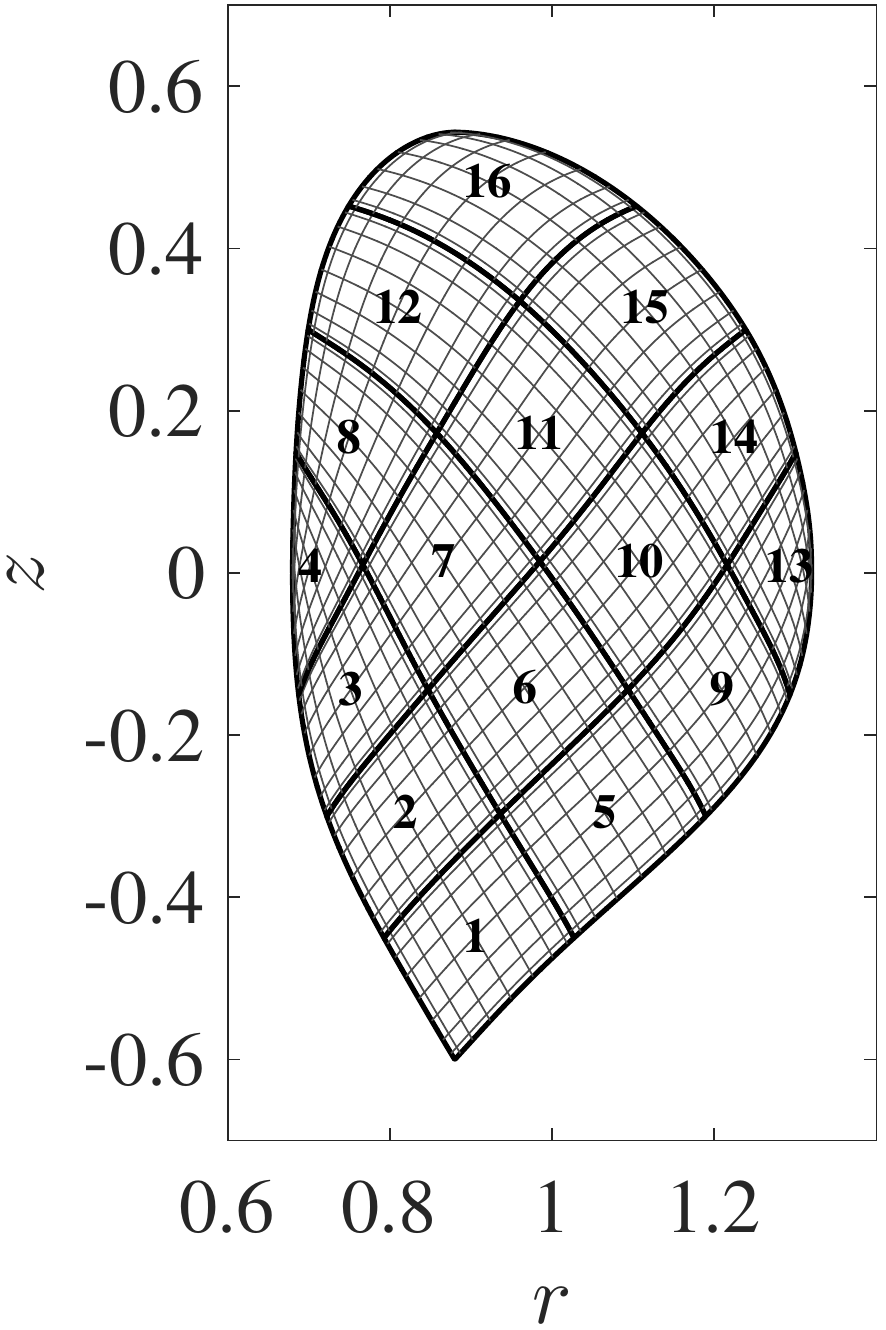}\hspace{0.75cm}
				\includegraphics[height=0.271\textwidth]{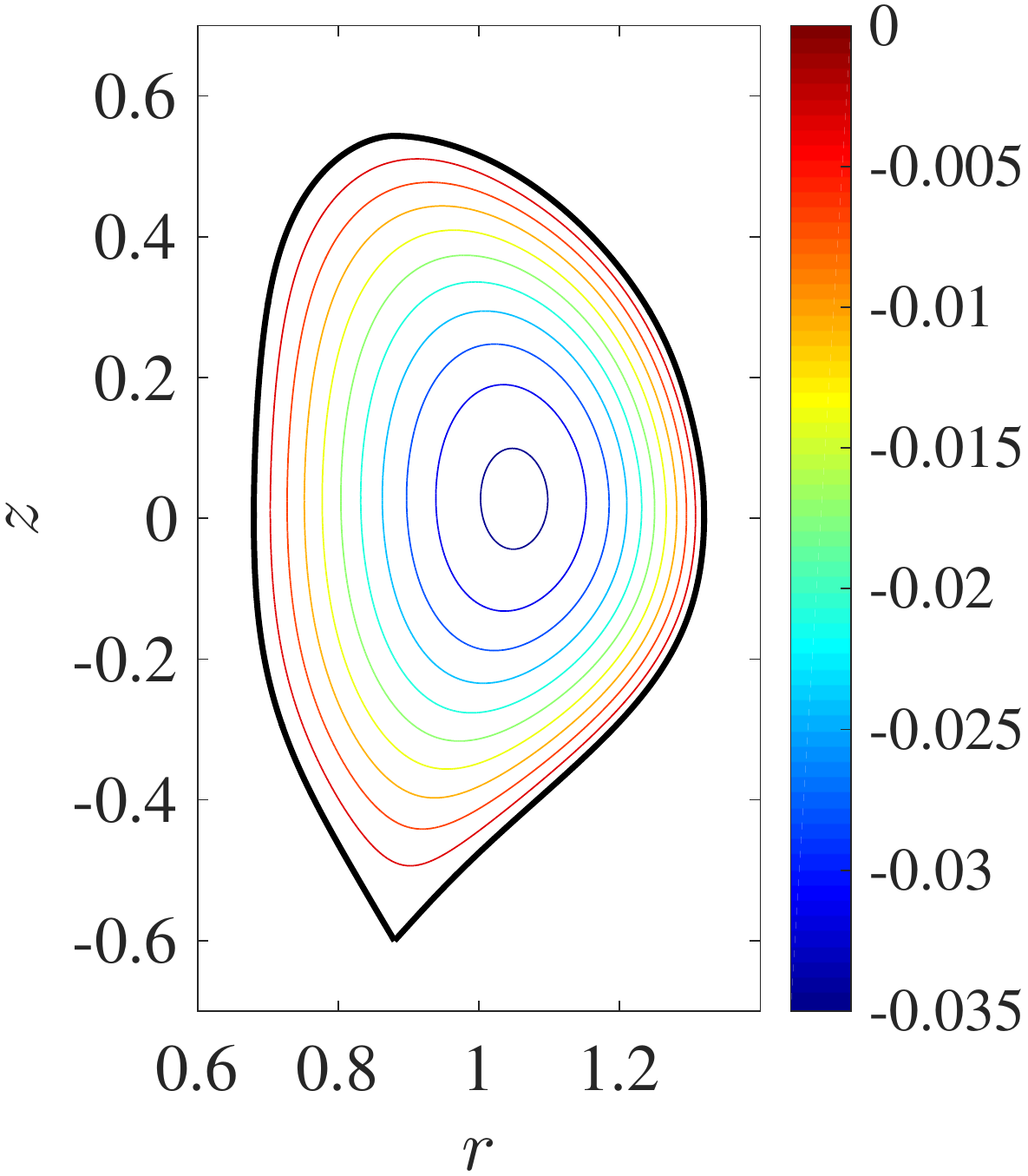}\hspace{0.75cm}
				\includegraphics[height=0.271\textwidth]{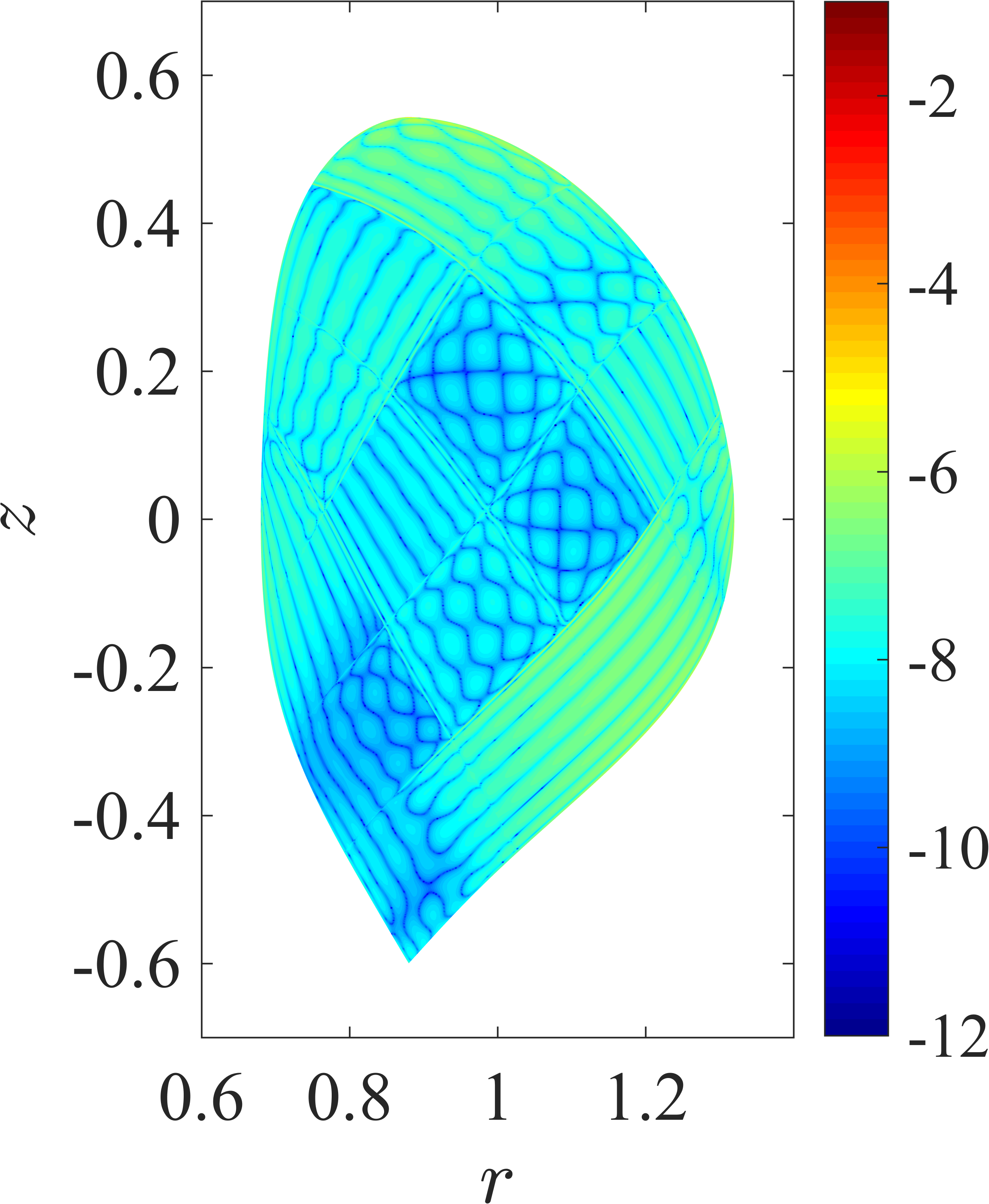}
				\end{center}
				\caption{\reviewerone{Numerical solution of the X-point Soloviev test case, \eqref{eq::soloviev_test_case_x_point}. From left to right: (i) computational mesh with $4\times 4$ elements of polynomial degree $p=8$, %(ii) analytical solution \eqref{eq::soloviev_test_case_analytical_solution}, $\psi_{a}$, 
				(ii) numerical solution using the mesh in (i), $\psi_{h}$, and (iii) logarithmic error between the analytical solution and the numerical one, $\log_{10}|\psi_{a}-\psi_{h}|$.}}
				\label{fig::soloviev_test_case_xpoint}
			\end{figure}
						
			The convergence tests for $h$-refinement and $p$-refinement show very similar results to the ITER and NSTX cases of \secref{sec::test_cases_soloviev},  both for $\psi_{h}$ and for $\vec{h}_{h}$, see \figref{fig::curved_soloviev_test_case_convergence_xpoint} and \figref{fig::curved_soloviev_test_case_convergence_h_x_point}. As seen before, accuracy up to machine precision is achieved.
			
			\begin{figure}[htb]
				\begin{center}
				\includegraphics[height=0.295\textwidth]{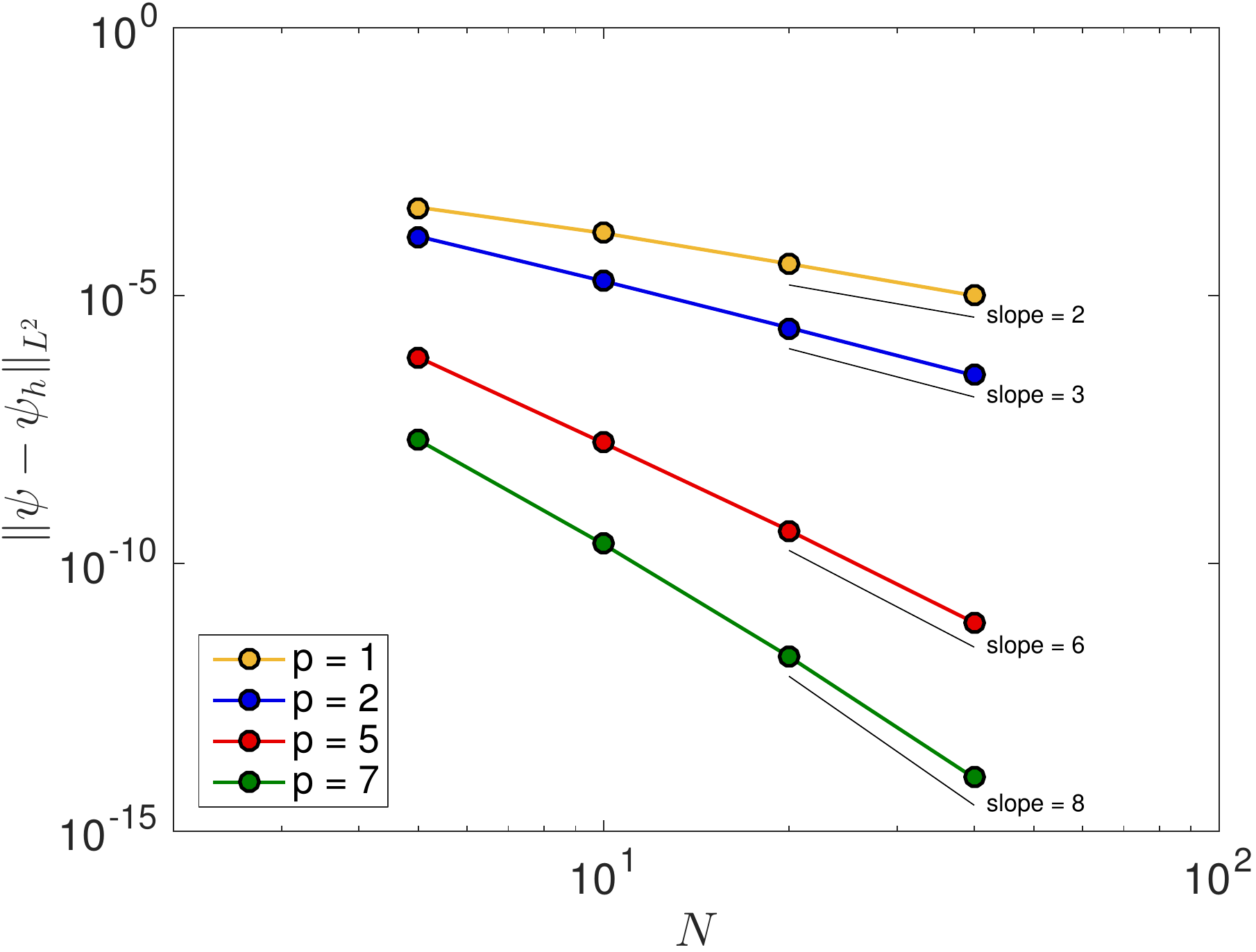} \hspace{1cm}
				\includegraphics[height=0.295\textwidth]{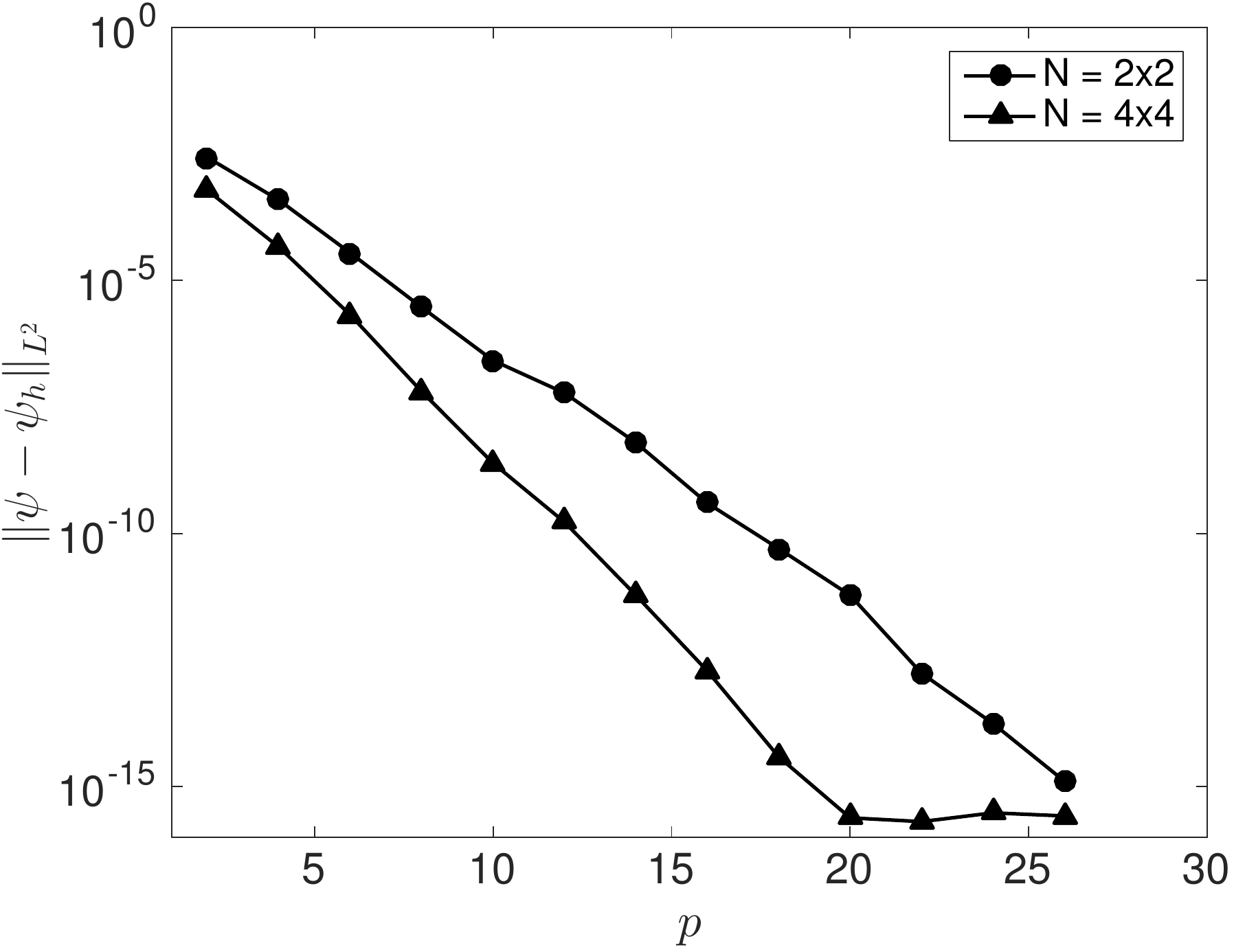}
				\end{center}
				\caption{\reviewerone{Convergence plots for the numerical solution of $\psi(r,z)$ of the X-point Soloviev test case \eqref{eq::soloviev_test_case_x_point}. Right: $p$-convergence plots.}}
				\label{fig::curved_soloviev_test_case_convergence_xpoint}
			\end{figure}
			
			\begin{figure}[htb]
				\begin{center}
				\includegraphics[height=0.295\textwidth]{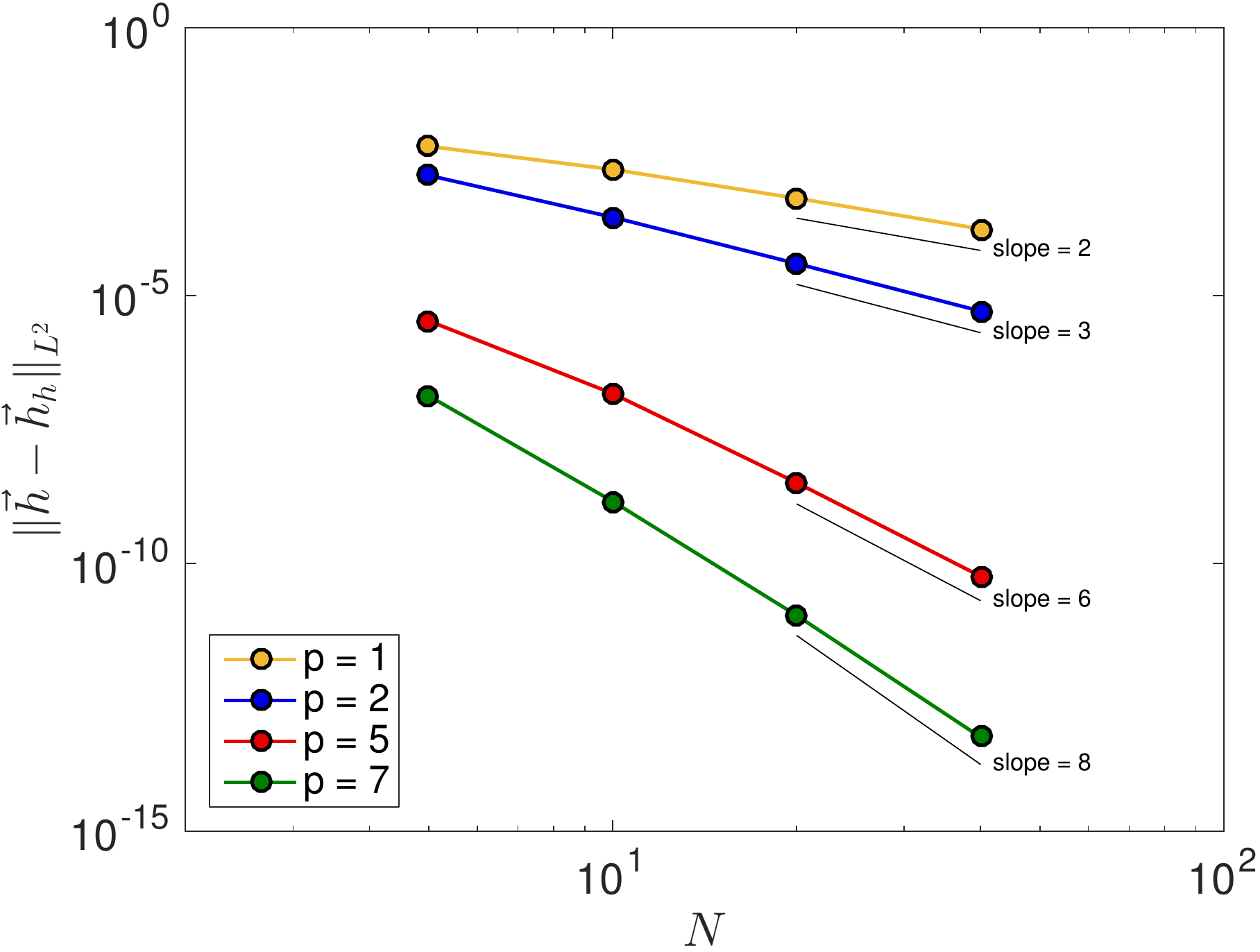} \hspace{1cm}
				\includegraphics[height=0.295\textwidth]{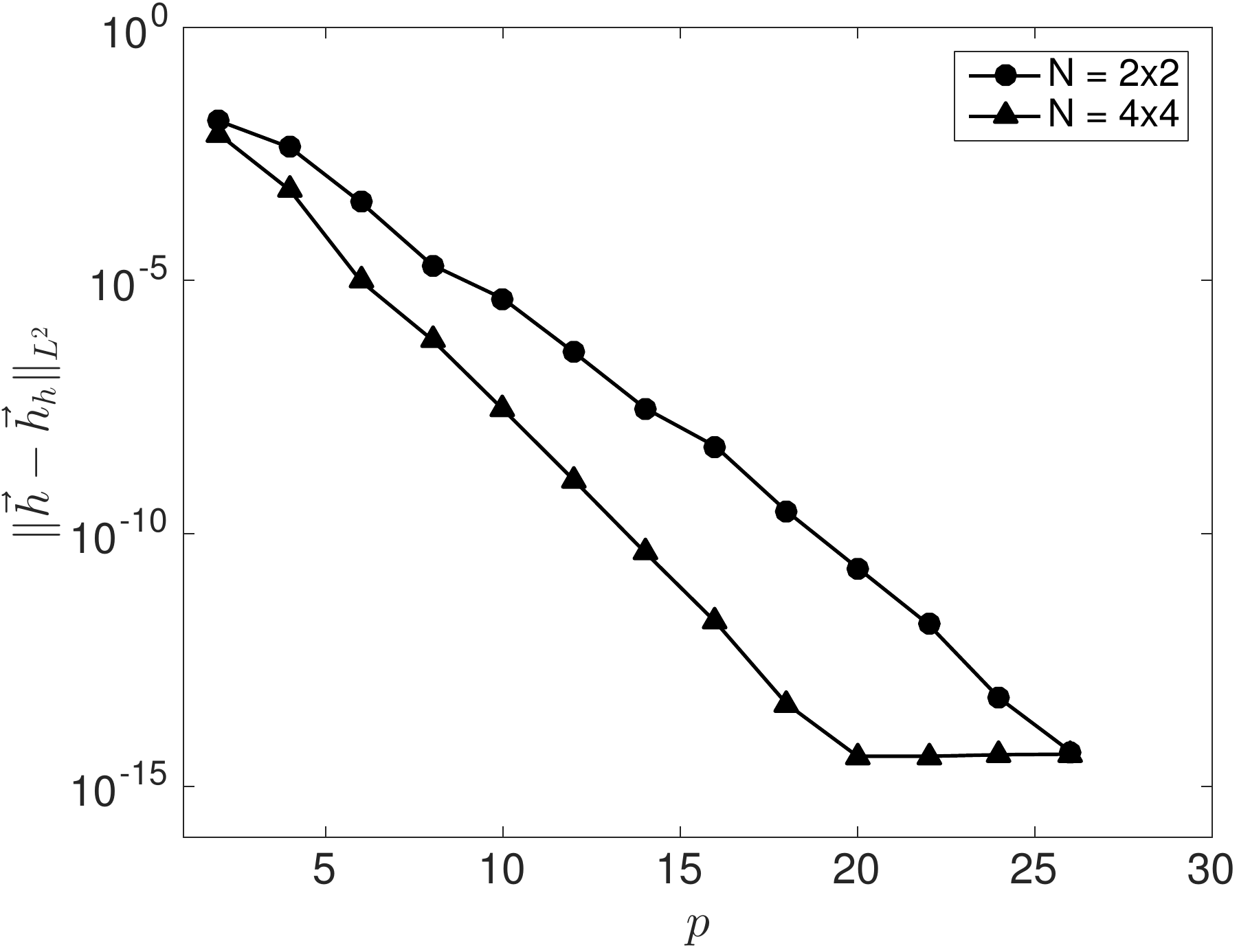}
				\end{center}
				\caption{\reviewerone{Convergence plots for the numerical solution of $\vec{h}_{h}(r,z)$ of the X-point Soloviev test case \eqref{eq::soloviev_test_case_x_point}. Left: $h$-convergence plots. Right: $p$-convergence plots.}}
				\label{fig::curved_soloviev_test_case_convergence_h_x_point}
			\end{figure}
			
			As in \secref{sec::test_cases_soloviev}, the error between the flux integral $\int_{\Omega_{p}} J_{h}\mathrm{d}V$ and the contour integral  $\int_{\partial\Omega_{p}}\vec{h}_{h}\mathrm{d}\vec{l}\,$ shows excellent results, see \figref{fig::curved_soloviev_test_case_convergence_integral_x_point}.
			
			\begin{figure}[htb]
				\begin{center}
				\includegraphics[height=0.295\textwidth]{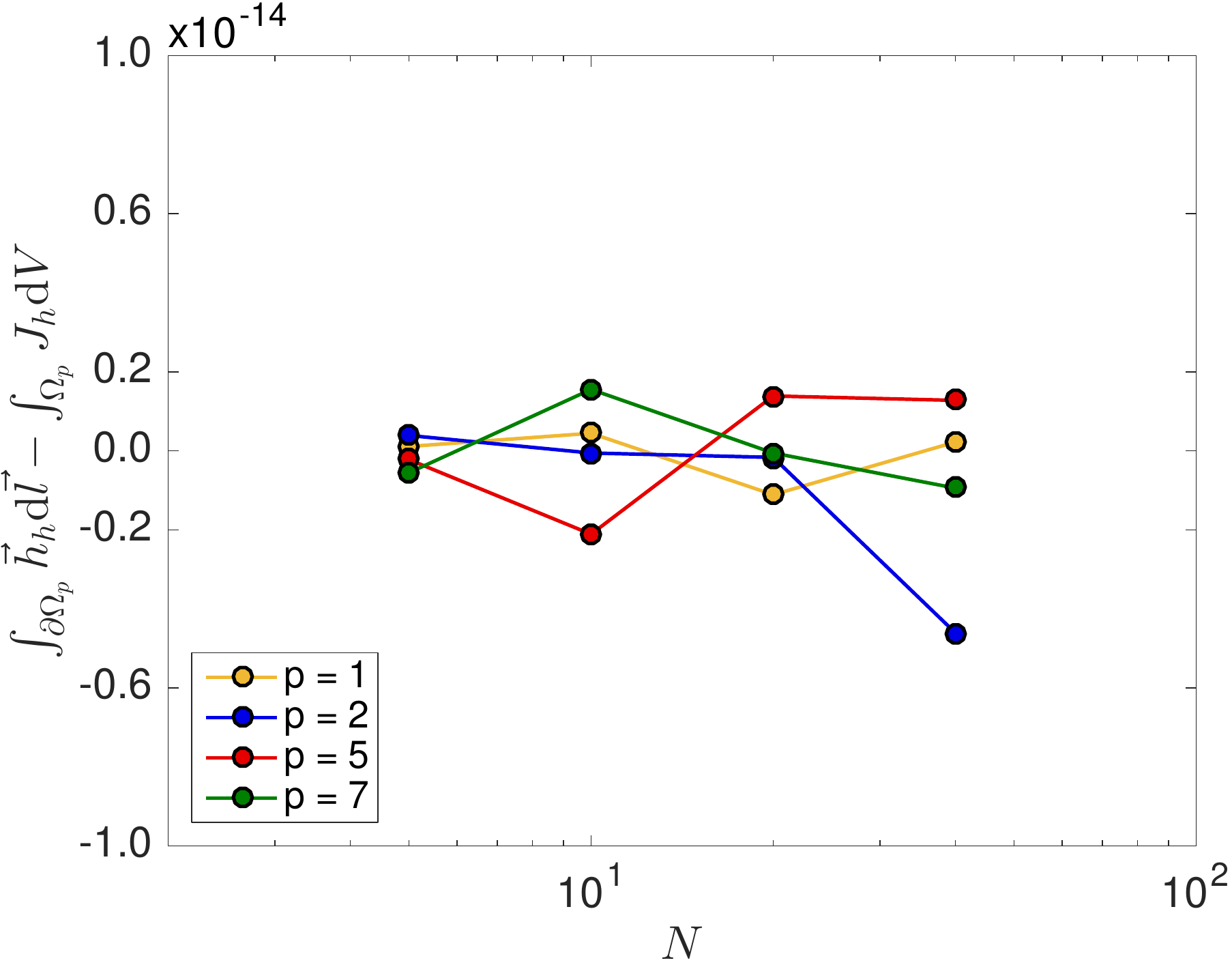} \hspace{1cm}
				\includegraphics[height=0.295\textwidth]{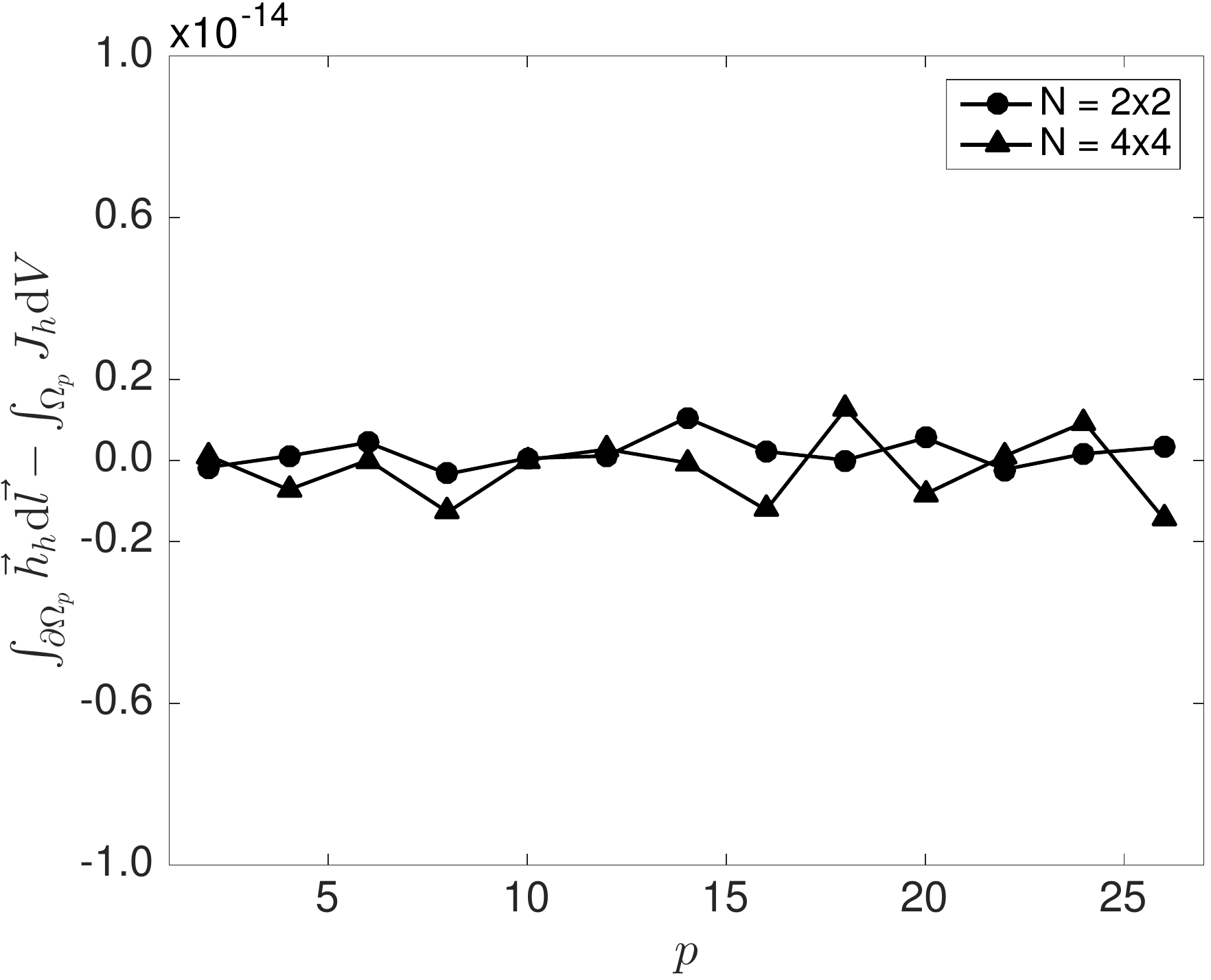}
				\end{center}
				\caption{\reviewerone{Error between the flux integral $\int_{\Omega_{p}} J_{h}\mathrm{d}V$ and the contour integral  $\int_{\partial\Omega_{p}}\vec{h}_{h}\mathrm{d}\vec{l}$, for the X-point Soloviev test case \eqref{eq::soloviev_test_case_x_point}. Left: $h$-convergence plots. Right: $p$-convergence plots.}}
				\label{fig::curved_soloviev_test_case_convergence_integral_x_point}
			\end{figure}
			
		\FloatBarrier

			\subsubsection{Linear eigenvalue problem}
				The second test for the X-point plasma configuration is the linear eigenvalue problem of \secref{sec::test_cases_linear_eigenvalue}. The same Grad-Shafranov problem in \eqref{eq::linear_eigenvalue_test_case} with $C_{1} = -1.0$ and $C_{2} = 2.0$ is solved with the plasma boundary $\Omega_{p}$ given by \eqref{eq:x_point_plasma_shape}.
				
				We first show in \figref{fig::linear_eigenvalue_xpoint_test_case} two example solutions obtained with a mesh of $4\times 4$ elements of polynomial degree $p=8$ and $p=16$. As can be seen the error $E$ given by \eqref{eq:nonlinear_error} considerably decreases when the polynomial degree of the elements is increased from $p=8$ to $p=16$.
				\begin{figure}[!ht]
				\begin{center}
				\includegraphics[height=0.271\textwidth]{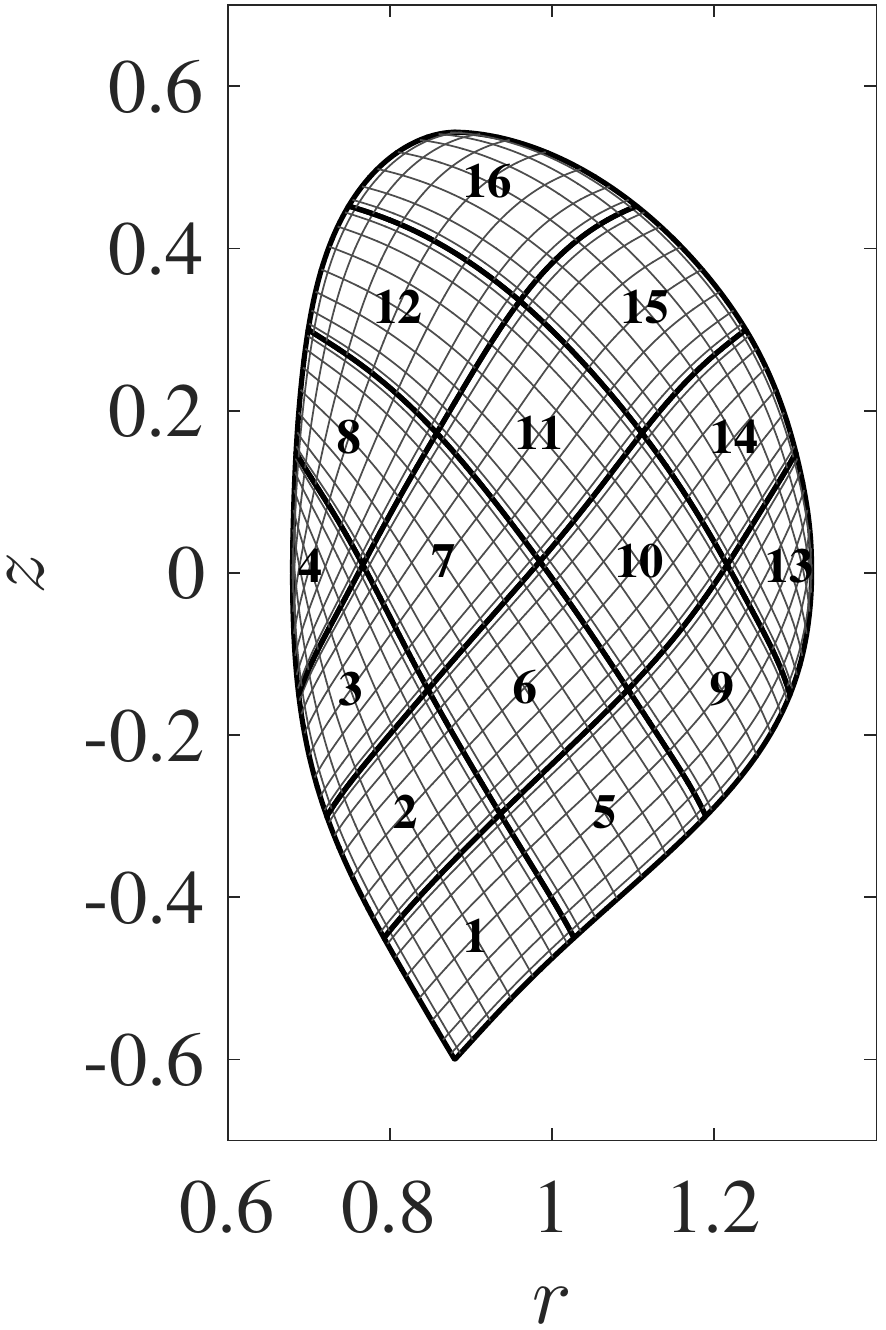}\hspace{0.75cm}
				\includegraphics[height=0.271\textwidth]{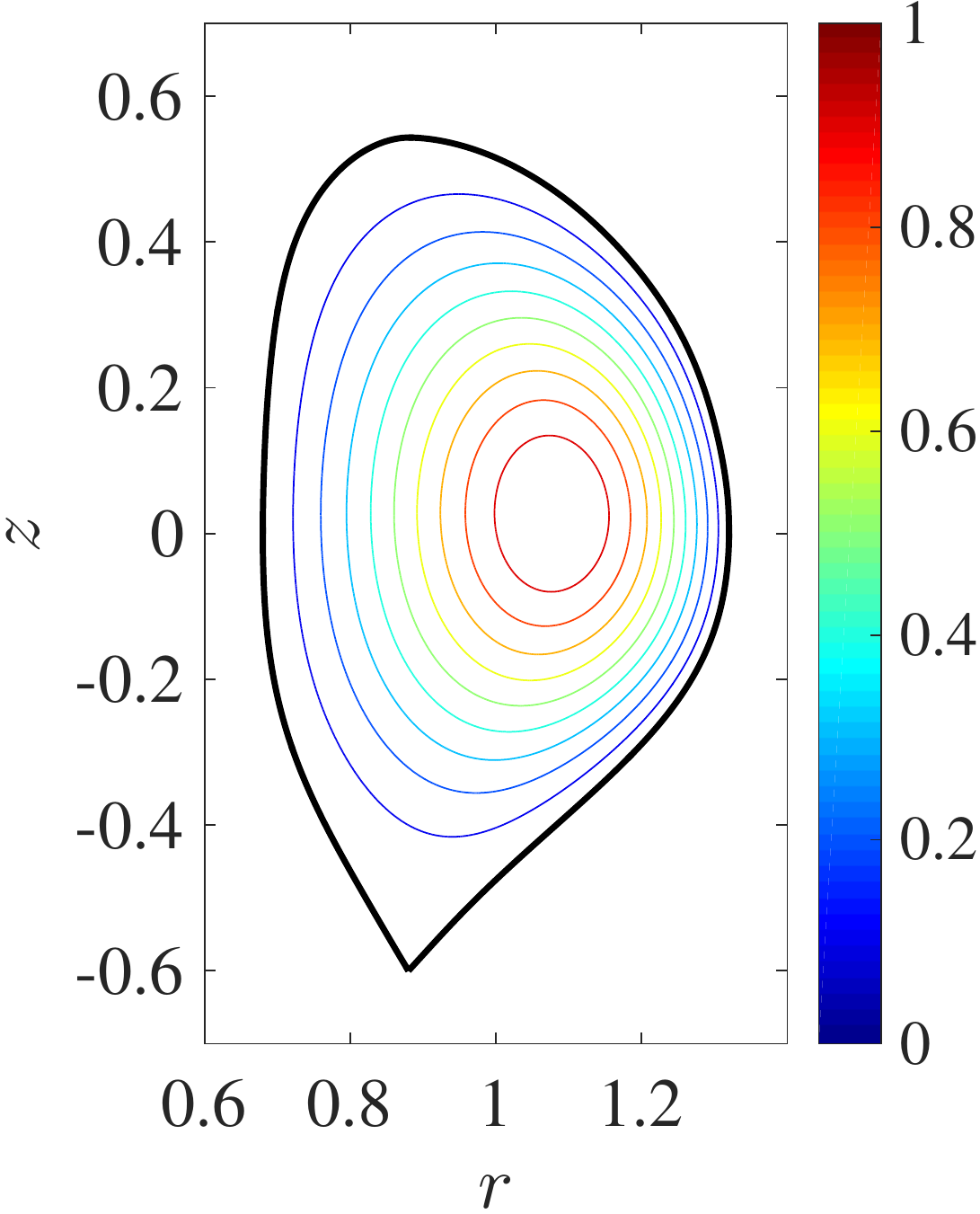}\hspace{0.75cm}
				\includegraphics[height=0.271\textwidth]{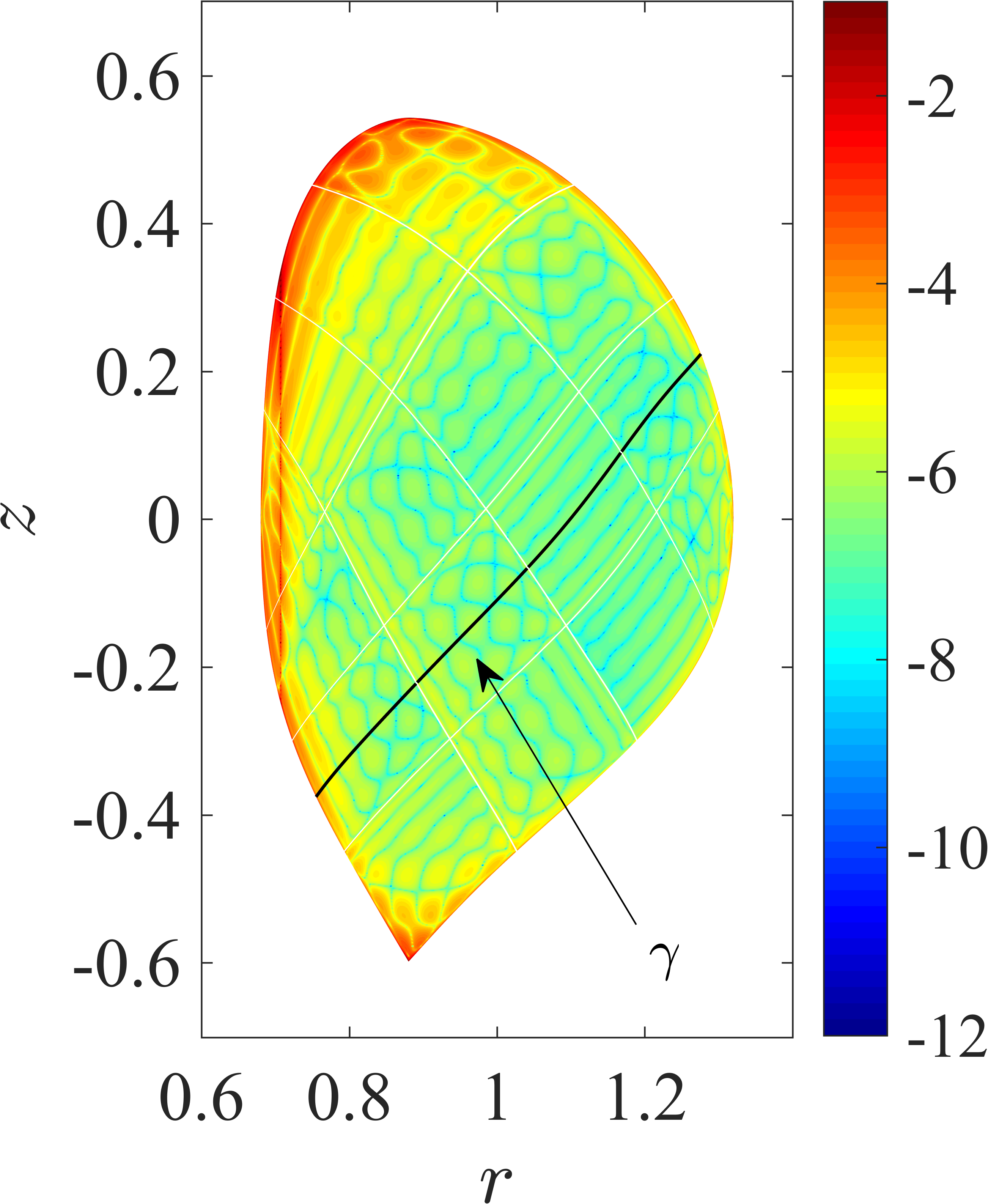}\\
				\includegraphics[height=0.271\textwidth]{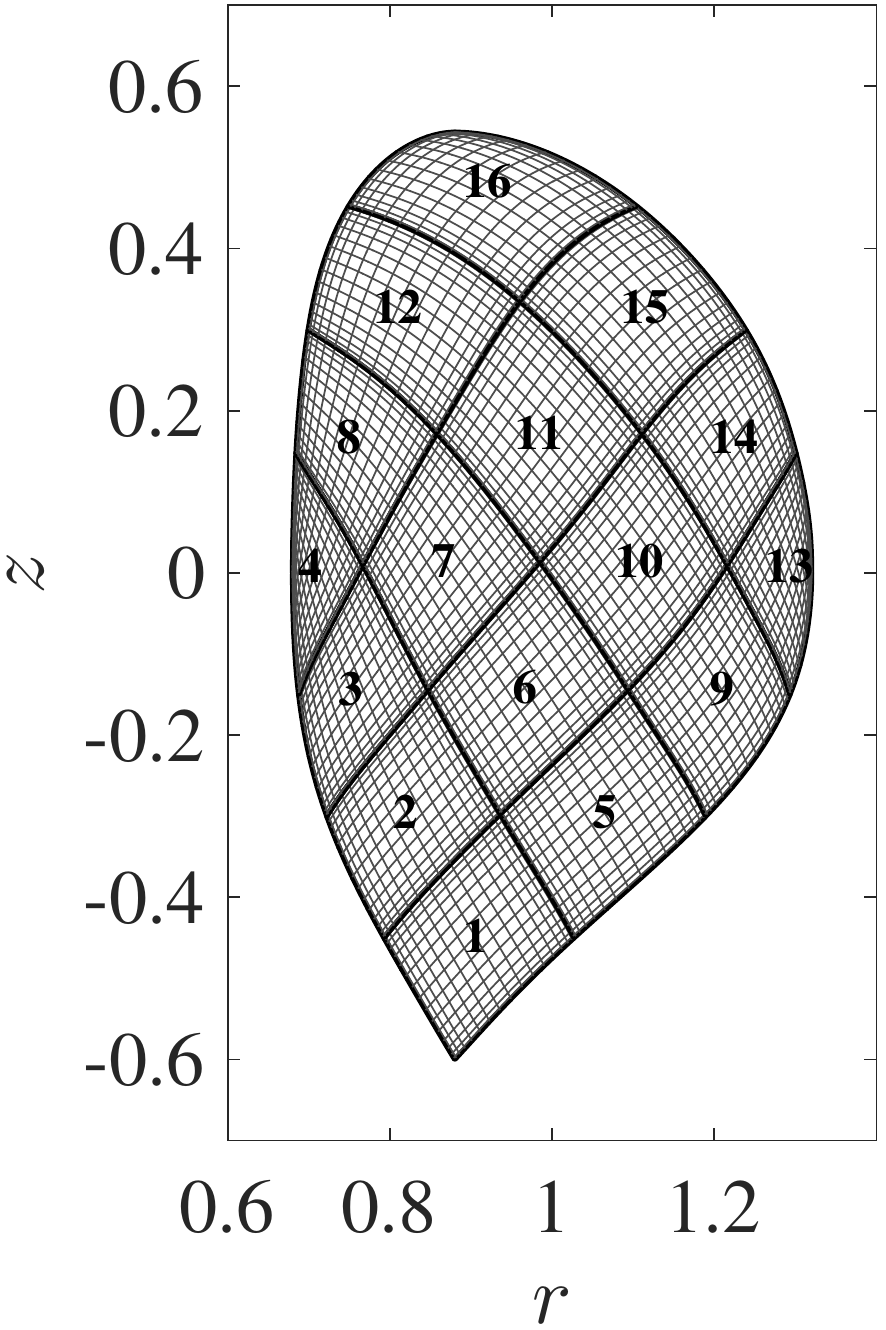}\hspace{0.75cm}
				\includegraphics[height=0.271\textwidth]{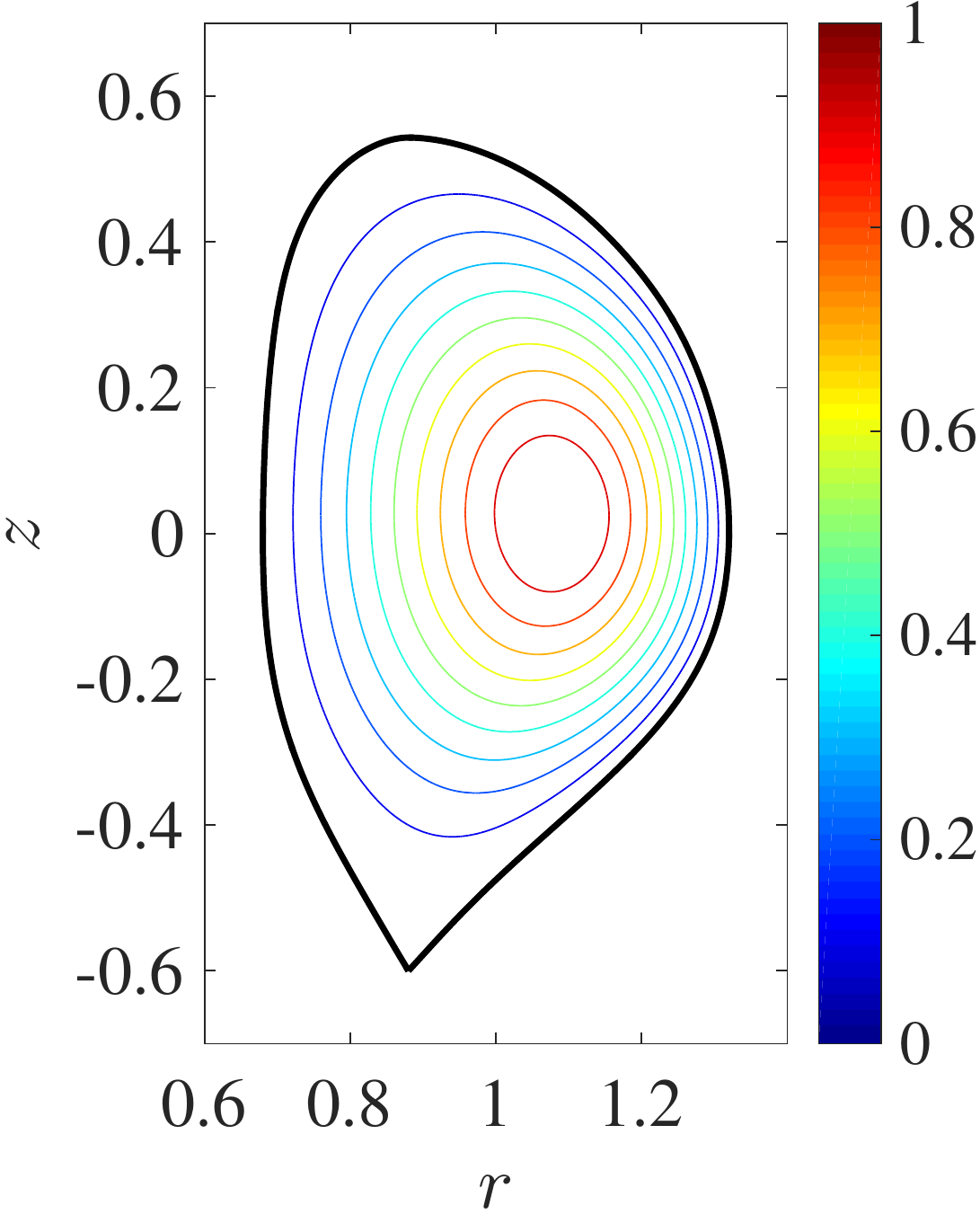}\hspace{0.75cm}
				\includegraphics[height=0.271\textwidth]{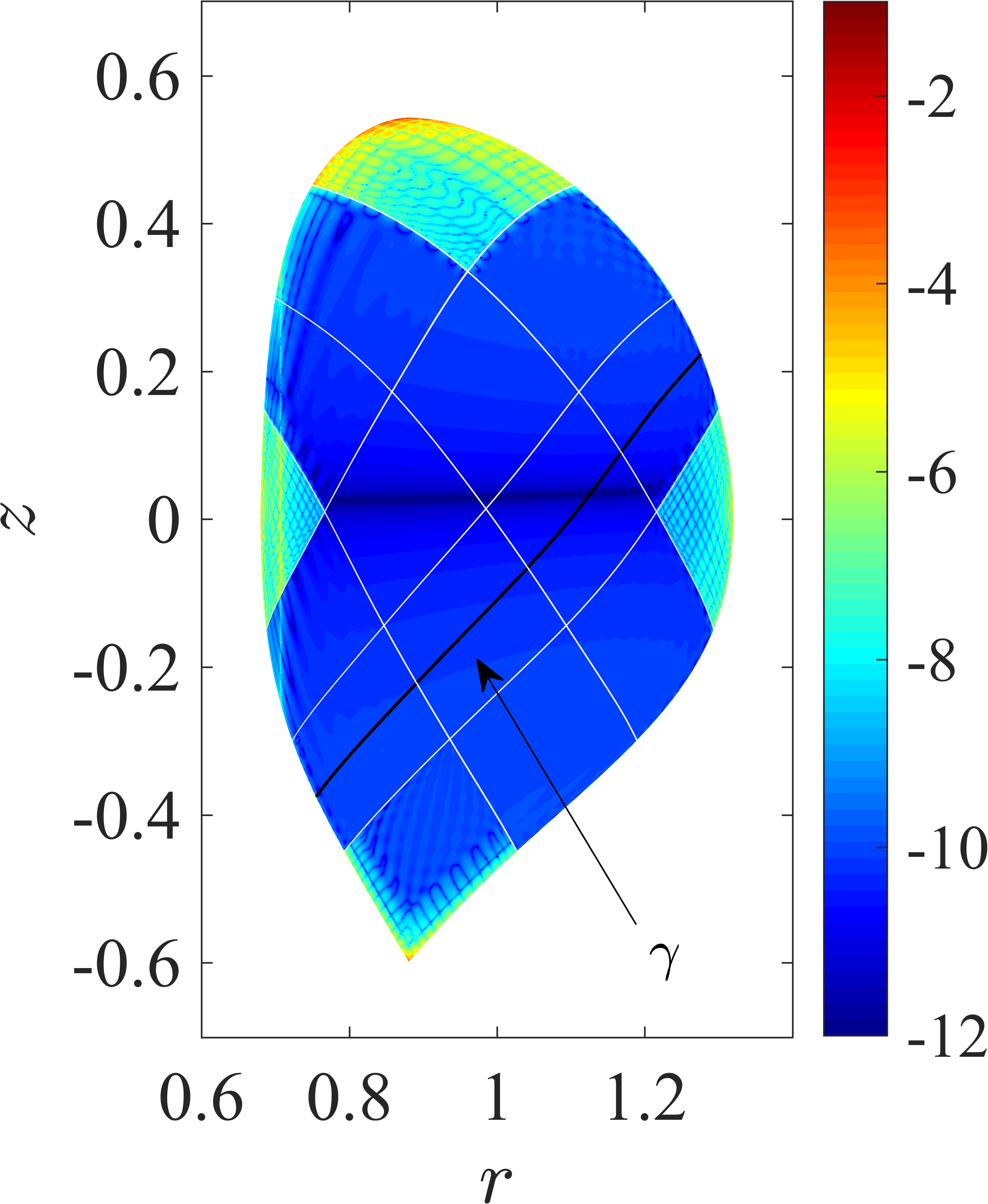}\\
				\end{center}
				\caption{\reviewerone{Numerical solution of the linear eigenvalue problem \eqref{eq::linear_eigenvalue_test_case} for the X-point plasma shape given in \eqref{eq:x_point_plasma_shape}, with $|\psi_{0}|=1.0$, $C_{1}=-1.0$ and $C_{2}=2.0$ (high Shafranov shift). Top: mesh of $4\times 4$ elements and elements of polynomial degree $p=8$. Bottom: mesh of $4\times 4$ elements and elements of polynomial degree $p=16$. From left to right: computational mesh, numerical solution, and error as given by \eqref{eq:nonlinear_error}.}}
				\label{fig::linear_eigenvalue_xpoint_test_case}
			\end{figure}
			
			For the ease of comparison, the error \eqref{eq:nonlinear_error} along the line $\gamma$ (see \figref{fig::linear_eigenvalue_xpoint_test_case}) is shown in \figref{fig::linear_eigenvalue_xpoint_test_case_error_along_line}. It is possible to see that the proposed method can accurately solve the linear eigenvalue Grad-Shafranov problem for a plasma with an X-point.
			
			\begin{figure}[!ht]
				\begin{center}
				\includegraphics[width=0.395\textwidth]{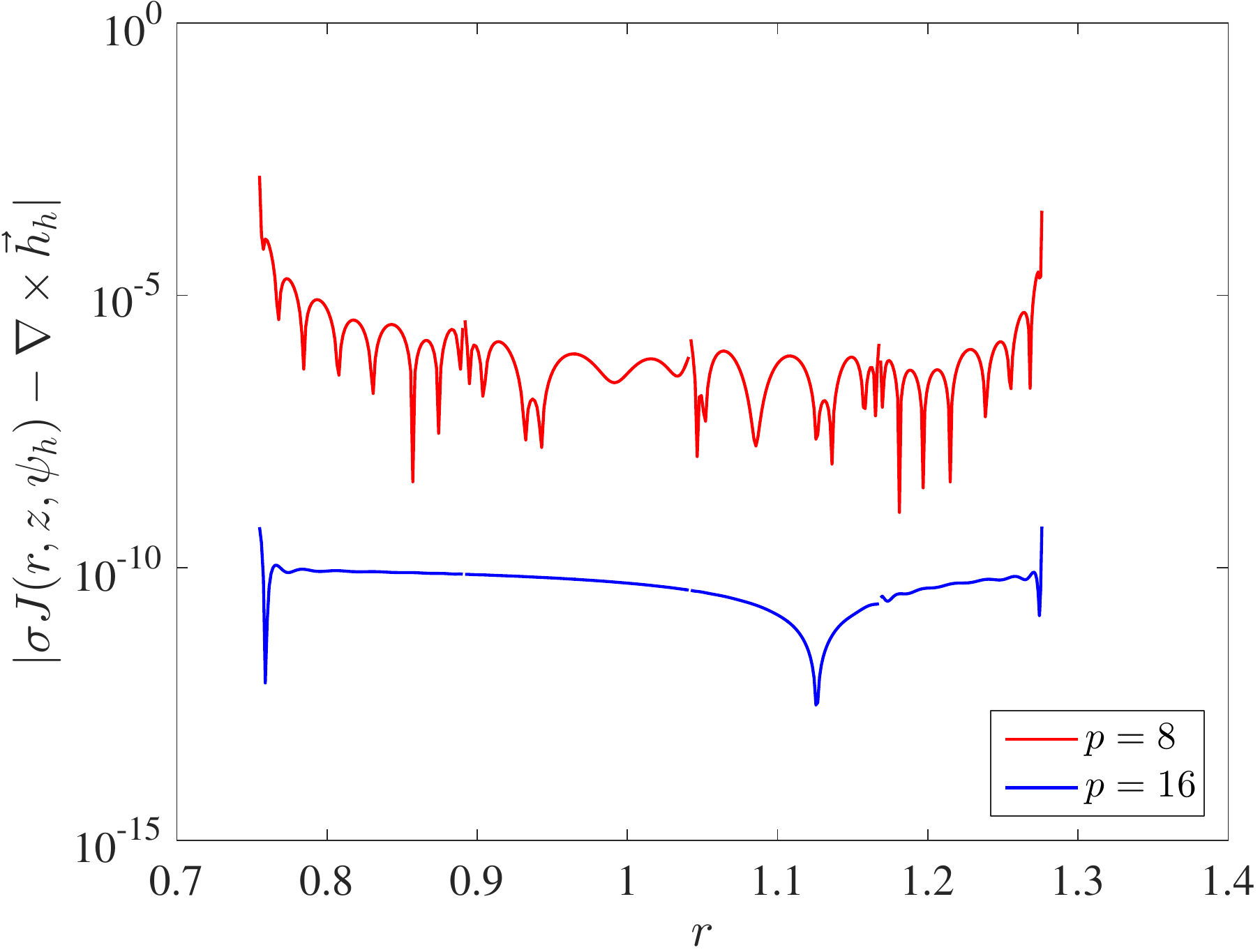}
				\end{center}
				\caption{\reviewerone{Error, as given by \eqref{eq:nonlinear_error}, along the curve $\gamma$ (see \figref{fig::linear_eigenvalue_xpoint_test_case}).}}
				\label{fig::linear_eigenvalue_xpoint_test_case_error_along_line}
			\end{figure}
			
			\FloatBarrier
				
			\subsubsection{Nonlinear eigenvalue problem}
				The third and final test case for the X-point plasma configuration is the non-linear Grad-Shafranov problem of \secref{sec::test_cases_nonlinear_eigenvalue} with $\eta=0.1$. The difference between this case and the previously discussed cases is the X-point plasma shape. 
			
				Numerical solutions corresponding to a mesh with $4\times 4$ elements of polynomial degree $p=8$ and $p=16$ are shown in \figref{fig::nonlinear_eigenvalue_xpoint_test_case}. As can be seen, the mimetic spectral element solver can adequately compute the non-linear equilibrium solution for a plasma with an X-point, showing a similar error to the cases discussed in \secref{sec::test_cases_nonlinear_eigenvalue}.
			
				\begin{figure}[!ht]
				\begin{center}
				\includegraphics[height=0.271\textwidth]{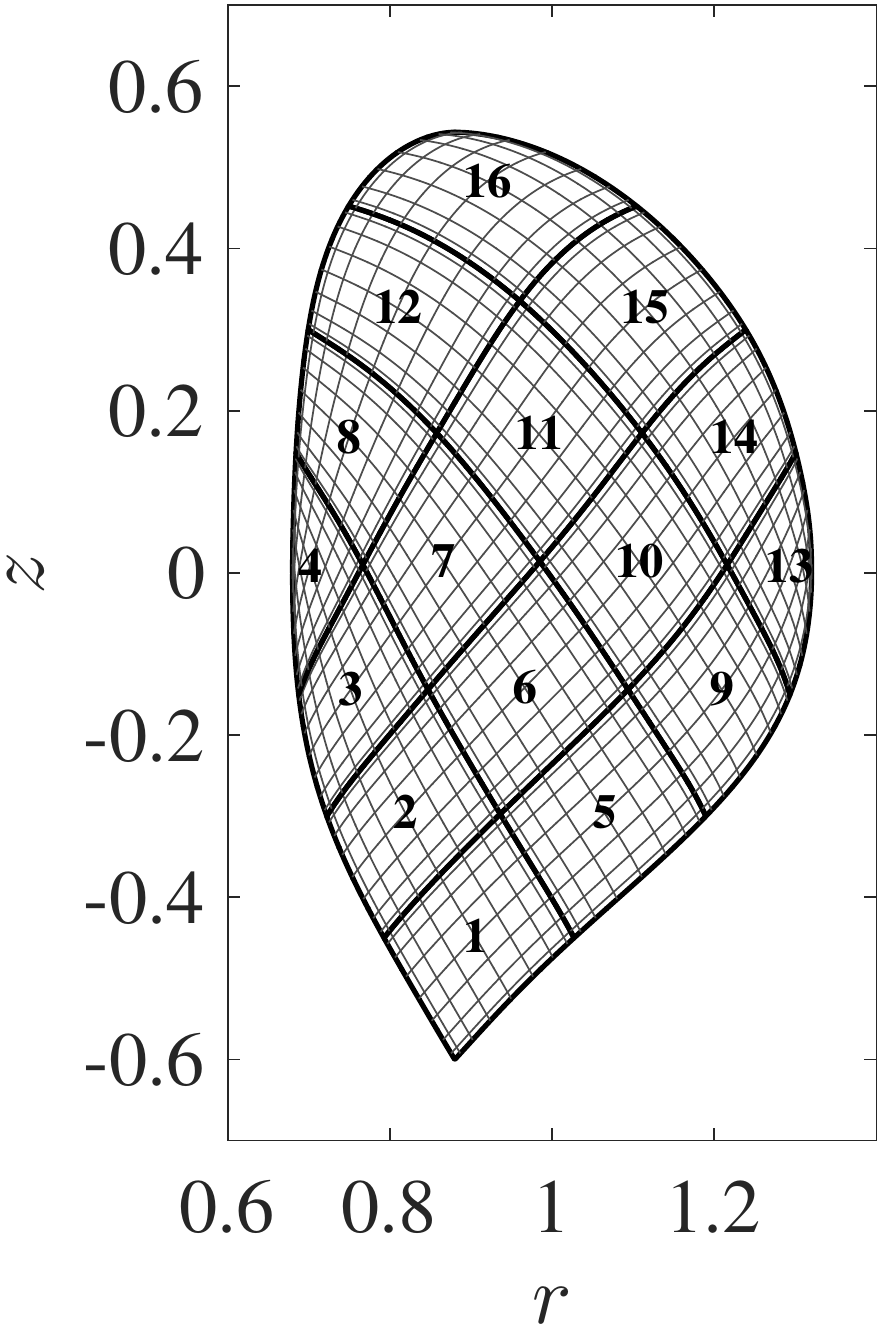}\hspace{0.75cm}
				\includegraphics[height=0.271\textwidth]{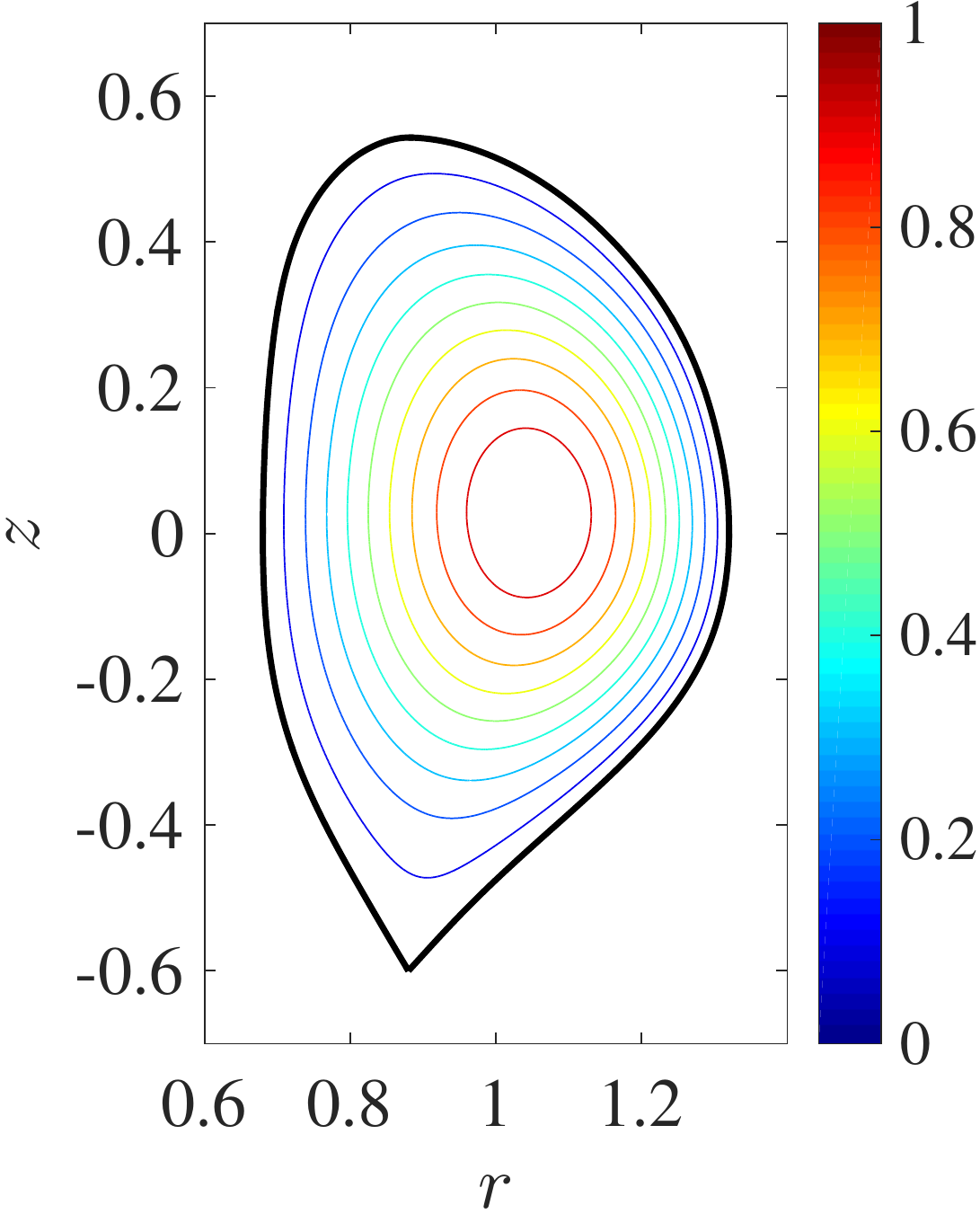}\hspace{0.75cm}
				\includegraphics[height=0.271\textwidth]{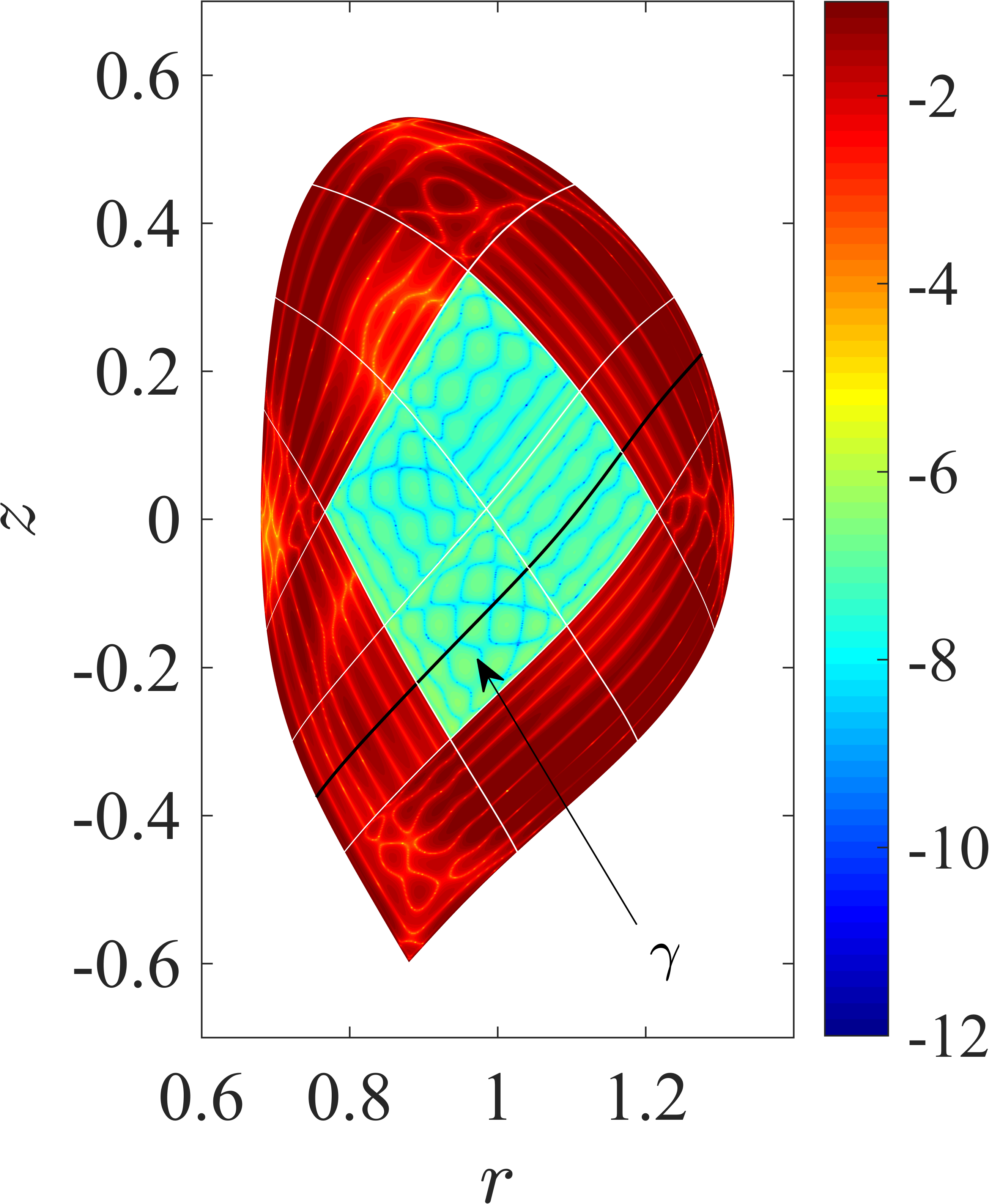}\\
				\includegraphics[height=0.271\textwidth]{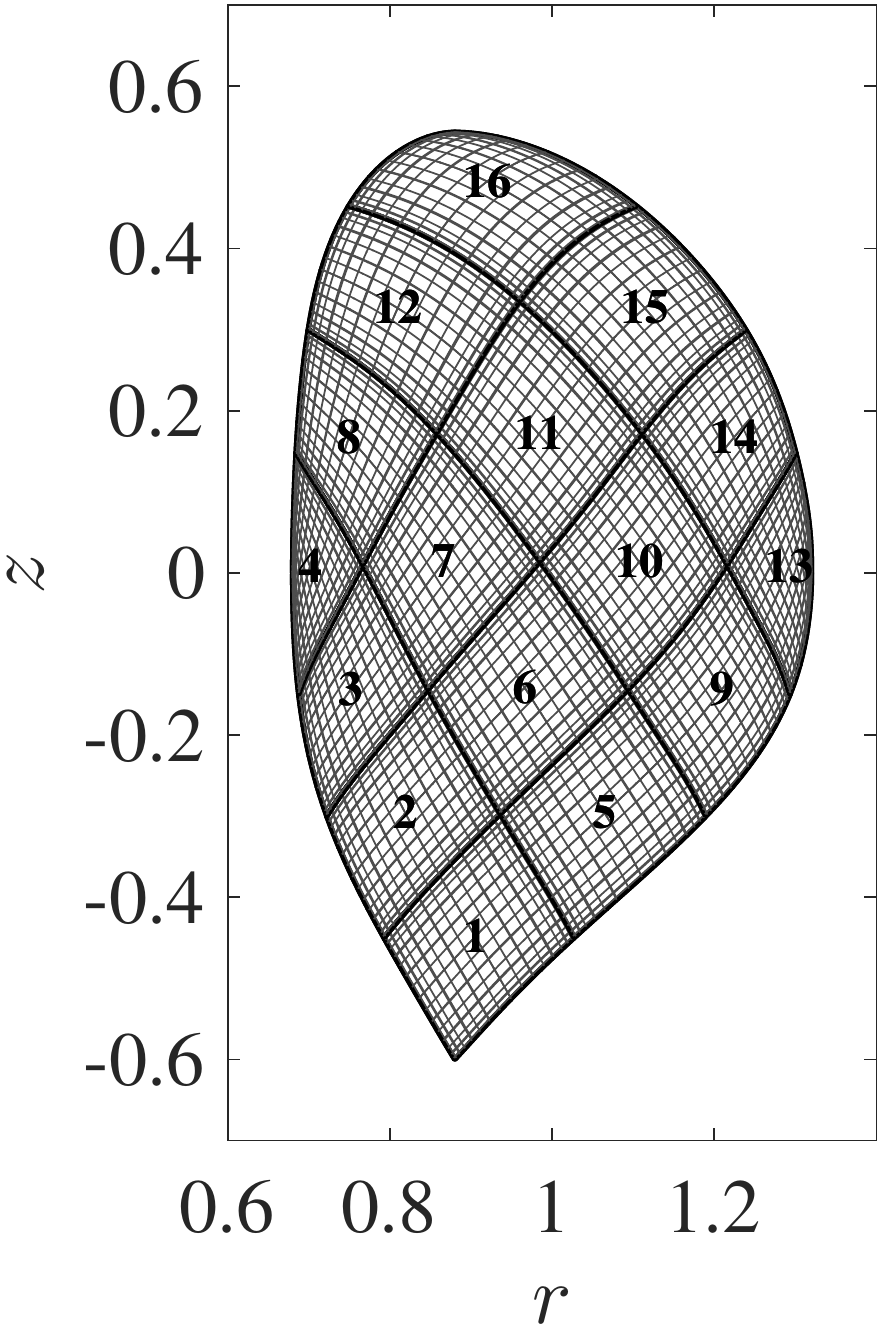}\hspace{0.75cm}
				\includegraphics[height=0.271\textwidth]{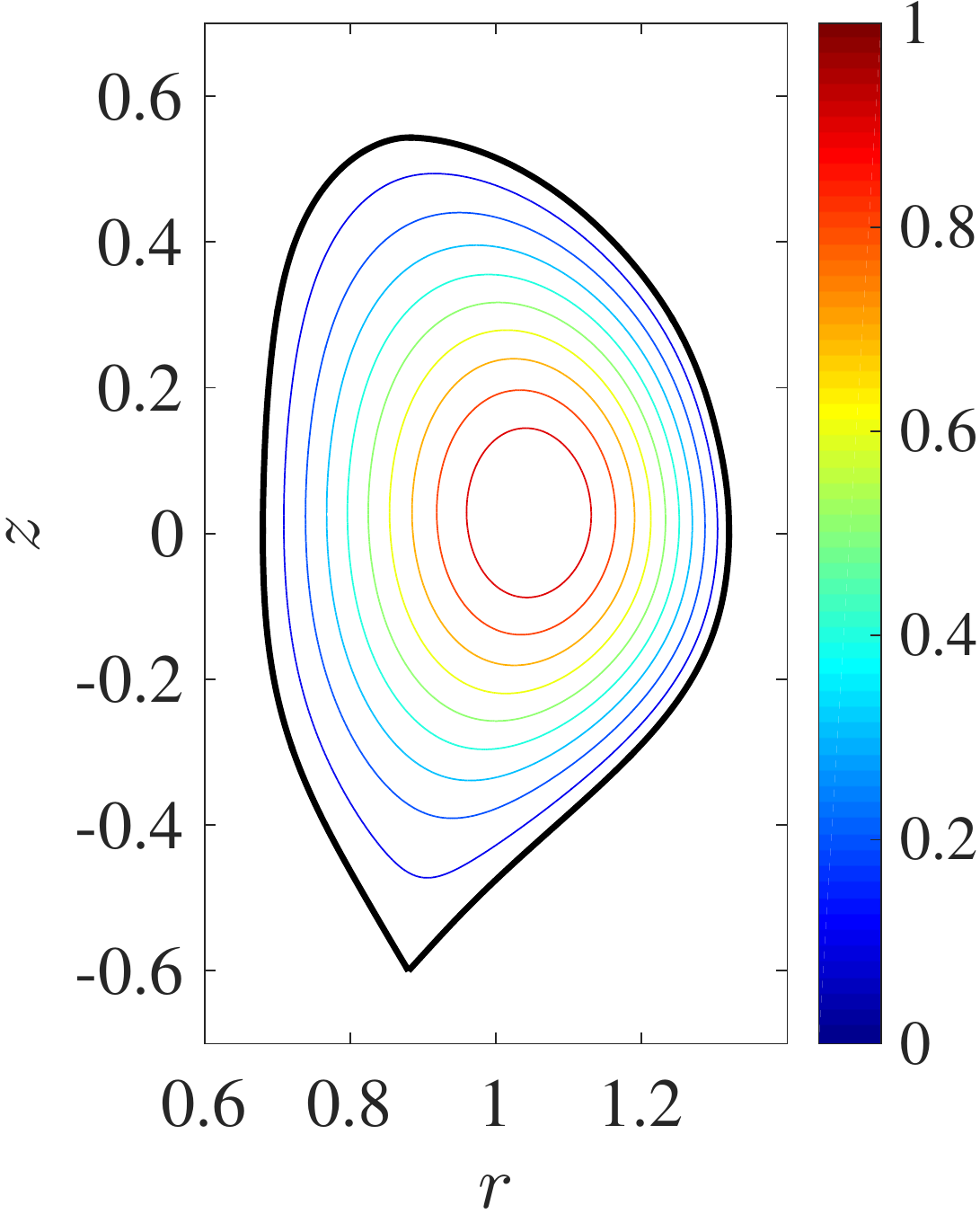}\hspace{0.75cm}
				\includegraphics[height=0.271\textwidth]{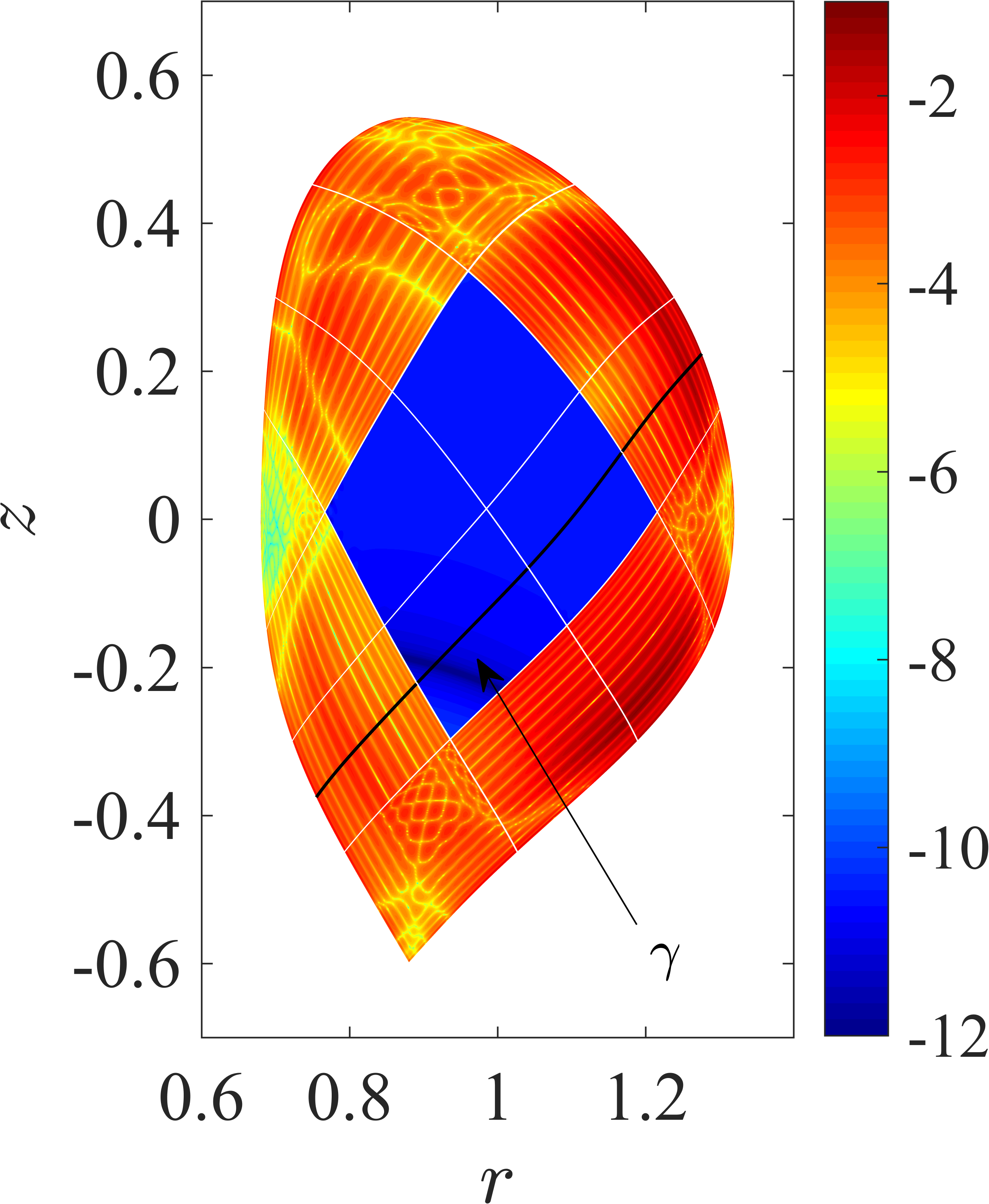}\\
				\end{center}
				\caption{\reviewerone{Numerical solution of the non-linear eigenvalue X-point test case, \eqref{eq::nonlinear_eigenvalue_test_case}, with $|\psi_{0}|=1.0$ and $\eta=0.1$. Top: mesh of $4\times 4$ elements and elements of polynomial degree $p=8$. Bottom: mesh of $4\times 4$ elements and elements of polynomial degree $p=16$. From left to right: computational mesh, numerical solution, and error as given by \eqref{eq:nonlinear_error}.}}
				\label{fig::nonlinear_eigenvalue_xpoint_test_case}
			\end{figure}
			
			As seen in \secref{sec::test_cases_nonlinear_eigenvalue}, for these meshes the elements at the boundary show a substantially larger error than the interior ones. This is again due to the large gradient in $\sigma J(r,z,\psi_{h})$. In \figref{fig::linear_eigenvalue_xpoint_test_case_error_along_line} we compare $\sigma J(r,z,\psi_{h})$ to $\nabla\times\vec{h}_{h}(r,z)$ for $p=8,16$ along the line $\gamma$, see \figref{fig::nonlinear_eigenvalue_xpoint_test_case} left. As can be seen, there is a sharp variation of $\sigma J(r,z,\psi_{h})$ close to the edge of the plasma. The error \eqref{eq:nonlinear_error} along the line $\gamma$ (see \figref{fig::nonlinear_eigenvalue_xpoint_test_case}) is shown in \figref{fig::linear_eigenvalue_xpoint_test_case_error_along_line} right. It is possible to see a substantial reduction in the error when the polynomial degree of the elements is increased from $p=8$ to $p=16$.
			
			\begin{figure}[!ht]
				\begin{center}
				\includegraphics[width=0.365\textwidth]{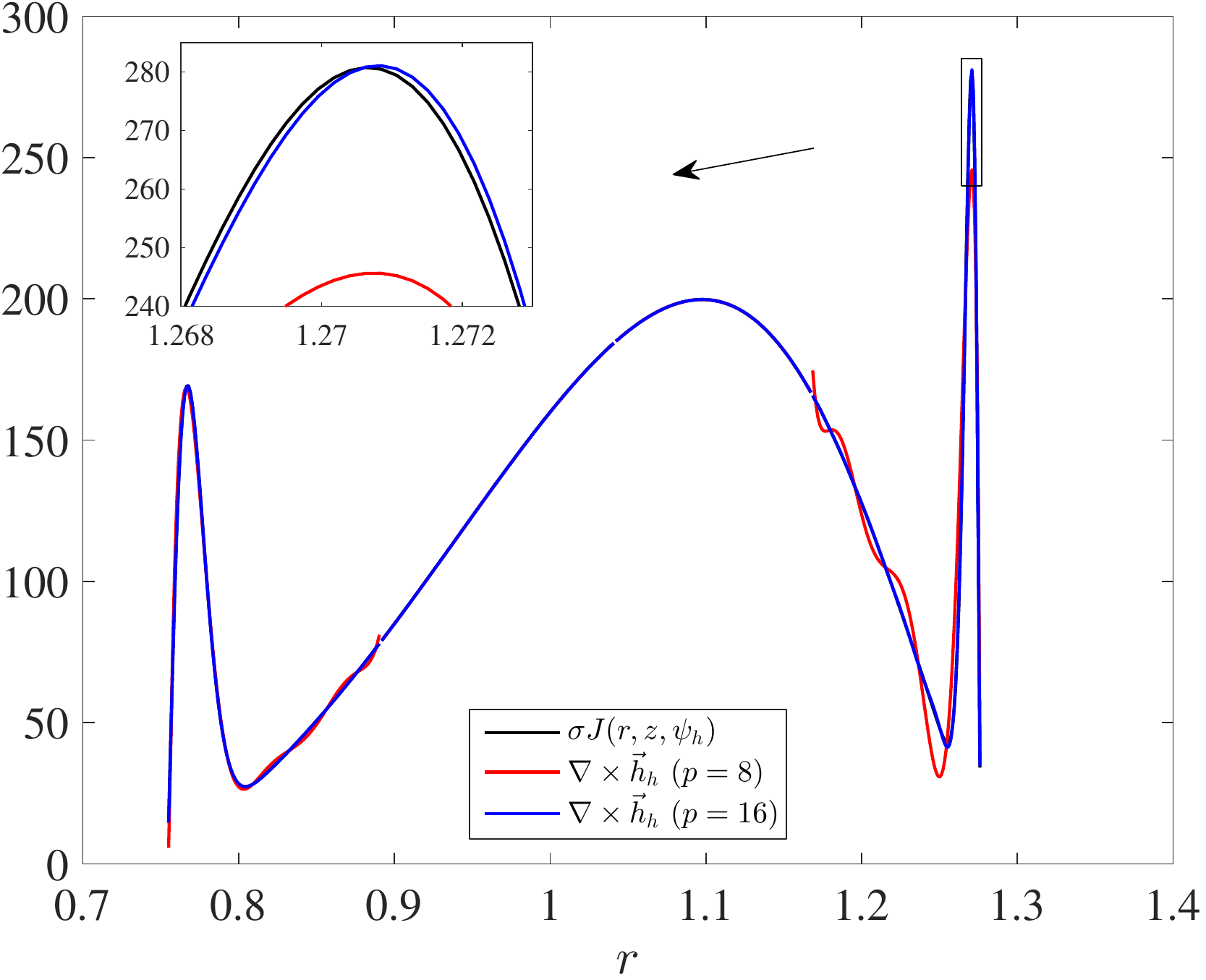}\hspace{0.5cm}
				\includegraphics[width=0.395\textwidth]{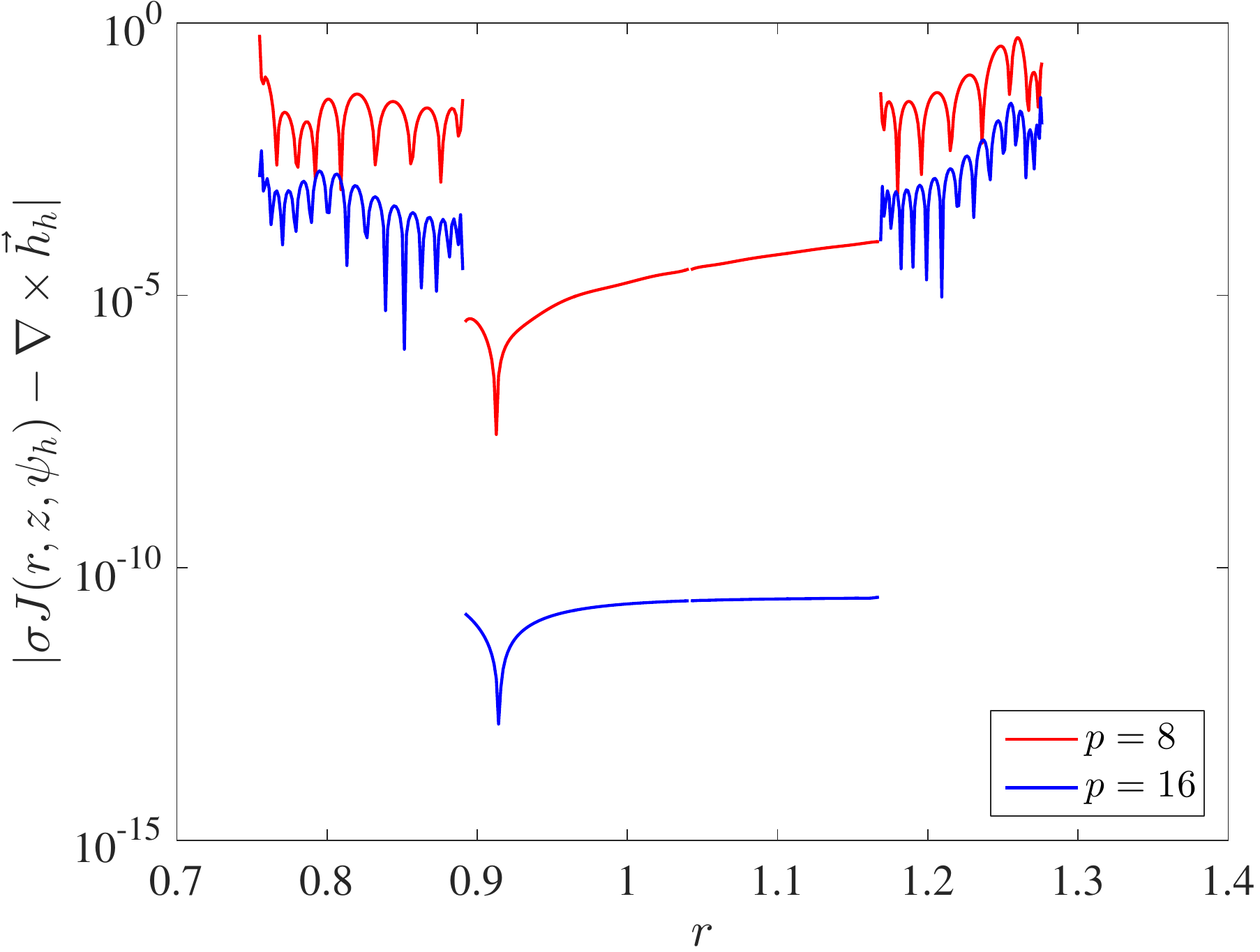}
				\end{center}
				\caption{\reviewerone{Left: Comparison between $\sigma J(r,z,\psi_{h})$ and $\nabla\times\vec{h}_{h}$ for $p=8,16$, along the curve $\gamma$ (see \figref{fig::nonlinear_eigenvalue_xpoint_test_case}). Right: Error, as given by \eqref{eq:nonlinear_error}, along the curve $\gamma$ (see \figref{fig::nonlinear_eigenvalue_xpoint_test_case}).}}
				\label{fig::linear_eigenvalue_xpoint_test_case_error_along_line}
			\end{figure}

			\end{reviewer1}
			\FloatBarrier
	
\section{Conclusions} \label{sec::conclusions}
	This article presents a new, fixed-boundary, Eulerian Grad-Shafranov solver based on the mimetic spectral element framework \cite{Palha2014,Kreeft2011}. \reviewerone{The advantages of this method are: (i) high order accuracy, enabling both spectral and geometric convergence, (ii) geometric flexibility, allowing for arbitrary plasma shapes (symmetric, asymmetric and  including X-points) and (iii) exact discretization of topological laws, resulting in an exact computation of the total plasma current.
	
	%The FRC and spheromak test cases, which include the geometric axis in their domain, show that this method can intrinsically account for regularity at the geometric axis. This is due to the special choice of basis functions with degrees of freedom associated to integral quantities, as opposed to the standard nodal degrees of freedom.
	
	We have shown for a wide range of profiles, including fusion relevant profiles with a pressure pedestal and high Shafranov shift, that the proposed method is capable of accurately computing equilibria.} Highly accurate solutions of the order $10^{-15}$ (machine precision) can be achieved with relatively coarse meshes, using high order basis functions. This characteristic represents one of the attractions of spectral element based discretizations: small numerical errors can be obtained with a much smaller number of degrees of freedom using higher order basis functions.
	
	Nevertheless, the proposed method still allows for further improvement, mainly in terms of speed. An important improvement to this method would be the use of a Newton solver instead of the Picard iteration presented. Another prossible improvement is the optimization of the discretization of the right-hand side, $J_{\phi}(r,z,\psi)$, required in each iteration. These improvements are the subject of ongoing and future research.

\appendix
\begin{reviewer1}
\section{X-point Soloviev solution} \label{ap:x_point}
	The X-point Soloviev solution used in this work corresponds to the up-down asymmetric ITER-like configuration presented in \cite{Cerfon2010}. The analytical solution is given by
	\begin{equation}
		\psi_{a}(r,z) = \frac{r^{4}}{8} + A \left(\frac{r^{2}}{2}\ln r - \frac{r^{4}}{8}\right) + \sum_{k=1}^{12}c_{k}\, \psi_{k} (r,z), \label{eq:x_point_soloviev_analytical}
	\end{equation}
	with
	\begin{align}
		\psi_{1}(r,z) & = 1, \nonumber \\
		\psi_{2}(r,z) & = r^{2}, \nonumber \\
		\psi_{3}(r,z) & = z^{2} - r^{2}\ln r, \nonumber \\
		\psi_{4}(r,z) & = r^{4} - 4r^{2}z^{2}, \nonumber \\
		\psi_{5}(r,z) & = 2z^{4} - 9z^{2}r^{2} + 3r^{4}\ln r - 12 r^{2} z^{2} \ln r, \nonumber \\
		\psi_{6}(r,z) & = r^{6} - 12 r^{4}z^{2} + 8r^{2}z^{4}, \nonumber \\
		\psi_{7}(r,z) & = 8z^{6} - 140z^{4}r^{2} + 75z^{2}r^{4} - 15r^{6} \ln r + 180r^{4}z^{2}\ln r - 120r^{2} z^{4}\ln r, \nonumber \\
		\psi_{8}(r,z) & = z, \nonumber \\
		\psi_{9}(r,z) & = zr^{2}, \nonumber \\
		\psi_{10}(r,z) & = z^{3} - 3zr^{2}\ln r, \nonumber \\
		\psi_{11}(r,z) & = 3zr^{4} - 4z^{3}r^{2}, \nonumber \\
		\psi_{12}(r,z) & = 8z^{5} - 45zr^{4} - 80z^{3}r^{2}\ln r + 60zr^{4} \ln r, \nonumber
	\end{align}
	and the values for the $c_{i}$ coefficients given in \tabref{tab:ci_coefficients}.
	\begin{table}[htp]
		\caption{Values for the $c_{i}$ coefficients for the up-down asymmetric ITER-like configuration of \eqref{eq:x_point_soloviev_analytical}.}
		\begin{center}
			\begin{tabular}{|cS[table-format=2.16]||cS[table-format=2.16]||cS[table-format=2.16]|}
				\hline
				$c_{1}$ : & 0.0864912785478807 & $c_{5}$ : & 0.3807375276922255 & $c_{9}$ : & 0.7401867427139835 \\
				$c_{2}$ : & 0.3236475999311713 & $c_{6}$ : & -0.3573346678775972 & $c_{10}$ : & -0.4397718916520960 \\
				$c_{3}$ : & -0.5227047152014734 & $c_{7}$ : & -0.0148740157319066 & $c_{11}$ : & -0.1071308624644806 \\
				$c_{4}$ : & -0.2319735789049367 & $c_{8}$ : & 0.1480149379993163 & $c_{12}$ : & 0.0127862151469652\\ \hline
				
			\end{tabular}
		\end{center}
		\label{tab:ci_coefficients}
	\end{table}%

\end{reviewer1}

% Bibliography 
\section*{References}
\bibliographystyle{elsart-num-sort} 
\def\url#1{}
\bibliography{./library}

\begin{thebibliography}{10}
\expandafter\ifx\csname url\endcsname\relax
  \def\url#1{\texttt{#1}}\fi
\expandafter\ifx\csname urlprefix\endcsname\relax\def\urlprefix{URL }\fi

\bibitem{abraham_diff_geom}
R.~Abraham, J.~E. Marsden, T.~Ratiu, {Manifolds, Tensor Analysis, and
  Applications}, vol.~75 of Applied Mathematical Sciences, Springer, 2001.

\bibitem{Albanese2015}
R.~Albanese, R.~Ambrosino, M.~Mattei, {CREATE-NL+: A robust control-oriented
  free boundary dynamic plasma equilibrium solver}, Fusion Engineering and
  Design.

\bibitem{Artaud2010b}
J.~F. Artaud, V.~Basiuk, F.~Imbeaux, M.~Schneider, J.~Garcia, G.~Giruzzi,
  P.~Huynh, T.~Aniel, F.~Albajar, J.~M. An{\'{e}}, A.~B{\'{e}}coulet,
  C.~Bourdelle, A.~Casati, L.~Colas, J.~Decker, R.~Dumont, L.~G. Eriksson,
  X.~Garbet, R.~Guirlet, P.~Hertout, G.~T. Hoang, W.~Houlberg, G.~Huysmans,
  E.~Joffrin, S.~H. Kim, F.~K{\"{o}}chl, J.~Lister, X.~Litaudon, P.~Maget,
  R.~Masset, B.~P{\'{e}}gouri{\'{e}}, Y.~Peysson, P.~Thomas, E.~Tsitrone,
  F.~Turco, {The CRONOS suite of codes for integrated tokamak modelling},
  Nuclear Fusion 50 (2010) 043001.

\bibitem{Bellan2002}
P.~M. Bellan, {Generalization of cylindrical spheromak solution to finite beta
  and large reversed shear}, Physics of Plasmas 9 (2002) 3050.

\bibitem{Blum1984}
J.~Blum, J.~{Le Foll}, {Plasma equilibrium evolution at the resistive diffusion
  timescale}, Computer Physics Reports 1 (1984) 465--494.

\bibitem{bossavit_japan_computational_1}
A.~Bossavit, {Computational electromagnetism and geometry: (1) Network
  equations}, Journal of the Japan Society of Applied Electromagnetics 7 (1999)
  150--159.

\bibitem{bossavit_japan_computational_2}
A.~Bossavit, {Computational electromagnetism and geometry: (2) Network
  constitutive laws}, Journal of the Japan Society of Applied Electromagnetics
  7 (1999) 294--301.

\bibitem{bossavit_japan_computational_4}
A.~Bossavit, {Computational electromagnetism and geometry: (4) From degrees of
  freedom to fields}, Journal of the Japan Society of Applied Electromagnetics
  8 (2000) 102--109.

\bibitem{bossavit_japan_computational_5}
A.~Bossavit, {Computational electromagnetism and geometry: (5) The ``Galerkin
  hodge''}, Journal of the Japan Society of Applied Electromagnetics 8 (2000)
  203--209.

\bibitem{bouman::icosahom2009}
M.~Bouman, A.~Palha, J.~Kreeft, M.~Gerritsma, {A Conservative Spectral Element
  Method for Curvilinear Domains}, in: Spectral and High Order Methods for
  Partial Differential Equations, vol.~76 of Lecture Notes in Computational
  Science and Engineering, Springer, 2011, pp. 111--119.

\bibitem{Brambilla1999}
M.~Brambilla, {Numerical simulation of ion cyclotron waves in tokamak plasmas},
  Plasma Physics and Controlled Fusion 41 (1999) 1--34.

\bibitem{brezzi1991mixed}
F.~Brezzi, M.~Fortin, {Mixed and Hybrid Finite Element Methods}, vol.~15 of
  Springer Series in Computational Mathematics, Springer, 1991.

\bibitem{Budny1992}
R.~V. Budny, M.~G. Bell, H.~Biglari, M.~Bitter, C.~E. Bush, C.~Z. Cheng, E.~D.
  Fredrickson, B.~Grek, K.~W. Hill, H.~Hsuan, A.~C. Janos, D.~L. Jassby, D.~W.
  Johnson, L.~C. Johnson, B.~LeBlanc, D.~C. McCune, D.~R. Mikkelsen, H.~K.
  Park, A.~T. Ramsey, S.~A. Sabbagh, S.~D. Scott, J.~F. Schivell, J.~D.
  Strachan, B.~C. Stratton, E.~J. Synakowski, G.~Taylor, M.~C. Zarnstorff,
  S.~J. Zweben, {Simulations of deuterium-tritium experiments in TFTR}, Nuclear
  Fusion 32 (1992) 429--447.

\bibitem{Cenacchi1988}
G.~Cenacchi, A.~Taroni, {JETTO a free boundary plasma transport code}, Tech.
  rep., ENEA (1988).

\bibitem{Cerfon2010}
A.~J. Cerfon, J.~P. Freidberg, {``One size fits al'' analytic solutions to the
  Grad-Shafranov equation}, Physics of Plasmas 17 (2010) 032502.

\bibitem{Coster2010a}
D.~P. Coster, V.~Basiuk, G.~Pereverzev, D.~Kalupin, R.~Zagorksi,
  R.~Stankiewicz, P.~Huynh, F.~Imbeaux, {The European Transport Solver}, IEEE
  Transactions on Plasma Science 38 (2010) 2085--2092.

\bibitem{CorsicaReport1997}
J.~A. Crotinger, L.~LoDestro, L.~D. Pearlstein, A.~Tarditi, T.~A. Casper, E.~B.
  Hooper, {Corsica: a comprehensive simulation of toroidal magnetic-fusion
  devices}, Tech. rep., Final Report to the LDRD program (1997).

\bibitem{Czarny2008}
O.~Czarny, G.~Huysmans, {B{\'{e}}zier surfaces and finite elements for MHD
  simulations}, Journal of Computational Physics 227 (2008) 7423--7445.

\bibitem{Fable2013a}
E.~Fable, C.~Angioni, A.~Ivanov, K.~Lackner, O.~Maj, S.~Yu, G.~Pautasso,
  G.~Pereverzev, {A stable scheme for computation of coupled transport and
  equilibrium equations in tokamaks}, Nuclear Fusion 53 (2013) 033002.

\bibitem{Felici2011}
F.~Felici, O.~Sauter, S.~Coda, B.~P. Duval, T.~P. Goodman, J.-M. Moret, J.~I.
  Paley, {Real-time physics-model-based simulation of the current density
  profile in tokamak plasmas}, Nuclear Fusion 51 (2011) 083052.

\bibitem{frankel}
T.~Frankel, {The Geometry of Physics}, Cambridge University Press, 2004.

\bibitem{gerritsma::edge_basis}
M.~Gerritsma, {Edge Functions for Spectral Element Methods}, in: Spectral and
  High Order Methods for Partial Differential Equations, vol.~76 of Lecture
  Notes in Computational Science and Engineering, Springer, 2011, pp. 199--207.

\bibitem{Gerritsma}
M.~Gerritsma, M.~Bouman, A.~Palha, {Least-Squares Spectral Element Method on a
  Staggered Grid}, in: Large-Scale Scientific Computing, vol. 5910 of Lecture
  Notes in Computer Science, Springer, 2010, pp. 653--661.

\bibitem{gerritsmaicosahom2012}
M.~Gerritsma, R.~Hiemstra, J.~Kreeft, A.~Palha, P.~P. Rebelo, D.~Toshniwal,
  {The Geometric Basis of Numerical Methods}, in: Spectral and High Order
  Methods for Partial Differential Equations, vol.~95 of Lecture Notes in
  Computational Science and Engineering, Springer, 2013, pp. 17--35.

\bibitem{Goedbloed1984}
J.~Goedbloed, {Some remarks on computing axisymmetric equilibria}, Computer
  Physics Communications 31 (1984) 123--135.

\bibitem{Goedbloed2010Book}
J.~P. Goedbloed, R.~Keppens, S.~Poedts, {Advanced Magnetohydrodynamics: With
  Applications to Laboratory and Astrophysical Plasmas}, Cambridge University
  Press, 2010.

\bibitem{gordon::transfinite_mapping}
W.~J. Gordon, C.~A. Hall, {Transfinite element methods: blending-function
  interpolation over arbitrary curved element domains}, Numerische Mathematik
  21 (1973) 109--129.

\bibitem{Gorler2011}
T.~G{\"{o}}rler, X.~Lapillonne, S.~Brunner, T.~Dannert, F.~Jenko, F.~Merz,
  D.~Told, {The global version of the gyrokinetic turbulence code GENE},
  Journal of Computational Physics 230 (2011) 7053--7071.

\bibitem{Gourdain2006}
P.-A. Gourdain, J.-N. Leboeuf, R.~Y. Neches, {High-resolution
  magnetohydrodynamic equilibrium code for unity beta plasmas}, Journal of
  Computational Physics 216 (2006) 275--299.

\bibitem{Grad1958}
H.~Grad, H.~Rubin, {Hydromagnetic equilibria and force-free fields}, Journal of
  Nuclear Energy 7 (1958) 284--285.

\bibitem{Gruber1981}
R.~Gruber, F.~Troyon, D.~Berger, L.~Bernard, S.~Rousset, R.~Schreiber,
  W.~Kerner, W.~Schneider, K.~Roberts, {ERATO stability code}, Computer Physics
  Communications 21 (1981) 323--371.

\bibitem{Heumann2015}
H.~Heumann, J.~Blum, C.~Boulbe, B.~Faugeras, G.~Selig, J.-M. An{\'{e}},
  S.~Br{\'{e}}mond, V.~Grandgirard, P.~Hertout, E.~Nardon, {Quasi-static
  free-boundary equilibrium of toroidal plasma with CEDRES++: Computational
  methods and applications}, Journal of Plasma Physics (2015) 1--35.

\bibitem{Hinton1976a}
F.~L. Hinton, R.~D. Hazeltine, {Theory of plasma transport in toroidal
  confinement systems}, Reviews of Modern Physics 48 (1976) 239--308.

\bibitem{Hirshman1979a}
S.~P. Hirshman, S.~C. Jardin, {Two-dimensional transport of tokamak plasmas},
  Physics of Fluids 22 (1979) 731.

\bibitem{Howell2014}
E.~Howell, C.~Sovinec, {Solving the Grad-Shafranov equation with spectral
  elements}, Computer Physics Communications 185 (2014) 1415--1421.

\bibitem{Humphreys2015}
D.~Humphreys, G.~Ambrosino, P.~de~Vries, F.~Felici, S.~H. Kim, G.~Jackson,
  A.~Kallenbach, E.~Kolemen, J.~Lister, D.~Moreau, A.~Pironti, G.~Raupp,
  O.~Sauter, E.~Schuster, J.~Snipes, W.~Treutterer, M.~Walker, A.~Welander,
  A.~Winter, L.~Zabeo, {Novel aspects of plasma control in ITER}, Physics of
  Plasmas 22 (2015) 021806.

\bibitem{Helena1990}
G.~T.~A. Huysmans, J.~P. Goedbloed, W.~Kerner, {Isoparametric bicubic Hermite
  elements for solution of the Grad-Shafranov equation}, International Journal
  of Modern Physics C 2 (1991) 371--376.

\bibitem{HymanShashkovSteinberg97}
J.~M. Hyman, M.~Shashkov, S.~Steinberg, {The numerical solution of diffusion
  problems in strongly heterogeous non-isotropic materials}, Journal of
  Computational Physics 132 (1997) 130--148.

\bibitem{Imazawa2015}
R.~Imazawa, Y.~Kawano, K.~Itami, {Meshless method for solving fixed boundary
  problem of plasma equilibrium}, Journal of Computational Physics 292 (2015)
  208--214.

\bibitem{IvanovSPIDER2005}
A.~A. Ivanov, R.~R. Khayrutdinov, S.~Y. Medvedev, Y.~Y. Poshekhonov, {New
  adaptive grid plasma evolution code SPIDER}, in: 32nd EPS Conference on
  Plasma Physics, 2005, pp. 2146--2149.

\bibitem{Jardin2004}
S.~Jardin, {A triangular finite element with first-derivative continuity
  applied to fusion MHD applications}, Journal of Computational Physics 200
  (2004) 133--152.

\bibitem{Jardin2010Book}
S.~C. Jardin, {Computational Methods in Plasma Physics}, Chapman \& Hall / CRC
  Computational Science, CRC Press, 2010.

\bibitem{Kikuchi1984}
F.~Kikuchi, K.~Nakazato, T.~Ushijima, {Finite element approximation of a
  nonlinear eigenvalue problem related to MHD equilibria}, Japan Journal of
  Applied Mathematics 1 (1984) 369--403.

\bibitem{Kreeft2011}
J.~Kreeft, A.~Palha, M.~Gerritsma, {Mimetic framework on curvilinear
  quadrilaterals of arbitrary order}, Arxiv preprint (2011) 69.

\bibitem{Lapillonne2009}
X.~Lapillonne, S.~Brunner, T.~Dannert, S.~Jolliet, A.~Marinoni, L.~Villard,
  T.~Görler, F.~Jenko, F.~Merz, {Clarifications to the limitations of the
  s-$\alpha$ equilibrium model for gyrokinetic computations of turbulence},
  Physics of Plasmas 16 (2009) 032308.

\bibitem{Lee2015}
J.~Lee, A.~Cerfon, {ECOM: A fast and accurate solver for toroidal axisymmetric
  MHD equilibria}, Computer Physics Communications 190 (2015) 72--88.

\bibitem{Li2014}
X.~Li, L.~E. Zakharov, V.~V. Drozdov, {Edge equilibrium code for tokamaks},
  Physics of Plasmas 21 (2014) 012505.

\bibitem{LoDestro1994}
L.~L. LoDestro, L.~D. Pearlstein, {On the Grad-Shafranov equation as an
  eigenvalue problem, with implications for q solvers}, Physics of Plasmas 1
  (1994) 90--95.

\bibitem{Ludwig1997a}
G.~O. Ludwig, {Direct variational solutions of the tokamak equilibrium
  problem}, Plasma Physics and Controlled Fusion 39 (1997) 2021--2037.

\bibitem{Lutjens1992}
H.~L{\"{u}}tjens, A.~Bondeson, A.~Roy, {Axisymmetric MHD equilibrium solver
  with bicubic Hermite elements}, Computer Physics Communications 69 (1992)
  287--298.

\bibitem{Lutjens1996}
H.~L{\"{u}}tjens, A.~Bondeson, O.~Sauter, {The CHEASE code for toroidal MHD
  equilibria}, Computer Physics Communications 97 (1996) 219--260.

\bibitem{mattiussi1997analysis}
C.~Mattiussi, {An analysis of finite volume, finite element, and finite
  difference methods using some concepts from algebraic topology}, Journal of
  Computational Physics 133 (1997) 289--309.

\bibitem{Neuman1977}
S.~P. Neuman, {Theoretical derivation of Darcy's law}, Acta Mechanica 25 (1977)
  153--170.

\bibitem{palha:lssc2009}
A.~Palha, M.~Gerritsma, {Mimetic Least-Squares Spectral/$hp$ Finite Element
  Method for the Poisson Equation}, in: Large-Scale Scientific Computing, vol.
  5910 of Lecture Notes in Computer Science, Springer, 2010, pp. 662--670.

\bibitem{palha::icosahom2009}
A.~Palha, M.~Gerritsma, {Spectral Element Approximation of the Hodge-$\star$
  Operator in Curved Elements}, in: Spectral and High Order Methods for Partial
  Differential Equations, vol.~76 of Lecture Notes in Computational Science and
  Engineering, Springer, 2010, pp. 283--291.

\bibitem{Palha2014}
A.~Palha, P.~P. Rebelo, R.~Hiemstra, J.~Kreeft, M.~Gerritsma,
  {Physics-compatible discretization techniques on single and dual grids, with
  application to the Poisson equation of volume forms}, Journal of
  Computational Physics 257 (2014) 1394--1422.

\bibitem{Parail2013a}
V.~Parail, R.~Albanese, R.~Ambrosino, J.-F. Artaud, K.~Besseghir, M.~Cavinato,
  G.~Corrigan, J.~Garcia, L.~Garzotti, Y.~Gribov, F.~Imbeaux, F.~Koechl,
  C.~Labate, J.~Lister, X.~Litaudon, A.~Loarte, P.~Maget, M.~Mattei,
  D.~McDonald, E.~Nardon, G.~Saibene, R.~Sartori, J.~Urban, {Self-consistent
  simulation of plasma scenarios for ITER using a combination of 1.5D transport
  codes and free-boundary equilibrium codes}, Nuclear Fusion 53 (2013) 113002.

\bibitem{Parks2003}
P.~B. Parks, M.~J. Schaffer, {Analytical equilibrium and interchange stability
  of single- and double-axis field-reversed configurations inside a cylindrical
  cavity}, Physics of Plasmas 10 (2003) 1411--1423.

\bibitem{Pataki2013}
A.~Pataki, A.~J. Cerfon, J.~P. Freidberg, L.~Greengard, M.~O’Neil, {A fast,
  high-order solver for the Grad-Shafranov equation}, Journal of Computational
  Physics 243 (2013) 28--45.

\bibitem{ASTRA2002}
G.~V. Pereverzev, P.~N. Yushmanov, {ASTRA Automated System for TRansport
  Analysis in a Tokamak}, Tech. rep., IPP (2002).

\bibitem{Rebelo2014}
P.~P. Rebelo, A.~Palha, M.~Gerritsma, {Mixed mimetic spectral element method
  applied to Darcy's problem}, in: Spectral and High Order Methods for Partial
  Differential Equations - ICOSAHOM 2012, vol.~95 of Lecture Notes in
  Computational Science and Engineering, Springer, 2014, pp. 373--382.

\bibitem{robidoux-polynomial}
N.~Robidoux, {Polynomial histopolation, superconvergent degrees of freedom, and
  pseudospectral discrete Hodge operators}, Unpublished:
  http://people.math.sfu.ca/$\sim$nrobidou/public\_html/prints/histogram/histogram.pdf.

\bibitem{Robidoux2011}
N.~Robidoux, S.~Steinberg, {A discrete vector calculus in tensor grids},
  Computational Methods in Applied Mathematics 11 (2011) 23--66.

\bibitem{SAITOH2012}
A.~Saitoh, T.~Itoh, N.~Matsui, A.~Kamitani, H.~Nakamura, {Application of
  collocation meshless method to eigenvalue problem}, Plasma and Fusion
  Research 7 (2012) 2406096.

\bibitem{Shafranov1958}
V.~D. Shafranov, {Magnetohydrodynamical equilibrium configurations}, Soviet
  Physics JETP 6 (1958) 545--554.

\bibitem{Sovinec2004}
C.~Sovinec, A.~Glasser, T.~Gianakon, D.~Barnes, R.~Nebel, S.~Kruger,
  D.~Schnack, S.~Plimpton, A.~Tarditi, M.~Chu, N.~Team, {Nonlinear
  magnetohydrodynamics simulation using high-order finite elements}, Journal of
  Computational Physics 195 (2004) 355--386.

\bibitem{Takeda1991}
T.~Takeda, S.~Tokuda, {Computation of MHD equilibrium of tokamak plasma},
  Journal of Computational Physics 93 (1991) 1--107.

\bibitem{tonti1975formal}
E.~Tonti, {On the formal structure of physical theories}, Tech. rep., Italian
  National Research Council (1975).

\bibitem{BookTonti2013}
E.~Tonti, {The Mathematical Structure of Classical and Relativistic Physics},
  Birkh\"{a}user, 2013.

\bibitem{Tonti2014}
E.~Tonti, {Why starting from differential equations for computational
  physics?}, Journal of Computational Physics 257 (2014) 1260--1290.

\bibitem{Zakharov1999}
L.~E. Zakharov, A.~Pletzer, {Theory of perturbed equilibria for solving the
  Grad-Shafranov equation}, Physics of Plasmas 6 (1999) 4693.

\end{thebibliography}
%\bibliography{library}

\end{document}